\documentclass[phd, final, twoside]{cornell}

\usepackage{amsmath, amssymb, bbm}
\usepackage{epsf}
\usepackage[final]{graphicx}
\usepackage{color}

\usepackage{hangcaption}
\newcommand{\CAP}[2]{\singlespacing\hangcaption[#1]{#2}\normalspacing}
\renewcommand{\caption}[1]{\singlespacing\hangcaption{#1}\normalspacing}

\usepackage{dcolumn}
\newcolumntype{d}[1]{D{.}{.}{#1}}

\usepackage{pictexwd}

\usepackage{rotating}

\newcommand{\gfour}{{^{(4)}{\mathbf g}}}
\newcommand{\Lie}{{\cal L}}
\newcommand{\CF}{\psi}
\newcommand{\trK}{K}

\newcommand{\g}{g}
\newcommand{\cg}{\tilde{g}}
\newcommand{\K}{K}
\newcommand{\A}{A}
\newcommand{\cA}{\tilde{A}}
\newcommand{\deriv}{\nabla}
\newcommand{\cderiv}{\tilde{\nabla}}
\newcommand{\Long}{\mathbb{L}}
\newcommand{\cLong}{\tilde{\mathbb{L}}}
\newcommand{\R}{R}
\newcommand{\cR}{\tilde{R}}
\newcommand{\N}{N}
\newcommand{\cN}{\tilde{N}}
\newcommand{\Christoffel}{\Gamma}
\newcommand{\cChristoffel}{\tilde{\Gamma}}
\renewcommand{\u}{u}
\newcommand{\cu}{\tilde{u}}
\newcommand{\s}{\sigma}
\newcommand{\cs}{\tilde{\sigma}}
\newcommand{\cj}{{\tilde{\jmath}\,}}
\newcommand{\crho}{\tilde{\rho}}

\newcommand{\omg}{(m\Omega)}
\newcommand{\Eadm}{E_{\mbox{\rm\tiny ADM}}}
\newcommand{\pps}{\mbox{\small$++$}}
\newcommand{\pms}{\mbox{\small$+-$}}
\newcommand{\mms}{\mbox{\small$--$}}

\newcommand{\tableline}{\hline}

\newcommand{\dnachd}[2]{\frac{\partial #1}{\partial #2}}
\newcommand{\ve}[1]{\ensuremath{\mathbf{#1}}}

\newcommand{\tildeLapLong}{\tilde\Delta_L}

\newcommand{\eqnref}[1]{(\ref{#1})}

\newcommand{\ComparingLong}{\mathbb{L}}
\newcommand{\LapLong}{\Delta_\Long}
\newcommand{\TF}{\mbox{TF}}

\newcommand{\ISCO}{ {\mbox{\small ISCO} }}

\renewcommand{\S}{ {\cal S}}
\newcommand{\BC}{ {\cal R}}
\newcommand{\fournabla}{\underline{\nabla}}

\begin{document}

\title{Initial data for black hole evolutions}
\author{Harald Paul Pfeiffer}
\conferraldate{August}{2003}
\maketitle

\copyrightholder{Harald Paul Pfeiffer}
\copyrightyear{2003}
\makecopyright

\begin{abstract}
We discuss the initial value problem of general relativity in its
recently unified Lagrangian and Hamiltonian pictures and present a
multi-domain pseudo-spectral collocation method to solve the resulting
coupled nonlinear partial differential equations.  Using this code, we
explore several approaches to construct initial data sets containing
one or two black holes: We compute quasi-circular orbits for spinning
equal mass black holes and unequal mass (nonspinning) black holes
using the effective potential method with Bowen-York extrinsic
curvature.  We compare initial data sets resulting from different
decompositions, and from different choices of the conformal metric
with each other.  Furthermore, we use the quasi-equilibrium method to
construct initial data for single black holes and for binary black
holes in quasi-circular orbits.  We investigate these binary black
hole data sets and examine the limits of large mass-ratio and wide
separation.  Finally, we propose a new method for constructing
spacetimes with superposed gravitational waves of possibly very large
amplitude.
\end{abstract}

\begin{biosketch}
Harald Pfeiffer was born on January 29 1974, and spent his childhood
in R\"uden\-hau\-sen, Germany.  Having done elementary school, high-school
and civil service just two miles away from home, he decided to
increase distance from home to study physics at the university of
Bayreuth, a sleepy town mostly known for the composer Wagner.  Being
adventurous, and somewhat bored with the pace the usual courses
proceeded, he ended up in a lecture, and subsequently in a seminar,
about general relativity, a topic which has had captured him in the
form of popular books since much earlier days.

Asking a few professors for opportunities to spend {\em one} year
abroad, one of them, Werner Pesch, simply answered ``Go to Cornell,''
an idea that caught his fancy immediately.  He applied for grad school
in the US and, as he wasn't sure he wanted to go so far away for such
a long time, he also applied to the University of Cambridge, UK.  When
he learned that he could defer admission to Cornell for a year, he
decided to take both and went to Cambridge in October 1997 to ``attend
diligently a Course in Advanced Study in Mathematics'' (so says the
Cambridge diploma), or, as it is known to the world, to do Part III.

Besides classes, Cambridge was great for punting, Backgammon and
enjoying good food at various formal dinners.  However, socially,
Cambridge turned out to differ only slightly from Germany; important
foods like Croissants were readily available, and there were just too
many Germans to hang out with.  Harald needed to go further away into
real alien territory and started graduate school at Cornell in 1998.
Ithaca has been a really nice place, and Cornell a great school to
learn physics, do physics, and meet a large variety of amazing people.

One year before he went to England, Sylke entered his life.  But she
left immediately to study in Ireland and when she returned, Harald was
getting ready for England.  It was finally in Ithaca that they could
spent an extended period of time together.  Sylke and Harald got
married in 2000, and a precious son was born to them in Spring 2003.

Harald has accepted a postdoctoral position at Caltech, the very
school that didn't receive any of his letters of recommendation
when he applied to its PhD program six years earlier.
\end{biosketch}

\begin{acknowledgements}
It is a pleasure to express my gratitude to all those people who have
helped me toward this point in my life.  First of all, I would like to
thank Saul Teukolsky, my thesis adviser, for his guidance and support
throughout my time at Cornell.  Saul was available for advice whenever
I needed it, but also gave me considerable freedom to pursue my own
ideas.  I continue to be amazed by his ability to answer very concisely
whatever question I put before him as well as his talent to ask
relevant questions.  I am very much indebted to Greg Cook with whom I
have worked on several projects over the years, and who taught me a
lot about the initial value problem.  The idea of developing a
spectral elliptic solver is his.  It is a pleasure to thank James York
for many discussions and a joint paper, which have refined my
understanding of general relativity and the initial value problem
tremendously.  I am also grateful to Jimmy for serving as a proxy on
my B-exam.  On a day to day basis, I worked with Larry Kidder, Mark
Scheel and, later, Deirdre Shoemaker.  I thank Mark and Larry for
sharing their code with me, and all three for putting up with my
sometimes too short patience and too high temper.  I enjoyed working
with them and look forward to continue to do so.  I am also indebted to
Dong Lai for guiding me through a completely different research
project.  Thanks go to {\'E}anna Flanagan and Eberhard Bodenschatz
who served on my special committee, and to Werner Pesch and Helmut
Brand for significant advice since my time at Bayreuth.

The development of my code depended crucially on an existing code base
for spectral methods written by Larry and Mark, on modern iterative
methods for linear systems made available by the PETSc-team, and on
visualization software developed by the Cornell undergraduates Adam
Bartnik, Hiro Oyaizu and Yor Limkumnerd;  I thank all these people.  I
also acknowledge the Department of Physics, Wake Forest University,
for access to their IBM SP2, on which most of the computations in this
thesis were performed.

I am very grateful for the many people who made life in Ithaca and in
Space Sciences so comfortable and fun: The sixth floor crowd, among
them Steve Drasco, {\'E}tienne Racienne, Marc Favata, Wolfgang Tichy,
John Karcz, Akiko Shirakawa, Wynn Ho, and in particular Jeandrew Brink
with her love for ice cream and deep and/or crazy ideas.  Furthermore
Gil Toombes, Eileen Tan (thanks for the car at several occasions),
Wulf Hofbauer, Horace \& Ileana Stoica and Shaffique Adam.  St. Luke
Lutheran church has always been a warm, welcoming and supportive
place.  Maybe even more important are friends back home, among them
Stephan Wildner, Uli Schwarz, G{\"u}nter Auernhammer, Elke G{\"o}tz,
Katharina Eisen, Uli Schwantag and Dirk Haderlein.

I wish to thank my parents Betty and Wilhelm Pfeiffer for teaching me
the truly important things in life and for supporting me throughout my
studies.  They provided the roots and the foundation that enabled me
to reach for things they never imagined to exist.  Finally, I wish to
thank my wife, although words cannot capture her impact on me: Thank
you for accompanying me through this amazing life.

\end{acknowledgements}

\contentspage
\tablelistpage
\figurelistpage

\normalspacing
\setcounter{page}{1}
\pagenumbering{arabic}
\pagestyle{cornell}


\chapter{Introduction}

General relativity is entering a very exciting phase with the
commissioning of several gravitational wave detectors.  The detection
of gravitational waves serves two main purposes.  It allows
fundamental tests of the theory of general relativity in the genuinely
nonlinear regime, and it advances astrophysics by opening a new
observational window which can be used for detailed studies of
individual sources and for source statistics.

One prime target for gravitational wave detectors is binary systems of
compact objects, black holes or neutron stars.  Such a binary system
emits energy and angular momentum through gravitational radiation, so
that the orbital separation slowly decreases.  The best-known example
is the ``Hulse-Taylor'' pulsar.  Gravitational radiation tends to
circularize orbits, leading to almost circular orbits during late
inspiral.  For binary black holes, the orbits become unstable at some
small separation, the so-called {\em innermost stable circular orbit},
where the slow adiabatic inspiral changes to a dynamical plunge.  The
two black holes merge and form a single distorted black hole, which
subsequently settles down to a stationary black hole by emission of
further gravitational radiation during the ring-down phase.

Early inspiral can be treated with post-Newtonian expansion, and the
late ring-down phase is accessible to perturbation theory.  It is
generally believed, however, that the late inspiral and the dynamical
merger can only be treated with a full numerical evolution of
Einstein's equations, which is therefore essential to obtain the
complete waveform of a binary black hole coalescence.  The knowledge
of the wave-form is essential for finding the gravitational wave
signal amidst the detector noise in the first place (cross-correlation
with known waveform templates greatly enhances the sensitivity of the
detectors), and to extract as much information as possible about
identified events.  One needs to know the prediction of general
relativity, especially in the nonlinear regime, to compare it to the
observations.

Besides the importance of numerical relativity for the experiments, it
also encompasses the intellectual challenge of solving the two-body
problem.  While the Newtonian two-body problem is treated completely
in freshman mechanics, the general relativistic analogue, the binary
black hole, is still unsolved.

Any evolution must start with {\em initial data}.  For an evolution of
Einstein's equations, setting initial data is difficult for several
reasons:

First, the initial data must satisfy constraints, analogous to the
divergence equations of electrodynamics,
\begin{equation}\label{eq:Maxwell}
\begin{aligned}
\nabla\cdot\vec E&=4\pi\rho,\\
\nabla\cdot\vec B&=0.
\end{aligned}
\end{equation}
Any initial data $\vec E$ and $\vec B$ for an evolution of Maxwell's
equations must satisfy Eqs.~(\ref{eq:Maxwell}).  Similarly, initial
data for Einstein's equations must satisfy constraint equations, which
are, however, much more complicated than Eqs.(\ref{eq:Maxwell}).
Thus, one is faced with the {\em mathematical} task of finding a
well-defined method to construct solutions to the constraint
equations of general relativity.  This problem has been an active
research area for almost sixty years.

Second, formalisms to solve the constraint equations usually lead to
sets of coupled nonlinear elliptic partial differential equations in
three dimensions, often on a computational domain which has excised
regions.  Solving such a set of equations is a formidable {\em
numerical} task.

Third, the mathematical formalisms to construct initial data sets
leave an enormous amount of freedom.  Some ten functions of space can
be freely specified.  These functions are generally called the ``free
data.''  Hence, even with an elliptic solver in hand, one is faced
with the {\em physical} question of which initial data sets from the
infinite parameter space are astrophysically relevant.  Initial data
representing a binary black hole, for example, must contain two black
holes which, if evolved, actually move on almost circular orbits.
Furthermore, the initial data should not contain unphysical
gravitational radiation; ideally, however, it should contain the
outgoing gravitational radiation emitted during the preceding
inspiral.  

This thesis attempts to contribute to each of these three aspects.  

In chapter~\ref{chapter:IVP} we present the latest formalisms to
decompose the constraint equations.  We outline the main ideas of the
two relevant papers by York~\cite{York:1999} and Pfeiffer \&
York~\cite{Pfeiffer-York:2003}.  We also comment on various issues
which are important in general, and/or will arise in later chapters of
the thesis.

Chapter~\ref{chapter:Code} develops a new code to solve elliptic
partial differential equations, a pseudospectral collocation method
with domain decomposition.  The exponential convergence of spectral
methods for smooth functions allows very accurate solutions.  The
domain decomposition makes it possible to handle complex topologies,
e.g. with excised spheres, and to distribute grid-points to improve
accuracy.  Last, but not least, the code is designed to be modular and
very flexible, making it easy to explore different partial
differential equations or different boundary conditions.  All three of
these characteristics have proved to be essential for the present
work.

The remaining chapters of the thesis explore initial data sets with a
variety of approaches.

Chapters~\ref{chapter:Spin} and~\ref{chapter:Testmass} compute binary
black holes in circular orbits for spinning black holes
(chapter~\ref{chapter:Spin}) and unequal mass black holes
(chapter~\ref{chapter:Testmass}).  We employ the effective potential
method based on conformally flat inversion-symmetric Bowen-York
initial data.  The most important result from these two chapters is,
perhaps, that this method is not optimal.  In the test-mass limit one
can assess the quality of many of the approximations used, and one can
compare the numerical method against an analytical result.  This makes
it possible to pin-point the problem of the numerical method with some
confidence.  We argue that the choice of the so-called extrinsic
curvature, namely the Bowen-York extrinsic curvature is problematic.
In the calculation for spinning equal-mass black holes, the symptoms
of failure are easily seen (disappearance of the ISCO for corotating
holes with moderate spin, and an unphysical $(\mbox{spin})^4$-effect);
however, the causes are less clear due to the large number of
assumptions made.

As there is widespread belief that the approximation of conformal
flatness limits the physical relevance of initial data sets (for
example, binary compact objects are not conformally flat at second
post-Newtonian order), we relax this approximation in
chapter~\ref{chapter:Comparing}.  We compute data sets that are not
conformally flat using different mathematical formalisms and different
choices for the free data.  We then compare these data sets with each
other.  For the data sets considered, we do not find evidence that the
non-conformally flat initial data sets are superior.  We find,
however, a sensitive dependence on the extrinsic curvature.  Two
different choices for the extrinsic curvature (``ConfTT'' and
``mConfTT'' in the language of chapter~\ref{chapter:Comparing}), which
are equally well motivated, result in drastically different initial
data sets.  While the approximation of conformal flatness will have to
be addressed eventually, especially for rotating black holes, I
believe that it is currently not the limiting factor.

One formalism to solve the constraints, the conformal thin sandwich
method, does not require specification of an extrinsic curvature.  It
thus might well avoid ambiguities like the one we just mentioned.  A
special case of this method was recently used in another approach to
compute quasi-circular orbits of binary black holes, the {\em helical
Killing vector approximation}, which has been generalized by Cook to
the {\em quasi-equilibrium method.}  Chapter~\ref{chapter:QE}
describes this method and examines the resulting set of partial
differential equations and boundary conditions.  We compute single
black hole spacetimes, and binary black holes in circular orbits with
two different choices of the remaining free data.  We further examine
the test-mass limit as well as the limit of widely separated black
holes.  In both limits, we show that the current choices for the
remaining free data are unsatisfactory.  We also discuss a proposal for
a lapse boundary condition put forth by Cook, and explain how to
perform a certain extrapolation required to use these data sets as
initial data for evolutions.

Chapter~\ref{chapter:GW}, finally, changes gear again and describes a
new and general way to construct spacetimes with superposed
gravitational radiation.  The attractive feature of this method is
that it admits a simple physical interpretation of the wave, and one
can specify directly the radial shape and angular dependence of the
wave, as well as the direction of propagation.  As examples, I
construct spacetimes without a black hole containing ingoing spherical
pulses of varying amplitude, as well as spacetimes containing one
black hole surrounded by a similar pulse.  The amplitude of these
pulses can be very large; in extreme cases most of the energy
contained in the spacetime is {\em outside} the black hole.  In this
chapter, we solve the conformal thin sandwich equations on very
general, nonflat conformal manifolds; the fact that this is easily
possible highlights the robustness of the conformal thin sandwich
formalism and of the spectral elliptic solver.

During the work described in this thesis, I have also improved an
apparent horizon finder that was developed at Cornell a few years ago.
Details can be found in sections \ref{sec:AHfinder} and
\ref{sec:Comparing:TestingCTS}.

The primary goal of this thesis is to construct initial data {\em for
evolutions}.  Indeed, data sets from chapters~\ref{chapter:Comparing},
\ref{chapter:QE} and \ref{chapter:GW} are used in evolutions by the
numerical relativity group at Cornell.  As the results from evolutions
are preliminary so far, I restrict discussion here to the initial data
sets.


\chapter{The initial value problem of general relativity}
\label{chapter:IVP}

Numerical relativity attempts to construct a spacetime with metric
$\gfour_{\mu\nu}$ ($\mu, \nu, \ldots=0,1,2,3$) which satisfies
Einstein's equations,
\begin{equation}\label{eq:Einstein}
G_{\mu\nu}=8\pi G T_{\mu\nu}.
\end{equation}

Within the standard 3+1
decomposition~\cite{Arnowitt-Deser-Misner:1962, York:1979} of
Einstein's equations, the first step is to single out a time
coordinate ``$t$'' by foliating the spacetime with spacelike
$t\!\!=$const.  hypersurfaces.  Each such hypersurface surface has a
future pointing unit-normal $n^\mu$, induced metric
$g_{\mu\nu}=\gfour_{\mu\nu}+n_\mu n_\nu$ and extrinsic curvature
$K_{\mu\nu} = -\frac{1}{2}\Lie_n g_{\mu\nu}$.  We use the label $t$ of
the hypersurfaces as one coordinate and choose 3-dimensional
coordinates $x^i$ within each hypersurface ($i, j, \ldots=1,2,3$).
The three dimensional metric $g_{\mu\nu}$ and $K_{\mu\nu}$ are purely
spatial tensors; we denote their spatial components by $\g_{ij}$ and
$\K_{ij}$.  The spacetime metric can be written as
\begin{equation}\label{eq:spacetime-ds}
ds^2=-\N^2dt+\g_{ij}\left(dx^i+\beta^i dt\right)
\left(dx^j+\beta^jdt\right),
\end{equation}
where $\N$ and $\beta^i$ denote the lapse function and shift vector,
respectively.  $\N$ measures the proper separation between neighboring
hypersurfaces along the surface normals and $\beta^i$ determines how the
coordinate labels move between hypersurfaces: Points along the
integral curves of the ``time''--vector $t^\mu=\N n^\mu + \beta^\mu$
(where $\beta^\mu=[0, \beta^i]$), have the same spatial coordinates
$x^i$.

Einstein's equations~(\ref{eq:Einstein}) decompose into evolution
equations and constraint equations for the quantities $\g_{ij}$
and $\K_{ij}$.  

The {\em evolution equations} determine how $\g_{ij}$ and $\K_{ij}$
are related between neighboring hypersurfaces,
\begin{align}
\label{eq:dtgij1}
\partial_t\g_{ij}
&=-2\N \K_{ij}+\deriv_i\beta_j+\deriv_j\beta_i\\
\partial_t\K_{ij}
&=\N\left(\R_{ij}-2\K_{ik}\K^k_j+\trK\K_{ij}-8\pi GS_{ij}+4\pi G\g_{ij}(S-\rho)
\right)\nonumber\\
\label{eq:dtKij1}
&\qquad-\deriv_i\deriv_j\N+\beta^k\deriv_k\K_{ij}+\K_{ik}\deriv_j\beta^k
+\K_{kj}\deriv_i\beta^k.
\end{align}
Here, $\deriv_i$ and $\R$ are the covariant derivative and the scalar
curvature (trace of the Ricci tensor) of $\g_{ij}$, respectively,
$\trK=\K_{ij}\g^{ij}$ is the trace of $\K^{ij}$, the so-called {\em
mean curvature}, $\rho$ and $S_{ij}$ are matter density and stress
tensor\footnote{In later chapters of this thesis, we deal exclusively
with vacuum spacetimes for which all matter terms vanish; we include
them in this chapter for completeness.}, respectively, and
$S=S_{ij}\g^{ij}$ denotes the trace of $S_{ij}$.

The {\em constraint equations} are conditions within each hypersurface
alone, ensuring that the three-dimensional surface can be embedded
into the four-dimensional spacetime:
\begin{align}
\label{eq:Ham1}
\R+\trK^2-\K_{ij}\K^{ij}&=16\pi G\rho,\\
\label{eq:Mom1}
\deriv_j\left(\K^{ij}-\g^{ij}\trK\right)&=8\pi G j^i,
\end{align}
with $j^i$ denoting the matter momentum density.
Equation~(\ref{eq:Ham1}) is called the {\em Hamiltonian constraint},
and Eq.~(\ref{eq:Mom1}) is the {\em momentum constraint}.

Cauchy initial data for Einstein's equations consists of $(\g_{ij},
\K^{ij})$ on one hypersurface satisfying the constraint
equations~(\ref{eq:Ham1}) and (\ref{eq:Mom1}).  After choosing lapse
and shift (which are arbitrary --- they merely choose a specific
coordinate system), Eqs.~(\ref{eq:dtgij1}) and (\ref{eq:dtKij1})
determine $(\g_{ij}, \K^{ij})$ at later times.  Analytically, the
constraints equations are preserved under the evolution.  (In practice
during numerical evolution of Eqs.~(\ref{eq:dtgij1}) and
(\ref{eq:dtKij1}), or any other formulation of Einstein's equations,
many problems arise.  However, we will focus on the initial value
problem here.)

The constraints~(\ref{eq:Ham1}) and (\ref{eq:Mom1}) restrict four of
the twelve degrees of freedom of $(\g_{ij}, \K^{ij})$.  As these
equations are not of any standard mathematical form, it is not obvious
which four degrees of freedom are restricted.  Hence, finding any
solutions is not trivial, and it is even harder to construct specific
solutions that represent certain astrophysically relevant situations
like a binary black hole.

Work on solving the constraint equation dates back almost sixty years
to Lichnerowicz \cite{Lichnerowicz:1944}, but today's picture emerged
only very recently in work due to York
\cite{York:1999,Pfeiffer-York:2003}.  I will describe two general
approaches to solving the constraint equations.  The first one is
based on the metric and its {\em time-derivative} on a hypersurface,
whereas the second one rests on the metric and the {\em extrinsic
curvature}.  Since the extrinsic curvature is essentially the
canonical momentum of the metric
(e.g.~\cite{Misner-Thorne-Wheeler:1973}), the latter approach belongs
to the Hamiltonian picture of mechanics whereas the former one is in
the spirit of Lagrangian mechanics.

\section{Preliminaries}
\label{sec:IVP:Preliminaries}

Both pictures make use of a conformal transformation on the metric,
\begin{equation}
\label{eq:g}
\g_{ij}=\CF^4\cg_{ij}
\end{equation}
with strictly positive {\em conformal factor} $\CF$.  $\cg_{ij}$ is
referred to as the {\em conformal metric}.  From (\ref{eq:g}) it
follows that the Christoffel symbols of the physical and conformal
metrics are related by
\begin{equation}\label{eq:Gamma-scaling}
\Christoffel^i_{jk}
=\cChristoffel^i_{jk}+2\CF^{-1}
        \left(\delta^i_j\partial_k\CF+\delta^i_k\partial_j\CF
              -\cg_{jk}\cg^{il}\partial_l\CF\right).
\end{equation}
Equation~(\ref{eq:Gamma-scaling}) implies that the scalar curvatures of
$\g_{ij}$ and $\cg_{ij}$ are related by
\begin{equation}\label{eq:R-scaling}
\R=\CF^{-4}\cR-8\CF^{-5}\cderiv^2\CF.
\end{equation}
Equations (\ref{eq:g})--(\ref{eq:R-scaling}) were already known to
Eisenhart~\cite{Eisenhart:1925}.  Furthermore, for any symmetric
tracefree tensor $\tilde S^{ij}$,
\begin{equation}\label{eq:div-scaling}
\deriv_j\left(\CF^{-10}\tilde S^{ij}\right)=\CF^{-10}\cderiv_j\tilde S^{ij},
\end{equation}
where $\cderiv$ is the covariant derivative of $\cg_{ij}$.
Lichnerowicz~\cite{Lichnerowicz:1944} used Eqs.~(\ref{eq:g}) to
(\ref{eq:div-scaling}) to treat the initial value problem on maximal
slices, $\trK=0$.  For non-maximal slices, we split the extrinsic
curvature into trace and tracefree parts,
\begin{equation}\label{eq:K-split}
\K^{ij}=\A^{ij}+\frac{1}{3}\g^{ij}\trK.
\end{equation}

With (\ref{eq:R-scaling}) and (\ref{eq:K-split}), the Hamiltonian constraint
(\ref{eq:Ham1}) becomes
\begin{equation}\label{eq:Ham2}
\cderiv^2\CF-\frac{1}{8}\CF\cR-\frac{1}{12}\CF^5\trK^2
+\frac{1}{8}\CF^5\A_{ij}\A^{ij}+2\pi G \CF^{5}\rho=0,
\end{equation}
a quasi-linear Laplace equation for $\CF$.  Local uniqueness proofs of
equations like (\ref{eq:Ham2}) usually linearize around an (assumed)
solution, and then use the maximum principle to conclude that ``zero''
is the only solution of the linearized equation.  However, the signs
of the last two terms of (\ref{eq:Ham2}) are such that the maximum
principle cannot be applied.  Consequently, it is not immediately
guaranteed that Eq.~(\ref{eq:Ham2}) has unique (or even locally
unique) solutions\footnote{As a physical illustration of possible
problems, consider the matter term $2\pi G\CF^5\rho$, a source which
pushes $\CF$ to larger values.  The physical volume element,
$dV=\sqrt{\g}d^3x=\CF^6\sqrt{\cg}d^3x$, expands as $\CF$ becomes
larger.  With {\em physical} matter energy density $\rho$ given, the
total matter content will grow like $\CF^6$ and will therefore become
{\em stronger}, pushing $\CF$ to even larger values.  Beyond some
critical value of $\rho$, a ``run-away'' might set in pushing $\CF$ to
infinity.  Indeed I observed this behavior while solving the
constraint equations coupled to a scalar field.}.  The term
proportional to $\A_{ij}\A^{ij}$ will be dealt with later; for the
matter terms we follow York~\cite{York:1979} and introduce conformally
scaled source terms:
\begin{align}
j^i&=\CF^{-10}\cj^i,  \label{eq:j-scaling}\\
\rho&=\CF^{-8}\crho.  \label{eq:rho-scaling}
\end{align}
The scaling for $j^i$ makes the momentum constraint below somewhat
nicer; the scalings of $\rho$ and $j^i$ are tied together such that
the dominant energy condition preserves sign:
\begin{equation}
\rho^2-\g_{ij}j^ij^j
=\CF^{-16}\left(\crho^2-\cg_{ij}\cj^i\cj^j\right)
\ge 0.
\end{equation}
The scaling of $\rho$, Eq.~(\ref{eq:rho-scaling}) modifies the matter
term in (\ref{eq:Ham2}) to $2\pi G\CF^{-3}\crho$ with
negative semi-definite linearization for $\crho\ge 0$.

The decomposition of $\K_{ij}$ into trace and tracefree part,
Eq.~(\ref{eq:K-split}), turns the momentum constraint~(\ref{eq:Mom1})
into
\begin{equation}\label{eq:Mom2}
\deriv_j\A^{ij}-\frac{2}{3}\deriv^i\trK=8\pi Gj^i.
\end{equation}
For time-symmetric vacuum spacetimes (where
$\A^{ij}\!=\!\trK\!=\!j^i\!=\!0$ solve the momentum constraint
(\ref{eq:Mom2}) trivially), only the first two terms of
(\ref{eq:Ham2}) remain.  This simplified equation was used in
beautiful early work on vacuum spacetimes, for example a positivity of
energy proof by Brill \cite{Brill:1959} and construction of multi
black hole spacetimes by Misner \cite{Misner:1963} and Brill~\&
Lindquist \cite{Brill-Lindquist:1963}.

The conformal transformation (\ref{eq:g}) implies one additional, very
simple, conformal scaling relation.  The longitudinal operator \cite{Deser:1967, York:1973, York:1974}
\begin{equation}\label{eq:L}
(\Long V)^{ij}\equiv\deriv^iV^j+\deriv^jV^i-\frac{2}{3}\g^{ij}\deriv_kV^k,
\end{equation}
satisfies~\cite{York:1973}
\begin{equation}\label{eq:L-scaling}
(\Long V)^{ij}=\CF^{-4}(\cLong V)^{ij}.
\end{equation}
Here $(\cLong V)^{ij}$ is given by the same formula (\ref{eq:L}) but
with quantities associated with the conformal metric $\cg_{ij}$.  (In
fluid dynamics $(\Long V)^{ij}$ is twice the shear of the velocity
field $V^i$).  In $d$ dimensions, the factor $2/3$ in
Eq.~(\ref{eq:L}) is replaced by $2/d$; Eq.~(\ref{eq:L-scaling})
holds for all $d$.

\section{Conformal thin sandwich formalism}
\label{sec:IVP:CTS}

\begin{figure}[b]
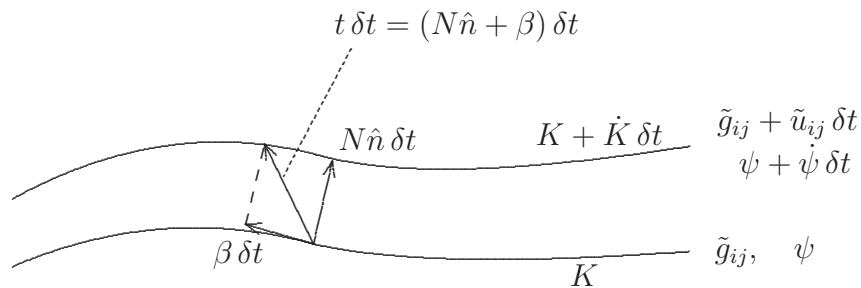

\hspace*{.8cm}
{
\beginpicture
\setcoordinatesystem units <10mm,10mm>
\setplotarea x from -1 to 12.2, y from -0.7 to 4.3
\setquadratic
\plot 0 0.9 2  1.6   4 1.5    6 1.3 9 1.6 /
\plot 0 0 2  0.5   4 0.3    6 0.1 9 0.2 /

\setlinear
\arrow <2mm> [0.375,0.75] from 4 0.3 to 4.25 1.4 
\put {$\N\hat n\,\delta t$} at 4.85 1.7

\arrow <2mm> [0.375,0.75] from 4 0.3 to 3.1 0.56
\put {$\beta\,\delta t$} at 3. 0.15

\arrow <2mm> [0.375,0.75] from 4 0.3 to 3.35 1.63 
\put {$t\,\delta t=(\N\hat n+\beta)\,\delta t$} at 5.9 3.2
\setdashes<0.5mm>
\setlinear
\plot 4.5 2.9 3.6 1.2 /

\setdashes<1.5mm>
\setlinear
\plot 3.1 0.56 3.35 1.63 /

\put {$\cg_{ij},\quad\CF$} at 10 0.2
\put {$\cg_{ij}+\cu_{ij}\,\delta t$} at 10.3 1.9

\put{$\CF+\dot\CF\,\delta t$} at 10.4 1.4

\put {$\trK$} at 7.6 -0.12
\put {$\trK+\dot \trK\,\delta t$} at 7.8 1.8
\endpicture
}
\caption{\label{fig:ConformalThinSandwich}Setup for conformal thin
sandwich equations}
\end{figure}

We now derive a formalism to solve the constraint equations
\cite{York:1999} which deals with the conformal metric and its {\em
time derivative}.  This formalism thus represents the {\em Lagrangian}
viewpoint.  Figure~\ref{fig:ConformalThinSandwich} illustrates the
basic setup; we deal with {\em two} hypersurfaces separated by the
infinitesimal $\delta t$ (thus the name ``thin sandwich''), and
connected by the lapse $\N$ and the shift $\beta^i$.  The mean
curvature of each hypersurface is $\trK$ and $\trK+\dot\trK\delta t$,
respectively.  The metric is split into conformal factor and conformal
metric; on the first hypersurface, this is simply Eq.~(\ref{eq:g}).
On the second hypersurface, the conformal factor and the conformal
metric will both be different from their values on the first
hypersurface.  The split into conformal factor and conformal metric is
not unique and is synchronized between the two surfaces by requiring
that the conformal metrics on both hypersurfaces have the same
determinant to first order in $\delta t$.  The variation of the
determinant of $\cg_{ij}$ is given by
\begin{equation}
\delta \cg=\cg\cg^{ij}\delta
\cg_{ij}=\cg\cg^{ij}\cu_{ij}\delta t
\end{equation}
(the first identity holds for any square matrix), so that
$\cu_{ij}=\partial_t\cg_{ij}$ must be traceless.  On the first
hypersurface the Hamiltonian constraint will eventually determine
$\CF$.

Besides the relationships indicated in Figure
\ref{fig:ConformalThinSandwich}, the conformal thin sandwich formalism
rests on the nontrivial scaling behavior of the lapse function:
\begin{equation}\label{eq:N-scaling}
\N=\CF^6 \cN.
\end{equation}
The scaling (\ref{eq:N-scaling}) appears in a bewildering variety of
contexts (see references in \cite{York:1999, Pfeiffer-York:2003}).
For example, studies of {\em hyperbolic} evolution systems for
Einsteins equations
(e.g. \cite{Choquet-Bruhat-York:1997,Kidder-Scheel-Teukolsky:2001})
find that hyperbolicity requires that the lapse anti-density
(``densitized lapse'') $\alpha\equiv\g^{-1/2}\N$ is freely
specifiable, not $\N$ directly (recall $\sqrt{\g}=\CF^6\sqrt{\cg}$, so
that $\alpha$ is essentially equivalent to $\cN$).  The scaling
(\ref{eq:N-scaling}) is crucial in the present context as well,
cf. Eq.~(\ref{eq:Aij-scaling}) below.

Substitution of Eq.~(\ref{eq:g}) into the evolution equation for
the metric, Eq.~(\ref{eq:dtgij1}), yields
\begin{equation}\label{eq:dtgij2}
4\CF^4(\partial_t\ln\CF)\cg_{ij}+\CF^4\partial_t\cg_{ij}
=-2\N\K_{ij}+\deriv_i\beta_j+\deriv_j\beta_i.
\end{equation}
Since $\partial_t\cg_{ij}=\cu_{ij}$ is traceless, the left hand side of
Eq.~(\ref{eq:dtgij2}) is already split into trace and tracefree parts.
Splitting the right hand side, too, gives
\begin{align}\label{eq:uij}
\CF^4\cu_{ij}&=-2\N\A_{ij}+(\Long\beta)_{ij}
\intertext{and}
\label{eq:dtphi}
\partial_t\ln\CF&=-\frac{1}{6}\N\trK+\frac{1}{6}\deriv_k\beta^k.
\end{align}
(The factor $\CF^4$ in (\ref{eq:dtphi}) cancels because the trace is
performed with the physical inverse metric
$\g^{ij}=\CF^{-4}\cg^{ij}$).  Equation~(\ref{eq:uij}) is the tracefree
piece of $\partial_t\g_{ij}$, thus for $\u_{ij}\equiv \CF^4\cu_{ij}$,
\begin{equation}\label{eq:dtg-tracefree}
\u_{ij}=\partial_t\g_{ij}-\frac{1}{3}\g_{ij}\g^{kl}\partial_t\g_{kl}.
\end{equation}
Now solve Eq.~(\ref{eq:uij}) for $\A^{ij}$,
\begin{equation}\label{eq:CTS-Aij}
\A^{ij}=\frac{1}{2\N}\left((\Long\beta)^{ij}-\u^{ij}\right),
\end{equation}
and rewrite with conformal quantities [using (\ref{eq:N-scaling}),
(\ref{eq:L-scaling}) and $\u^{ij}=\CF^{-4}\cu^{ij}$]:
\begin{equation}\label{eq:Aij-scaling}
\A^{ij}
=\frac{1}{2\CF^6\cN}\Big(\CF^{-4}(\cLong\beta)^{ij}-\CF^{-4}\cu^{ij}\Big)
=\CF^{-10}\frac{1}{2\cN}\Big((\cLong\beta)^{ij}-\cu^{ij}\Big)
\equiv\CF^{-10}\cA^{ij}
\end{equation}
with the conformal tracefree extrinsic curvature 
\begin{equation}\label{eq:cAij}
\cA^{ij}=\frac{1}{2\cN}\Big((\cLong\beta)^{ij}-\cu^{ij}\Big).
\end{equation}
Equation~(\ref{eq:Aij-scaling}) shows that the formula for $\A^{ij}$
is form invariant under conformal transformations; this hinges on the
scaling of $\N$ in Eq.~(\ref{eq:N-scaling}).  Substitution of
Eq.~(\ref{eq:Aij-scaling}) into the momentum constraint
(\ref{eq:Mom2}) and application of Eq.~(\ref{eq:div-scaling}) yields
\begin{equation}\label{eq:Mom3}
\cderiv_j\left(\frac{1}{2\cN}(\cLong\beta)^{ij}\right)
-\cderiv_j\left(\frac{1}{2\cN}\cu^{ij}\right)-\frac{2}{3}\CF^6\cderiv^i\trK
=8\pi G\cj^i,
\end{equation}
whereas Eq.~(\ref{eq:Aij-scaling}) modifies the Hamiltonian constraint
(\ref{eq:Ham2}) to\footnote{Indices on conformal quantities are raised
and lowered with the conformal metric, for example
$$
\A_{ij}
=\g_{ik}\g_{jl}\A^{kl}=\CF^4\cg_{ik}\CF^4\cg_{jl}\CF^{-10}\cA^{kl}
=\CF^{-2}\cA_{ij}.
$$} 
\begin{equation}\label{eq:Ham3}
\cderiv^2\CF-\frac{1}{8}\CF \cR-\frac{1}{12}\CF^5\trK^2
+\frac{1}{8}\CF^{-7}\cA_{ij}\cA^{ij}=-2\pi G\CF^{-3}\crho.
\end{equation}
Equations~(\ref{eq:Mom3}) and~(\ref{eq:Ham3}) constitute elliptic
equations for $\beta^i$ and $\CF$.  We thus find the following procedure
to compute a valid initial data set:
\renewcommand{\theenumi}{{\bf \Alph{enumi}}}
\begin{enumerate}
\item Choose the free data 
\begin{equation}\label{eq:CTS-freedata1}
(\cg_{ij},\; \cu_{ij},\;\;\;\trK,\; \cN)
\end{equation}
(and matter terms if applicable).
\item Solve Eqs.~(\ref{eq:Mom3}) and (\ref{eq:Ham3}) for $\beta^i$ and $\CF$.
\item Assemble $\g_{ij}=\CF^4\cg_{ij}$ and
$\K^{ij}=\CF^{-10}\cA^{ij}+\frac{1}{3}\g^{ij}\trK$.
\end{enumerate}

The conformal thin sandwich formalism does not involve
transverse-traceless decompositions, and is therefore somewhat more
convenient than the extrinsic curvature decomposition discussed in
section \ref{sec:IVP:ExCurv}.  We now comment on several issues
related to the conformal thin sandwich formalism.

\subsection[Fixing $\cN$ via $\dot\trK$]{Fixing \boldmath$\cN$ via $\dot\trK$}

The free data so far is given by~(\ref{eq:CTS-freedata1}).  Whereas
$\cg_{ij}$ and $\cu_{ij}=\partial_t\cg_{ij}$ constitute a
``variable~\&~velocity pair'' $(q, \dot q)$ in the spirit of
Lagrangian mechanics, the remaining free data does not.

As we show starting with Eq.~(\ref{eq:dtK}) below, specification of
$\partial_t\trK=\dot\trK$ fixes $\cN$ through an elliptic equation.
If we take $\dot\trK$ as the ``free'' quantity instead of $\cN$, then
the free data for the conformal thin sandwich formalism becomes
\begin{equation}\label{eq:CTS-freedata2}
(\cg_{ij},\; \cu_{ij},\;\;\;\trK,\;\dot\trK)
\end{equation}
plus matter terms if applicable.  These free data consists completely
of $(q, \dot q)$ pairs as appropriate for a Lagrangian viewpoint.  The
free data (\ref{eq:CTS-freedata2}) is also useful practically for
computations of quasi-equilibrium initial data sets in chapter
\ref{chapter:QE}.  In quasi-equilibrium, $\dot\trK=0$ is a natural and
simple choice, whereas it is not obvious at all which conformal lapse
$\cN$ one should use.  In particular, $\cN$ will depend on the slicing
of the spacetime.

We now derive the elliptic equation for $\cN$~(see, e.g.,
\cite{Smarr-York:1978b}).  The trace of the evolution equation for
$\K_{ij}$, Eq.~(\ref{eq:dtKij1}), gives after some calculations
\begin{equation}\label{eq:dtK}
\partial_tK-\beta^k\partial_k\trK
=\N\left(\R+\trK^2+4\pi G(S-3\rho)\right)-\deriv^2\N.
\end{equation}
Elimination of $\R$ with the Hamiltonian constraint (\ref{eq:Ham1}) 
yields 
\begin{equation}
\deriv^2\N-\N\left(\K_{ij}\K^{ij}+4\pi
G(S+\rho)\right)=-\partial_tK+\beta^k\partial_k\trK.
\end{equation}
Rewriting the Laplace operator with conformal derivatives, 
\begin{align*}
\deriv^2\N
&=\frac{1}{\sqrt{\g}}\partial_i\left(\sqrt{\g}
\g^{ij}\partial_j\N\right)\\
&=\frac{\CF^{-6}}{\sqrt{\cg}}\partial_i\left(\CF^2\sqrt{\cg}\cg^{ij}
\partial_j\N\right)\\
&=\CF^{-4}\cderiv^2\N+2\CF^{-4}\cderiv_{\!i}\ln\CF\cderiv^i\N,
\end{align*}
together with Eqs.~(\ref{eq:N-scaling}) and (\ref{eq:Ham3}) gives
\begin{align}\label{eq:dtK3}
\cderiv^2\cN+14\cderiv^i\ln\CF\cderiv_{\!i}&\cN
+\cN\bigg[\frac{3}{4}\cR+\frac{1}{6}\CF^4\trK^2
          \!-\!\frac{7}{4}\CF^{-8}\cA_{ij}\cA^{ij}\\
+&42\cderiv_{\!i}\ln\CF\cderiv^i\ln\CF\!-\!4\pi G\CF^4(S\!+\!4\rho)\bigg]
=-\CF^{-2}\left(\partial_t\trK\!-\!\beta^k\partial_k\trK\right),
\nonumber
\end{align}
which can also be rewritten as
\begin{align}\label{eq:dtK4}
\cderiv^2(\cN\CF^7)-(\cN\CF^7)\bigg[\frac{1}{8}\cR+\frac{5}{12}\CF^4\trK^2
+\frac{7}{8}\CF^{-8}\cA_{ij}\cA^{ij}&+2\pi G\CF^4(\rho+2S)\bigg]\\
&=-\CF^5\left(\partial_t\trK\!-\!\beta^k\partial_k\trK\right).
\nonumber
\end{align}
Equation (\ref{eq:dtK3}) determines the freely specifiable $\cN$
directly, whereas Eq.~(\ref{eq:dtK4}) is shorter and therefore
computationally somewhat more convenient.

We now have a system of {\em five} coupled partial differential
equations, (\ref{eq:Mom3}), (\ref{eq:Ham3}) and (\ref{eq:dtK3}) [or
(\ref{eq:dtK4})] with freely specifiable data (\ref{eq:CTS-freedata2}).

\subsection{Invariance to conformal transformations of the free data}
\label{sec:IVP:CTS:invariance}

Assume we specify free data (\ref{eq:CTS-freedata1}), solve the
conformal thin sandwich equations for conformal factor $\CF$ and shift
$\beta^i$, and assemble the physical initial data $(\g_{ij},
\K^{ij})$.  We now pick a function $\Psi>0$, and perform a conformal
transformation on the free data (\ref{eq:CTS-freedata1}) by setting
\begin{equation}\label{eq:freedata-scaled}
\cg'_{ij}=\Psi^{-4}\cg_{ij},\quad
{\cu{'}{\,}}^{ij}=\Psi^{4}\cu^{ij},\quad
K'=K,\quad
\cN'=\Psi^{-6}\cN,
\end{equation}
plus the scalings $\crho'=\Psi^8\crho,\quad
{\cj'}{\,}^i=\Psi^{10}\cj^i$ for matter terms if applicable.  These
free data, together with conformal factor $\CF'=\Psi\CF$ and the shift
$\beta'{\,}^i=\beta^i$ (where $\CF$ and $\beta^i$ are the solutions to
the conformal thin sandwich equations using the {\em original} free data)
lead to the {\em same} physical initial data $(\g_{ij}, \K^{ij})$:
\begin{align}
\g'_{ij}
&=\CF'{\,}^4\;\cg'_{ij}=\CF^4\cg_{ij}=\g_{ij},\\
\A'{\,}^{ij}
&=\CF'{\,}^{-10}\frac{1}{2\cN'}
\left(\big(\cLong'\,\beta'\big)^{ij}-\cu'{\,}^{ij}\right)
\nonumber\\
\label{eq:Aij-conformal-invariance}
&=(\Psi\CF)^{-10}\frac{1}{2\Psi^{-6}\cN}
\left(\Psi^4\big(\cLong\,\beta\big)^{ij}-\Psi^4\cu^{ij}\right)
=\A^{ij}.
\end{align}
Here, we used Eqs.~(\ref{eq:g}),~(\ref{eq:L-scaling})
and~(\ref{eq:Aij-scaling}).  Adding the trace of the extrinsic
curvature to (\ref{eq:Aij-conformal-invariance}) is trivial.

We thus find that only the {\em conformal equivalence class} of
$\cg_{ij}$ is relevant for the physical solution.  This is a very
desirable property; we introduced $\cg_{ij}$ as a {\em conformal}
metric, so its overall scaling should not matter.  Fixing $\cN$ by
specification of $\dot\trK$ preserves this invariance, as the
$\dot\trK$-equation, (\ref{eq:dtK3}), is derived from {\em physical}
quantities in the first place.

The extrinsic curvature decomposition introduced in the next section
is also invariant under conformal transformations of the free data.
We note that invariance under conformal transformations of the free
data is not trivial; earlier variants of the constraint decompositions
did not possess it, giving rise to ambiguities in the free
data~\cite{Tichy-Bruegmann-etal:2003}.

\subsection{Gauge degrees of freedom}

The physical initial data $(\g_{ij}, \K^{ij})$ has twelve degrees of
freedom\footnote{Matter just adds four additional degrees of freedom
in $\crho$ and $\cj^i$ which determine the four physical matter
variables $\rho$ and $j^i$.  It therefore does not influence the
counting of degrees of freedom of the geometrical objects $\g_{ij}$
and $\K^{ij}$.}.  It is constrained by four constraint equations, so
there should be eight degrees of freedom in the freely specifiable
data.  However, taking into account that only the conformal
equivalence class of $\cg_{ij}$ is relevant and that $\cu_{ij}$ is
traceless, one sees that the free data (\ref{eq:CTS-freedata1}) or
(\ref{eq:CTS-freedata2}) consists of {\em twelve} quantities, not
eight.  To examine this further, we consider the substitutions
\begin{equation}\label{eq:beta-gauge}
\begin{aligned}
\cu^{ij}\;&\to\;\cu^{ij}+(\cLong W)^{ij},\\
\beta^i\;&\to\;\beta^i+W^i.
\end{aligned}
\end{equation}
The vector $W^i$ disappears from
Eqs.~(\ref{eq:cAij})--(\ref{eq:Ham3}), therefore the substitution
(\ref{eq:beta-gauge}) will not change the physical initial data set
$(\g_{ij}, \K^{ij})$; it merely tilts the time-axis and changes the
coordinate labels on the second hypersurface.  Thus, $\cu_{ij}$
contains three gauge degrees of freedom associated with the shift.
($W^i$ enters into Eq.~(\ref{eq:dtK3}) as an advection term, though,
because $\partial_tK$ is the derivative {\em along} the time-vector,
which of course must be sensitive to the direction {\em of} the
time-vector).

The fourth ``missing'' degree of freedom is hidden in the lapse
function $\cN$:  One can construct every possible initial data set
$(\g_{ij}, \K^{ij})$ with {\em any} (non-pathologic) choice of $\cN$.
This is easiest seen by going backward from the physical initial data
$(\g_{ij}, \K^{ij})$ (satisfying the constraints) to the free data.
Given $(\g_{ij}, \K^{ij})$ {\em and} any $\cN$ and $\beta^i$, there
exists free data such that solving the conformal thin sandwich
equations gives back the original physical initial data set $(\g_{ij},
\K^{ij})$.  These free data are:
\begin{equation}
\cg_{ij}=\g_{ij},\quad \cu^{ij}=(\Long\beta)^{ij}-2\cN \A^{ij},\quad 
\trK=\K^{ij}\g_{ij}
\end{equation}
plus the given $\cN$ (whatever it may be).  With these choices for the
free data, $\CF\equiv 1$ and the given $\beta^i$ will reconstruct the
physical spacetime $(\g_{ij}, \K^{ij})$ as can be seen from
Eqs.~(\ref{eq:g}) and (\ref{eq:Aij-scaling}).  Therefore, $\CF\equiv
1$ and the given $\beta^i$ will solve the conformal thin sandwich
equations (\ref{eq:Mom3}) and (\ref{eq:Ham3}).  

The fact that we were free to choose $\beta^i$ reflects again the
gauge-symmetry illustrated in Eq.~(\ref{eq:beta-gauge}), but in
addition, we showed that the choice of $\cN$ does not restrict the set
of ``achievable'' initial data sets.

Note that the physical initial data contain further gauge freedom:
Covariance under spatial transformations implies that $\g_{ij}$ (and
$\cg_{ij}$) contain three gauge degrees of freedom associated with the
choice of coordinates.  Furthermore, $\trK$ can be interpreted as {\em
time}~\cite{York:1972}, fixing the temporal gauge.  Thus, an initial
data set has only four physical degrees of freedom---in perturbed flat
space they are simply the two polarizations of gravitational waves.

\subsection{Implications for an evolution of the initial data}
\label{sec:IVP:ImplicationsForEvolution}

During the solution of the conformal thin sandwich equations, one finds a
shift $\beta^i$ and a lapse $N$.  {\em If} one uses this lapse and
shift in a subsequent evolution of the initial data $(\g_{ij},
\K^{ij})$ ---recall that any lapse and shift can be used--- then 
Eq.~(\ref{eq:dtg-tracefree}) implies that {\em initially},
\begin{equation}\label{eq:dtg-tracefree2}
\partial_t\g_{ij}-\frac{1}{3}\g_{ij}\g^{kl}\partial_t\g_{kl}
=\CF^4\cu_{ij}.
\end{equation}
The freely specifiable piece $\cu_{ij}$ thus directly controls
the tracefree part of the time-derivative of the metric. 
If we specified $\dot\trK$ as part of the free data, then, of course,
the evolution will initially have $\partial_t\trK=\dot\trK$, too.
Finally, from (\ref{eq:dtphi}), we find 
\begin{align}
\partial_t\ln\CF&=\frac{1}{6}\left(\deriv_k\beta^k-\N\trK\right)\nonumber\\
\label{eq:dtphi2}
&=\frac{1}{6}\cderiv_k\beta^k+\beta^k\partial_k\ln\CF-\frac{1}{6}\CF^6\cN\trK.
\end{align}
This is essentially the trace of $\partial_t\g_{ij}$.  

Since we have been very successful so far with specification of
time-derivatives ($\partial_t\cg_{ij}$ and $\partial_t\trK$), one
might be tempted to turn (\ref{eq:dtphi2}) around and use it as the
{\em definition} of $\trK$ in terms of $\partial_t\ln\CF$.  This idea
certainly comes to mind when looking for quasi-equilibrium solutions,
for which as many time-derivatives as possible should vanish.  From
(\ref{eq:dtphi2}) we find
\begin{equation}\label{eq:K-attempt}
K=\frac{1}{\CF^6\cN}\left(\cderiv_k\beta^k
       -6\left(\partial_t-\beta^k\partial_k\right)\ln\CF\right).
\end{equation}
We see that $\trK$ contains first derivatives of the shift.  The
gradient of $\trK$ enters the momentum constraint (\ref{eq:Mom3}),
therefore (\ref{eq:K-attempt}) will modify the second derivatives of
the elliptic operator in (\ref{eq:Mom3}).  The terms containing second
derivatives of $\beta^i$ in Eq.~(\ref{eq:Mom3}) become
\begin{equation}
\cderiv_j\left(\frac{1}{2\cN}(\cLong\beta)^{ij}\right)
-\frac{2}{3}\CF^6\cderiv^i\left(\frac{1}{\CF^6\cN}\cderiv_k\beta^k\right).
\end{equation}
Using the product rule, discarding first derivatives of $\beta^i$, and 
using the definition of $\Long$, Eq.~(\ref{eq:L}), we find
\begin{equation}
\frac{1}{2\cN}\left(
\cderiv_j\cderiv^j\beta^i+\cderiv_j\cderiv^i\beta^j-2\cderiv^i\cderiv_k\beta^k
\right).
\end{equation}
Commuting the derivatives in the second term yields a term
proportional to the Riemann tensor, which is discarded as it contains
no derivatives of $\beta^i$ at all.  After multiplication by $\cN$,
the highest-order derivatives of $\beta^i$ in the momentum constraint
become
\begin{equation}
\partial_j\partial^j\beta^i-\partial^i\partial_k\beta^k.
\end{equation}
This operator is non-invertible; for example it maps the functions
\begin{equation}
f^j(x^l) = k^j \exp\left(i\,k^lx_l\right)
\end{equation}
to zero for any choice of the wave-vector $k^i$.  Thus, the attempt to
fix $\trK$ via (\ref{eq:K-attempt}) makes the momentum constraint
non-invertible.  

In practice I encountered a consequence of this fact when constructing
quasi-equilibrium slices of spherically symmetric spacetimes.  For
that case, we show in section~\ref{sec:QE:SphericalSymmetricSolves}
that the conformal thin sandwich equations with $\cu_{ij}=0$ and
$\dot\trK=0$ (and appropriate boundary conditions) are so successful
in picking out the time-like Killing vector that for {\em any} choice
of $\trK$, the solution satisfies
$\partial_t\ln\CF=0$.  If $\partial_t\ln\CF=0$ for any choice of
$\trK$, then $\partial_t\ln\CF=0$ obviously cannot determine a unique
$\trK$.

In contrast to the trace-free part (\ref{eq:dtg-tracefree2}) we can
{\em not} easily control $\partial_t\ln\CF$ by choices of the free
data.  We can only evaluate (\ref{eq:dtphi2}) {\em after} solving the
conformal thin sandwich equations.

\section{Extrinsic curvature decomposition}
\label{sec:IVP:ExCurv}

The second method to construct solutions of the constraint equations
if based on a decomposition of the extrinsic curvature.  Early
variants of this approach \cite{Murchadha-York:1974b, York:1979} have
been available for almost thirty years, but the final version was
developed only very recently \cite{Pfeiffer-York:2003}.  
We will make use of the equations and results from section
\ref{sec:IVP:Preliminaries}, in particular, we use a conformal
metric, $\g_{ij}=\CF^4\cg_{ij}$, and split the extrinsic curvature
into trace and trace-free parts, $\K^{ij}=\A^{ij}+1/3\g^{ij}\trK$,
cf. Eqs.~(\ref{eq:g}) and (\ref{eq:K-split}).

We start with a {\em weighted transverse traceless} decomposition of
$\A^{ij}$,
\begin{equation}\label{eq:barAij-TT}
\A^{ij}=\A^{ij}_{TT}+\frac{1}{\s}\left(\Long V\right)^{ij}.
\end{equation}
Here, $\A^{ij}_{TT}$ is transverse, $\deriv_{\!j}\A^{ij}_{TT}=0$, and
traceless, $\g_{ij}\A^{ij}_{TT}=0$, and $\s$ is a strictly positive
and bounded function.  Appearance of the weight function $\s$ is a key
point in the extrinsic curvature formulation; its inclusion is the
major improvement of~\cite{Pfeiffer-York:2003} over the older variants.

Given a symmetric tracefree tensor like $\A^{ij}$, the decomposition
(\ref{eq:barAij-TT}) can be obtained by taking the divergence of
Eq.~(\ref{eq:barAij-TT}),
\begin{equation}\label{eq:div-Aij-TT}
\deriv_j\A^{ij}=\deriv_j\left[\s^{-1}(\Long V)^{ij}\right].
\end{equation}
The right hand side, $\deriv_j\left[\s^{-1}(\Long\,.\,)^{ij}\right]$,
is a well-behaved elliptic operator in divergence form, so no problem
should arise when solving (\ref{eq:div-Aij-TT}) for $V^i$.
Substitution of the solution $V^i$ back into (\ref{eq:barAij-TT})
yields the transverse traceless part $\A^{ij}_{TT}$.  In the presence
of boundaries, Eq.~(\ref{eq:div-Aij-TT}) requires boundary conditions
which will influence the solution $V^i$ and the decomposition
(\ref{eq:barAij-TT}).  For closed manifolds, existence and uniqueness
of the decomposition (\ref{eq:barAij-TT}) for the case $\s\equiv 1$
was shown in \cite{York:1974}.

We now conformally scale the quantities on the right hand side of
(\ref{eq:barAij-TT}) with the goal of rewriting the momentum
constraint in conformal space.

First, we set 
\begin{equation}\label{eq:AijTT-scaling}
\A^{ij}_{TT}\equiv \CF^{-10}\cA^{ij}_{TT}.
\end{equation}
Equation~(\ref{eq:div-scaling}) ensures that $\cA^{ij}_{TT}$ is
transverse with respect to $\cg_{ij}$ if and only if $\A^{ij}_{TT}$ is
transverse with respect to the physical metric $\g_{ij}$.

Because of Eq.~(\ref{eq:L-scaling}), and because $\Long$ is the
conformal Killing operator, we will not scale the vector $V^i$.

The conformal scaling of the weight function is given by
\begin{equation}\label{eq:sigma-scaling}
\s=\CF^6\cs.
\end{equation}
The most immediate reason for this scaling is that it allows
Eq.~(\ref{eq:Aij-scaling2}) below; several more reasons will be
mentioned in the sequel.

Using the scaling relations (\ref{eq:L-scaling}),
 (\ref{eq:AijTT-scaling}) and (\ref{eq:sigma-scaling}), we can now
 rewrite Eq.~(\ref{eq:barAij-TT}) as
\begin{equation}\label{eq:Aij-scaling2}
\A^{ij}
=\CF^{-10}\left(\cA^{ij}_{TT}+\frac{1}{\cs}(\cLong V)^{ij}\right)
\equiv\CF^{-10}\cA^{ij},
\end{equation}
where
\begin{equation}\label{eq:Aij-TT}
\cA^{ij}=\cA^{ij}_{TT}+\cs^{-1}(\cLong V)^{ij}
\end{equation}
is a weighted transverse traceless decomposition in the conformal
space.  By virtue of the scaling of the weight function $\sigma$,
Eq.~(\ref{eq:sigma-scaling}), the weighted transverse traceless
decomposition thus commutes with the conformal transformation.

Equations~(\ref{eq:div-scaling}) and (\ref{eq:Aij-scaling2}) allow us to 
rewrite the momentum constraint (\ref{eq:Mom2}) as
\begin{equation}\label{eq:Mom4}
\cderiv_j\left(\frac{1}{\cs}(\cLong V)^{ij}\right)
-\frac{2}{3}\CF^6\cderiv^i\trK
=8\pi G\cj^i,
\end{equation}
an elliptic equation for $V^i$.  The Hamiltonian constraint
Eq.~(\ref{eq:Ham2}) reads
\begin{equation}\label{eq:Ham4}
\cderiv^2\CF-\frac{1}{8}\cR\CF
-\frac{1}{12}\CF^5\trK^2+\frac{1}{8}\CF^{-7}\cA_{ij}\cA^{ij}
=-2\pi G\CF^{-3}\crho,
\end{equation}
with $\cA^{ij}$ given by (\ref{eq:Aij-TT}).  Equation~(\ref{eq:Ham4})
is identical to Eq.~(\ref{eq:Ham3}) since in both formulations
$\A^{ij}=\CF^{-10}\cA^{ij}$, however, the definitions of $\cA^{ij}$
differ.

Starting from the {\em physical} initial data $(\g_{ij}, \K^{ij})$, we
have now rewritten the constraints (\ref{eq:Ham1}) and (\ref{eq:Mom1})
as elliptic equations (\ref{eq:Mom4}) and (\ref{eq:Ham4}).  In order
to construct a valid initial data set $(\g_{ij}, \K^{ij})$,
one performs this program backward:
\renewcommand{\theenumi}{\bf\Alph{enumi}}
\begin{enumerate}
\item\label{step-1} Choose the free data 
\begin{equation}\label{eq:ExCurv-freedata}
\left(\cg_{ij},\;\; \trK,\;\; \cA^{ij}_{TT},\;\;\cs\right)
\end{equation}
and matter terms if applicable.
\item\label{step-2} Solve Eqs.~(\ref{eq:Mom4}) and (\ref{eq:Ham4}) for
$V^i$ and $\CF$.
\item\label{step-3} Assemble the physical solution by Eqs.~(\ref{eq:g}),
 (\ref{eq:K-split}), and (\ref{eq:Aij-scaling2}):
\begin{align}
\g_{ij}&=\CF^4\cg_{ij},\\
\K^{ij}&=\CF^{-10}\Big(\cA^{ij}_{TT}+\cs^{-1}(\cLong V)^{ij}\Big)
+\frac{1}{3}\g^{ij}\trK.
\end{align}
\end{enumerate}

\renewcommand{\theenumi}{\arabic{enumi}}

The next subsection contains a few brief remarks, whereas later
subsections comment on specific issues in more detail.

\subsection{Remarks on the extrinsic curvature decomposition}

\subsubsection{Invariance to conformal transformations of the free data}

Similar to section \ref{sec:IVP:CTS:invariance}, one can show that the
physical initial data $(\g_{ij}, \K^{ij})$ is {\em invariant} to 
a conformal transformation of the free data.  For $\Psi>0$, the relevant
transformations are [cf. Eq.~(\ref{eq:freedata-scaled})]:
\begin{equation}\label{eq:freedata-scaled2}
\cg'_{ij}=\Psi^{-4}\cg_{ij},\quad
\cA'{\,}^{ij}_{TT}=\Psi^{10}A^{ij}_{TT},\quad
K'=K,\quad
\cs'=\Psi^{-6}\cs,
\end{equation}
plus the scalings $\crho'=\Psi^8\crho,\quad
{\cj'}{\,}^i=\Psi^{10}\cj^i$ for matter terms if applicable.  The
calculation is straightforward, the key-point being that the scaling of
the weight-function~(\ref{eq:sigma-scaling}) synchronizes the
conformal scaling of the transverse-traceless and longitudinal parts
of the weighted transverse traceless decomposition.

\subsubsection{Gauge degrees of freedom}

Because of the invariance to conformal scalings of the free data,
$\cg_{ij}$ supplies only five degrees of freedom, so that the free
data Eq.~(\ref{eq:ExCurv-freedata}) contains nine degrees of freedom.
The weight $\s$ (or $\cs$) merely parametrizes the transverse
traceless decomposition (\ref{eq:barAij-TT}).  For any choice of $\s$,
the decomposition (\ref{eq:barAij-TT}) can be performed, therefore
with any choice of $\cs$, all initial data sets can be generated for
appropriate choices of the free data.

\subsubsection{Construction of \protect\boldmath$\cA^{ij}_{TT}$}

To construct a transverse traceless tensor $\cA^{ij}_{TT}$ compatible
with the metric $\cg_{ij}$, one decomposes a general symmetric
tracefree tensor $\tilde M^{ij}$.  Write
\begin{equation}\label{eq:Mij-decomposition}
\tilde M^{ij}=\cA^{ij}_{TT}+\frac{1}{\cs}(\cLong W)^{ij}.
\end{equation}
The divergence of this equation, 
\begin{equation}\label{eq:divergence-Mij-decomposition}
\cderiv_j\tilde M^{ij}=\cderiv_j\left[\cs^{-1}(\cLong W)^{ij}\right],
\end{equation}
represents an elliptic equation for $W^i$.  Solving this equation, and
substituting $W^i$ back into (\ref{eq:Mij-decomposition}) yields
\begin{equation}
\cA^{ij}_{TT}=\tilde M^{ij}-\frac{1}{\cs}(\cLong W)^{ij}.
\end{equation}
The formula for $\cA^{ij}$, Eq.~(\ref{eq:Aij-TT}), now reads
\begin{equation}\label{eq:Aij-from-Mij}
\cA^{ij}=\tilde M^{ij}+\frac{1}{\cs}\big[\cLong(V-W)\big]^{ij},
\end{equation}
which depends only on the {\em difference} $V^i-W^i$.  On the other hand, 
subtraction of (\ref{eq:divergence-Mij-decomposition}) from 
the momentum constraint (\ref{eq:Mom4}) yields
\begin{equation}\label{eq:Mom5}
\cderiv_j\left(\frac{1}{\cs}\big[\cLong(V-W)\big]^{ij}\right)
+\cderiv_j\tilde M^{ij}-\frac{2}{3}\CF^6\cderiv^i\trK
=8\pi G\cj^i,
\end{equation}
which is an equation for the {\em difference} $V^i-W^i$.  Thus one can
combine the construction of $\cA^{ij}_{TT}$ from $\tilde M^{ij}$ with the
solution of the momentum constraint.  Instead of solving
(\ref{eq:divergence-Mij-decomposition}) for $W^i$ and then
(\ref{eq:Mom4}) for $V^i$, one can directly solve (\ref{eq:Mom5}) for
$V^i-W^i$. 

In the presence of boundaries, solutions of elliptic equations like
(\ref{eq:Mom5}) or (\ref{eq:divergence-Mij-decomposition}) will depend on
{\em boundary conditions}.  In chapter \ref{chapter:Comparing} of this
thesis, for example, we encounter the situation that we know boundary
conditions for the combined solve for $V^i-W^i$, but we do not know
boundary conditions for the individual solves for $W^i$ and $V^i$ (see
section \ref{Comparing:sec:mConfTT}).

\subsection[Identification of $\s$ with the lapse $\N$]{
Identification of \boldmath$\s$ with the lapse $\N$}

The extrinsic curvature formulation of the initial value problem as
presented so far is perfectly adequate for the mathematical task of
rewriting the constraints as well-defined equations.  However, it is
very natural to further identify the weight-function $\s$ with the
lapse function $\N$,
\begin{equation}\label{eq:sigma=2N}
\s=2\N,\qquad\cs=2\cN.
\end{equation}
One reason for this identification is that $\s$ and $\N$ have the same
conformal scaling behavior, cf. Eqs.~(\ref{eq:N-scaling}) and
(\ref{eq:sigma-scaling}).  A second reason is that with this
identification, the conformal thin sandwich equations become
equivalent to the extrinsic curvature formulation.  To see this,
note that by virtue of (\ref{eq:sigma=2N}), Eqs.~(\ref{eq:Aij-from-Mij}) 
and (\ref{eq:Mom5}) become
\begin{gather}\label{eq:cA5}
\cA^{ij}=\tilde M^{ij}+\frac{1}{2\cN}\left[\cLong(V-W)\right]^{ij},
\intertext{and}
\label{eq:Mom6}
\cderiv_j\left(\frac{1}{2\cN}\big[\cLong(V-W)\big]^{ij}\right)
+\cderiv_j\tilde M^{ij}-\frac{2}{3}\CF^6\cderiv^i\trK
=8\pi G\cj^i.
\end{gather}
With the identifications
\begin{equation}
\tilde M^{ij}\;\leftrightarrow\;-\frac{1}{2\cN}\cu^{ij},\qquad\quad
V^i-W^i\;\leftrightarrow\; \beta^i,
\end{equation}
Eqs.~(\ref{eq:cA5}) and (\ref{eq:Mom6}) are {\em
identical} to Eqs.~(\ref{eq:cAij}) and (\ref{eq:Mom3}) of the conformal
thin sandwich formalism.  The Lagrangian picture agrees completely
with the Hamilton\-ian picture.  A third reason for (\ref{eq:sigma=2N}) is
given in the immediately following section.

\subsection[Stationary spacetimes have $\A^{ij}_{TT}=0$]
{Stationary spacetimes have \boldmath$\A^{ij}_{TT}=0$}

Consider a stationary solution of Einstein's equations with timelike
Killing vector $l$.  Given a spacelike hypersurface $\Sigma$, there is
a preferred gauge so that the time-vector of an evolution coincides
with $l$, namely $\N=-n\cdot l$, $\beta=\,\perp\! l$, where $n$ is the
unit normal to $\Sigma$, and $\perp$ is the projection operator into
$\Sigma$.  With this choice of lapse and shift, $\g_{ij}$ and
$\K^{ij}$ will be time-independent.  Using $\partial_t\g_{ij}=0$ in
Eq.~(\ref{eq:dtgij1}) yields
\begin{equation}
\K_{ij}
=\frac{1}{2\N}\left(\deriv_i\beta_j + \deriv_j\beta_i\right).
\end{equation}
The tracefree part of this equation is
\begin{equation}
  \A^{ij}=\frac{1}{2\N}(\Long\beta)^{ij},
\end{equation}
which is a weighted transverse traceless decomposition with
$\A^{ij}_{TT}\equiv 0$.  Thus, with the appropriate weight factor
$\s=2\N$, the extrinsic curvature has {\em no transverse traceless
piece} for any spacelike slice in any spacetime with timelike Killing
vector (A similar argument is applicable in the ergosphere of a Kerr
black hole; however, one must be more careful with the choice of
$\Sigma$ relative to $l$).

This is a great result!  One generally identifies the transverse
traceless piece of the extrinsic curvature with {\em radiative degrees
of freedom}.  For stationary spacetimes which do not radiate, 
$\A^{ij}_{TT}$ should therefore vanish.

A transverse-traceless decomposition of $\A^{ij}$ without the
weight-factor, however, will in general lead to a nonzero transverse
traceless piece.  Thus, such a decomposition is incompatible with the
identification of $\A^{ij}_{TT}$ with ``gravitational radiation.''

These considerations provide another argument for the introduction of
the weight-function in Eq.~(\ref{eq:barAij-TT}), and the identification of
$\s$ with the lapse-function in Eq.~(\ref{eq:sigma=2N}).

\subsection{Earlier decompositions of the extrinsic curvature}

The scaling $\s=\CF^6\cs$ in Eq.~(\ref{eq:sigma-scaling}) {\em
synchronizes} the conformal scalings of the transverse traceless and
longitudinal parts of $\A^{ij}$.  Consequently, a transverse-traceless
decomposition {\em without} the weight-function does {\em not} commute
with the conformal transformation giving rise to two inequivalent
extrinsic curvature decompositions~\cite{York:1973, York:1979,
Cook:2000}.  They differ in whether one performs first the transverse
traceless decomposition or the conformal scalings.  Both these
variants are inequivalent to the Lagrangian viewpoint, the conformal
thin sandwich formalism.

Nonetheless these earlier decompositions, especially the ``Conformal
transverse traceless decomposition'' \cite{York:1979}, have been used
very successfully in a wide variety of contexts.  Indeed, since part
of this thesis was done before the modern decompositions were
developed, we use them as well in chapters~\ref{chapter:Spin} to
\ref{chapter:Comparing}.

\section{Black hole initial data}

So far we were concerned with describing mathematically well-defined
procedures for constructing initial data $(\g_{ij}, \K^{ij})$.  Both
the conformal thin sandwich equations and the the extrinsic curvature
decomposition can generate every possible initial data set, and in
chapter~\ref{chapter:Code} of this thesis we will present a numerical
code to solve these equations.

Having these mathematical and numerical tools, we now face the {\em
physical} question {\em how} to choose the free data such that the
final physical initial data sets have certain properties.  Moreover,
the presence of black holes often leads to {\em excised} regions from
the computational domain, which require {\em boundary
conditions}.  For elliptic equations, boundary conditions influence
the solution everywhere, so one must choose them with great care.

Both questions --how to choose the free data and how to choose
boundary conditions at the excised regions\footnote{Boundary
conditions at {\em infinity} are easily derived from asymptotic
flatness.}-- are difficult, because we do not know exactly how these
choices influence the physical solution, neither do we know what the
physical solution ``should'' be for certain astrophysically relevant
situations, for example an inspiraling binary black hole.

The approach taken in this thesis is to compute initial data sets with
both formalisms and with several different choices of free data.  Then
we look for consistency among these initial data sets.  Particularly
helpful are {\em sequences} of initial data sets.  For example, we
compute initial data representing binary black holes in quasi-circular
orbits as a function of the separation between the black holes.
Analysis of such a sequence, and verification that certain properties
along the sequence seem physically ``reasonable,'' can increase of
decrease the confidence one has in the relevance of the particular
approach used to construct the sequences.

In the rest of this section, we briefly review the choices made in
each of the subsequent chapters to put them into context.  From now
on, we focus exclusively on vacuum spacetimes with vanishing matter
terms, $\rho=j^i=S_{ij}=0$, containing one or two black holes.

The project presented in chapter \ref{chapter:Spin} was begun before
the modern constraint decompositions were discovered.  Therefore it
uses one of the older methods, the ``Conformal transverse traceless
decomposition'' \cite{York:1979, Cook:2000}.  It is a special case of
the extrinsic curvature decomposition outlined in section
\ref{sec:IVP:ExCurv} which is obtained by setting $\cs\equiv 1$.  It
further assumes the simplest possible free data: conformal flatness,
$\cg_{ij}=\mbox{flat}$, maximal slicing, $\trK\!=\!0$, and 
$\cA^{ij}_{TT}\!=\!0$.  The momentum constraint (\ref{eq:Mom4})
decouples from the Hamiltonian constraint (\ref{eq:Ham4}), and
simplifies to
\begin{equation}\label{eq:Mom7}
\cderiv_j(\cLong V)^{ij}=0,
\end{equation}
where in Cartesian coordinates, the derivatives are simple partial
derivatives.  Bowen and York~\cite{Bowen:1979,Bowen-York:1980} found
{\em analytic} solutions, $V_{BY}^i$, to (\ref{eq:Mom7}) describing
one or multiple black holes carrying linear and angular momenta.

The initial value problem is thus reduced to solving the Hamiltonian
constraint (\ref{eq:Ham4}), which becomes
\begin{equation}\label{eq:Ham7}
\Delta\CF+\frac{1}{8}\CF^{-7}(\cLong V_{BY})_{ij}(\cLong V_{BY})^{ij}=0,
\end{equation}
a quasi-linear flat space Laplace equation.

The boundary conditions at the black holes are derived by requiring
{\em inversion symmetry} with respect to the throat of each black
hole.  This leads to a two-sheeted topology, where each black hole
connects ``our'' universe through an Einstein-Rosen bridge to the same
``mirror'' universe \cite{Misner:1963}.  To satisfy inversion
symmetry, $\cA^{ij}_{BY}=(\cLong V_{BY})^{ij}$ must be modified by an
ingenious imaging
process~\cite{Kulkarni-Shepley-York:1983}\footnote{The {\em puncture
method} \cite{Brandt-Bruegmann:1997} makes the same assumptions on the
free data, but uses a three-sheeted topology instead, with each black
hole connecting to ``its own'' universe
(cf. \cite{Brill-Lindquist:1963}).  This approach does not require
excised regions in the computational domain.  A very recent paper
\cite{Hannam-Evans-etal:grqc0306028} indicates that the puncture
method cannot be easily extended to the conformal thin sandwich
equations.}.  This method is described in detail by
Cook~\cite{Cook:1990, Cook:1991, Cook:1994}.  Cook also developed an
elliptic solver for this problem~\cite{Cook:1994}, which I use in
chapter~\ref{chapter:Spin}.

Chapter \ref{chapter:Testmass} is a variation on
chapter~\ref{chapter:Spin}, exploring quasi-circular orbits for
unequal mass black holes (as opposed to spinning black holes).  It
uses the inversion symmetric Bowen-York data to construct
sequences of quasi-circular orbits extending toward the test-mass
limit.  This limit is known analytically, so one can compare directly
the computations against the correct results.

The ``Bowen-York initial data'' is relatively easy to construct (one
flat space Laplace equation instead of four or five coupled equations;
moreover, the puncture method~\cite{Brandt-Bruegmann:1997} does not
even require internal boundaries).  However, it makes very special
assumptions about the free data, $\cg_{ij}=\mbox{flat}$, $\trK=0$,
$\cA^{ij}_{TT}=0$, $\cs=1$.  Only a small subset of all initial data
sets can be reached with these restrictive assumptions.  As it is not
clear that realistic binary black hole data belongs into this class,
one needs to go beyond this approach.

In chapter \ref{chapter:Comparing}, we compare different constraint
decompositions, namely the two previous variants of the extrinsic
curvature decomposition, as well as the conformal thin sandwich
equations with free data~(\ref{eq:CTS-freedata1}) (specification
of $\cN$, not $\dot\trK$).  We also explore different choices for some
of the free data by choosing $\cg_{ij}$ and $\trK$ (and $\cN$ for
conformal thin sandwich) based on superposed Kerr-Schild data.

Chapter~\ref{chapter:QE} employs the conformal thin sandwich formalism
with specification of $\dot\trK$, i.e. we solve the five coupled
partial differential equations (\ref{eq:Mom3}), (\ref{eq:Ham3}),
(\ref{eq:dtK4}).  We also explore different boundary conditions at the
horizons of the black holes.

Chapter \ref{chapter:GW}, finally, explores spacetimes without black
holes, or with just one.  We use the conformal thin sandwich equations
with $\dot\trK$-equation to construct initial data for a spacetime
with superposed gravitational radiation.  In this chapter, we solve
the conformal thin sandwich equations on very general, nonflat
conformal manifolds $\cg_{ij}$.






\renewcommand{\thefootnote}{\fnsymbol{footnote}}

\chapter[A multidomain spectral method for solving elliptic equations]
{A multidomain spectral method for solving elliptic equations\footnote[1]{H. P. Pfeiffer, L. E. Kidder, M. A. Scheel and S. A. Teukolsky, Comput. Phys. Commun. {\bf 152}, 253 (2003).}}
\label{chapter:Code}

\renewcommand{\thefootnote}{\arabic{footnote}}

\section{Introduction}

Elliptic partial differential equations (PDE) are a basic and
important aspect of almost all areas of natural science.  Numerical
solutions of PDEs in three or more dimensions pose a formidable
problem, requiring a significant amount of memory and CPU time.
Off-the-shelf solvers are available; however, it can be difficult to
adapt such a solver to a particular problem at hand, especially when
the computational domain of the PDE is nontrivial, or when one deals
with a set of coupled PDEs.

There are three basic approaches to solving PDEs: Finite differences,
finite elements and spectral methods. Finite differences are easiest
to code. However, they converge only algebraically and therefore need
a large number of grid points and have correspondingly large memory
requirements.  Finite elements and spectral methods both expand the
solution in basis functions. Finite elements use many subdomains and
expand to low order in each subdomain, whereas spectral methods use
comparatively few subdomains with high expansion orders.  Finite
elements are particularly well suited to irregular geometries
appearing in many engineering applications.  For sufficiently regular
domains, however, spectral methods are generally faster and/or more
accurate.

Multidomain spectral methods date back at least to the work of
Orszag\cite{Orszag:1980}.  In a multidomain spectral method, one has
to match the solution across different subdomains.  Often this is
accomplished by a combination of solves on individual subdomains
together with a global scheme to find the function values at the
internal subdomain boundaries.  Examples of such global schemes are
relaxational iteration\cite{Funaro-Quarteroni-Zanolli:1988}, an
influence matrix\cite{Macaraeg-Streett:1986,Boyd:2001}, or the
spectral projection decomposition
method\cite{Gervasio-Ovtchinnikov-Quarteroni:1997}.  For simple PDEs
like the Helmholtz equation, fast solvers for the subdomain solves are
available. For more complicated PDEs, or for coupled PDEs, the
subdomain solves will typically use an iterative solver.  One drawback
of these schemes is that information from the iterative subdomain
solves is not used in the global matching procedure until the
subdomain solves have completely converged.
The question arises whether efficiency can be improved by avoiding
this delay in communication with the matching procedure.
\\

In this paper we present a spectral method for coupled
nonlinear PDEs based on pseudospectral collocation with domain
decomposition.  This method does not split subdomain solves and
matching into two distinct elements.  Instead it combines satisfying
the PDE on each subdomain, matching between subdomains, and satisfying
the boundary conditions into one set of equations.  This system of
equations is then solved with an iterative solver, typically
GMRES\cite{Templates}.  At each iteration, this solver thus has
up-to-date information about residuals on the individual subdomains
and about matching and thus can make optimal use of all information.

The individual subdomains implemented are rectangular blocks {\em and}
spherical shells.  Whereas either rectangular blocks (see e.g.
\cite{Demaret-Deville:1991,Ku:1995,Pinelli-Vacca-Quarteroni:1997})
or spherical shells\cite{Grandclement-Bonazzola-etal:2001} have been
employed before, we are not aware of work using both.  The code
supports an arbitrary number of blocks and shells that can touch each
other and/or overlap.

Moreover, the operator $\cal S$ at the core of the method (see section
\ref{sec:OperatorS}) turns out to be modular, i.e. the code fragments
used to evaluate the PDE, the boundary conditions, and the matching
conditions are independent of each other. Thus the structure of the
resulting code allows for great flexibility, which is further enhanced
by a novel point of view of the mappings that are used to map
collocation coordinates to the physical coordinates. This flexibility
is highlighted in the following aspects:
\begin{itemize}
\item The user code for the particular PDE at hand is completely
  independent from the code dealing with the spectral method and
  domain decomposition. For a new PDE, the user has to supply only the
  code that computes the residual and its linearization.
\item {\em Mappings} are employed to control how collocation points
  and thus resolution are distributed within each subdomain.  New
  mappings can be easily added which are then available for {\em all}
  PDEs that have been coded.
\item The solver uses standard software packages for the
  Newton-Raphson step, the iterative linear solvers, and the
  preconditioning.  Thus one can experiment with many different linear
  solvers and different preconditioners to find an efficient
  combination.  The code will also automatically benefit from
  improvements to these software packages.
\item The code is dimension independent (up to three dimensions).
\item Many properties of a particular solve can be chosen at runtime,
  for example the domain decomposition, the mappings used in each
  subdomain, as well as the choice of the iterative solver.  The user
  can also choose among differential operators and boundary conditions
  previously coded at runtime.
\end{itemize}

In the next section we recall some basics of the pseudo-spectral
collocation method. In section \ref{sec:Implementation_Code} we describe
our approach of combining matching with solving of
the PDE. For ease of exposition, we interweave the new method with
more practical issues like mappings and code modularity. The central
piece of our method, the operator $\cal S$, is introduced in section
\ref{sec:OperatorS} for a one-dimensional problem and then extended to
higher dimensions and spherical shells.  In section \ref{sec:Examples}
we solve three example problems. Many practical issues like
preconditioning and parallelization are discussed in this section, and
we also include a detailed comparison to a finite difference code.

\section{Spectral Methods}
\label{sec:SpectralMethods}

Spectral methods have been widely used for many years, and several
good books are available
(e.g.\cite{Gottlieb-Orszag:1977,Canuto-Hussaini,Boyd:2001}).
Therefore we present here only the most important ideas relevant to
our problem.  We deal with a second order nonlinear elliptic partial
differential equation or system of equations,
\begin{equation}\label{eq:PDE}
  ({\cal N}u)(x)=0,\qquad x\in {\cal D},
\end{equation}
in some domain  ${\cal D}\subset\mathbbmss{R}^d$
with boundary conditions
\begin{equation}\label{eq:bc}
  g(u)(x)=0,\qquad x\in \partial{\cal D}.
\end{equation}
The function $u$ can be a single variable giving rise to a single PDE,
or it can be vector-valued giving rise to a coupled set of PDEs.
Throughout this paper we assume that the problem has a unique
solution.  We also absorb a possible right-hand side into the elliptic
operator $\cal N$.

The fundamental idea of spectral methods is to approximate functions,
like the solution to the PDE \eqnref{eq:PDE}, as a truncated series in
some basis functions $\Phi_k(x)$:
\begin{equation}
  \label{eq:Expansion}
  u(x) \approx u^{(N)}(x)\equiv \sum_{k=0}^{N}\tilde u_{k}\Phi_k(x).
\end{equation}
The coefficients $\tilde u_k$ are called the {\em spectral
coefficients}. The choice of basis functions depends on the boundary
conditions; for periodic problems, a Fourier series or an expansion in
spherical harmonics are suitable, for nonperiodic problems,
eigenfunctions of singular Sturm-Liouville operators are used.  With
the appropriate $\Phi_k$, the series \eqnref{eq:Expansion} converges
exponentially in $N$, provided $u(x)$ is smooth.

Derivatives of $u^{(N)}$ are computed via
\begin{equation}\label{eq:deriv}
  \frac{du^{(N)}(x)}{dx}=\sum_{k=0}^N\tilde u_k\frac{d\Phi_k(x)}{dx},
\end{equation}
and similarly for higher derivatives.  As $d\Phi_k/dx$ are known
analytically, Eq.~\eqnref{eq:deriv} can be used to evaluate
derivatives exactly (i.e. to numerical roundoff) at any point
$x$\footnote{In practice, derivatives, Eq.~\eqnref{eq:deriv}, are
computed via a truncated series in $\Phi_k$.  In this case,
derivatives will be exact only if $d\Phi_k/dx$ can be represented by a
truncated series~\eqnref{eq:Expansion}.}.  No special treatment of
boundary points is necessary.  Thus $(\cal N u^{(N)})(x)$ can be
evaluated exactly at any $x$, too.  The resulting function $(\cal
Nu^{(N)})(x)$, however, can in general {\em not} be represented
exactly as a truncated series \eqnref{eq:Expansion}.  Nonlinearities
in $\cal N$ will introduce higher frequency modes, and expressing
$({\cal N} u^{(N)})(x)$ by the series \eqnref{eq:Expansion} will fold
these modes back into the retained $N+1$ modes.  This is called {\em
aliasing}; the errors are exponentially small, however, with
sufficient resolution.

In order to compute the spectral coefficients $\tilde u_k$ we use 
{\em pseudo-spectral collocation} where one requires
\begin{equation}\label{eq:PSC}
  ({\cal N}u^{(N)})(x_i)=0, \qquad i=0, \ldots, N.
\end{equation}
The points $x_i$ are called {\em collocation points}, and are chosen
as the abscissas of the Gaussian quadrature associated with the basis
set $\Phi_k$.  Eq.~\eqnref{eq:PSC} is essentially equivalent to
approximating the residual $({\cal N}u^{(N)})(x)$ as a truncated series
\eqnref{eq:Expansion}, and requiring that this approximation vanishes.

Although we concentrate on elliptic problems, we note that spectral
methods are also widely used for hyperbolic and parabolic PDEs.  The
spectral coefficients become functions of time, $\tilde u_k(t)$, and
one rewrites the hyperbolic or parabolic PDE as a system of ordinary
differential equations for $\tilde u_k(t)$, which can be solved by
various time stepping methods (e.g. Runge-Kutta).  For details, see
e.g. \cite{Gottlieb-Orszag:1977,Canuto-Hussaini,Boyd:2001}.

\subsection{Chebyshev polynomials}

Chebyshev polynomials are widely used as basis functions for spectral
methods. They satisfy the convenient analytical properties of
``classical'' orthogonal polynomials. Their defining differential
equation is a {\em singular} Sturm-Liouville problem, and so Chebyshev
expansions converge exponentially for smooth functions $u$ {\em
  independent} of the boundary conditions satisfied by $u$
\cite{Gottlieb-Orszag:1977,Boyd:2001}.

Chebyshev polynomials are defined by
\begin{equation}
  \label{eq:Chebyshev}
  T_k(X)=\cos(k \arccos X), \qquad X\in [-1,1].
\end{equation}
They are defined on the interval $X\in [-1,1]$ only; usually one needs
to map the {\em collocation coordinate} $X\in[-1,1]$ to the physical
coordinate of the problem, $x\in [a,b]$. We use the convention that
the variable $X$ varies over the interval $[-1,1]$, whereas $x$ is
defined over arbitrary intervals.  We will describe our approach to
mappings below in the implementation section.

For an expansion up to order $N$ (i.e. having a total of
$N+1$ basis functions) the associated collocation points are
\begin{equation}\label{eq:ChebyshevCollocationPoints}
  X_i=\cos\left(\frac{i\pi}{N}\right),\quad i=0, \ldots, N.
\end{equation}
Define the real space values 
\begin{equation}\label{eq:Chebyshev-Expansion}
  u_i\equiv u^{(N)}(X_i)=\sum_{k=0}^N\tilde u_k T_k(X_i).
\end{equation}
Using the discrete orthogonality relation
\begin{equation}
  \label{eq:Chebyshev-Orthogonality}
  \delta_{jk}=\frac{2}{N\bar c_k}\sum_{i=0}^N
  \frac{1}{\bar c_i}T_j(X_i)T_k(X_i)
\end{equation}
with 
\begin{equation}
  \bar c_i=\left\{\begin{aligned}&2\qquad k=0\mbox{ or }k=N\\
      &1\qquad k=1,\ldots, N-1,\end{aligned}\right.
\end{equation}
we can invert \eqnref{eq:Chebyshev-Expansion} and find
\begin{equation}\label{eq:Chebyshev-inverseExpansion}
  \tilde u_j=\frac{2}{N\bar c_j}\sum_{i=0}^{N}\frac{u_i}{\bar c_i}T_j(X_i).
\end{equation}

Both matrix multiplications \eqnref{eq:Chebyshev-Expansion} and
\eqnref{eq:Chebyshev-inverseExpansion} can be performed with a fast
cosine transform in ${\cal O}(N\ln N)$ operations, another reason for
the popularity of Chebyshev basis functions. 

There are the same number of real space values $u_i$ and spectral
coefficients $\tilde u_k$, and there is a one-to-one mapping between
$\{u_i\}$ and $\{\tilde u_k\}$. Hence one can represent the function
$u^{(N)}$ by either $\{u_i\}$ or $\{\tilde u_k\}$.

The spectral coefficients of the derivative, 
\begin{equation}\label{eq:ExpansionDerivative}
  \frac{du^{(N)}}{dX}(X)=\sum_{k=0}^{N}\tilde u_k'T_k(X),
\end{equation}
are given by the recurrence relation
\begin{equation}\label{eq:DerivativeRecurrence}
\begin{aligned}
  \tilde u_i'&=\tilde u_{i+2}'+2(i+1)\tilde u_{i+1},\qquad i=1,\ldots,N-1,\\
  \tilde u_0'&=\frac{1}{2}\tilde u_2'+\tilde u_1,
\end{aligned}
\end{equation}
with $\tilde u_{N+1}=\tilde u_N=0$.  The coefficients of the second
derivative,
\begin{equation}
  \frac{d^2u^{(N)}}{dX^2}(X)=\sum_{k=0}^{N-1}\tilde u_k''T_k(X),
\end{equation}
are obtained by a similar recurrence relation, or by applying
\eqnref{eq:DerivativeRecurrence} twice.

\subsection{Basis functions in higher dimensions}
\label{sec:SpecMethods-Basisfunctions}

In higher dimensions one can choose tensor grids of lower dimensional
basis functions.  For example, a $d$-dimensional cube $[-1, 1]^d$ can
be described by Chebyshev polynomials along each coordinate axis.  For
a three-dimensional sphere or spherical shell, tensor products of
spherical harmonics for the angles and a Chebyshev series for the
radial coordinate\cite{Grandclement-Bonazzola-etal:2001} are used.
It is also possible to expand the angular piece in a double
Fourier-series\cite{Orszag:1974}.

\subsection{Domain Decomposition}



If the computational domain $\cal D$ has a different topology than the
basis functions, then an expansion in the basis functions cannot cover
$\cal D$ completely. Moreover, the particular problem at hand might
require different resolution in different regions of the computational
domain which will render a single overall expansion inefficient.

One circumvents these problems with domain decomposition. The computational
domain $\cal D$ is covered by $N_{\cal D}$ subdomains 
\begin{equation}
  {\cal D}=\bigcup_{\mu=1}^{N_{\cal D}}{\cal D}_\mu,
\end{equation}
each having its own set of basis functions and expansion coefficients:
\begin{equation}
  u^{(\mu)}(x)=\sum_{k=0}^{N_\mu}\tilde u^{(\mu)}_k\Phi^{(\mu)}_k(x),
  \qquad x\in {\cal D}_\mu,\quad \mu=1, \ldots N_{\cal D}.
\end{equation}
Here $u^{(\mu)}$ denotes the approximation in the $\mu$-th domain, and
we have dropped the additional label $N$ denoting the expansion order
of $u^{(\mu)}$.  The individual subdomains ${\cal D}_\mu$ can touch
each other or overlap each other. To ensure that the functions
$u^{(\mu)}$ ---each defined on a single subdomain ${\cal D}_\mu$
only--- actually fit together and form a smooth solution of the PDE
\eqnref{eq:PDE} on the full domain ${\cal D}$, they have to satisfy
matching conditions.  In the limit of {\em infinite} resolution, we must
have that
\begin{itemize}
\item for touching subdomains ${\cal D}_\mu$ and ${\cal D}_\nu$ the
  function and its normal derivative must be smooth on the surface
  where the subdomains touch:
\begin{equation}
  \label{eq:Matching-touch}
\begin{aligned}
  u^{(\mu)}(x)&=u^{(\nu)}(x)\\
  \dnachd{u^{(\mu)}}{n}(x)&=-\dnachd{u^{(\nu)}}{n}(x)
  \end{aligned}
  \qquad
   x\in \partial{\cal D}_\mu\cap\partial{\cal D}_\nu
\end{equation}
(The minus sign in the second equation of \eqnref{eq:Matching-touch} 
occurs because we use the outward-pointing normal in each subdomain.)

\item for overlapping subdomains ${\cal D}_\mu$ and ${\cal D}_\nu$ the
  functions $u^{(\mu)}$ and $u^{(\nu)}$ must be identical in ${\cal
    D}_\mu\cap{\cal D}_\nu$. By uniqueness of the solution of the PDE,
  it suffices to require that the functions are identical on the
  boundary of the overlapping domain:
  \begin{equation}\label{eq:Matching-overlap}
  u^{(\mu)}(x)=u^{(\nu)}(x),
  \qquad x\in \partial\left({\cal D}_\mu\cap{\cal D}_\nu\right).
  \end{equation}
\end{itemize}
For {\em finite} resolution, Eqs.~\eqnref{eq:Matching-touch} and
\eqnref{eq:Matching-overlap} will in general not hold for all $x$, but
only at a discrete set of points.  Between these points, these
equations will be violated by an exponentially small amount.
We will see in the next section how these conditions are actually
implemented in the code.

\section{Implementation} 
\label{sec:Implementation_Code}

In this section we describe our specific approaches to several aspects
of multi-dimensional pseudo-spectral collocation with domain
decomposition.

\subsection{One-dimensional Mappings} 

Chebyshev polynomials are defined for $X\in [-1, 1]$.
Differential equations in general will be defined on a different
interval $x\in [a,b]$. In order to use Chebyshev polynomials, one
introduces a mapping
\begin{equation}
X: [a, b]\to [-1, 1],\quad x\to X=X(x)
\end{equation}
that maps the {\em physical coordinate} $x$ onto the {\em collocation
coordinate} $X$.

One could explicitly substitute this mapping into the PDE under
consideration. Derivatives would be multiplied by a Jacobian, and we would
obtain the PDE on the interval $[-1, 1]$.  For example, the
differential equation in the variable $x$
\begin{equation}\label{eq:Example}
  \dnachd{^2u}{x^2}+u=0,\qquad x\in [a,b],
\end{equation}
becomes the following differential equation in the variable $X$:
\begin{equation}\label{eq:ExampleMapped}
  X'^2\dnachd{^2u}{X^2}+X''\dnachd{u}{X}+u=0,\qquad X\in [-1,1],
\end{equation}
where $X'=\partial X/\partial x$ and $X''=\partial^2X/\partial x^2$.
Now one could expand $u(X)$ in Chebyshev polynomials, compute
derivatives $\partial/\partial X$ via the recurrence relation
\eqnref{eq:DerivativeRecurrence} and code
Eq.~\eqnref{eq:ExampleMapped} in terms of $\partial u/\partial X$.
This approach is common in the literature
\cite{Boyd:2001,Kidder-Finn:2000}. However, it has several
disadvantages: As one can already see from this simple example, the
equations become longer and one has to code and debug more terms.
Second, and more important, it is inflexible, since for each different
map one has to derive and code a mapped equation
\eqnref{eq:ExampleMapped}.  {\em A priori} one might not know the
appropriate map for a differential equation, and in order to try
several maps, one has to code the mapped equation several times.
Also, for domain decomposition, a different map is needed for each
subdomain.

We propose a different approach.  We still expand in terms of
Chebyshev polynomials on $X\in [-1, 1]$ and obtain the physical
solution via a mapping $X(x)$,
\begin{equation}
  u(x)=\sum_{k=0}^N\tilde u_kT_k(X(x)),
\end{equation}
and we still compute $\partial u(X)/\partial X$ and
$\partial^2u(X)/\partial X^2$ via the recurrence relation
\eqnref{eq:DerivativeRecurrence}. However, now we do {\em not}
substitute $\partial u(X)/\partial X$ and
$\partial^2u(X)/\partial X^2$ into the mapped differential
equation, Eq.~\eqnref{eq:ExampleMapped}.  Instead we compute first
numerically
\begin{align}\label{eq:FirstDeriv-Mapped}
  \dnachd{u(x)}{x}&=X'\dnachd{u(X)}{X}\\
\label{eq:SecondDeriv-Mapped}
  \dnachd{^2u(x)}{x^2}&
  =X'^2\dnachd{^2u(X)}{X^2}+X''\dnachd{u(X)}{X}
\end{align}
and substitute these values into the original physical differential
equation \eqnref{eq:Example}.  The collocation points are thus mapped
to the physical coordinates
\begin{equation}
  \label{eq:PhysicalGridPoints}
  x_i=X^{-1}(X_i).
\end{equation}

This approach separates the code into three distinct parts: 
\begin{enumerate}
\item Code dealing with the basis functions: transforms between
  collocation space $X$ and spectral space, evaluation of derivatives
  $\partial/\partial X$ via recurrence relations. This code depends
  only on the collocation coordinates $X\in [-1,1]$ (and on the
  angular coordinates $\theta, \phi$ for spherical shells).
\item Mappings that map between collocation coordinate $X$ and
  physical coordinates $x$.
\item The ``user code'' implementing the physical PDE [in our example
  Eq.~\eqnref{eq:Example}] that deals only with the physical coordinate
  $x$.
\end{enumerate}

These three elements are independent of each other:
\begin{itemize}
\item A user who wants to code another differential equation has
  only to write the code that evaluates the differential operator
  ${\cal N}$ in the physical space with physical derivatives. Then
  immediately all previously coded mappings are available for this
  new differential equation, as well as all basis functions.

\item In order to introduce a new mapping, one has to code only four
  functions, namely $X(x)$, its inverse $x(X)$, as well as the
  derivatives $X'(x)$ and $X''(x)$.  This new map can then be used 
  for any differential equation already coded or to be coded later.
  
\item In order to switch to a different set of basis functions, one
  has only to code the transforms and the recurrence relations for
  the derivatives. 
\end{itemize}

In practice we use three different mappings
\begin{equation}\label{eq:Mappings}
\begin{aligned}
  &\mbox{linear:}& X(x) &= Ax+B\\
  &\mbox{log:} &   X(x) &= A\log(Bx+C)\\
  &\mbox{inverse:}& X(x)&= \frac{A}{x}+B
\end{aligned}
\end{equation}
In each case the constants $A, B$ are chosen such that $[a, b]$ is
mapped to $[-1, 1]$.  The log mapping has one additional parameter
which is used to fine-tune the relative density of collocation points
at both ends of the interval $[a, b]$.  We show the effects of
different mappings in our first example in section \ref{sec:Example1}.

\subsection{Basis functions and Mappings in higher Dimensions} 

\subsubsection{Rectangular Blocks}

In order to expand in a $d$-dimensional rectangular block, 
\begin{equation}
{\cal D}=[a_1,
b_1]\times[a_2,b_2]\times\ldots\times[a_d, b_d],  
\end{equation}
we use a tensor product of Chebyshev polynomials with a 1-$d$ mapping
along each coordinate axis:
\begin{equation}\label{eq:Block-nD}
  u(x_1, \ldots, x_d)
  =\sum_{k_1=0}^{N_1}\sum_{k_2=0}^{N_2}\cdots\sum_{k_d=0}^{N_d}
    \tilde u_{k_1\cdots k_d}
T_{k_1}\!\!\left(X^{(1)}(x_1)\right)\cdots
T_{k_d}\!\left(X^{(d)}(x_d)\right).
\end{equation}
We use $d$ mappings
\begin{equation}
  X^{(l)}: [a_l,b_l]\to [-1, 1],\quad l=1, \ldots d,
\end{equation}
and the collocation points in physical space are the mapped
collocation points along each dimension,
\begin{equation}
  x_{i_1\cdots i_d}=\Big(x_{i_1}^{(1)}, \ldots, x_{i_d}^{(d)}\Big),
\end{equation}
where the coordinate along the $l$-th dimension $x^{(l)}_{i_l}$ is
given by Eq.~\eqnref{eq:PhysicalGridPoints} using $X^{(l)}$.

Note that such a $d$-dimensional rectangle has as many spectral
coefficients $\tilde u_{k_1\cdots k_d}$ as grid point values
$u_{i_1\ldots i_d}=u(x_{i_1}, \ldots, x_{i_d})$.  Therefore we can
equivalently solve for the spectral coefficients or the real space
values.  We will solve for the real space values $u_{i_1\ldots i_d}$.

\subsubsection{Spherical Shell}

In a spherical shell with inner and outer radii $0<R_1<R_2$ we use a
mapping for the radial coordinate.  A function $u(r,\theta,\phi)$ is
thus expanded as
\begin{equation}\label{eq:ExpansionSphere}
  u(r, \theta, \phi)
  =\sum_{k=0}^{N_r}\sum_{l=0}^L\sum_{m=-l}^l
    \tilde u_{klm}T_k(X(r))Y_{lm}(\theta,\phi),
\end{equation}
where real-valued spherical harmonics are used:
\begin{equation}\label{eq:Ylm}
  Y_{lm}(\theta, \phi)\equiv\left\{
\begin{aligned}
&P_l^{m}(\cos\theta)\cos(m\phi), &&m\ge 0\\
&P_l^{|m|}(\cos\theta)\sin(|m|\phi),&&m<0
\end{aligned}\right.
\end{equation}
$P_l^m(\cos\theta)$ are the associated Legendre polynomials.
Associating the $\sin$-terms with negative $m$ is not standard, but
eliminates the need to refer to two sets of spectral coefficients, one
each for the $\cos$-terms and the $\sin$-terms.  The radial mapping
$X:[R_1,R_2]\to[-1,1]$ can be any of the choices in
Eq.~\eqnref{eq:Mappings}.  The radial collocation points $r_i, i=0,
\dots, N_r$ are given by Eq.~\eqnref{eq:PhysicalGridPoints}.

For the angle $\phi$, Eq.~\eqnref{eq:Ylm} leads to a Fourier series
with equally spaced azimuthal collocation points
\begin{equation}
  \phi_i=\frac{2\pi i}{N_\phi},\qquad i=0, 1, \ldots, N_\phi-1.
\end{equation}
There is a total of $N_\theta=L+1$ angular collocation points
$\theta_i$, which are the abscissas of Gauss-Legendre integration.
All $\theta_i$ are in the interior of the interval $[0, \pi]$, hence
no collocation point will be placed on the $z$-axis of the polar
coordinate system.

In order to resolve the full Fourier series in $\phi$ up to $m=L$, one
needs $N_\phi\ge 2L+1$, since for $N_\phi=2L$, the term $\sin(L\phi)$
vanishes at all collocation points $\phi_i$.  We use $N_\phi=2(L+1)$
since FFTs are more efficient with an even number of points.

The expansion \eqnref{eq:ExpansionSphere} has a total of
$(N_r+1)(L+1)^2$ spectral coefficients but a total of $(N_r+1)
N_\theta N_\phi=2(N_r+1)(L+1)^2$ collocation points.  This means a
spherical shell has {\em more} collocation points than spectral
coefficients and the expansion~\eqnref{eq:ExpansionSphere}
approximates the grid point values in a least-square sense
only\cite{Swarztrauber:1979}.  Performing a spectral transform and its
inverse will thus project the grid point values into a subspace with
dimension equal to the number of spectral coefficients.
The implications of this fact for our code are discussed below in
section~\ref{sec:S-in-Spheres}.


\subsubsection{Representation of vectors, tensors and derivatives}

In both kinds of subdomains, rectangular blocks and spherical shells,
we expand the {\em Cartesian components} of vectors and tensors, and
we compute {\em Cartesian} derivatives, $\partial/\partial x,
\partial/\partial y, \partial/\partial z$.  These quantities are
smooth everywhere, and thus can be expanded in scalar spherical
harmonics.  By contrast, the spherical components of a vector field in
a spherical shell are discontinuous at the poles and cannnot be
expanded in scalar spherical harmonics. One would have to use
e.g. vector spherical
harmonics\cite{Swarztrauber:1979,Swarztrauber:1981}.  

For a spherical shell we provide an additional wrapper around the
basis functions and the radial mapping that transforms polar
derivatives $ \partial/\partial r, \partial/\partial\theta,
\partial/\partial\phi$ to Cartesian derivatives\footnote{The polar
singularity at $\theta=0,\pi$ is not a problem, as no collocation points
are located there.}.  This involves multiplications by sines and
cosines of the angles $\theta,\phi$ which can be performed grid point
by grid point in real space.  Alternatively, it can be done spectrally
by expressing, e.g.  $\sin\theta$ in spherical harmonics, and then
using addition theorems to reduce products of spherical harmonics to
simple sums.  Carrying out the transformation spectrally is slightly
better in practice.

Representing vectors and derivatives in Cartesian coordinates in both
kinds of subdomains increases flexibility, too. We can use the {\em
same} code to evaluate the residual in {\em both} kinds of subdomains.

We remark that for given order $l$ exact representation of derivatives
of $Y_{lm}$ in Cartesian coordinates requires spherical harmonics up
to order $l+1$.  For example, the function $f(\ve x)=|\ve x|$ has only
a $Y_{00}$ component.  However, the $z$-component of its gradient is
$\cos\theta$ which is proportional to $Y_{10}$.  Thus the derivatives
of the highest retained $l$-modes cannot be represented accurately.
This is not important, however, as the amplitude of these modes
decreases exponentially with resolution.

\subsection[The operator ${\cal S}$]{The operator \protect\boldmath${\cal S}$}
\label{sec:OperatorS}

We now introduce the operator $\cal S$, the centerpiece of our method.
It combines the solution of the PDE, the boundary conditions and
matching between different subdomains.

\begin{figure}
\begin{centering}
\includegraphics[scale=1.1]{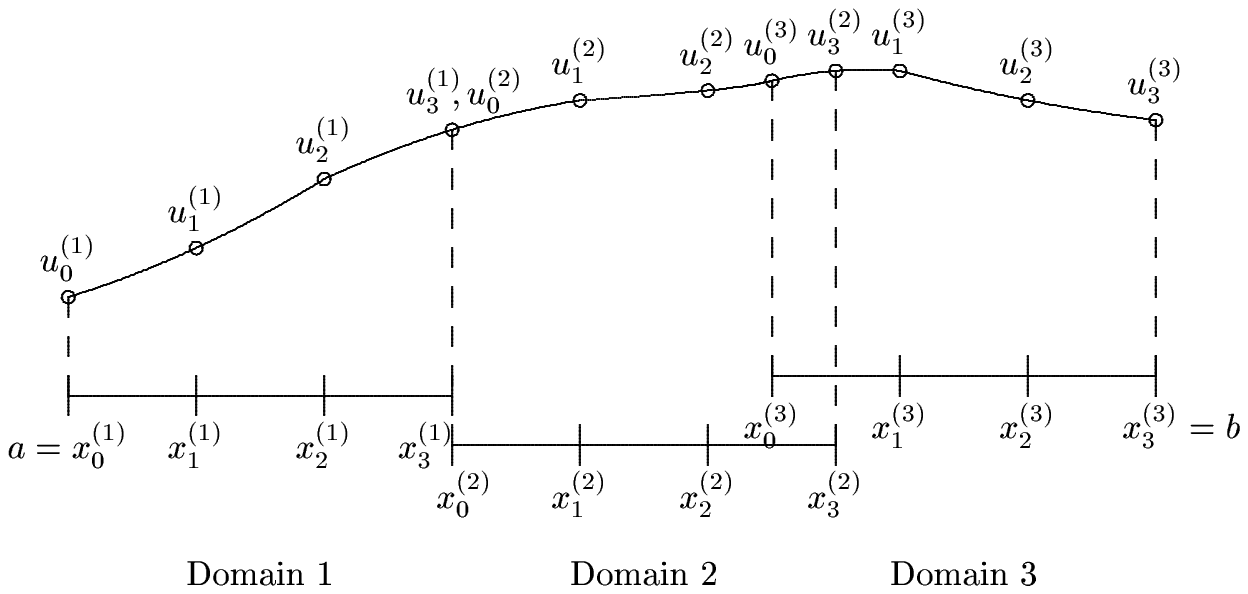}
\CAP{Illustration of matching with three subdomains in one
  dimension}{\label{fig:Illustration}Illustration of matching with
  three subdomains in one dimension.  Subdomains 1 and 2 touch each
  other, and subdomains 2 and 3 overlap.  $x_i^{(\mu)}$ denotes the
  coordinate of the $i$-th collocation point of the $\mu$-th
  subdomain, and $u_i^{(\mu)}$ denotes the function values at the grid
  points.}
\end{centering}
\end{figure}
  
We introduce $\cal S$ first with a simple case, a one-dimensional
differential equation with a Dirichlet boundary condition at one end
and a von Neumann boundary condition at the other end:

\begin{align}
  ({\cal N}u)(x)&=0, \qquad a<x<b,\\
  u(a)&=A,\\
  \dnachd{u}{x}(b)&=B.
\end{align}
To explain our procedure for matching, we assume three domains as
depicted in Figure \ref{fig:Illustration}.  $N_\mu$ denotes the
highest retained expansion order in domain $\mu$; here
$N_\mu=3$ for all domains.  Domains 2 and 3 overlap.  Domains 1 and 2
touch so that the collocation points $x^{(1)}_3$ and $x^{(2)}_0$
represent the same physical point. The function value, however, is
represented twice, once assigned to domain 1, $u^{(1)}_3$, and once
belonging to domain 2: $u^{(2)}_0$. Using just the grid point values
within one subdomain, we can expand the function in that subdomain and
can evaluate derivatives. We can also interpolate the function to
arbitrary positions $x$. Thus, given values
$\{u^{(\mu)}_i, i=0, \ldots, N_\mu\}$, we can compute $u^{(\mu)}(x)$
for $x\in [x^{(\mu)}_0, x^{(\mu)}_{N_\mu}]$.

In order to determine the unknowns $u^{(\mu)}_i$, we need one equation
per unknown. We will first write down these equations and then explain
where they come from.
\begin{subequations}\label{eq:S3}
\begin{align}
  \label{eq:S3_A}
  \big({\cal N}u^{(\mu)}\big)\big(x^{(\mu)}_i\big)&=0,
  \quad  i=1,\ldots, N_\mu-1, \quad\mu=1,\ldots ,N_{\cal D}\\
  \label{eq:S3_B}
  u^{(1)}_0-A&=0\\
  \label{eq:S3_C}
  \dnachd{u^{(3)}}{x}\big(x^{(3)}_3\big)-B&=0\\
  \label{eq:S3_D}
  u^{(1)}_3-u^{(2)}_0&=0\\
  \label{eq:S3_E}
  \dnachd{u^{(1)}}{n}\big(x^{(1)}_3\big)
  +\dnachd{u^{(2)}}{n}\big(x^{(1)}_3\big)&=0\\
  \label{eq:S3_F}
  u^{(2)}_3 - u^{(3)}\big(x^{(2)}_3\big)&=0\\
  \label{eq:S3_G}
  u^{(3)}_0 - u^{(2)}\big(x^{(3)}_0\big)&=0
\end{align}
\end{subequations}
  
Eq.~\eqnref{eq:S3_A} represents the actual pseudo-spectral
equation~\eqnref{eq:PSC}.  It is enforced only for collocation points
that are {\em not} on the boundary of a subdomain.
Eqs.~\eqnref{eq:S3_B} and \eqnref{eq:S3_C} encode the boundary
conditions.  Eqs.~\eqnref{eq:S3_D} and \eqnref{eq:S3_E} describe the
value and derivative matching at touching subdomain boundaries. These
equations follow from Eq.~\eqnref{eq:Matching-touch}.
Eqs.~\eqnref{eq:S3_F} and \eqnref{eq:S3_G} perform matching between
overlapping subdomains as given by Eq.~\eqnref{eq:Matching-overlap}.

We will view the left-hand side of Eqs.~\eqnref{eq:S3} as a non-linear
operator $\cal S$. This operator acts on the set of grid point values
for {\em all} subdomains $\big\{u^{(\mu)}_i\big\}$ ($\mu=1,2,3,
i=0,\ldots, N_\mu$ in the example) and returns a residual that
incorporates the actual pseudo-spectral condition Eq.~\eqnref{eq:PSC},
the boundary conditions, and the matching conditions between different
subdomains.  If we denote the vector of {\em all} grid point values by
$\underline{\bf u}$, then the discretized version of the partial
differential equation becomes
\begin{equation}\label{eq:Su=0}
  {\cal S}\underline{\bf u}=0.
\end{equation}
The solution of Eq.~\eqnref{eq:Su=0} clearly is the solution of the
partial differential equation we want to obtain. By virtue of
Eq.~\eqnref{eq:Su=0} we thus have condensed the PDE, the boundary
conditions and matching into one set of nonlinear equations. 

We comment on some implementation issues:
\begin{itemize}
\item The action of the operator $\cal S$ can be computed very easily:
  Given grid point values $\underline{\bf u}$, every subdomain is
  transformed to spectral space and derivatives are computed.  Using
  the derivatives we can compute Eqs.~\eqnref{eq:S3_A},
  \eqnref{eq:S3_E} and any boundary conditions that involve
  derivatives like Eq.~\eqnref{eq:S3_C}. The interpolations necessary
  in Eqs.~\eqnref{eq:S3_F} and \eqnref{eq:S3_G} are done by summing
  up the spectral series.
  
\item $\cal S\underline{\bf u}$ can be computed in parallel:
  Everything except the matching conditions depends only on the set of
  grid point values within {\em one} subdomain. Therefore the natural
  parallelization is to distribute subdomains to different processors.
  
\item The code fragments implementing the nonlinear operator $\cal N$,
  the boundary conditions and the matching conditions are independent
  of each other. In order to change boundary conditions, one has
  only to modify the code implementing Eqs.~\eqnref{eq:S3_B} and
  \eqnref{eq:S3_C}.  In particular, the code for the matching-equations
  \eqnref{eq:S3_D}-\eqnref{eq:S3_G} can be used for {\em any}
  differential operator $\cal N$ and for {\em any} boundary condition.
\end{itemize}

We have now introduced the operator $\cal S$ in one dimension. Next we
address how to solve Eq.~\eqnref{eq:Su=0}, and then we generalize
$\cal S$ to higher dimensions. We present our method in this order
because the generalization to higher dimensions depends on some
details of the solution process.

\subsection[Solving ${\cal S}u=0$]{Solving \protect\boldmath${\cal S}u=0$}

In this section we describe how we solve the system of nonlinear
equations~\eqnref{eq:Su=0}.  Our procedure is completely standard and
requires three ingredients: A Newton-Raphson iteration to reduce the
nonlinear equations to a linear solve at each iteration, an iterative
linear solver, and the preconditioner for the linear solver.  For
these three steps we employ the software package
PETSc\cite{petsc-home-page}. We now comment on each of these three
stages.

\subsubsection{Newton-Raphson with line searches}

PETSc\cite{petsc-home-page} implements a Newton-Raphson method with
line searches, similar to the method described in
\cite{NumericalRecipes}.  Given a current guess $\underline{\bf
  u}_{\mbox{\footnotesize old}}$ of the solution, a Newton-Raphson
step proceeds as follows: Compute the residual
\begin{equation}
  \underline{\bf r}\equiv {\cal S}\underline{\bf u}_{\mbox{\footnotesize old}}
\end{equation}
and linearize $\cal S$ around the current guess $\underline{\bf
  u}_{\mbox{\footnotesize old}}$ of the solution:
\begin{equation}
  \label{eq:Jacobian}
  {\cal J}\equiv \frac{\partial\cal S}{\partial\underline{\bf u}}
(\underline{\bf u}_{\mbox{\footnotesize old}}).
\end{equation}
The Jacobian ${\cal J}$ is a $N_{DF}\times N_{DF}$-dimensional
matrix, $N_{DF}$ being the number of degrees of freedom. Next compute
a correction $\delta\underline{\bf u}$ by solving the linear system
\begin{equation}
\label{eq:Linearization}
 {\cal J}\delta\underline{\bf u}=-\underline{\bf r}. 
\end{equation}
Finally a line-search is performed in the direction of
$\delta\underline{\bf u}$. Parametrize the new solution by
\begin{equation}
  \underline{\bf u}_{\mbox{\footnotesize new}}
  =\underline{\bf u}_{\mbox{\footnotesize old}}
  +\lambda\;\,\delta\underline{\bf u}
\end{equation}
and determine the parameter $\lambda>0$ such that the residual of the
new solution,
\begin{equation}
  ||{\cal S}(\underline{\bf u}_{\mbox{\footnotesize new}})||,
\end{equation}
has sufficiently decreased. Of course, close enough to the true
solution, the full Newton-Raphson step $\lambda=1$ will lead to
quadratic convergence.  PETSc offers different algorithms to perform
this line-search.  The default method which employs cubic backtracking
worked very well in all our tests.
The line search ensures that in each iteration the
residual does indeed decrease, which is not guaranteed in
Newton-Raphson without line searches.

\subsubsection{Linear Solve}

In each Newton-Raphson iteration one has to solve
Eq.~\eqnref{eq:Linearization}, a linear system of $N_{DF}$ equations.
For large systems of linear equations, iterative linear
solvers\cite{Templates} are most efficient. Such iterative solvers
require solely the ability to compute the matrix-vector product ${\cal
  J}\underline{\bf v}$ for a given vector $\underline{\bf v}$.  Since
spectral derivatives and spectral interpolation lead to {\em full}
(i.e. non-sparse) matrices it is impractical to set up the matrix
${\cal J}$ explicitly.  One can compute these matrix-vector products
instead with the linearized variant of the code that computes the
operator $\cal S$, i.e.  equations \eqnref{eq:S3_A}-\eqnref{eq:S3_G}
and their multidimensional generalizations.  Thus our method requires
the linearizations of the operator $\cal N$ [Eq.~\eqnref{eq:S3_A}] and
of the boundary conditions [Eqs.~\eqnref{eq:S3_B} and
\eqnref{eq:S3_C}].  The matching equations
\eqnref{eq:S3_D}-\eqnref{eq:S3_G} are linear anyway, so one can reuse
code from $\cal S$ for these equations.  The linearizations are merely
Frechet derivatives\cite{Boyd:2001} of the respective operators
evaluated at the collocation points, and therefore the Newton-Raphson
iteration applied to the discretized equations is equivalent to the
Newton-Kantorovitch iteration applied to the PDE.

PETSc includes several different linear iterative solvers (GMRES,
TFQR, ...)  that can be employed for the linear solve inside the
Newton-Raphson iteration.  The choice of linear solver and of options
for the linear solver and for the Newton-Raphson iteration are made at
runtime. This allows one to experiment with different linear solvers
and with a variety of options to find an efficient combination.  Note
that the matching conditions \eqnref{eq:Matching-touch} and
\eqnref{eq:Matching-overlap} lead to a nonsymmetric matrix $\cal J$.
Therefore only iterative methods that allow for nonsymmetric matrices
can be used.

\subsubsection{Preconditioning}\label{sec:Preconditioning}

In practice one will find that the Jacobian ${\cal J}$ is
ill-conditioned and thus the iterative method employed to solve
Eq.~\eqnref{eq:Linearization} will need an increasing number of
iterations as the number of collocation points is increased.  The {\em
  spectral condition number} $\kappa$ of a matrix is the ratio of
largest to smallest eigenvalue of this matrix,
\begin{equation}
  \kappa=\frac{\lambda_{max}}{\lambda_{min}}.
\end{equation}
For second order differential equations discretized with Chebyshev
polynomials, one finds $\kappa\propto N^4$, $N$ being the number of
grid points per dimension.  Solving a linear system to given accuracy
will require\cite{Axelsson:1994,Templates} ${\cal O}(\kappa)$ iterations of
the Richardson method, and ${\cal O}(\sqrt{\kappa})$ iterations of modern
iterative methods like conjugate gradients or GMRES.  Although modern
methods are better than Richardson iteration, it is still vital to
keep $\kappa$ close to $1$.

This is achieved with {\em preconditioning}. Instead of solving
Eq.~\eqnref{eq:Linearization} directly, one solves
\begin{equation}\label{eq:BJ}
 {\cal B}{\cal J}\;\delta\underline{\bf u}
=-{\cal B}\underline{\bf r},
\end{equation}
with the preconditioning matrix ${\cal B}$.  Now the iterative solver
deals with the matrix ${\cal B}\cal J$. If ${\cal B}$ is a good
approximation to ${\cal J}^{-1}$, then ${\cal B}{\cal J}$ will be
close to the identity matrix, the condition number will be close to
unity, and the linear solver will converge quickly. 

Hence the problem reduces to finding a matrix $\cal B$ that
approximates ${\cal J}^{-1}$ sufficiently well and that can be
computed efficiently.
There exist many different approaches, most notably finite difference
preconditioning\cite{Orszag:1980} and finite element
preconditioning\cite{Deville-Mund:1985}; we will follow a two-stage
process proposed by Orszag\cite{Orszag:1980}.  First, initialize a
matrix ${\cal A}_{FD}$ with a finite difference approximation of the
Jacobian $\cal J$.  Second, approximately invert ${\cal A}_{FD}$ to
construct ${\cal B}$,
\begin{equation}\label{eq:PC}
  {\cal B}\approx {\cal A}_{FD}^{-1}.
\end{equation}
In one spatial dimension ${\cal A}_{FD}$ is tridiagonal and direct
inversion ${\cal B}\equiv{\cal A}_{FD}^{-1}$ is feasible.  In two or
more dimensions, direct inversion of ${\cal A}_{FD}$ is too expensive;
for problems in one two-dimensional subdomain, hardcoded incomplete
LU-factorizations have been developed\cite{Canuto-Hussaini}.  In our
case we have to deal with the additional complexity that the Jacobian
and therefore ${\cal A}_{FD}$ contains matching conditions.  Since we
choose the domain decomposition at runtime, nothing is known about the
particular structure of the subdomains.

We proceed as follows: We initialize ${\cal A}_{FD}$ with the finite
difference approximation of ${\cal J}$.  It is sufficient to include
those terms of the Jacobian in ${\cal A}_{FD}$ that cause the
condition number to increase with the expansion order. These are the
second spatial derivatives and the first derivatives from matching
conditions and boundary conditions, Eqs.~\eqnref{eq:S3_E}
and~\eqnref{eq:S3_C}. Including the value matching conditions
\eqnref{eq:S3_D}, \eqnref{eq:S3_F}, \eqnref{eq:S3_G} in ${\cal
  A}_{FD}$ improves the ability of the preconditioner to represent
modes extending over several subdomains and thus decreases the number
of iterations, too.  In the first example in section
\ref{sec:Example1} we demonstrate that preconditioning is indeed
necessary, and that one should precondition not only the second order
derivatives, but also the matching conditions.  Some details about the
finite difference approximations are given in appendix
\ref{sec:FD-details}.

Having set up ${\cal A}_{FD}$ we then use the software package
PETSc\cite{petsc-home-page} for the approximate inversion of
Eq.~\eqnref{eq:PC}.  PETSc provides many general purpose
preconditioners that perform the step~\eqnref{eq:PC} either explicitly
or implicitly, most notably ILU and the overlapping Schwarz method.
With PETSc we can explore these to find the most efficient one.  We
will describe our particular choices for preconditioning below for
each example.

\subsection[${\cal S}$ in higher dimensions]{\protect\boldmath${\cal S}$ in higher dimensions}

Generalizing $\cal S$ to multiple dimensions is conceptually
straightforward, since Eqs.~\eqnref{eq:S3} generalize nicely to higher
dimensions.  In order to simplify the matching between touching
subdomains, we require that on a surface shared by touching
subdomains, the collocation points are {\em identical}.  If, for
example, two three-dimensional rectangular blocks touch along the
$x$-axis, then both blocks must have identical lower and upper bounds
of the blocks along the $y$ and $z$ axis and both blocks must use the
same mappings and the same number of collocation points along the $y$-
and $z$-axis.  For concentric spherical shells, this restriction
implies that all concentric shells must have the same number of
collocation points in the angular directions.  With this restriction,
matching between touching subdomains remains a point-by-point
operation.

For overlapping domains, no restriction is needed. If a boundary point
of one subdomain happens to be within another subdomain, then an
equation analogous to \eqnref{eq:S3_F} is enforced using spectral
interpolation.

The actual implementation of the operator $\cal S$ involves
bookkeeping to keep track of which subdomains overlap or touch, or of
what equation to enforce at which grid point. When running in
parallel, matching conditions have to be communicated across
processors.

\subsection[Extension of $\cal S$ to Spherical Shells]{Extension of \protect\boldmath$\cal S$ to Spherical Shells}
\label{sec:S-in-Spheres}

Spherical shells have the additional complexity of having more
collocation points than spectral coefficients, $N_{col}>N_{spec}$, at
least in our formulation.  Transforming to spectral space and back to
real space projects the real-space values into a
$N_{spec}$-dimensional subspace.  Since spectral transforms are used
for derivatives and interpolation, a sphere has effectively only
$N_{spec}$ degrees of freedom.  If we naively try to impose $N_{col}$
equations, one at each collocation point, and if we try to solve for
real space values at each collocation point, we find that the linear
solver does not converge.  This happens because more equations are
imposed than degrees of freedom are available. Thus we cannot solve
for the real space values $u_{ijk}$ in a spherical shell.

The next choice would be to solve for the spectral coefficients
$\tilde u_{klm}$ as defined in Eq.~\eqnref{eq:ExpansionSphere} This is
also problematic as it prohibits finite-difference preconditioning.
One finds guidance on how to proceed by considering the prototypical
elliptic operator, the Laplacian. Application of $\nabla^2$ to
an expansion in spherical harmonics yields
\begin{equation}
  \nabla^2\;\sum_{l,m}a_{lm}(r)Y_{lm}
=\sum_{l,m}\!\left[\!-\frac{l(l\!+\!1)a_{lm}(r)}{r^2}
  \!+\!\frac{1}{r^2}\dnachd{}{r}\!\left(\!r^2\dnachd{a_{lm}(r)}{r}\right)
\right]Y_{lm}.
\end{equation}
We see that the different $(lm)$-pairs are uncoupled. The angular
derivatives will therefore be diagonal in spectral space (with
diagonal elements $-l(l+1)/r^2$).  However, one has to precondition
the radial derivatives in order to keep the spectral conditioning
number low and must therefore keep real-space information about the
radial direction. We therefore solve for the coefficients $\hat
u_{ilm}$ of an expansion defined by
\begin{equation}\label{eq:ExpansionSphere2}
  u(r_i, \theta, \phi)
  =\sum_{l=0}^L\sum_{m=-l}^l
    \hat u_{ilm}Y_{lm}(\theta, \phi).
\end{equation}
This mixed real/spectral expansion has $N_{spec}$ coefficients $\hat
u_{ilm}$ and retains real space information about the radial
coordinate necessary for finite difference preconditioning.  In order
to precondition the flat space Laplacian in a spherical shell, ${\cal
  A}_{FD}$ is initialized with the diagonal matrix
$\mbox{diag}(-l(l+1)/r_i^2)$ for the angular piece of $\nabla^2$ and
with finite differences for the radial derivatives.  More general
differential operators are discussed in the last example, 
section~\ref{sec:Example3}, and in appendix
\ref{sec:Nonflat-Preconditioning}.

In order to evaluate ${\cal S}\underline{\bf u}$ for a spherical
shell, we proceed as follows.  $\underline{\bf u}$ contains the
coefficients $\hat u_{ilm}$.  Transform these coefficients to real
space values.  This involves only an angular transform.  Compute
boundary conditions, matching conditions, and the residual of the
nonlinear elliptic operator $\cal N$ at each collocation point as in
rectangular blocks.  At this stage we have $N_{col}$ collocation point
values, all of which should vanish for the desired solution. We
transform these values back into the coefficients of
Eq.~\eqnref{eq:ExpansionSphere2} and return these coefficients as the
residual of the operator $\cal S$.

\subsection{Implementation Details}

Our code is written in C++ and utilizes various software packages
extensively.  PETSc\cite{petsc-home-page} is used for linear and
nonlinear solves as well as for preconditioning.  The software package
KeLP \cite{kelp-home-page} is used to implement domain decomposition.
It provides functionality to iterate over boundary points of a
specific subdomain as well as routines for handling the interprocessor
communication needed for matching between subdomains.
Spherepack\cite{spherepack-home-page} provides routines to handle
spherical harmonics such as computation of collocation points and
spectral transforms.  For Fourier transforms we employ
DFFTPACK\cite{dfftpack-home-page}.

The object oriented programming language C++ supports the modularity
of our code very well.  The different elements, e.g. mappings, are
represented by polymorphic classes.  In total, the code is about 50000
lines long.  It shares its infrastructure, i.e. domain decomposition,
spectral transforms, mappings, IO, etc., with a separate spectral
evolution code that we have written.  The code is currently not well
documented.  However, anyone interested in using it should contact us.

\section{Examples} 
\label{sec:Examples}

\subsection[$\nabla^2u=0$ in 2-D]{\protect\boldmath $\nabla^2u=0$ in 2-D}
\label{sec:Example1}
\begin{figure}
\begin{centering}
 \includegraphics[scale=0.65]{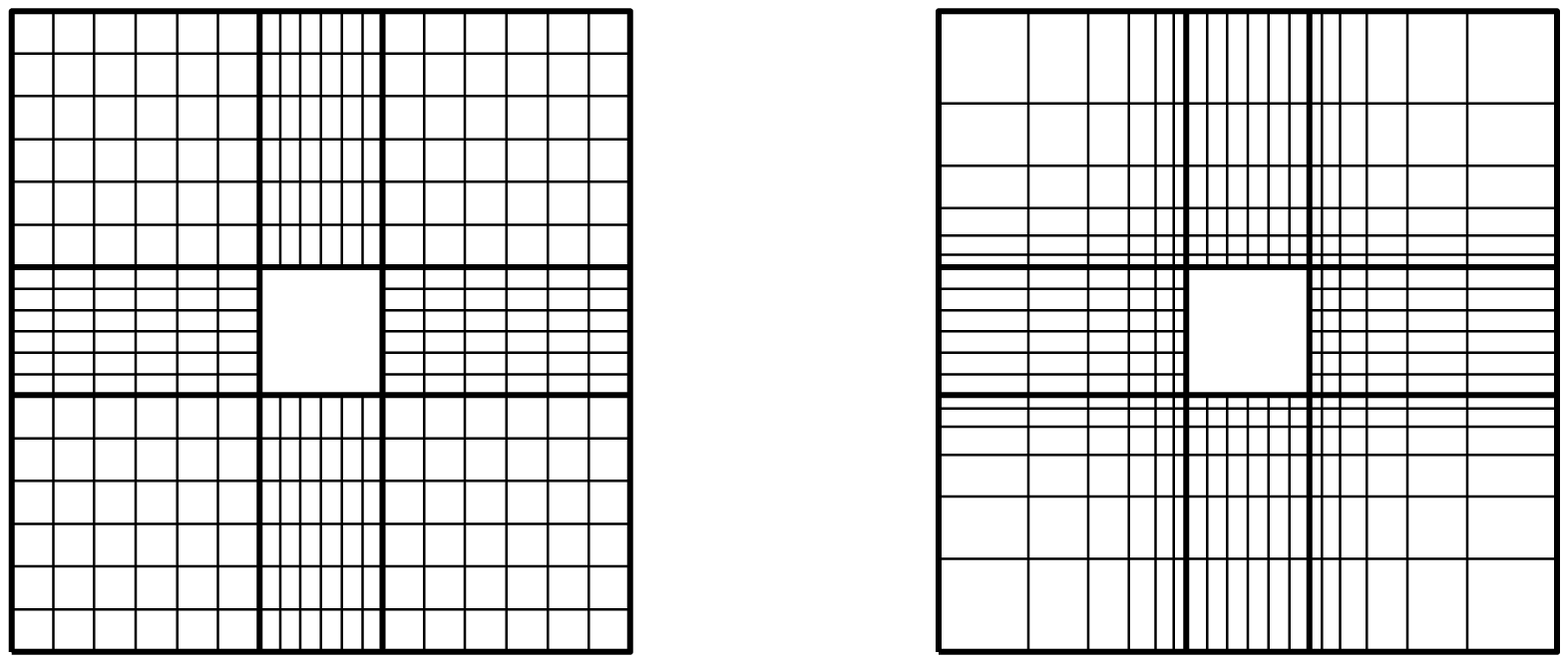}
\CAP{Domain decomposition for Laplace equation in a
  square}{\label{fig:SketchRectangles}Domain decomposition for Laplace
  equation in a square. The left plot illustrates linear mappings in
  all subdomains and the right plot shows log-linear-log mappings
  along each axis.}
\end{centering}
\end{figure}

As a first test, we solve the Laplace equation in two dimensions with
Dirichlet boundary conditions:
\begin{align}\label{eq:Test1a}
  \nabla^2 u(x,y)&=0 &&(x,y)\in {\cal D}\\
  \label{eq:Test1b}
  u(x,y)&=\ln(x^2+y^2)&&(x,y)\in \partial{\cal D}
\end{align}
The computational domain ${\cal D}$ is a square with side $2L$
centered on the origin with a smaller square with side $2$ excised:
\begin{equation}
    {\cal D}=\{(x,y) | -L\le x,y\le L\} - \{(x,y)| -1<x,y<1\}
\end{equation}
This domain is decomposed into 8 touching rectangles as shown in
figure \ref{fig:SketchRectangles}. This figure also illustrates the
difference between linear mappings and logarithmic mappings. The right
plot of figure \ref{fig:SketchRectangles} shows that logarithmic
mappings move grid points closer to the excised rectangle.  For
clarity, both plots neglect the fact that the Chebyshev collocation
points given in Eq.~\eqnref{eq:ChebyshevCollocationPoints} are
clustered toward the boundaries.

\begin{figure}
\begin{centering}
\includegraphics[scale=0.4]{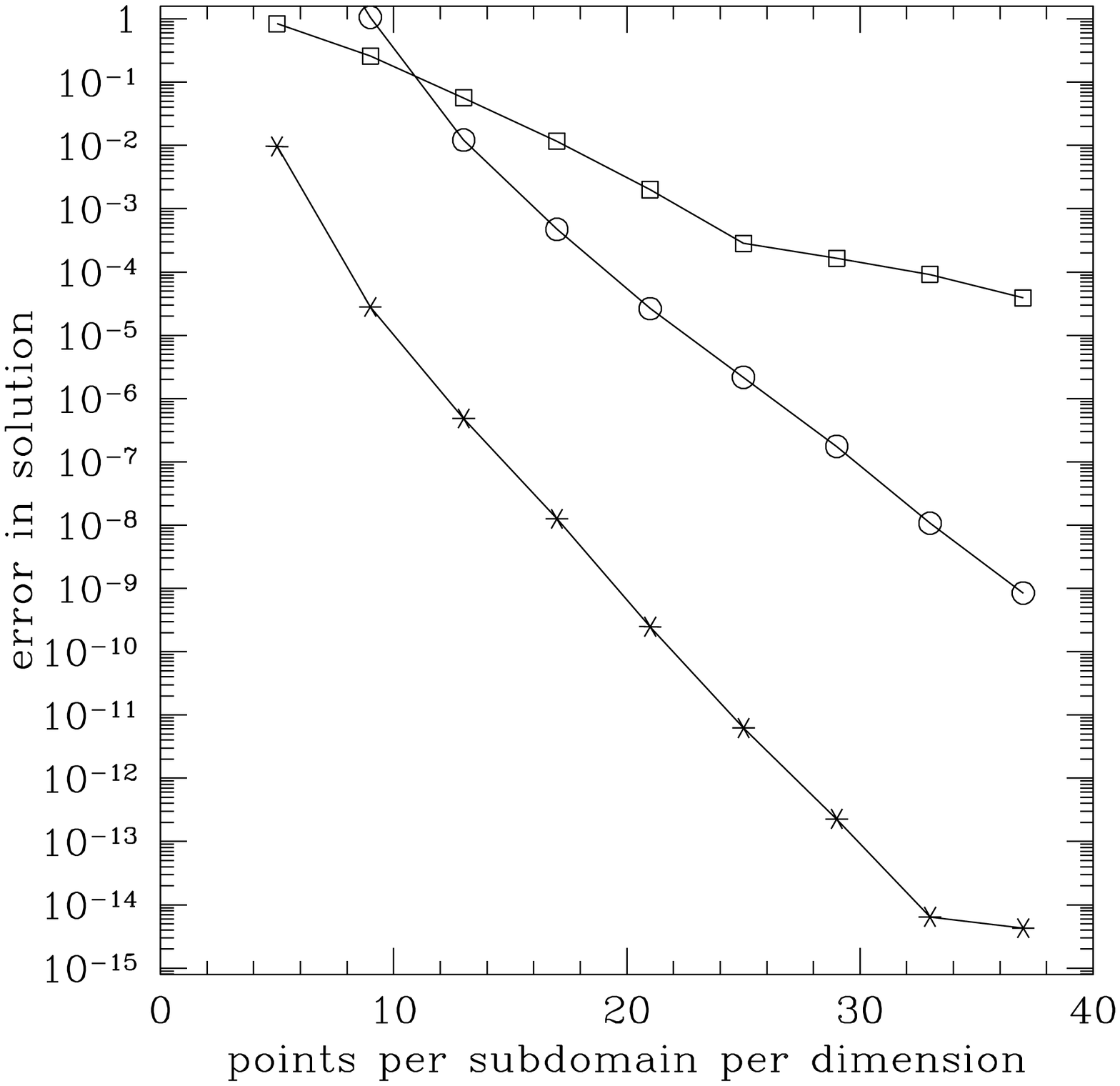}
\CAP{Convergence of solution of Laplace equation in a
  square}{\label{fig:Laplace2D}Convergence of solution of Laplace
  equation in a square (rms errors). Stars denote $L=5$ with linear
  mappings, squares $L=100$ with linear mappings, and circles $L=100$
  with log mappings.}
\end{centering}
\end{figure}

We solve Eqs.~\eqnref{eq:Test1a} and~\eqnref{eq:Test1b} for three
cases:
\begin{itemize}
\item $L=5$ with linear mappings
\item $L=100$ with linear mappings
\item $L=100$ with logarithmic mappings.
\end{itemize}
Equation~\eqnref{eq:Test1a} is linear, therefore only one
Newton-Raphson iteration with one linear solve is necessary.  The
numerical solution is compared to the analytic solution
$u(x,y)=\ln(x^2+y^2)$.  The errors are shown in figure
\ref{fig:Laplace2D}. In the small computational domain extending only
to $x,y=\pm 5$, the accuracy of the solution quickly reaches numerical
roundoff.  In the larger domain extending to $L=100$ the achievable
resolution with the same number of collocation points is of course
lower.  With linear mappings we achieve an absolute accuracy of
$10^{-4}$ with a total of $\sim 10000$ collocation points. This is
already better than finite difference codes. However this accuracy can
be increased with better adapted mappings. Since the solution
$\ln(x^2+y^2)$ changes much faster close to the origin than far away,
one expects better convergence if more collocation points are placed
close to the origin. This can be achieved with logarithmic mappings.
Figure \ref{fig:Laplace2D} shows that logarithmic mappings roughly
double the convergence rate. At the highest resolution the difference
is over four orders of magnitude.

\begin{figure}[tb]
\centerline{\includegraphics[scale=0.40]{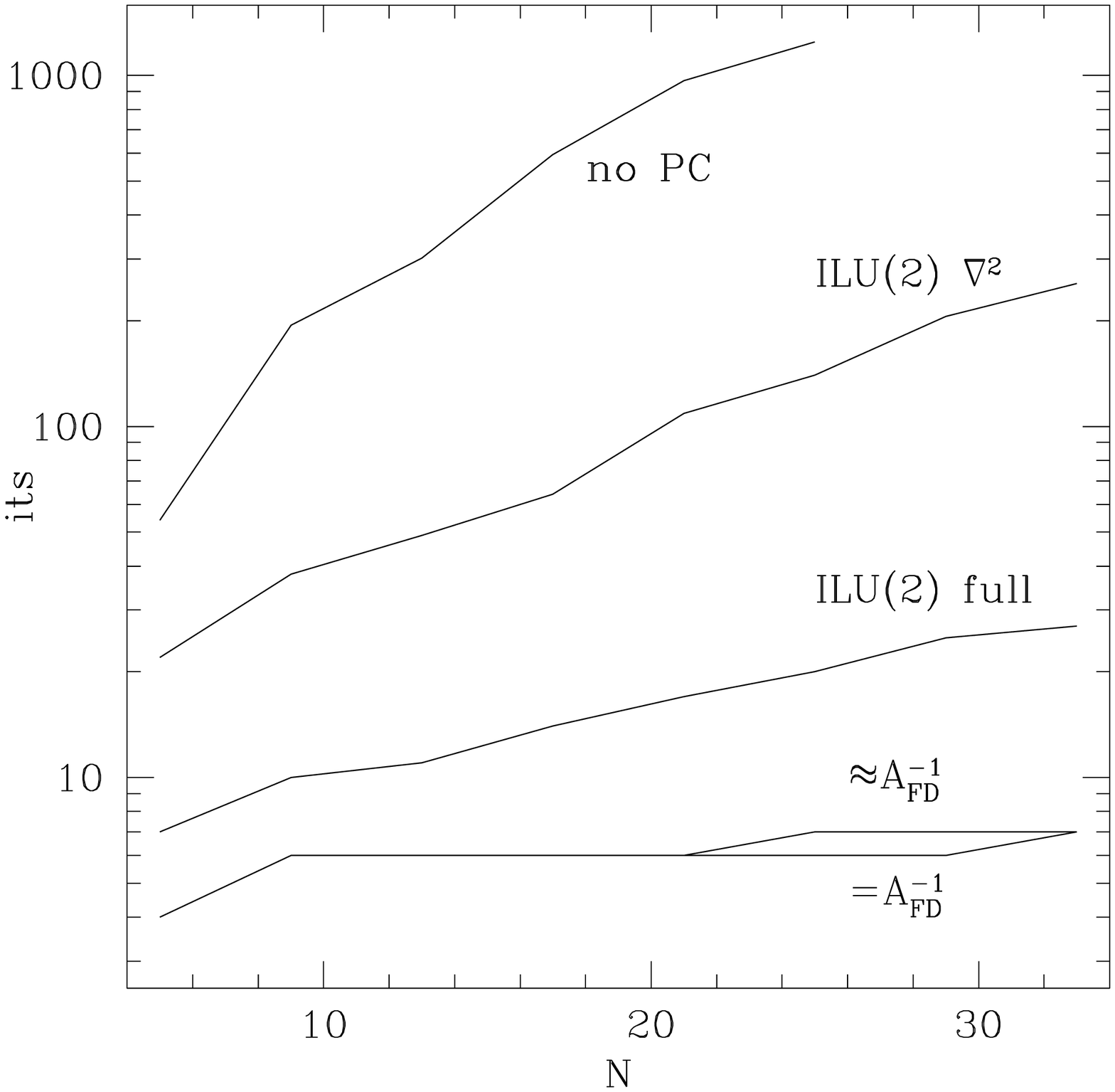}}
\CAP{Iteration count of the linear solver for different types of
   preconditioning}{\label{fig:tab:Laplace2D}Number of iterations in
   the linear solver for various kinds of preconditioning as a
   function of the number of collocation points $N$ per dimension per
   subdomain.  ``ILU(2) $\nabla^2$'' denotes ILU(2) preconditioning of
   only the Laplace operator, whereas in ``ILU(2) full'' the
   derivative matching conditions are preconditioned also. ``$={\cal
   A}_{FD}^{-1}$'' and ``$\approx{\cal A}_{FD}^{-1}$'' denote exact
   and approximate inversion of the preconditioning matrix based on
   Laplace operator and matching conditions.}
\end{figure}

Figure \ref{fig:tab:Laplace2D} compares the number of iterations
$N_{its}$ in the linear solver for different choices of the finite
difference preconditioning matrix ${\cal A}_{FD}$ [section
\ref{sec:Preconditioning}]. Without any preconditioning, ${\cal
A}_{FD}=\mathbf{1}$, $N_{its}$ increases very quickly with the number
of collocation points.  If only second derivative terms are included
in ${\cal A}_{FD}$ then $N_{its}$ grows more slowly. Inclusion of both
second derivatives and matching conditions \eqnref{eq:S3_D} and
\eqnref{eq:S3_E} improves the convergence further (see curve labeled
``ILU(2) full'').  In this case ILU(2) preconditioning completely
controls the largest eigenvalue of the preconditioned operator $\cal B
J$ [cf. Eq.~\eqnref{eq:BJ}], $\lambda_{max}\lesssim 2.6$; however, the
smallest eigenvalue $\lambda_{min}$ approaches zero as $N$ is
increased.  Hence, the condition number and thus the required number
of iteration still increases with resolution.  It is typical that ILU
has difficulties controlling the long wavelength modes, and the
problem is aggravated because the subdomains are only weakly
coupled. Figure \ref{fig:tab:Laplace2D} also contains results for
exact and approximate inversion of ${\cal A}_{FD}$. These methods
control $\lambda_{min}$, too, and lead to an iteration count
independent of resolution.  Direct inversion is computationally
expensive and is only feasible for small problems like this one.
Approximate inversion of ${\cal A}_{FD}$, our preferred method for
more complex geometries in 3 dimensions, will be explained in detail
in the next example.

\subsection{Quasilinear Laplace equation with two excised spheres}

This solver was developed primarily for elliptic problems in numerical
relativity. Accordingly we now solve an equation that has been of
considerable interest in that field over the last few years (see e.g.
\cite{Cook-Choptuik-etal:1993,Cook:2000} and references therein).
Readers not familiar with relativity can simply view this problem as
another test example for our new solver\footnote{The solution of this
  problem describes two black holes. The surfaces of the spheres
  $S_{1,2}$ correspond to the horizons of the black holes, the
  function $A^2$ encodes information about spins and velocities of the
  black holes, and the solution $\psi$ measures the deviation from a
  flat spacetime. Far away from the black holes one has $\psi\approx
  1$ with an almost Minkowski space, close to the holes we will find
  $\psi\sim 2$ with considerable curvature of spacetime}.  We solve
\begin{equation}\label{eq:BBH}
  \nabla^2\psi+\frac{1}{8}A^2\psi^{-7}=0
\end{equation}
for the function $\psi=\psi(x,y,z)$. $A^2=A^2(x,y,z)$ is a known,
positive function, and the computational domain is $\mathbbmss{R}^3$ with
two excised spheres,
\begin{equation}\label{eq:D-BBH}
  {\cal D}=\mathbbmss{R}^3-S_1-S_2.
\end{equation}
The radii $r_{1/2}$ and centers of the spheres are given.
The function $\psi$ must satisfy a Dirichlet boundary condition at
infinity, and Robin boundary conditions at the surface of each excised
sphere:
\begin{align}\label{eq:InftyBC-BBH}
  \psi\to 1&&&\mbox{as }r\to\infty\\
\label{eq:RobinBC-BBH}
  \dnachd{\psi}{r}+\frac{\psi}{2r_i}=0&&&\vec r\in\partial S_i,\;\; i=1,2
\end{align}
$\partial/\partial r$ in Eq.~\eqnref{eq:RobinBC-BBH} denotes the
radial derivative in a coordinate system centered at the center of
sphere $i$.

\begin{figure}
\centerline{
  \includegraphics[angle=90,scale=0.35]{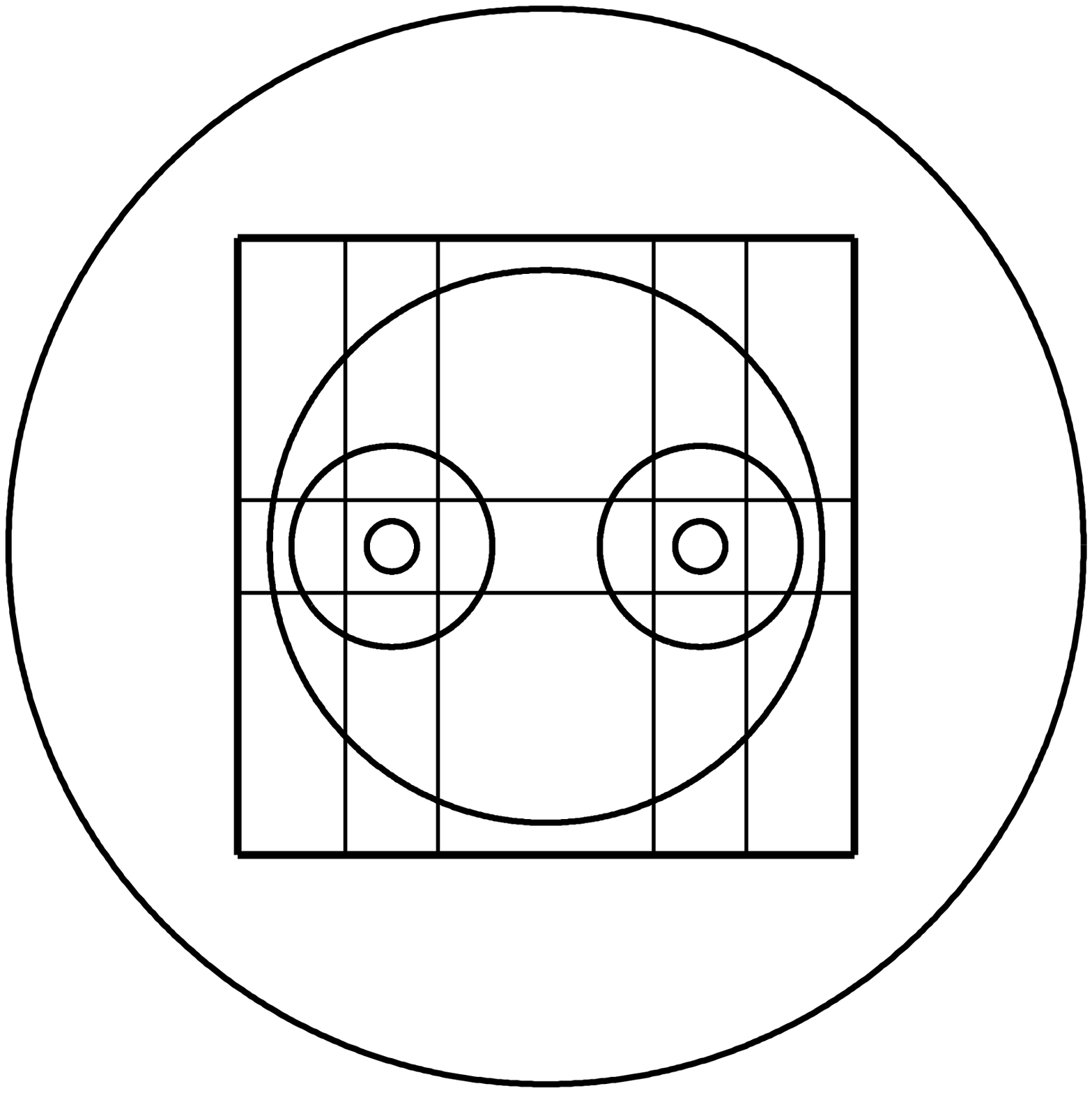}}
\CAP{Domain decomposition for the domain $\mathbbmss{R}^3$ with
  two excised spheres}{\label{fig:Domains-BBH}Cut through the domain
  decomposition for the computational domain Eq.~(\ref{eq:D-BBH}).}
\end{figure}
Figure \ref{fig:Domains-BBH} sketches the domain decomposition used
for the computational domain $\cal D$. We surround each excised sphere
with a spherical shell. These two spherical shells are matched
together with $5\times 3\times 3$ rectangular blocks, where the two
blocks that contain the excised spheres $S_{1,2}$ are removed.
Finally, we surround this structure with a third spherical shell
extending to very large outer radius. This gives a total of 46
subdomains, namely 3 shells and 43 rectangular blocks.

In the inner spheres we use a log mapping for the radial coordinate.
In the rectangular blocks, a combination of linear and logarithmic
mappings is used similar to the 2D example in figure
\ref{fig:SketchRectangles}. In the outer sphere an inverse mapping is
used which is well adapted to the fall-off behavior $\psi\sim
1+a\,r^{-1}+\cdots$ for large radii $r$. The outer radius of the outer
spherical shell is chosen to be $10^9$ or $10^{10}$ and a Dirichlet
boundary condition $\psi=1$ is used to approximate
Eq.~\eqnref{eq:InftyBC-BBH}.  For this particular problem, we could
also place the outer boundary {\em at} infinity without impact on
convergence or runtime.  We did not do this, because more specialized
analysis tools described in \cite{Pfeiffer-Cook-Teukolsky:2002}
currently require a finite outer radius\footnote{One can also
approximate \eqnref{eq:InftyBC-BBH} by a Robin boundary condition at
smaller outer radius.  This leads to slower convergence, probably
because the preconditioning necessary for a Robin boundary condition
is less effective in the stretched outer sphere.}.

We now present two solutions with different sizes and locations of the
excised spheres. In sections \ref{sec:Example2-Preconditioning} to
\ref{sec:Example2-ParallelExecution}, we then discuss several topics
including preconditioning and parallelization.

\subsubsection{Equal sized spheres}

First we choose two equal sized spheres with radii $r_1=r_2=1$.  The
separation between the centers of the spheres is chosen to be 10, the
outer radius of the outer sphere is $10^9$.

We solve equation \eqnref{eq:BBH} at several resolutions. The highest
resolution uses $29^3$ collocation points in each rectangular block,
$29\times 21\times 42$ collocation points (radial, $\theta$ and $\phi$
directions) in the inner spherical shells and $29\times 16\times 32$
in the outer spherical shell. We use the difference in the solutions
at neighboring resolutions as a measure of the error. We denote the
pointwise maximum of this difference by $L_{inf}$ and the
root-mean-square of the grid point values by $L_2$.  We also compute
at each resolution the quantity
\begin{equation}
\label{eq:M}
  M=-\frac{1}{2\pi}\int_{\infty}\dnachd{\psi}{r}d^2S
\end{equation}
which is the total mass of the binary black hole system. $M$ will be
needed in the comparison to a finite difference code below.  The
difference $\Delta M$ between $M$ at neighboring resolutions is again
a measure of the error of the solution.

\begin{figure}
\begin{centering}
  \includegraphics[scale=0.4]{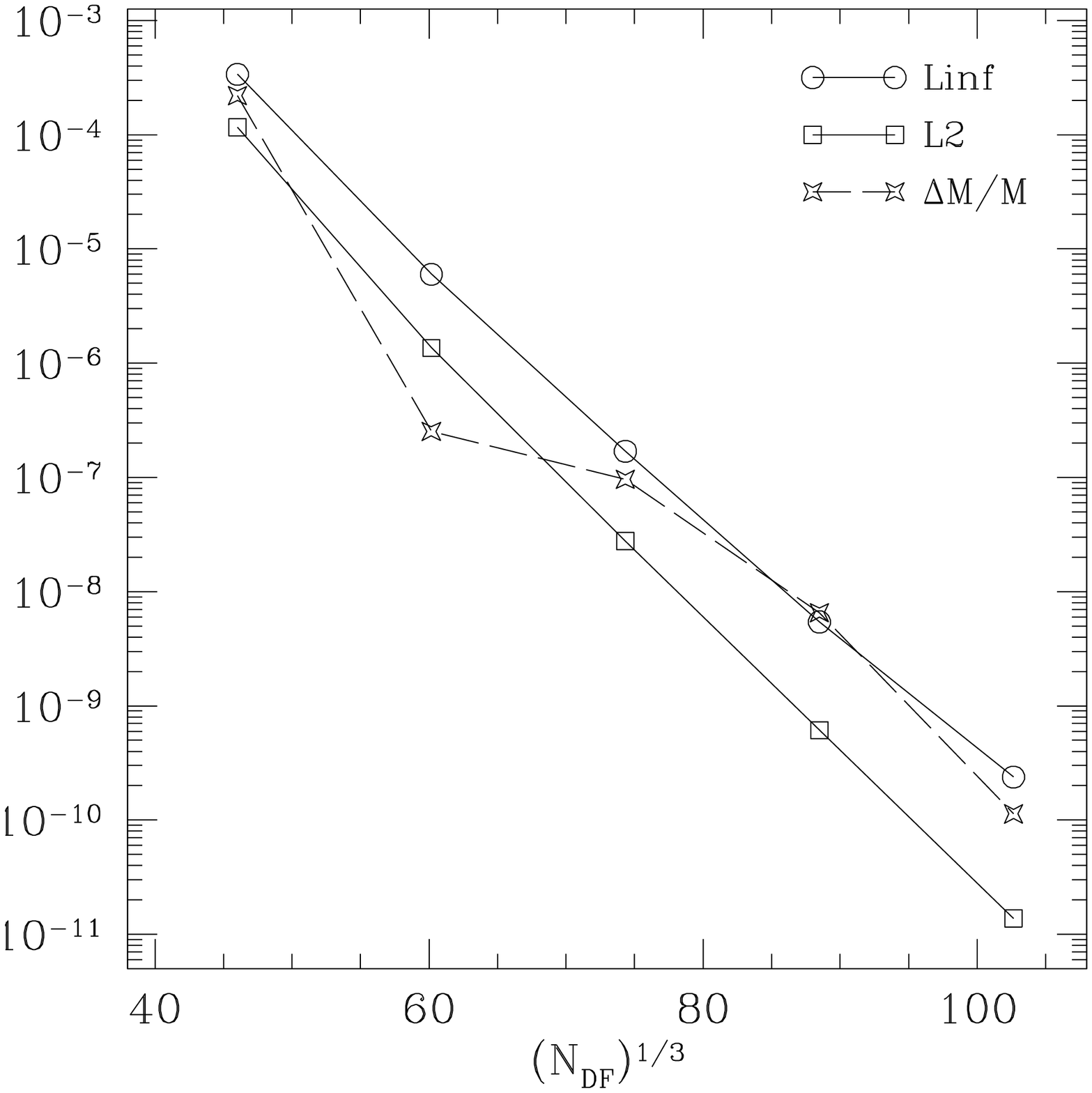}
\CAP{Convergence of solution of
Eqs.~\eqnref{eq:BBH}-\eqnref{eq:RobinBC-BBH} for equal-sized
holes}{\label{fig:Convergence-BBH}Convergence of solution of
Eqs.~\eqnref{eq:BBH}-\eqnref{eq:RobinBC-BBH} with radii of excised spheres
$r_1=r_2=1$ at separation $10$.  $N_{DF}$, $L_{inf}$, $L_2$, $M$ and
$\Delta M$ are defined in the text immediately before and after
Eq.~\eqnref{eq:M}.}
\end{centering}
\end{figure}

Figure \ref{fig:Convergence-BBH} shows the convergence of the solution
$\psi$ with increasing resolution. Since the rectangular blocks and
the spheres have different numbers of collocation points, the cube
root of the total number of degrees of freedom, $N_{DF}^{1/3}$ is used
to label the $x$-axis.  The exponential convergence is apparent.
Because of the exponential convergence, and because Linf, L2 and
$\Delta M$ utilize differences to the next lower resolution, the
errors given in figure \ref{fig:Convergence-BBH} are essentially the
errors of the next {\em lower} resolution. Note that at the highest
resolutions the approximation of the outer boundary condition
{\eqnref{eq:InftyBC-BBH}} by a Dirichlet boundary condition at finite
outer radius $10^9$ becomes apparent: If we move the outer boundary to
$10^{10}$, $M$ changes by $2\cdot 10^{-9}$ which is of order $1/10^9$
as expected.

On the coarsest resolution $\psi=1$ is used as the initial guess.
Newton-Raphson then needs six iterations to converge. On the finer
resolutions we use the result of the previous level as the initial guess.
These initial guesses are so good that one Newton-Raphson iteration is
sufficient on each resolution.

\subsubsection{Nonequal spheres --- Different length scales}
\label{sec:Example2-LengthScales}

With the multidomain spectral method it is possible to distribute
resolution differently in each subdomain. This allows for geometries
with vastly different length scales. As an example, we again solve 
equations \eqnref{eq:BBH}-\eqnref{eq:RobinBC-BBH}. The radii of the
spheres are now $r_1=1$ and $r_2=0.05$, and the separation of the centers
of the spheres is $100$.  The separation of the holes is thus 2000
times the radius of the smaller sphere.  A finite difference code
based on a Cartesian grid for this geometry would have to use adaptive
mesh refinement.

\begin{figure}
\begin{centering}
  \includegraphics[scale=0.4]{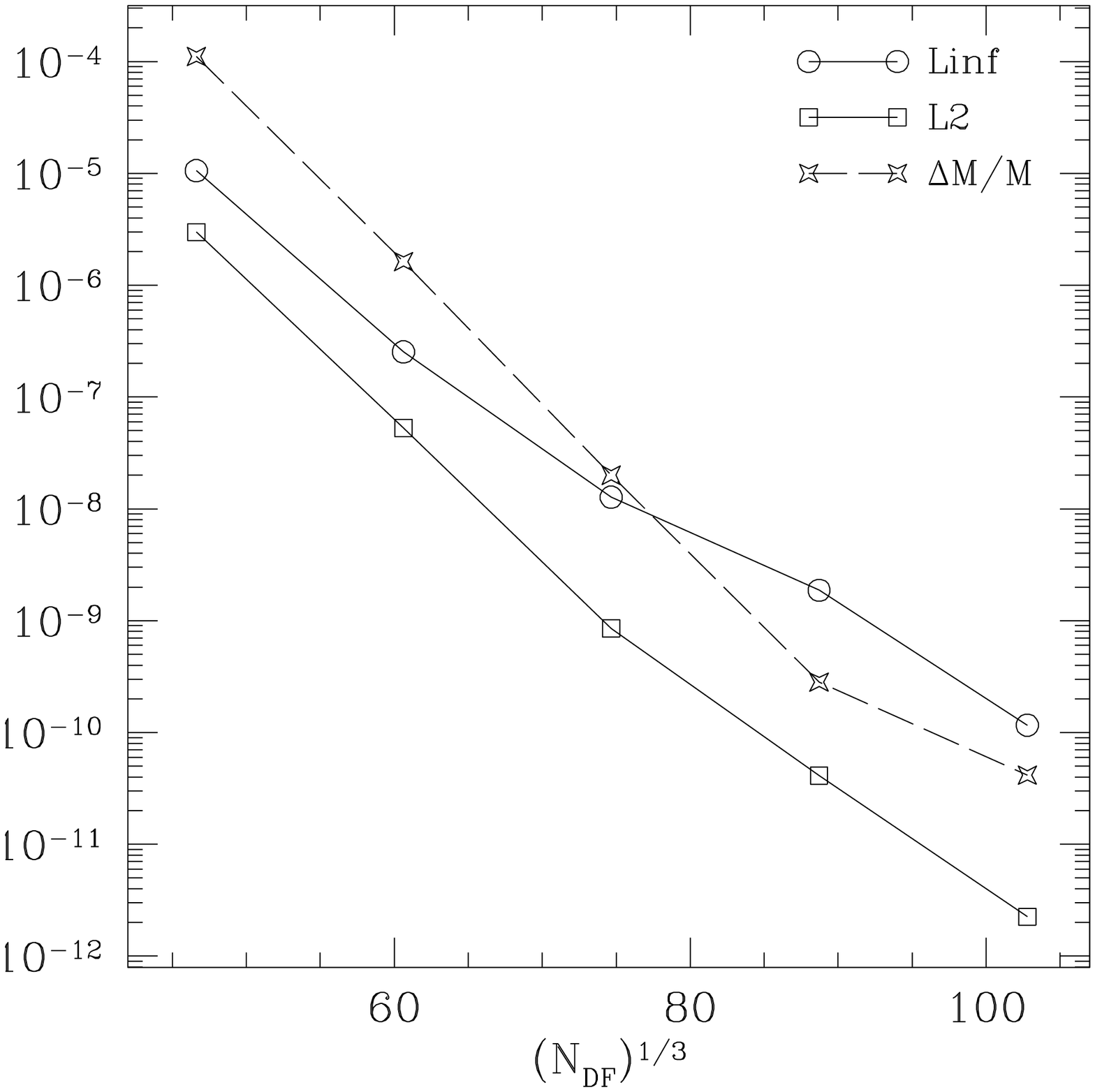}
\CAP{Convergence of solution of
Eqs.~\eqnref{eq:BBH}-\eqnref{eq:RobinBC-BBH} for widely separated, unequal
sized holes}{\label{fig:Convergence-MassRatio}Convergence of solution
of Eqs.~\eqnref{eq:BBH}-\eqnref{eq:RobinBC-BBH} for excised spheres of
radii $r_1=1$, $r_2=0.05$ at separation $100$.  Symbols as in Figure
\ref{fig:Convergence-BBH}.}
\end{centering}
\end{figure}

With the spectral solver, we still use the domain decomposition
depicted in figure \ref{fig:Domains-BBH}, but now the inner radii of
the two inner spherical shells are different. The outer boundary of
the outer sphere is at $10^{10}$.  The number of collocation points in
each subdomain is almost identical to the case with equal sized
spheres of figure \ref{fig:Convergence-BBH}, except we add 8
additional radial collocation points to the shell around the small
excised sphere.  As before we solve on different resolutions and
compute the norms of the differences of the solution between different
resolutions, as well as of the total mass $M$. The results are shown
in figure \ref{fig:Convergence-MassRatio}. The exponential convergence
shows that the solver can handle the different length scales involved
in this problem.

\subsubsection{Preconditioning}
\label{sec:Example2-Preconditioning}

The finite difference approximation ${\cal A}_{FD}$ is initialized
with the second derivative terms, the matching conditions in touching
domains [cf. Eqs.~\eqnref{eq:S3_D} and \eqnref{eq:S3_E}], and with a FD
approximation of the Robin boundary condition
Eq.~\eqnref{eq:RobinBC-BBH}.  Running on a single processor, we could
again define the preconditioner ${\cal B}$ via an ILU decomposition of
${\cal A}_{FD}$.  However, when running on multiple processors, an ILU
decomposition requires a prohibitive amount of communication, and
block ASM preconditioning\cite{Smith-Bjorstad-Gropp:1996} with one
block per processor becomes favorable.  After considerable
experimentation, we settled on an implicit definition of ${\cal B}$
via its action on vectors. ${\cal B}\underline{\bf v}$ is {\em
  defined} to be the approximate solution $\underline{\bf w}$ of
\begin{equation}\label{eq:PC-LinearSystem}
  {\cal A}_{FD}\underline{\bf w}=\underline{\bf v}.
\end{equation}

Equation \eqnref{eq:PC-LinearSystem} is solved using a second, inner
iterative solver, usually GMRES preconditioned with ILU (on a single
processor) or a block ASM method (on multiple processors).  The inner
solver is restricted to a fixed number of iterations.  Applying a
fixed number of iterations of an iterative solver is {\em not} a
linear operation, hence ${\cal B}$ represents no longer a matrix, but
a nonlinear operator.  In the outer linear solve we therefore use
FGMRES\cite{Saad:1993}, a variant of GMRES that does not require that
the preconditioner $\cal B$ is linear.  With this preconditioning the
outer linear solver needs about 20 iterations to reduce the residual
of the linear solve by $10^{-5}$.

More inner iterations reduce the number of iterations in the outer
linear solve, but increase the computations per outer iteration.  We
found the optimal number of inner iterations to be between 15-20.  In
all the computations given in this paper we use 20 inner iterations,
except for the 2-D example in Figure \ref{fig:tab:Laplace2D} where 10
inner iterations sufficed.

\subsubsection{Multigrid}

We also experimented with multigrid
algorithms\cite{Canuto-Hussaini,NumericalRecipes} to improve the
runtime.  The potential for multigrid is fairly small, since the
number of collocation points is so low. In this particular problem, an
accuracy of better than $10^{-6}$ can be achieved with $17^3$ grid
points per domain, which limits multigrid to at most two coarse
levels.

In addition it is not trivial to construct a restriction operator. The
obvious and canonical choice for a restriction operator is to
transform to spectral space, discard the upper half of the spectral
coefficients, and transform back to real space on a coarser grid. This
does not work here because the operator $\cal S$ uses the boundary
points of each subdomain to convey information about matching between
subdomains and about boundary conditions. Since these boundary points
are filled using different equations than the interior points, the
residual will typically be {\em discontinuous} between boundary points
of a subdomain and interior points.  Information about discontinuities
is mainly stored in the high frequency part of a spectral expansion
and discarding these will thus result in a loss of information about
matching between grids.  However, the coarse grid correction of a
multigrid algorithm is supposed to handle long wavelength modes of the
solution. In our case these extend over several subdomains and thus
information about matching is essential. Hence the simple approach of
discarding the upper half of the frequencies discards the most vital
parts of the information required by the coarse grid solver.

Thus one seems to be compelled to use a real space restriction
operator.  We examined straight injection\cite{NumericalRecipes}
which performed fairly well. The execution speed was comparable to the
preconditioning with an inner linear solve as described in section
\ref{sec:Example2-Preconditioning}. Since we did not achieve a
significant code speed-up, there was no reason to keep the increased
complexity of the multigrid algorithm.

\subsubsection{Comparison to a Finite Difference Code} 

The computational domain Eq.~\eqnref{eq:D-BBH} is challenging for 
standard finite difference codes based on a regular Cartesian grids
for two reasons:
\begin{enumerate}
\item The boundaries of the excised spheres do not coincide with
  coordinate boundaries, so complicated interpolation/extrapolation is
  needed to satisfy the boundary condition \eqnref{eq:RobinBC-BBH}
  (This problem led to a reformulation of the underlying physical
  problem without excised spheres\cite{Brandt-Bruegmann:1997}).
\item Resolving both the rapid changes close to the excised spheres
  {\em and} the fall-off behavior toward infinity requires a large
  number of grid points.
\end{enumerate}

\begin{figure}[tb]
\begin{centering}
  \includegraphics[scale=0.43]{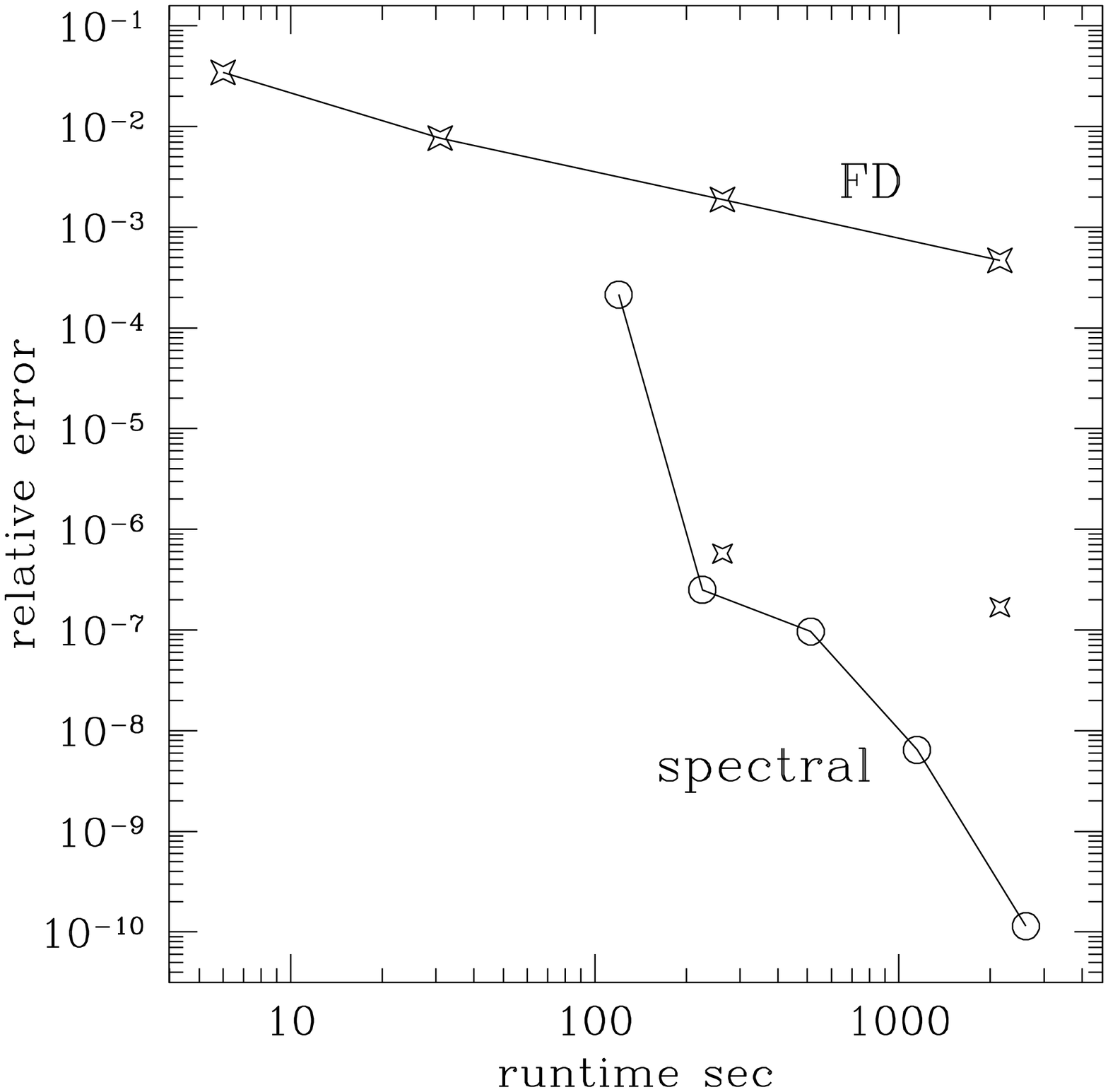} \CAP{Comparison
of runtime vs. achieved accuracy for the new spectral solver and the
Cad{\'e}z code}{\label{fig:CompareCadez}Comparison of runtime
vs. achieved accuracy for the new spectral solver and the Cad{\'e}z
code. Plotted is the achieved accuracy of the total mass $M$
vs. runtime needed to solve
Eqs.~\eqnref{eq:BBH}-\eqnref{eq:RobinBC-BBH} for both codes.}
\end{centering}
\end{figure}

In \cite{Cook-Choptuik-etal:1993} three different methods were
developed to solve the boundary value problem
\eqnref{eq:BBH}-\eqnref{eq:RobinBC-BBH}.  The best one turned out to
be a finite difference code based on a specially adapted coordinate
system, the so-called Cad{\'e}z coordinates. This code is an
FAS-multigrid algorithm developed specifically for the differential
equation \eqnref{eq:BBH}. Care was taken that the truncation error is
strictly even in grid-spacing $h$, thus allowing one to take two or
three solutions at different resolutions and Richardson extrapolate to
$h\to 0$.  The Cad{\'e}z code is thus specially built for this
equation in this geometry and it is unlikely that it can be
significantly improved upon by any finite difference method.

On the other hand, our spectral solver is general purpose. The domain
decomposition is not restricted to $\mathbbmss{R}^3$ with two excised
spheres and we do not employ any specific optimizations for this
particular problem.

We compare these two codes for the configuration with equal sized
spheres.  Figure \ref{fig:CompareCadez} shows a plot of runtime vs.
achieved accuracy for both codes.  These runs were performed on a
single RS6000 processor; the highest resolution runs needed about 1GB
of memory.  The solid line labeled FD represents the results of the
finite difference code without Richardson extrapolation. This line
converges quadratically in grid spacing.  The two stars represent
Richardson extrapolated values. The superiority of the spectral code
is obvious.  In accuracy, the spectral method outperforms even the
finite difference code with Richardson extrapolation by orders of
magnitude.  Only very few finite difference codes allow for Richardson
extrapolation, hence one should also compare the finite difference
code without Richardson extrapolation to the spectral code: Then the
{\em lowest} resolution of the spectral code is as accurate as the
{\em highest} resolution of the finite difference code and faster by a
factor of 20.  Note also that the Cad{\'e}z code cannot handle excised
spheres of very different sizes or spheres that are widely separated.
In particular, it cannot be used for the configuration in
section~\ref{sec:Example2-LengthScales}, which is readily solved by
our method.

\subsubsection{Parallelization}
\label{sec:Example2-ParallelExecution}

Most computations during a solve are local to each subdomain; the
operator $\cal S$ and the Jacobian $\cal J$ need communicate only
matching information across subdomains.  The inner linear solve is a
completely standard parallel linear solve with an explicitly known
matrix ${\cal A}_{FD}$. The software package PETSc has all the
necessary capabilities to solve this system of equations efficiently
in parallel. Hence parallelization by distributing different
subdomains to different processors is fairly straightforward.

However, different elements of the overall solve scale with different
powers of the number of collocation points per dimension.  If we
denote the number of collocation points per dimension by $N$, the
following scalings hold in three dimensions (the most interesting
case): A spectral transform in a rectangular domain requires
${\cal O}(N^3\log N)$ operations; the transform in a sphere ---where no
useful fast transform for the Legendre polynomials is available---
requires ${\cal O}(N^4)$ operations; interpolation to {\em one} point is
${\cal O}(N^3)$, so interpolation to {\em all} ${\cal O}(N^2)$ boundary points
scales like ${\cal O}(N^5)$. Thus the optimal assignment of subdomains to
processors is a function of $N$.  Moreover, assignment of subdomains to
processors is a discrete process --- it is not possible to move an
arbitrary fraction of computations from one processor to the another.
One always has to move a whole subdomain with all the computations
associated with it.  This makes efficient load balancing difficult.

At high resolution, the ${\cal O}(N^5)$ interpolation consumes most of
the runtime. Note that the outer spherical shell interpolates to $78$
block surfaces, whereas the inner shells each interpolate to $6$ block
surfaces. These interpolations are parallelized by first distributing
the data within each sphere to all processors. Then each processor
interpolates a fraction of the points and the results are gathered
again.

We present scaling results in table~\ref{tab:ParallelScaling}. These
results were obtained on the SP2 of the physics department of Wake
Forest University, and on NCSA's Platinum cluster, whose nodes have
two Intel Pentium processors each.  The listed times are cumulative
times for solving at five different resolutions, each solve using the
next lower solution as initial guess.  Not included in these times is
the set up in which the code determines which subdomain is responsible
for interpolating which ``overlapping'' boundary point. Also not
included is input/output.

\begin{table}
\begin{centering}
\CAP{Runtime and scaling efficiency of the spectral code.}{\label{tab:ParallelScaling}Runtime and scaling efficiency. Three processors host one shell and
$n_1$ blocks each, the remaining processors host $n_2$ blocks each. 
  The last four columns refer to the Platinum cluster.
}
\centerline{
\begin{tabular}{|c|cc|rc|rc|rc|}\hline
   & &
   & \multicolumn{2}{c|}{SP2} 
   & \multicolumn{2}{c|}{2 procs/node} 
   & \multicolumn{2}{c|}{1 proc/node} \\
   Nprocs & $n_1$ & $n_2$ & t[sec] & eff. & t[sec] & eff. & t[sec] & eff.\\\hline
    1 &     &    & 2344 &      & 1654 &      & 1654 &      \\
    4 & 10  & 13 &  786 & 0.75 &  764 & 0.54 &  643 & 0.64 \\
    8 & 4-5 &  6 &  384 & 0.76 &  381 & 0.54 &  304 & 0.68 \\
   18 & 0   &  3 &      &      &  198 & 0.46 &  156 & 0.59 \\
   26 & 0   &  2 &      &      &  140 & 0.45 &  111 & 0.57 \\
   46 & 0   &  1 &      &      &   87 & 0.41 &   73 & 0.49 \\\hline
  \end{tabular}
}
\end{centering}
\end{table}

On the SP2 we achieve a scaling efficiency of 75\%, whereas the Intel
cluster has a lower scaling efficiency between around 54\% (8
processors), and 41\% (46 processors).  Given all the limitations
mentioned above these numbers are very encouraging.

Changing from serial to parallel execution degrades performance in two
ways: First, the ILU preconditioner used within the approximate inner
linear solve is replaced by an overlapping block ASM preconditioner.
Since this preconditioner is less efficient than ILU, the approximate
inner linear solve is less accurate after its fixed number of
iterations. Therefore the outer linear solve needs more iterations to
converge to the required accuracy of $10^{-5}$. The single processor
code needs 19 outer linear iterations, whereas the parallel codes need
23 or 24.  Thus the maximally achievable scaling efficiency is
limited to $19/23\approx 0.83$. The scaling efficiency on the SP2 is
close to this limit.

The second reason for low scaling efficiency is that we have not
optimized the MPI calls in any way. The fact that the scaling
efficiency on the cluster is much better if only one processor per node
is used, suggests that the MPI calls are a bottleneck. Using both
processors on a node doubles the communication load on that node which
doubles the waiting time for MPI communication. The higher scaling
efficiency on the SP2 which has faster switches also suggests that
the runs on the PC cluster are communication limited.

\subsection[Coupled PDEs in nonflat geometry with excised spheres]
{Coupled PDEs in nonflat geometry with excised\,\,\,\,\,\,\,\,\,\,\, spheres}
\label{sec:Example3}

So far we have been considering only PDEs in a single variable.
However, the definition of the operator $\cal S$ is not restricted to
this case. In this section we present a solution of four coupled
nonlinear PDEs.  These equations are

\begin{figure}
\begin{centering}
  \includegraphics[scale=0.4]{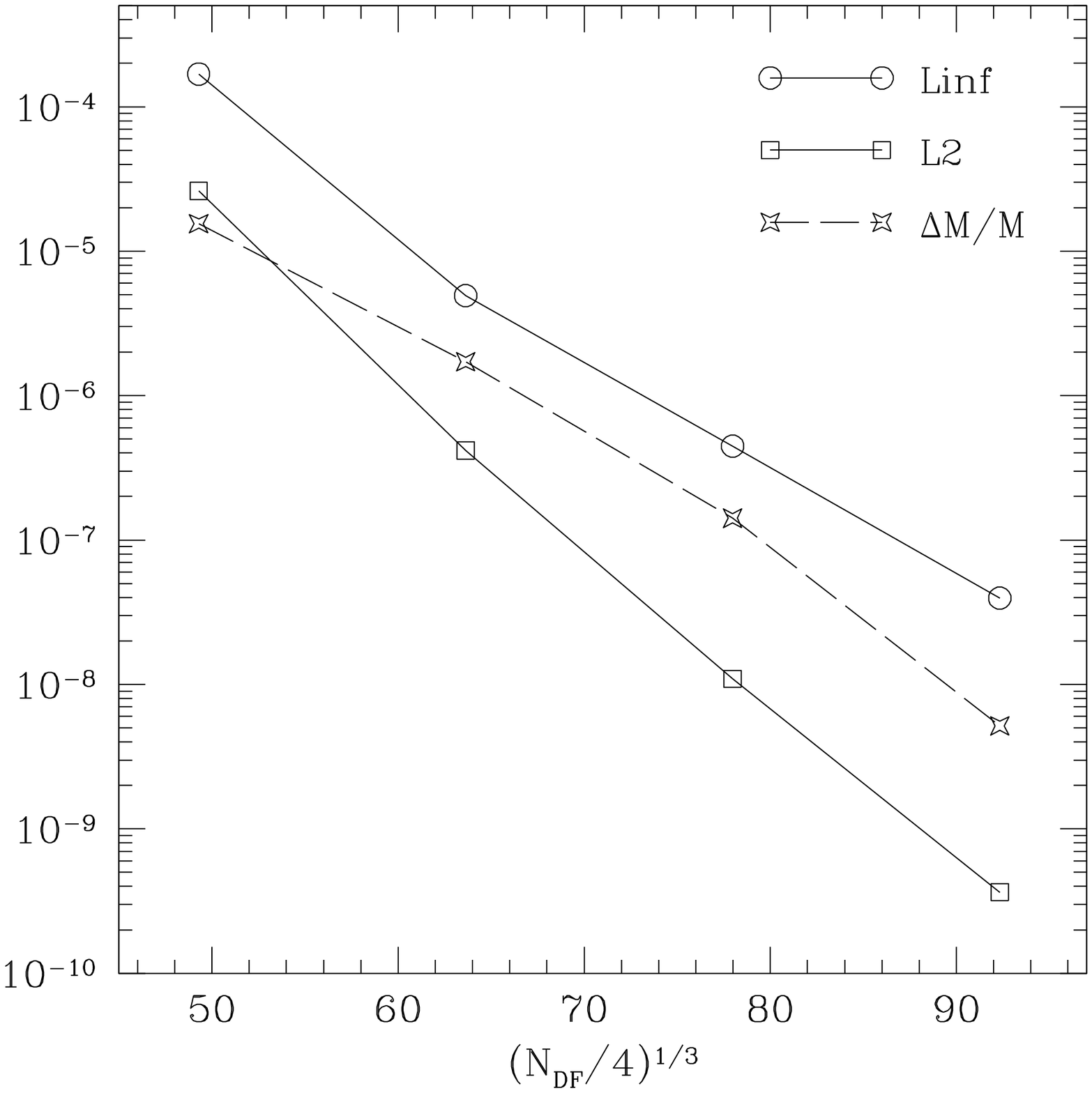}
\CAP{Convergence of solution to coupled PDEs \eqnref{eq:ConfTT-1} and
  \eqnref{eq:ConfTT-2}}{\label{fig:Convergence-ConfTT}Convergence of solution to coupled PDEs \eqnref{eq:ConfTT-1} and
  \eqnref{eq:ConfTT-2}.  Definitions as in figure \ref{fig:Convergence-BBH}.}
\end{centering}
\end{figure}

\begin{align}
\label{eq:ConfTT-1}
    &\tilde\nabla^2\psi-\frac{1}{8}\psi\tilde R-\frac{1}{12}\psi^5K^2
    +\frac{1}{8}\psi^{-7}\sum_{i,j=1}^3\tilde A_{ij}\tilde A^{ij}=0,\\
\label{eq:ConfTT-2}
    &\qquad\tildeLapLong V^i-\frac{2}{3}\psi^6\tilde\nabla^iK
    +\sum_{j=1}^3\tilde\nabla_{\!j}\tilde M^{ij}=0,&\hspace*{-2cm}i=1,2,3
\end{align}
These equations are important for the binary black hole problem.  The
exact definitions of the various terms can be found in
\cite{Pfeiffer-Cook-Teukolsky:2002}.  For this paper, only the
following information is necessary: $\tilde\nabla^2$ is the Laplace
operator on a nonflat three-dimensional manifold, hence
Eq.~\eqnref{eq:ConfTT-1} is an elliptic equation for $\psi$.
$\tildeLapLong$ is a variant of the vector Laplacian, thus
Eq.~\eqnref{eq:ConfTT-2} is an elliptic equation for the vector $V^i,
i=1,2,3$. The variables $\tilde A_{ij}$ and $\tilde A^{ij}$ are
functions of $V^i$, so that Eqs.~\eqnref{eq:ConfTT-1} and
\eqnref{eq:ConfTT-2} have to be solved simultaneously. The functions
$\tilde R$, $K$ and $\tilde M^{ij}$ are given.

The computational domain again has two spheres $S_{1,2}$ excised, 
\begin{equation}
  {\cal D}=\mathbbmss{R}^3-S_1-S_2.
\end{equation}
The radii of the excised spheres are $r_1=r_2=2$ and the separation
between the centers is 10.

We have Dirichlet boundary conditions on all boundaries:
\begin{align}
  \psi&=1,\\
  V^i&=0,\qquad i=1,2,3.
\end{align}

We solve the problem again at several different resolutions. On the
coarsest level two Newton-Raphson iterations are necessary, whereas
the finer levels need only one Newton-Raphson iteration. The linear
solver needs 30 iterations to reduce the residual by $10^5$.  Details
about constructing ${\cal A}_{FD}$ for the nonflat differential
operators $\tilde\nabla^2$ and $\tildeLapLong$ are given in appendix
\ref{sec:Nonflat-Preconditioning}.

The convergence of the solutions is shown in figure
\ref{fig:Convergence-ConfTT}. We again find smooth exponential
convergence. Recall that the plotted quantities essentially give the
error of the next lower resolution. Hence the next-to-highest
resolution run with a total of $78^3\approx 500000$ collocation points
has a maximum pointwise error of $\sim 0.5\cdot 10^{-7}$. The wall
clock time for that run is less than 2 hours on four RS 6000
processors.

This problem has also been attacked with a finite difference
code\cite{Marronetti-Matzner:2000}.  The finite difference code required a
runtime of 200 CPU hours (on 16 nodes of an Origin 2000). The accuracy
of the finite difference code seems to be comparable to the lowest
resolution solve of our spectral solver, which took 16 minutes CPU
time.  Compared to the finite difference code the spectral code is
almost embarrassingly fast.

\section{Improvements}

The fact that spherical harmonics have fewer spectral coefficients
than collocation points causes a host of complications. We have to
solve for mixed real-spectral coefficients,
Eq.~\eqnref{eq:ExpansionSphere2}. This complicates the operator $\cal
S$, and severely complicates real space finite difference
preconditioning. A double Fourier series\cite{Orszag:1974} for the
angular variables might be superior to the expansion in spherical
harmonics, since this avoids the necessity for mixed real-spectral
coefficients. Moreover one can then use fast transforms for both
$\phi$ and $\theta$ which might improve runtime in the spherical
shells.

We are working on cylinders as a third possible subdomain type.  We
also hope to use more general mappings that are no longer restricted
to acting on each dimension separately.

In terms of pure runtime, one should try to optimize the interpolation
to boundary points of overlapping subdomains. This is the part of the
code that has the worst scaling with the number of unknowns. Replacing
the explicit summation of the series by one of the faster methods
discussed in \cite{Boyd:2001} should speed up the code tremendously.
As was seen in the discussion of parallelization in section
\ref{sec:Example2-ParallelExecution}, the code seems to be communication
limited on a PC cluster. One should therefore also optimize the MPI
calls. For example, one could overlap communication with subdomain
internal computations.

Even without all these improvements our code is already very fast.
This indicates the potential of our approach.

\section{Conclusion}

We have presented a new elliptic solver based on pseudo-spectral
collocation.  The solver uses domain decomposition with spherical
shells and rectangular blocks, and can handle nonlinear coupled
partial differential equations.

Our method combines the differential operator, the
boundary conditions and matching between subdomains in {\em one}
operator $\cal S$. The equation ${\cal S}\underline{\bf u}=0$ is then
solved with Newton-Raphson and an iterative linear solver.  We show
than one can employ standard software packages for nonlinear and
linear solves and for preconditioning.

The operator $\cal S$ has the added benefit that it is modular.
Therefore adaption of the method to a new PDE or to new boundary
conditions is easy; the user has only to code the differential
operator and/or the new boundary conditions.  We also discuss our
treatment of mappings which decouples mappings from the actual code
evaluating the differential operator, and from the code dealing with
basis functions and details of spectral transforms. This modularity
again simplifies extension of the existing code with e.g. new
mappings.

We demonstrated the capabilities of the new method with three examples
on non-simply-connected computational domains in two and three
dimensions and with one and four variables.  We also demonstrated that
the domain decomposition allows for vastly different length scales in
the problem.  During the examples we discussed various practical
details like preconditioning and parallelization.  Two of these
examples were real applications from numerical relativity.  We found
the spectral code at the {\em coarsest} resolution to be as accurate
as finite difference methods, but faster by one to two orders of
magnitude.

\section*{Acknowledgements}
 We thank Gregory Cook for helpful discussions.  This work was
 supported in part by NSF grants PHY-9800737 and PHY-9900672 to
 Cornell University. Computations were performed on the IBM SP2 of the
 Department of Physics, Wake Forest University, with support from an
 IBM SUR grant, as well as on the Platinum cluster of NCSA.

\section{Appendix A: Preconditioning of inverse mappings}
\label{sec:FD-details}

In a subdomain with inverse mapping that extends out to (almost)
infinity, the outermost grid points are distributed very unevenly in
physical space.  This causes finite-difference approximations of
derivatives to fail if they are based on the physical coordinate
positions.  Therefore we difference in the collocation coordinate $X$
and apply the mapping via Eq.~\eqnref{eq:SecondDeriv-Mapped}. At the
collocation grid point $X_i$ with grid spacing $h_-=X_i-X_{i-1}$ and
$h_+=X_{i+1}-X_i$ we thus use
\begin{align}
\label{eq:FD-dudX}
\left(\dnachd{u}{X}\right)_{\!i}
=&-\frac{h_+u_{i-1}}{h_-(h_-+h_+)}
+\frac{(h_+-h_-)u_i}{h_+h_-}
+\frac{h_-u_{i+1}}{h_+(h_-+h_+)},\\
\label{eq:FD-du2dX2}
\left(\dnachd{^2u}{X^2}\right)_{\!i}
=& \frac{2u_{i-1}}{h_-(h_-+h_+)}
  -\frac{2u_i}{h_-h_+}
  +\frac{2u_{i+1}}{h_+(h_-+h_+)},\\
\label{eq:FD-du2dx2}
\left(\dnachd{^2u}{x^2}\right)_{\!i}
=&\;X'^2_i\left(\dnachd{^2u}{X^2}\right)_{\!i}
  +X''_i\left(\dnachd{u}{X}\right)_{\!i}.
\end{align}
If one substitutes Eqs.~\eqnref{eq:FD-dudX} and \eqnref{eq:FD-du2dX2} into
\eqnref{eq:FD-du2dx2}, then the coefficients of $u_{i-1}, u_i$ and
$u_{i+1}$ are the values that have to be entered into the
FD-approximation matrix ${\cal A}_{FD}$.

Even with this trick, preconditioning of the radial derivatives in an
extremely stretched outer sphere is not yet sufficiently good.  The
preconditioned Jacobian ${\cal BJ}$ still contains eigenvalues of size
$\sim 40$. The eigenmodes are associated with the highest radial mode
in the outer sphere. We found that we can suppress these eigenvalues
by damping this highest radial mode by a factor of 10 after the PETSc
preconditioning is applied.

\section{Appendix B: Preconditioning the nonflat Laplacian}
\label{sec:Nonflat-Preconditioning}

In a nonflat manifold, the Laplace operator of Eq.~\eqnref{eq:ConfTT-1}
contains second and first derivatives of $\psi$,
\begin{equation}\label{eq:nonflat-Laplace}
  \tilde\nabla^2\psi
=\sum_{i,j=1}^3g^{ij}\frac{\partial^2\psi}{\partial x^i\partial x^j}
 +\sum_{i=1}^3f^i\dnachd{\psi}{x^i}.
\end{equation}
The coefficients $g^{ij}$ and $f^i$ are known functions of position.
Since our particular manifold is almost flat, we have $g^{ii}\approx
1$, and $g^{ij}\approx 0$ for $i\neq j$.  We base our
preconditioning only on the diagonal part of \eqnref{eq:nonflat-Laplace}, 

\begin{equation}\label{eq:nonflat-Laplacian-diagonal}
\sum_{i=1}^3g^{ii}\frac{\partial^2}{\partial {x^i}^2}.
\end{equation}

In rectangular blocks, Eq.~\eqnref{eq:nonflat-Laplacian-diagonal} can
be preconditioned without additional fill-in in ${\cal A}_{FD}$.
Inclusion of the mixed second derivatives from
Eq.~\eqnref{eq:nonflat-Laplace} in ${\cal A}_{FD}$ leads to a large
fill-in of ${\cal A}_{FD}$. The increase in computation time due to
the larger fill-in outweighs the improvement of convergence of the
iterative solver in our problems.

For the spherical shells, matters are complicated by the fact that we
use mixed real-space/spectral space coefficients [recall
Eq.~\eqnref{eq:ExpansionSphere2}].  It is easy to precondition the
angular piece of the flat space Laplacian, since our basis functions
$Y_{lm}$ are the eigenfunctions of this operator.  Derivative
operators with angle-dependent coefficients lead to convolutions in
spectral space and lead thus to a large fill-in in the preconditioning
matrix.  Therefore we can only precondition radial derivatives with
coefficients that are {\em independent} of the angles $\theta, \phi$.
We thus need to approximate Eq.~\eqnref{eq:nonflat-Laplacian-diagonal}
by a flat space angular Laplacian and constant coefficient radial
derivatives. 
We proceed as follows.

In spherical coordinates Eq.~\eqnref{eq:nonflat-Laplacian-diagonal} 
takes the form
\begin{equation}\label{eq:Cart2Sph}
\begin{aligned}
\sum_{i=1}^3g^{ii} \frac{\partial^2}{\partial{x^i}^2}
=&G^{\theta\theta}\dnachd{^2}{\theta^2}
  +G^{\phi\phi}\frac{\partial^2}{\partial\phi^2}
  +G^{rr}\frac{\partial^2}{\partial r^2}
+G^{\theta\phi}\dnachd{^2}{\theta\partial\phi}\\
  &+G^{\theta r}\frac{\partial^2}{\partial\theta\partial r}
  +G^{\phi r}\frac{\partial^2}{\partial\phi\partial r}
+F^{\theta}\dnachd{}{\theta}
  +F^{\phi}\dnachd{}{\phi}
  +F^{r}\dnachd{}{r}.
\end{aligned}
\end{equation}
The various functions $G$ and $F$ are related to $g^{ii}$ by standard
Cartesian-to-polar coordinate transformations.  These transformations
are singular for $\theta=0$ and $\theta=\pi$; however, no grid points
are located on this axis.  Moreover, only $G^{\theta\theta}$, $G^{rr}$
and $F^r$ will be used, and these three quantities are continuous at
the pole. We compute $G^{\theta\theta}$, $G^{rr}$ and $F^r$ at each
grid point.  For each radial grid point $r_i$, we average over angles
to obtain $\bar G^{\theta\theta}_i$, $\bar G^{rr}_i$ and $\bar F^r_i$.
Now precondition as if $\bar G^{\theta\theta}_i$ were part of an
angular piece of the flat space Laplacian, i.e. enter $-l(l+1)\bar
G^{\theta\theta}_i/r_i^2$ as the diagonal element belonging to the
unknown $\hat u_{ilm}$.  Further, precondition $\bar
G^{rr}_i\partial^2/\partial r^2+\bar F^r_i\partial/\partial r$ with
finite differences as described in appendix \ref{sec:FD-details}.
Ignore all other terms in Eq.~\eqnref{eq:Cart2Sph}.

The operator $\tildeLapLong$ in Eq.~\eqnref{eq:ConfTT-2} is defined by
\begin{equation}
  \tildeLapLong V^i
\equiv
\tilde\nabla^2 V^i+\frac{1}{3}\sum_{k=1}^3\tilde\nabla^i \tilde\nabla_kV^k
+\sum_{k=1}^3\tilde R^i_kV^k, 
\end{equation}
$\tilde\nabla$ and $\tilde R_{ij}$ being the covariant derivative
operator and Ricci tensor associated with the metric of the manifold.
$\tildeLapLong V^i$ contains thus the nonflat Laplace operator acting
on each component $V^i$ separately, plus terms coupling the
different components which involve second derivatives, too.  We
precondition only the Laplace operator $\tilde\nabla^2 V^i$ for each
component $V^i$ separately as described above and ignore the coupling
terms between different components.



%


\renewcommand{\thefootnote}{\fnsymbol{footnote}}

\chapter[Quasi-circular orbits for spinning binary black holes]{Quasi-circular orbits for spinning binary black holes\footnote[1]{H. P. Pfeiffer, S. A. Teukolsky, and G. B. Cook, Phys. Rev. D {\bf 62}, 104018 (2000).}}
\label{chapter:Spin}
\renewcommand{\thefootnote}{\arabic{footnote}}







\section{Introduction}
\label{sec:Spin:Introduction}

The inspiral and coalescence of binary black hole systems is a prime
target for upcoming gravitational wave detectors such as LIGO.  Such
systems will be circularized by the emission of gravitational waves,
and will evolve through a quasi-equilibrium sequence of circular
orbits. At the innermost stable circular orbit (ISCO) we expect a
transition to a dynamically plunging orbit. It is anticipated that this
transition will impart a characteristic signature on the gravitational
waveform. It is therefore important to know the orbital frequency at
the ISCO, since the corresponding gravitational wave frequency is
predominantly just twice this frequency.

Predicting the waveform in detail from the transition at the ISCO to
the final merger requires the full machinery of numerical
relativity. These calculations require appropriate initial data. Out of
the large space of solutions of the initial-value equations of general
relativity, we need an algorithm to select solutions corresponding to
black holes in quasi-circular orbits. The effective potential method
\cite{Cook:1994} allows one to construct such solutions, and to
determine the properties of the ISCO.

The effective potential is based on the fact that minimizing the
energy of a system yields an equilibrium solution. This follows from
the Hamiltonian equations of motion: If the Hamiltonian $\cal H$ is
minimized with respect to a coordinate $q$ and a momentum $p$, then
$\dot{q}=\partial{\cal H}/\partial p=0$ and $\dot{p} =
-\partial {\cal H}/\partial q=0$.  The energy of two objects in
orbit about each other can be lowered by placing the objects at rest
at their center of mass. Therefore minimizing the energy with respect
to all coordinates and momenta will not yield a circular orbit. To
find circular orbits in Newtonian gravity, one can minimize the energy
while holding the angular momentum constant. This procedure works as
well for a test-mass orbiting a Schwarzschild black hole, where one
minimizes the ADM energy. This can be seen as follows. For geodesic
motion, one finds \cite{Wald:1984}
\begin{equation}\label{eqn:Schwarzschild}
\frac12\dot r^2 + \frac12\left(1-\frac{2M}{r}\right)
	\left(\frac{\tilde{L}^2}{r^2}+1\right) = \frac12\tilde{E}^2.
\end{equation}
Here $M$ is the mass of the black hole, $\tilde{E}$ is the energy per
unit rest mass of the test-particle as seen from infinity and
$\tilde{L}$ its orbital angular momentum per unit rest mass. Denote
the rest-mass of the test-particle by $M'$. Then the ADM energy is simply
$\Eadm=M+\tilde{E}M'$, and minimizing $\Eadm$ is equivalent to
minimizing $\tilde{E}$. Hence minimizing the left hand side of
(\ref{eqn:Schwarzschild}) with respect to $r$ yields the radius of
circular orbits as a function of angular momentum. Minimization of
(\ref{eqn:Schwarzschild}) with respect to $\dot r$ yields $\dot r=0$,
which is necessary for a circular orbit. From the minimum one finds
the energy of the test-particle as a function of angular momentum.
Obviously, one needs to keep $M$ and $M'$ constant during the
minimization, so the prescription to compute circular orbits becomes:
Minimize $\Eadm$ while keeping the angular momentum and the rest
masses constant.

These ideas have been formalized as variational principles for finding
equilibria for rotating and binary stars in Newtonian gravity. There
is also a similar variational principle for rotating stars in general
relativity \cite{Hartle-Sharp:1967}. Binary systems in general
relativity are not strictly in equilibrium because they emit
gravitational waves. However, for orbits outside the innermost stable
circular orbit, the gravitational radiation reaction time scale is
much longer than the orbital period. It is therefore a good
approximation to treat the binary as an equilibrium system.

In this paper we apply this minimization principle to rotating binary
black hole systems.  Let the masses of the holes be $M_1$ and $M_2$,
the spins be ${\bf S}_1$ and ${\bf S}_2$, and the total angular
momentum of the system be ${\bf J}$.  We exploit the invariance under
rescaling of the mass by using dimensionless quantities $M_1/M_2$,
${\bf S}_1/M_1^2$, ${\bf S}_2/M_2^2$, and ${\bf J}/\mu m$, where
$m=M_1+M_2$ denotes the total mass and $\mu=M_1M_2/m$ the reduced
mass.  Then we adopt the following straightforward prescription to
locate quasi-circular orbits: Minimize the scaled ADM energy $\Eadm/m$
with respect to the separation of the holes, while keeping $M_1/M_2$,
${\bf S}_1/M_1^2$, ${\bf S}_2/M_2^2$, and ${\bf J}/\mu m$ constant.

It is somewhat involved to carry out this simple prescription. The
computation of the ADM energy becomes more difficult than for the
Schwarzschild example above. More importantly, however, no rigorous
definitions exist for the mass or spin of an individual black hole in
a spacetime containing two black holes. We will address these issues
in Sec.~\ref{sec:Implementation}. Ultimately, we must use numerical
methods to generate and search among the solutions.  Our numerical
approach involves rootfinding, which is also described in
Sec.~\ref{sec:Implementation}.

In Sec.~\ref{sec:Results} we present the results of the effective
potential method. For the interpretation of these results, we need to
search for common apparent horizons in our binary black hole
data sets. These results are included in Sec.~\ref{sec:Results}, too.
We discuss our results and conclusions in Secs.~\ref{sec:Discussion}
and \ref{sec:conclusion}. The appendix contains details of the
apparent horizon searches.

\section{Implementation}\label{sec:Implementation}

In order to minimize the ADM energy while keeping $M_1/M_2$, ${\bf
J}/\mu m$, ${\bf S}_1/M_1^2$ and ${\bf S}_2/M_2^2$ constant, we need a
method to compute the ADM energy as a function of angular momentum,
masses and spins of the holes and separation. As a first step we
construct initial data $(\gamma_{ij},K_{ij})$ on a hypersurface as
described in \cite{Cook:1991,Cook-Choptuik-etal:1993,Cook:1994}. Our
particular approach assumes conformal flatness of the 3-metric
$\gamma_{ij}$, maximal embedding of the hypersurface, as well as
inversion symmetry conditions on the 3-metric $\gamma_{ij}$ and on the
extrinsic curvature $K_{ij}$.  The effective potential method is
independent of these assumptions and works with all methods that
compute initial data. For example, in \cite{Baumgarte:2000}, the
effective potential method was used without assuming inversion
symmetry.  In particular, the assumptions of maximal embedding and
conformal flatness are not essential but merely convenient---maximal
embedding decouples the Hamiltonian and momentum constraints within
the initial-data formalism we use, and conformal flatness allows for
an analytic solution of the momentum constraints. One disadvantage of
conformal flatness is that Kerr black holes do not admit conformally
flat 3-metrics, at least for the simple time slicings we are aware of.
In \cite{Garat-Price:2000} it was shown that the Kerr metric is not
conformally flat at second order in the spin parameter $S/M^2$.
Indeed, in Sec.~\ref{sec:PN} we identify this deviation in our
results.

Because we assume that the initial hypersurface is maximal, the
momentum and Hamiltonian constraints decouple. We follow the Bowen and
York \cite{Bowen-York:1980} prescription to solve the momentum
constraint analytically.  Then we need only solve one
three-dimensional quasi-linear elliptic differential equation, the
Hamiltonian constraint.  It is solved on a so-called {\v C}ade{\v z}
grid using a multigrid algorithm\cite{Cook-Choptuik-etal:1993}. The constructed
data sets depend on several input parameters, namely the radii and the
positions of the throats of the holes in the flat background space,
$a_i$ and ${\bf C}_i$, $i=1,2$, respectively, and their linear momenta
and spins, ${\bf P}_i$ and ${\bf S}_i$, $i=1,2$, respectively. We note
that in this initial-data prescription, ${\bf P}_i$ and ${\bf S}_i$
represent the \emph{physical} linear and angular momentum of the black
hole if it is isolated. We work in the zero momentum frame, where
${\bf P}_2=-{\bf P}_1$, and choose ${\bf P}_i$ perpendicular to ${\bf
C}_2-{\bf C}_1$ in order to realize a circular orbit. Then the
magnitude $P\equiv P_1=P_2$ is sufficient to describe the linear
momenta. Choosing $a_1$ as the fundamental length scale, we are left
with the following dimensionless input parameters: the ratio of the
throat radii $\alpha=a_1/a_2$, the dimensionless background separation
$\beta=|{\bf C}_1-{\bf C}_2|/a_1$, and the dimensionless linear momentum
and spins, $P/a_1$ and ${\bf S}_i/a_1^2$, $i=1,2$, respectively.

From the initial data we can rigorously compute the ADM energy
$\Eadm$, the total angular momentum ${\bf J}$ and the proper
separation between the apparent horizons of each hole, $\ell$.  The
total angular momentum is evaluated as in Ref.~\cite{Cook:1994}:
\begin{equation}\label{eqn:L_cook}
{{\bf J}} \equiv \left({\bf C}_1 - {\bf O}\right)\times{\bf P}_1
		+ \left({\bf C}_2 - {\bf O}\right)\times{\bf P}_2
		+ {\bf S}_1 + {\bf S}_2.
\end{equation}
Here ${\bf O}$ represents the point about which the angular momentum is
defined; it drops out immediately because ${\bf P}_1=-{\bf P}_2$.  When
orbiting black holes have spin, neither the individual spins of the
holes nor their orbital angular momentum ${\bf L}$ are rigorously
defined. We simply take ${\bf L}$ to be defined by
\begin{equation}
\label{eqn:L_def}
	{\bf L} \equiv {\bf J} - {\bf S}_1 - {\bf S}_2,
\end{equation}
with ${\bf S}_1$ and ${\bf S}_2$ defining the individual spins.

Finally, we need to define the masses of the individual holes. 
As in Ref.~\cite{Cook:1994}, we define the mass of each hole
via the Christoudoulou formula:
\begin{equation}\label{eqn:M}
M_i^2=M_{ir,i}^2+\frac{S_i^2}{4M_{ir,i}^2},
\end{equation}
\begin{equation}\label{eqn:M_ir}
M_{ir,i}^2=\frac{A_i}{16\pi},
\end{equation}
where $A_i$ is the area of the event horizon of the $i^{\mbox{th}}$
hole.  Clearly this definition is only rigorous for a stationary
spacetime.  Moreover, we cannot locate the event horizon from the
initial data slice alone. Therefore we must resort to using the
apparent horizons areas in equations (\ref{eqn:M}) and
(\ref{eqn:M_ir}) instead. Apparent horizons can be determined from
initial data and in the present case their positions are known to
coincide with the throats of the holes \cite{Cook:1991}.
For a stationary spacetime, apparent horizons and event horizons
coincide, and in a general, well-behaved spacetime, the event horizon
must coincide with or lie outside of the apparent horizon.  In the
latter case we will underestimate the mass of the black hole by using
the apparent horizon area. Some of the results of this work indicate
that this happens for very small separations of the holes.

With the individual masses we can finally define the {\em effective
potential} as the non-dimensional binding energy of the system:
\begin{equation}\label{eqn:Eb}
\frac{E_b}\mu \equiv (\Eadm-M_1-M_2)/\mu.
\end{equation}
Since the mass-ratio $M_1/M_2$ is kept constant during the minimization,
minimizing $E_b/\mu$ is equivalent to minimizing $\Eadm/m$.

We construct initial data sets starting from the {\em input}
parameters $\alpha$, $\beta$, $P/a_1$ and ${\bf S}_i/a_1^2$, and
compute the {\em physical} parameters $E_b/\mu$, $M_1/M_2$, $J/\mu m$
and ${\bf S}_i/M_i^2$. In order to construct an initial data set with
certain physical parameters we have to choose the input parameters
appropriately. This requires nonlinear rootfinding.

Within our effective potential approach, we will search for minima in
the binding energy as a function of the separation of the black holes.
Fortunately, it is not necessary to solve for a specific proper
separation $\ell/m$. It is sufficient to keep $\beta$ constant during
rootfinding and thus find a binary black hole configuration with some
separation $\ell/m$.  Our goal is to solve the following set of
equations [cf. Eqns.~(10a-d) of Ref.~\cite{Cook:1994}]:
\begin{subequations}
\label{eqn:roots}
\begin{eqnarray}
\label{eqn:X_root}
	\frac{M_1}{M_2}&=&\left[\frac{M_1}{M_2}\right]\\
\label{eqn:S1_root}
	\frac{S_1}{M_1^2}&=&\left[\frac{S_1}{M_1^2}\right]\\
\label{eqn:S2_root}
	\frac{S_2}{M_2^2}&=&\left[\frac{S_2}{M_2^2}\right]\\
\label{eqn:J_root}
	\frac{J}{\mu m}&=&\left[\frac{J}{\mu m}\right].
\end{eqnarray}
\end{subequations}
The bracketed quantities on the right hand sides of
(\ref{eqn:X_root}-\ref{eqn:S2_root}) denote the physical values to be
reached, and the expressions on the left-hand side represent functions
of the background parameters $\alpha$, $P/a_1$, $S_1/a_1^2$ and
$S_2/a_1^2$ as well as the fixed $\beta$.

For non-rotating holes, equations (\ref{eqn:S1_root}) and
(\ref{eqn:S2_root}) are trivially satisfied by $S_1=S_2=0$. For
spinning holes this is no longer the case.  Hence, it seems one has to
solve the complete set of Eqns.~(\ref{eqn:X_root}--\ref{eqn:J_root}).
However, in any initial data scheme where the physical spins of the
black holes are directly parameterized, Eqns.~(\ref{eqn:S1_root}) and
(\ref{eqn:S2_root}) can be eliminated. First, we note again that if
the physical spins are directly parameterized, from
Eqn.~(\ref{eqn:L_def}) we find that we can replace rootfinding in
$J/\mu m$ by rootfinding in $L/\mu m$. Thus Eqn.~(\ref{eqn:J_root}) is
replaced by
\begin{equation}\label{eqn:L_root}
\frac{L}{\mu m}=\left[\frac{L}{\mu m}\right].
\end{equation}
In the zero momentum frame, Eqns.~(\ref{eqn:L_cook}) and
(\ref{eqn:L_def}) simplify to
\begin{equation}\label{eqn:L}
\frac{L}{a_1^2} = \beta \frac{P}{a_1}.
\end{equation}
Thus we can rewrite $S_1$ as
\begin{equation}\label{eqn:S1}
\frac{S_1}{a_1^2} =\frac{S_1}{M_1^2}\cdot\frac{M_1}{M_2}
	\cdot\frac{M_1M_2}{L}
	\cdot \beta \frac{P}{a_1}.
\end{equation}
For a solution of
Eqns.~(\ref{eqn:X_root}--\ref{eqn:S2_root},\ref{eqn:L_root}), the
first three terms on the right hand side of (\ref{eqn:S1}) take the
values of the desired physical parameters, so we can replace them by
these parameters throughout the rootfinding.  A similar result holds
for $S_2$.  We perform only two-dimensional rootfinding, in $\alpha$
and $P/a_1$, and set in each iteration
\begin{subequations}
\begin{eqnarray}
\label{eqn:S1_soln}
	\frac{S_1}{a_1^2} &=& \left[\frac{S_1}{M_1^2}\right]
		\left[\frac{M_1}{M_2}\right]
		\left[\frac{L}{\mu m}\right]^{-1} \beta \frac{P}{a_1},\\
\label{eqn:S2_soln}
	\frac{S_2}{a_1^2} &=& \left[\frac{S_2}{M_2^2}\right]
		\left[\frac{M_1}{M_2}\right]^{-1}
		\left[\frac{L}{\mu m}\right]^{-1} \beta \frac{P}{a_1}.
\end{eqnarray}
\end{subequations}

For an important subset of spin configurations, even one-dimensional
rootfinding is sufficient as can be seen as follows: Consider
equal-sized holes with equal spin magnitudes on both holes.  If both
spins are parallel to the orbital angular momentum, or both spins are
antiparallel, there exists a symmetry under exchange of the two
holes. Therefore $\alpha$ must be equal to $1$ and we are left with
one free parameter, $P/a_1$. If one spin is parallel to the orbital
angular momentum and the other spin is antiparallel, however, this
property is lost.  One hole is co-rotating with the orbital motion and
the other hole is counter-rotating. The choice $\alpha=1$ would result
in holes with slightly different masses. We thus need two-dimensional
rootfinding in $\alpha$ and $P/a_1$ for this case.

Each ``function evaluation'' for the rootfinding involves the
computation of an initial data set $(\gamma_{ij}, K_{ij})$. High
resolution solutions take between 30 minutes and several hours of CPU
time on one RS6000 processor.  For maximum efficiency, we first
perform rootfinding with a Newton-Raphson method\cite{NumericalRecipes} on low
resolution data sets.  The numerical values for $M_1/M_2$ and $J/\mu
m$ differ slightly between low resolution and high resolution
solutions, therefore we solve on low resolution for adjusted values of
$[M_1/M_2]$ and $[J/\mu m]$.  With the input parameters found in the
low resolution rootfinding, a high resolution computation is performed
to verify that equations (\ref{eqn:X_root}) and (\ref{eqn:J_root}) are
indeed satisfied at high resolution, and to adjust the offset used in
the next low resolution rootfinding. If necessary, this procedure is
repeated. On average each complete rootfinding takes fewer than two
high resolution computations.

Following our prescription, we now minimize the binding energy with
respect to separation while keeping $M_1/M_2$, $L/\mu m$ and ${\bf
S}_i/M_i^2$ constant.  The binding energy of a sequence of solutions
with these quantities held constant represents a contour of the
effective potential.  Our code starts at large separation $\beta$ and
reduces $\beta$ until a minimum in $E_b/\mu$ is bracketed.  Then the
minimum is located with Brent's method \cite{NumericalRecipes}, yielding a
quasi-circular orbit for the prescribed values of $J/\mu m$,
$M_1/M_2$, and ${\bf S}_i/M_i^2$.  Note that each computation of
$E_b/\mu$ during the minimization along an effective potential contour
requires rootfinding.

By computing quasi-circular orbits for different $J/\mu m$, but fixed
$M_1/M_2$ and ${\bf S}_i/M_i^2$, a {\em sequence} of quasi-circular
orbits is obtained. A binary black hole that radiates away energy and
angular momentum will follow such a sequence approximately, assuming
that the spin on each hole remains constant.  We step towards smaller
$J/\mu m$, and compute only as many points along each effective
potential contour as are required for the minimization.  As soon as we
do not find a minimum in the effective potential contours anymore we
expect to be beyond the innermost stable circular orbit.  We trace out
some complete effective potential contours around the last value of
$J/\mu m$ to check the behavior of these curves.

Finally, from the binding energy $E_b/\mu$ and the angular
momentum $J/\mu m$ along the sequence, we compute the orbital angular
frequency as
\begin{equation}\label{eqn:Omega}
\Omega=\left.\frac{\partial E_b}{\partial J}
\right|_{\mbox{\footnotesize{sequence}}}
\end{equation}

\section{Results}
\label{sec:Results}
The parameter space of spinning binary black holes is large -- one
can vary the mass ratio of the holes as well as spin directions and
magnitudes.  Astrophysically most interesting are holes that co-rotate
with the orbital motion, i.e.\ with both spins ${\bf S}_i$ parallel to
the orbital angular momentum ${\bf L}$. In addition to these co-rotating
configurations, we examine configurations with one co-rotating hole
and one counter-rotating hole, and configurations with two
counter-rotating holes. We have the following three families of
sequences:
\begin{itemize}
\item The ``\pps\ sequences'' with two co-rotating holes.
\item The ``\pms\ sequences'' with one co-rotating and one counter-rotating
	 hole.
\item The ``\mms\ sequences'' with two counter-rotating holes.
\end{itemize}

We restrict ourselves to equal mass holes, $M_1=M_2\equiv M$ with
equal spin magnitudes $S_1=S_2\equiv S$.  As we will see, the
assumption of conformal flatness becomes questionable at high spins,
so we consider only spin magnitudes $S/M^2\le 0.50$.  We denote a spin
configuration by two plus or minus signs together with a number
specifying the spin magnitude on the holes. Thus ``$\pps0.25$''
denotes a configuration with two co-rotating holes and spin magnitudes
$S_1/M^2=S_2/M^2=0.25$.

\begin{figure}[tb]
\centerline{\epsfxsize=3.75in\epsffile{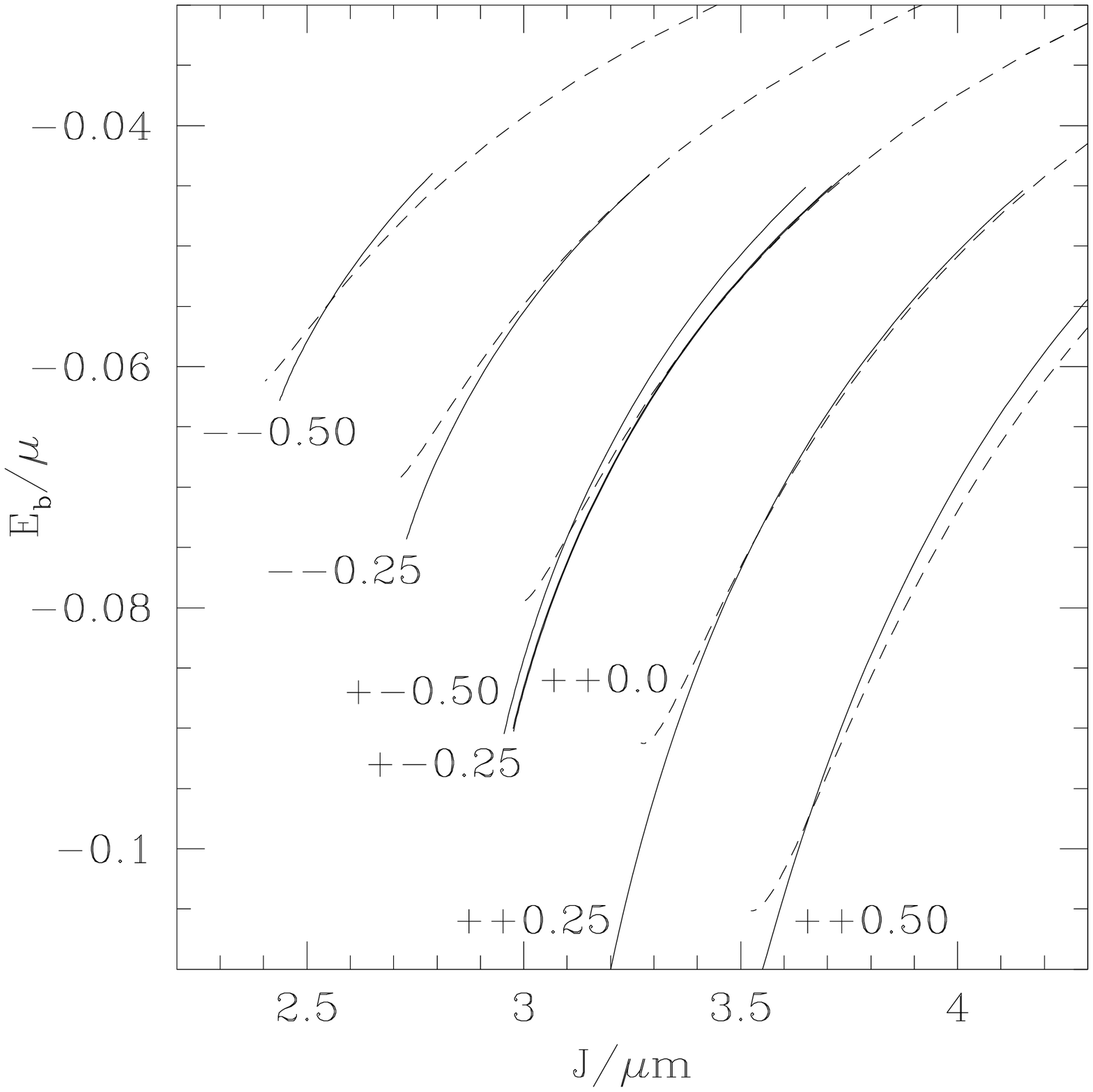}}
\CAP{Sequences of quasi-circular orbits for different spin
configurations: binding energy $E_b/\mu$ vs. angular momentum $J/\mu
m$}{\label{fig:seq_EbJ}Sequences of quasi-circular orbits for
different spin configurations.  Plotted is the binding energy
$E_b/\mu$ vs. the angular momentum $J/\mu m$ along the sequences.  The
solid lines represent the data, the dashed lines are the results based
on $\mbox{(post)}^2$-Newtonian theory.  As discussed later in this
paper, the effective potential method could not locate an ISCO for the
$\pps0.25$ and $\pps0.50$ sequences, although we believe each sequence
should terminate in one. }
\end{figure}

Quasi-circular orbits were computed for various values of $J/\mu m$
along each sequence. In Fig.~\ref{fig:seq_EbJ} the binding energy
$E_b/\mu$ along each sequence is plotted as a function of the angular
momentum $J/\mu m$. A binary black hole that loses energy and angular
momentum through gravitational radiation moves along such a sequence
if the spins of the individual holes remain constant.  The dashed
lines in Fig.~\ref{fig:seq_EbJ} represent the results of
$(\mbox{post})^2$-Newtonian theory which we describe in Sec.~\ref{sec:PN}.

\begin{figure}
\centerline{\epsfxsize=3.75in\epsffile{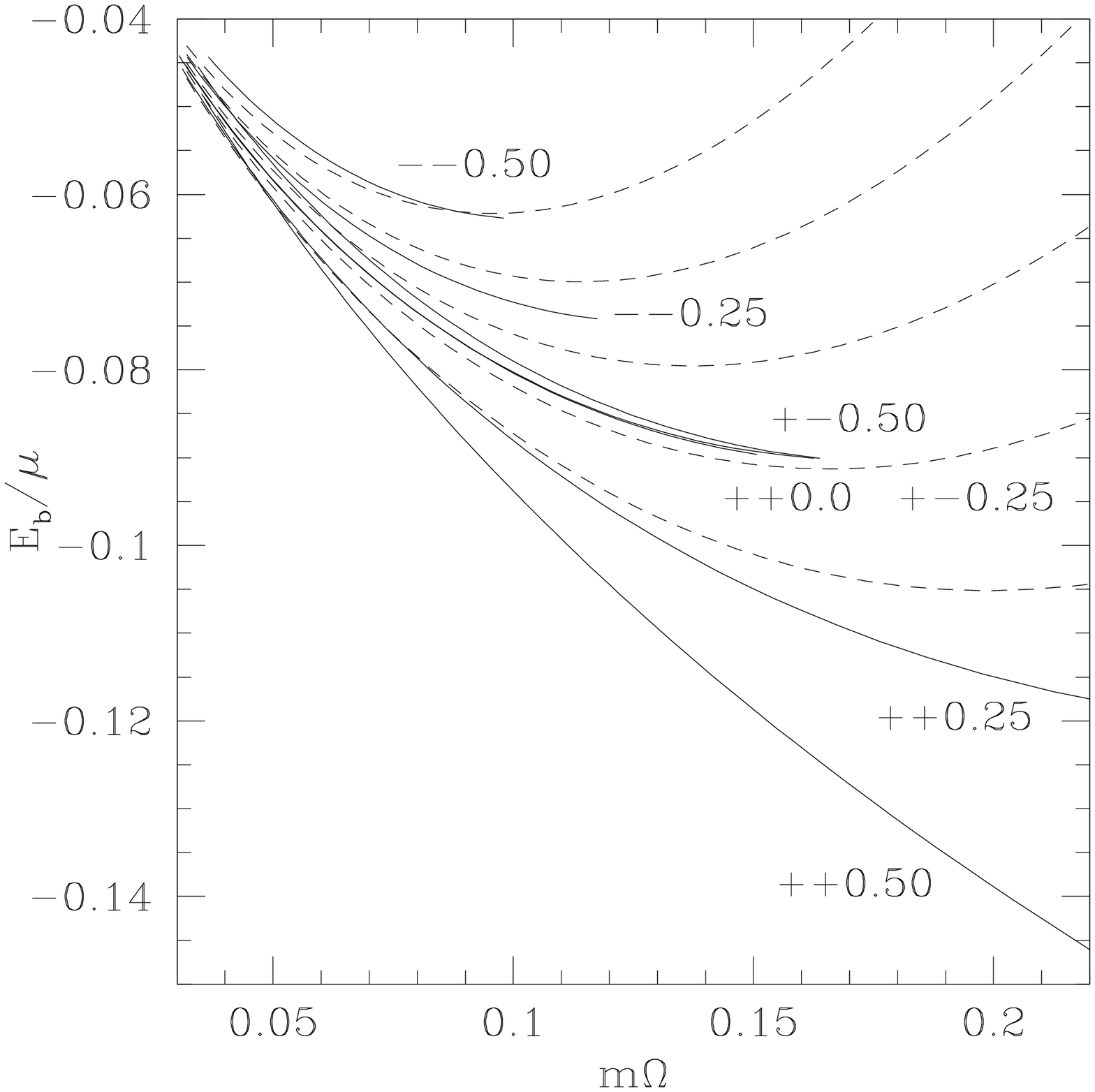}}
\CAP{Sequences of quasi-circular orbits for different spin
configurations: binding energy $E_b/\mu$ vs. orbital angular frequency
$m\Omega$}{\label{fig:seq_EbOmg}Sequences of quasi-circular orbits
for different spin configurations.  Plotted is the binding energy
$E_b/\mu$ vs. the orbital angular frequency $m\Omega$ along the
sequences.  The solid lines represent the data, the dashed lines are
the results based on $\mbox{(post)}^2$-Newtonian theory.  As discussed
later in this paper, the effective potential method could not locate
an ISCO for the $\pps0.25$ and $\pps0.50$ sequences, although we
believe each sequence should terminate in one. }
\end{figure}

\begin{figure}[bt]
\centerline{\epsfxsize=3.75in\epsffile{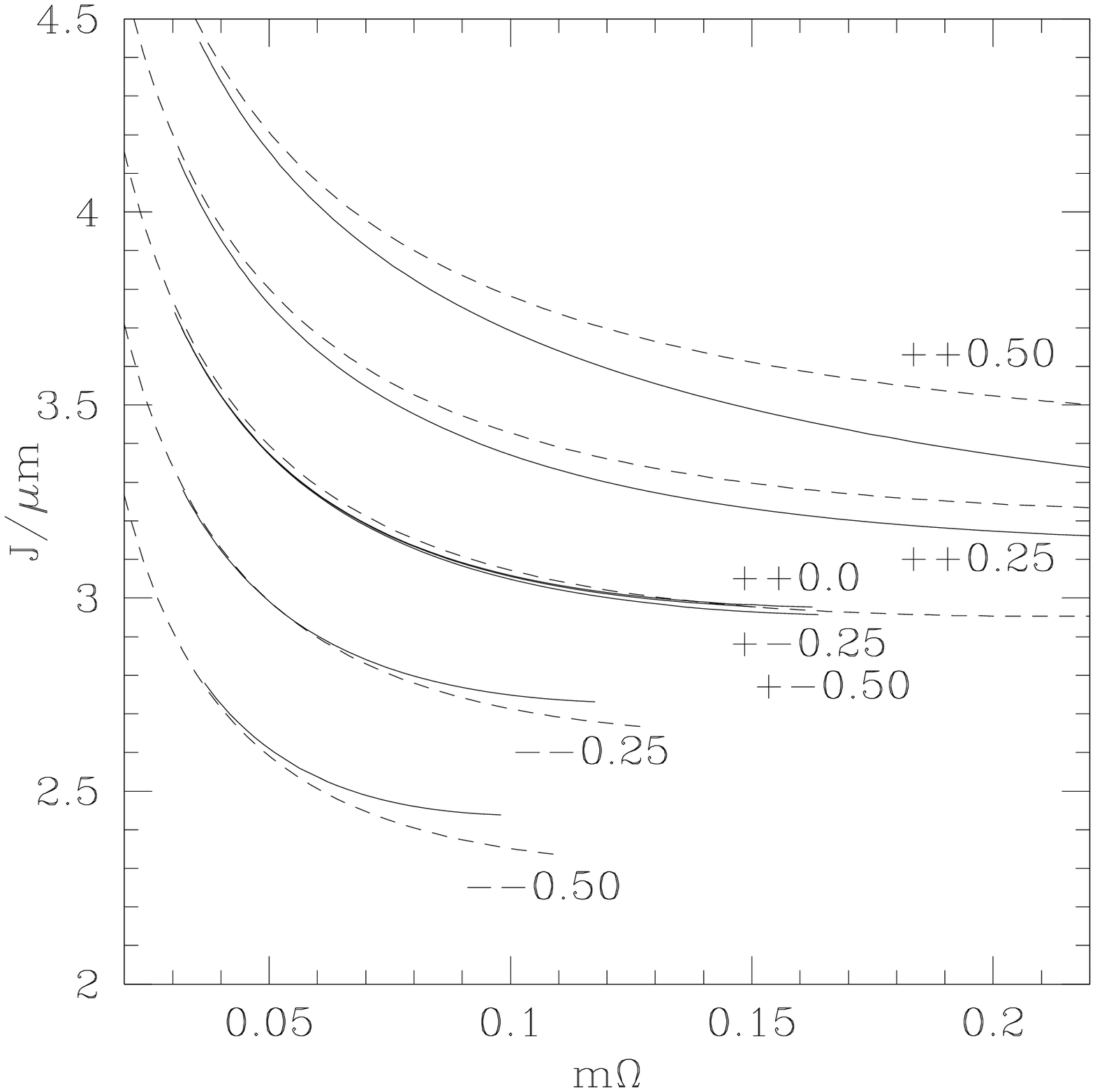}}
\CAP{Sequences of quasi-circular orbits for
different spin configurations: angular momentum $J/\mu
m$ vs. orbital angular frequency $m\Omega$}
{\label{fig:seq_JOmg}Sequences of quasi-circular orbits for
different spin configurations.  Plotted is the angular momentum $J/\mu
m$ vs. the orbital angular frequency $m\Omega$ along the sequences.
The solid lines represent the data, the dashed lines are the results
based on $\mbox{(post)}^2$-Newtonian theory.  As discussed later in
this paper, the effective potential method could not locate an ISCO
for the $\pps0.25$ and $\pps0.50$ sequences, although we believe each
sequence should terminate in one. }
\end{figure}

Using equation (\ref{eqn:Omega}) we compute the orbital angular
frequency.  In Figs.~\ref{fig:seq_EbOmg} and \ref{fig:seq_JOmg}, the
binding energy and the angular momentum along the sequences are plotted
as a function of orbital frequency.

\subsection{Behavior at large separations}
\label{sec:PN}

We compare our results to the $\mbox{(post)}^2$-Newtonian expansions
for spinning holes in quasi-circular orbit that were kindly provided
by L.\ Kidder.  The expressions for arbitrary spins and masses are
lengthy.  If one restricts attention to equal-mass holes, $M_1=M_2=M$,
$m=2M$, $\mu=M/2$, it turns out that only the {\em sum} of the spins
enters the $\mbox{(post)}^2$-Newtonian expansions.  In terms of
\begin{equation}\label{eqn:s_defn}
{\bf s}\equiv\frac{{\bf S}_1+{\bf S}_2}{M^2},
\end{equation}
and with $\hat{\bf L}$ being the unit-vector parallel to ${\bf L}$,
the $\mbox{(post)}^2$-Newtonian expansions become
\begin{subequations}
\begin{align}
\frac{E_b}{\mu}\label{eqn:PN_Eb}
&=-\frac{1}{2}\omg^{2/3}
\Bigg\{1-\frac{37}{48}\omg^{2/3}+\frac{7}{6}(\hat{\bf L}\cdot{\bf s})\omg\\
&\hspace*{3.2cm}
-\left(\frac{1069}{384}+\frac{1}{8}
		\left[3(\hat{\bf L}\cdot{\bf s})^2-{\bf s}^2\right]
	\right)\omg^{4/3}
\Bigg\}, \nonumber
\\
\left(\frac{J}{\mu m}\right)^2\!\!
\label{eqn:PN_J}
&=\omg^{-2/3}
\Bigg\{1+2(\hat{\bf L}\cdot{\bf s})\omg^{1/3}
+\left(\frac{37}{12}+{\bf s}^2\right)\omg^{2/3} \\
&\hspace{2.5cm   }\mbox{}
+\frac{1}{6}(\hat{\bf L}\cdot{\bf s})\omg
	+\left(\frac{143}{18}-\frac{37}{24}(\hat{\bf L}\cdot{\bf s})^2
		-\frac{7}{8}{\bf s}^2\right)
\omg^{4/3}\Bigg\}.\nonumber
\end{align}
\end{subequations}
\noindent
These expressions are plotted in
Figs.~\ref{fig:seq_EbJ}--\ref{fig:seq_JOmg} together with our results
from the effective potential method.  There is remarkable agreement.

The sum ${\bf S}_1+{\bf S}_2$ is zero for all \pms\ sequences with equal
spin magnitudes, so $\mbox{(post)}^2$-Newtonian theory predicts that
the \pms\ sequences are identical to the non-rotating sequence.  This
is remarkable, and indeed, in
Figs.~\ref{fig:seq_EbJ}--\ref{fig:seq_JOmg} the \pms\ sequences are
close to the $\pps0.0$ sequence.  However, a closer look reveals a
systematic behavior from which we can gain some insight into our
assumptions. For fixed angular momentum $J/\mu m$, consider the
difference in binding energy between a point on a \pms\ sequence
and a point on the non-rotating $0.0$ sequence,
\begin{equation}\label{eqn:DeltaE}
\Delta E_b/\mu(S)=\frac{E_b}{\mu}(\pms{S})-\frac{E_b}{\mu}(0).
\end{equation}

\begin{figure}[bt]
\centerline{\epsfxsize=3.5in\epsffile{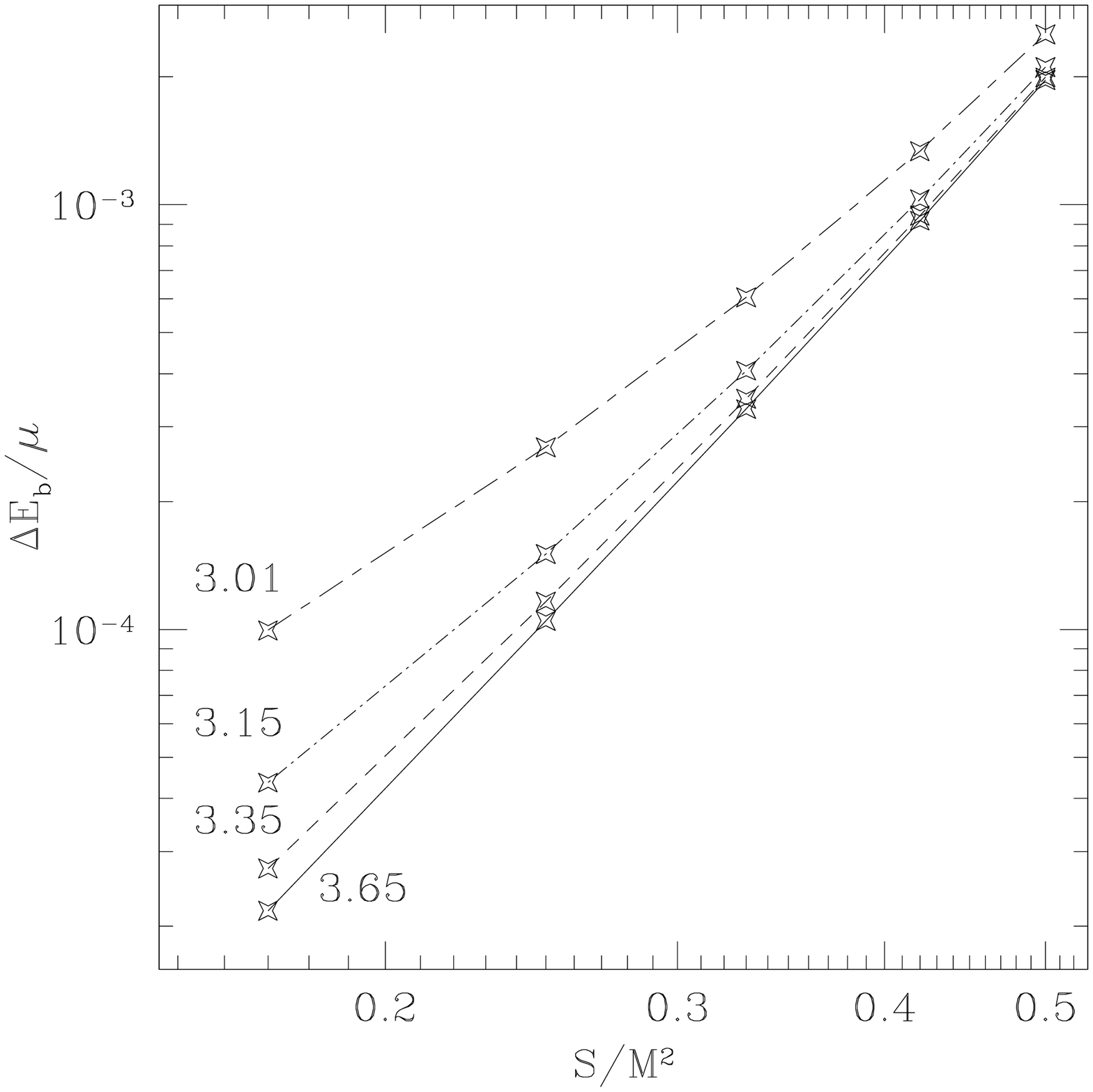}}
\CAP{Difference in binding energy $\Delta
E_b/\mu$ between \pms\ sequences and non-rotating sequence as a
function of spin of the \pms\ sequence for fixed angular momentum
$J/\mu m$
}
{\label{fig:spin4}Difference in binding energy $\Delta
E_b/\mu$ between \pms\ sequences and non-rotating sequence as a
function of spin of the \pms\ sequence for fixed angular momentum
$J/\mu m$.  Each curve is labeled by its value of $J/\mu m$.  $J/\mu
m=3.01$ is very close to the ISCOs that have $J/\mu m\approx 2.98$.
$J/\mu m=3.65$, $3.35$, $3.15$ and $3.01$ correspond to a separation
of $\ell/m \approx 12.3$, $9.6$, $7.7$ and $6.1$, respectively. }
\end{figure}

In Fig.~\ref{fig:spin4}, $\Delta E_b/\mu(S)$ is plotted as a function
of spin for several values of angular momentum $J/\mu m$.  $\Delta
E_b/\mu$ varies as the fourth power of spin.  This might be a physical
effect beyond $\mbox{(post)}^2$-Newtonian expansions, but for the
following reason it seems likely that one of our assumptions
introduces a non-physical contribution to $\Delta E_b/\mu$, too.
Figure~\ref{fig:spin4} strongly suggests that $\Delta E_b/\mu$ is
converging to a non-zero value as $J/\mu m$ (and thus separation)
increase, indicating
that there is a contribution to $\Delta E_b$ that is independent of
the separation of the holes.  For all spin configurations, $E_b$ must
approach zero in the limit of large separation, therefore any physical
contribution to $\Delta E_b$ should decrease with separation.
Moreover, a coupling between the holes, physical or unphysical, will
give rise to a separation-dependent contribution to $\Delta
E_b/\mu$. Therefore the separation-independent contribution must be a
non-physical effect due to properties of each {\em isolated} hole.  A
likely candidate is the underlying assumption of conformal flatness.
At large separations each hole should resemble a Kerr black hole,
which is {\em not} conformally flat.

Since the Kerr metric is the unique stationary state for a spinning
black hole, if the conformally flat initial data for a single hole
were evolved, the metric would relax to the Kerr metric and emit some
gravitational radiation.  Therefore the total energy contained in our
initial data slices is larger than in a more faithful conformally
non-flat data slice and $\Delta E_b/\mu$ should be positive, which it
indeed is.

We conclude that at large separations $\Delta E_b$ is contaminated by
an unphysical contribution because of the conformal flatness
assumption.  At small separation there might be additional physical
contributions beyond the $(\mbox{post})^2$-Newtonian order.

\subsection{Behavior at small separations -- ISCO}
\label{sec:results:ISCO}

In this section we report the key results of this work -- the
spin dependence of the innermost stable circular orbit. As we will
see, the interpretation of our data at small separations is somewhat
complicated. At large separations, the assumptions and approximations
we have used are reasonable, except for the assumption of conformal
flatness when the holes are spinning.  At small separations, the
interaction between the two black holes becomes relatively strong,
and our approximations begin to break down. Near the ISCO, we must
evaluate the quality of our assumptions to determine how reliable our
results are.

\begin{figure}
\centerline{\epsfxsize=5.5in\epsffile{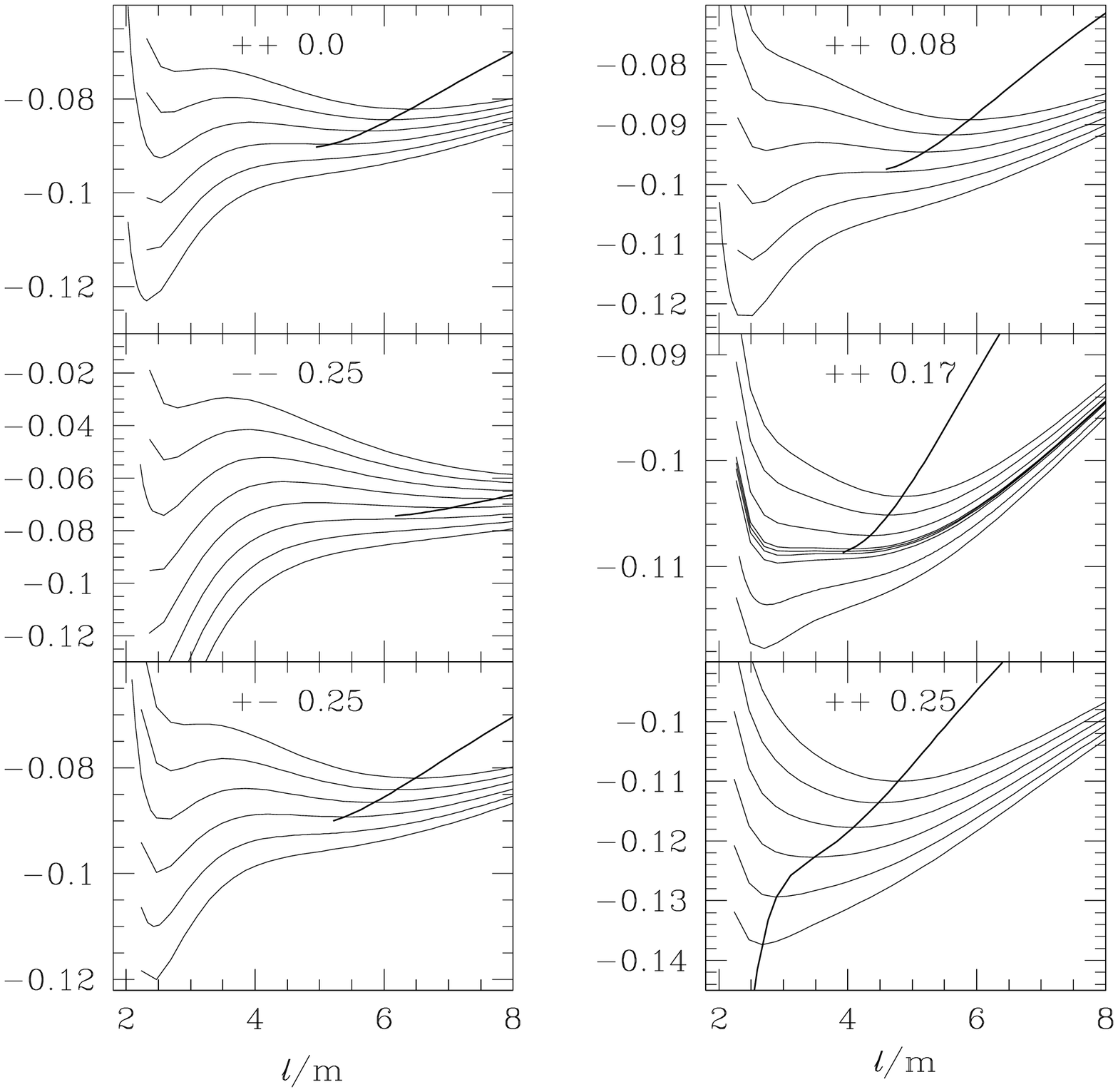}}
\CAP{Constant $J/\mu{m}$ contours of the
effective potential $E_b/\mu$ as a function of separation $\ell/m$ for
various spin configurations
}
{\label{fig:Eb_seq}Constant $J/\mu{m}$ contours of the
effective potential $E_b/\mu$ as a function of separation $\ell/m$ for
various spin configurations. The curves are spaced in steps of
$\Delta{J}/\mu{m}=0.02$ except for the $\mms0.25$ and the $\pps0.17$
configurations, which have steps of $0.04$ and $0.01$,
respectively. Also plotted is the sequence of quasi-circular orbits
connecting the minima of the effective potential. }
\end{figure}

In the neighborhood of each tentative ISCO, we compute a set of
complete effective potential contours. These are shown in
Fig.~\ref{fig:Eb_seq}. In each plot, the binding energy $E_b/\mu$
is shown as a function of separation $\ell/m$ for several different
values of angular momentum $J/\mu m$. Also plotted is the sequence of
quasi-circular orbits passing through the minima of the effective
potential.  Figure~\ref{fig:Eb_seq} shows the non-rotating sequence
$\pps0.0$, one example each of a \mms\ and a \pms\ sequence, and three
\pps\ sequences with different spin magnitudes.

Examining the constant $J$ contours of the effective potential for
{\em fixed} spin configurations, we find that they fall into three
regimes separated by critical values that we will label $J_A$ and
$J_B$. Contours with $J>J_A$ exhibit a single minimum positioned at
large separation $\ell/m$. This minimum moves inward as the angular
momentum decreases, i.e.\ the holes approach each other as angular
momentum and energy are radiated away.  We call this the ``outer''
minimum. As $J$ passes through the critical value $J_A$, a new
``inner'' minimum appears inside the outer minimum. In this region,
contours of the effective potential have two minima separated by a
local maximum. The maximum corresponds to the well known unstable
circular orbit of a Schwarzschild black hole. As $J$ decreases
further, $J_A>J>J_B$, the maximum moves outward whereas the outer
minimum continues to move inward -- the quasi-circular orbit associated
with the outer minimum continues to shrink. As $J$ passes through the
second critical value $J_B$ the outer minimum and the maximum meet in
an inflection point and disappear. The quasi-circular orbit associated
with the outer minimum disappears and this inflection point is
identified with the ISCO.  For $J<J_B$, only the inner minimum
remains.

This behavior for the non-rotating sequence was already found in
\cite{Cook:1994}. There, the inner minimum was dismissed as
unphysical, since the underlying assumptions become weaker at small
separations of the holes, and since a common event horizon might form.
We will discuss this ``unphysical'' region and the possibility and
consequences of the formation of a common event horizon below.
But first we continue discussing the behavior of the effective
potential for different spin configurations. 

As we increase the spin magnitude for the \mms\ configurations, the two
critical angular momentum values $J_A$ and $J_B$ move away from each
other. We see a more pronounced local maximum and the $E_b$ curves
look similar to the effective potential of Schwarzschild for a larger
interval of angular momenta. The ISCO moves outward to larger
separations as spin increases.

\begin{figure}
\centerline{\epsfxsize=3.5in\epsffile{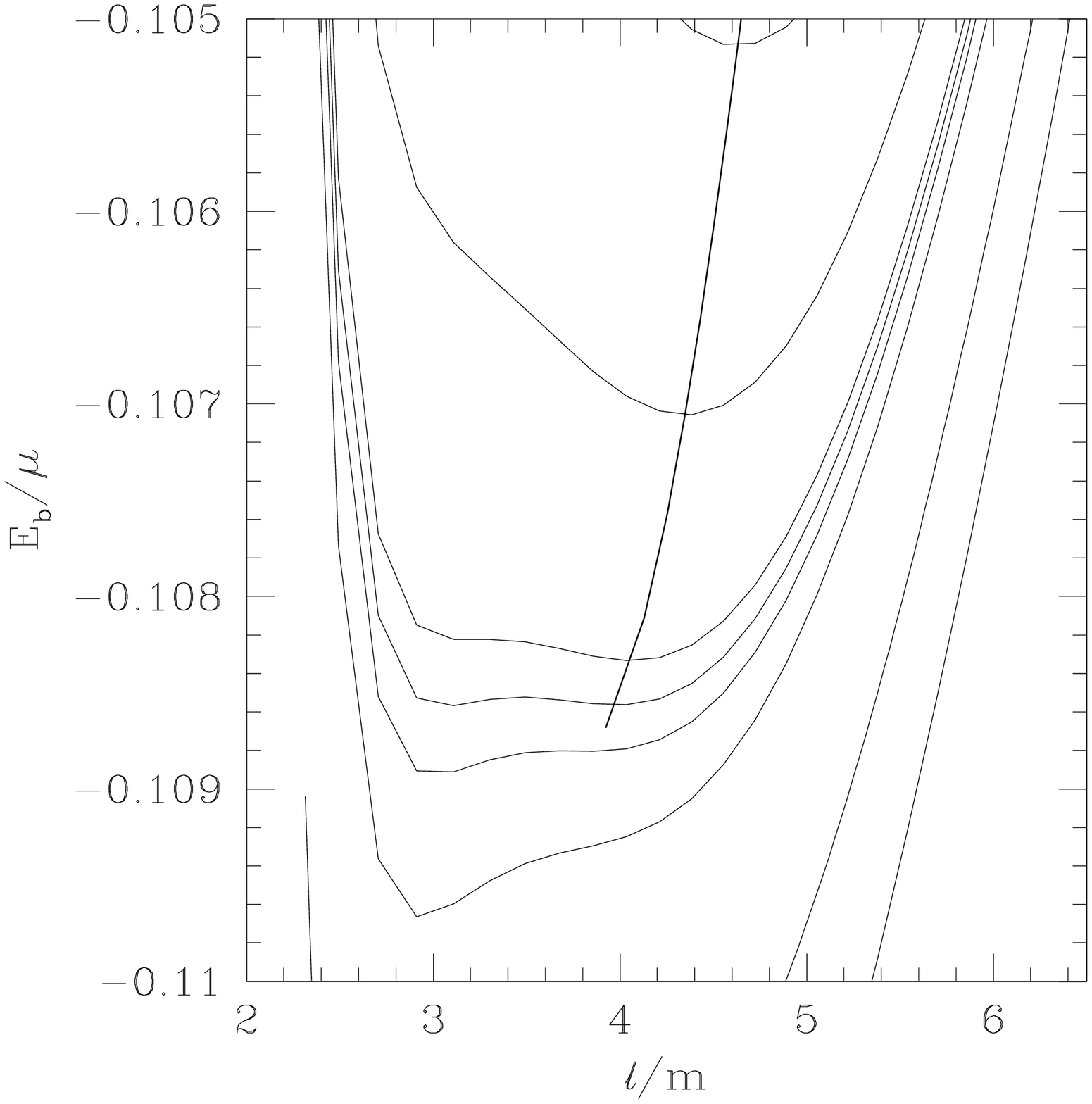}}
\CAP{
Enlargement of the $\pps0.17$
sequence of Fig.~\ref{fig:Eb_seq}
}
{\label{fig:PP017enlarge}Enlargement of the $\pps0.17$
sequence of Fig.~\ref{fig:Eb_seq}.  The displayed effective potential
contours (top to bottom) correspond to angular momenta $J/\mu m=3.12$,
$3.11$, $3.104$, $3.103$, $3.102$, $3.10$, $3.09$ and $3.08$. Also
shown is the sequence of quasi-circular orbits.}
\end{figure}

Conversely, as we increase the spin magnitude for the \pps\
configurations the interval $(J_B, J_A)$, where two minima and a local
maximum exist becomes smaller.  Slightly above $S/M^2=0.17$, $J_A$ and
$J_B$ merge and for $S/M^2\gtrsim0.17$, the regime with two minima and a
maximum is {\em not} present.  Figure~\ref{fig:PP017enlarge}
illustrates the small interval $(J_B, J_A)$ with an enlargement of the
$\pps0.17$ sequence.  As long as the regime with two minima and a
maximum is present, we can still define the ISCO by the inflection
point. It moves towards smaller separation of the holes as the spin is
increased.  However, since the inflection point ceases to exist at
some spin magnitude, we cannot define an ISCO for all $S/M^2$.
Therefore the \pps\ sequences displayed in
Figs.~\ref{fig:seq_EbJ}--\ref{fig:seq_JOmg} do not terminate.
Furthermore, we need a more careful analysis to determine whether the
ISCO properties for spin magnitudes close to the critical value
$S/M^2\approx 0.17$ are reliable.

The \pms\ configurations are very similar to the non-rotating
one. Given the weak dependence on spin within the \pms\ sequences, this
is not surprising.  We do not consider the \pms\ configurations
further.

\begin{figure}
\centerline{\epsfxsize=4in\epsffile{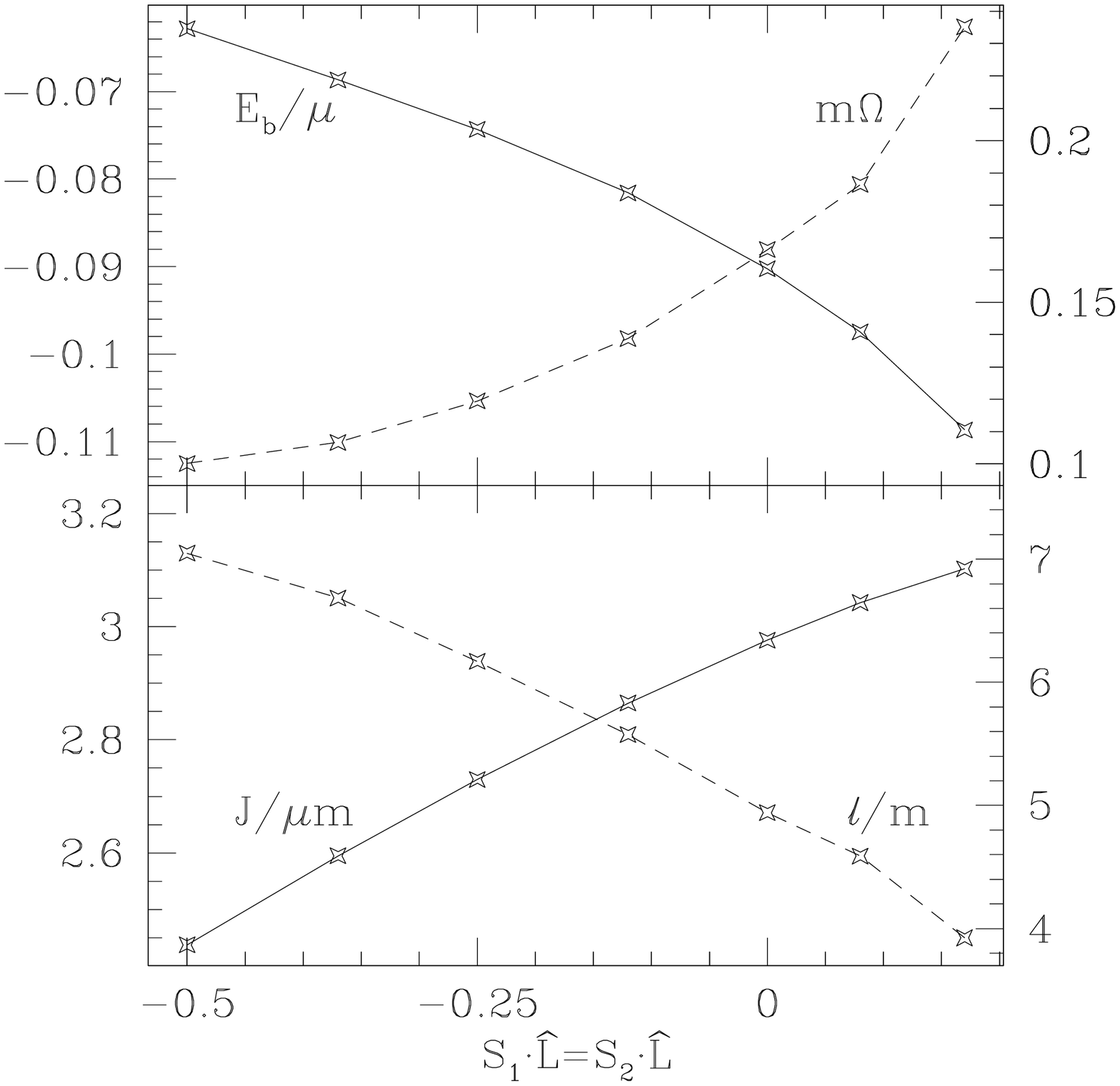}}
\CAP{
Values of several physical parameters at the
ISCO of the \pps\ and \mms\ sequences.
}{\label{fig:ISCO}Values of several physical parameters at the
ISCO of the \pps\ and \mms\ sequences.  Plotted are the binding energy
$E_b/\mu$, the orbital angular frequency $m\Omega$, the total angular
momentum $J/\mu m$ and the proper separation between the holes,
$\ell/m$ as a function of spin $S/M^2$ on the holes. The \pps\
sequences are plotted along the positive part of the horizontal axis,
the \mms\ sequences along the negative part as $-S/M^2$. The vertical
axes on the left side belong to $E_b/\mu$ and $J/\mu m$. }
\end{figure}

\begin{table}
\CAP{
Orbital parameters of the innermost stable
circular orbit for equal-mass spinning holes
}{\label{tab:ISCO}Orbital parameters of the innermost stable
circular orbit for equal-mass spinning holes. The second through sixth
columns give the data obtained in this work, the three columns to the
right give the data for a test mass orbiting a Kerr black hole.  The
results for the $\pps0.08$ and $\pps0.17$ sequences
will have larger systematic errors than the other cases (see text).}
\begin{tabular}{d{-1}|d{2}d{4}d{3}d{4}d{3}|d{5}d{4}d{5}}
\multicolumn{1}{c|}{Sequence} & 
\multicolumn{1}{c}{$\ell/m$}   & 
\multicolumn{1}{c}{$E_b/\mu$}  &  
\multicolumn{1}{c}{$m\Omega$}  &  
\multicolumn{1}{c}{$J/\mu m$}  & 
\multicolumn{1}{c|}{$L/\mu m$}  &
\multicolumn{1}{c}{$E_b/\mu$}  & 
\multicolumn{1}{c}{$L/\mu m$}  & 
\multicolumn{1}{c}{$m\Omega$}
\\
\tableline
\mms0.50 & 7.05 & -0.0628 & 0.100 & 2.438 & 3.438 & -0.04514 & 3.8842 
& 0.04935\\
\mms0.37 & 6.68 & -0.0687 & 0.107 & 2.595 & 3.335 & -0.04767 & 3.7834 
& 0.05319\\
\mms0.25 & 6.17 & -0.0743 & 0.120 & 2.730 & 3.230 & -0.05032 & 3.6856 
& 0.05727\\
\mms0.12 & 5.58 & -0.0815 & 0.139 & 2.865 & 3.105 & -0.05363 & 3.5738 
& 0.06242\\
\pps0.0   & 4.94 & -0.0901 & 0.166 & 2.976 & 2.976 & -0.05719 & 3.4641 
& 0.06804\\
\pps0.08  & 4.59 & -0.0975 & 0.186 & 3.042 & 2.882 & -0.05991 & 3.3870 
& 0.07237\\
\pps0.17  & 3.93 & -0.1087 & 0.235 & 3.103 & 2.763 & -0.06337 & 3.2957 
& 0.07793\\
\end{tabular}
\end{table}

Figure~\ref{fig:ISCO} and Table~\ref{tab:ISCO} summarize the orbital
parameters at the ISCO as a function of spin for the \mms\ sequences
and the \pps\ sequences.  The numerical errors in $E_b/\mu$, $L/\mu m$ and
$J/\mu m$ are less than 1 per cent, while $m\Omega$ and $\ell/m$ are
accurate to a few percent. However, for the \pps\ sequences the
systematic errors of our approach might be much larger.  The table
also includes ISCO parameters for a test mass orbiting a Kerr black
hole obtained from formulas in \cite{Bardeen-Press-Teukolsky:1972}.

\subsection{Common apparent horizons}

A common event horizon might be responsible for the strange
behavior of the effective potential at small separations, because once
a common event horizon forms, there are no longer two distinct
black holes.  It would be helpful to know the critical separation
where a common event horizon first forms.  However, in order to
locate the event horizon, knowledge of the complete spacetime is
needed. In the present case, only data on one time-slice is available,
and so we can only search for common apparent horizons. Since
the event horizon must lie outside the apparent horizon, the formation
of a common apparent horizon places a firm bound on the
formation of an event horizon.
 
\begin{table}\CAP{
Summary of common apparent horizon searches}{\label{tab:AH}Summary of
the common apparent horizon searches.  Listed are the sequences and
values of orbital angular momentum for which an apparent horizon
search was carried out.  The apparent horizon was found to form at a
separation $\ell_1/m<\ell/m<\ell_2/m$.  {\footnotesize(a): From
\cite{Cook-Abrahams:1992}, which found a critical separation
$\beta=4.17$. This corresponds to a proper separation of
$\ell/m\approx 1.89$.  }}
\centerline{\begin{tabular}{l|d{-1}d{-1}d{-1}}
\multicolumn{1}{c|}{Sequence} & \multicolumn{1}{c}{$L/\mu m$} & \multicolumn{1}{c}{$\ell_1/m$} & \multicolumn{1}{c}{$\ell_2/m$}\\
\hline
\mms0.37 & 3.38 & 2.32 & 2.38 \\
\mms0.25 & 3.10 & 2.20 & 2.25 \\
\mms0.25 & 3.34 & 2.24 & 2.29 \\
\pps0.0& 0.0 & \multicolumn{2}{c}{1.89\mbox{$^{\mbox{\footnotesize (a)}}$}}\\
\pps0.0  & 2.94 & 2.08 & 2.13 \\
\pps0.0  & 3.00 & 2.08 & 2.13 \\
\pps0.08 & 2.84 & 2.03 & 2.08 \\
\pps0.08 & 2.92 & 2.03 & 2.08 \\
\pps0.17 & 2.79 & 1.98 & 2.03 \\
\pps0.25 & 2.70 & 1.96 & 2.01 \\
\end{tabular}}
\end{table}

Searches for a common apparent horizon were carried out for
several spin configurations.  Details of the apparent horizon finder
and the method used to discern the formation of a common
apparent horizon are given in the Appendix.  In Table~\ref{tab:AH},
the results of the apparent horizon searches are listed.

For fixed spin configurations the common apparent horizon forms at
larger separation for larger angular momentum.  This can be seen from
the $\mms0.25$ and $\pps0.0$ sequences.  For varying spins and angular
momentum close to the ISCO values, the proper separation between the
throats at the formation of the common apparent horizon depends weakly
on the spin.  It decreases from $\ell/m\approx 2.3$ for the $\mms0.37$
sequence down to $\ell/m\approx 2.0$ for the $\pps0.17$ sequence.

{\em Notice that the segment of parameter space where common apparent
horizons form does {\em not} include the sequence of quasi-circular
orbit configurations.}  Indeed, the common apparent horizons form at a
separation inside the inner minimum where the effective potential {\em
increases} with decreasing separation.

\begin{figure}[bt]
\centerline{\epsfxsize=3.5in\epsffile{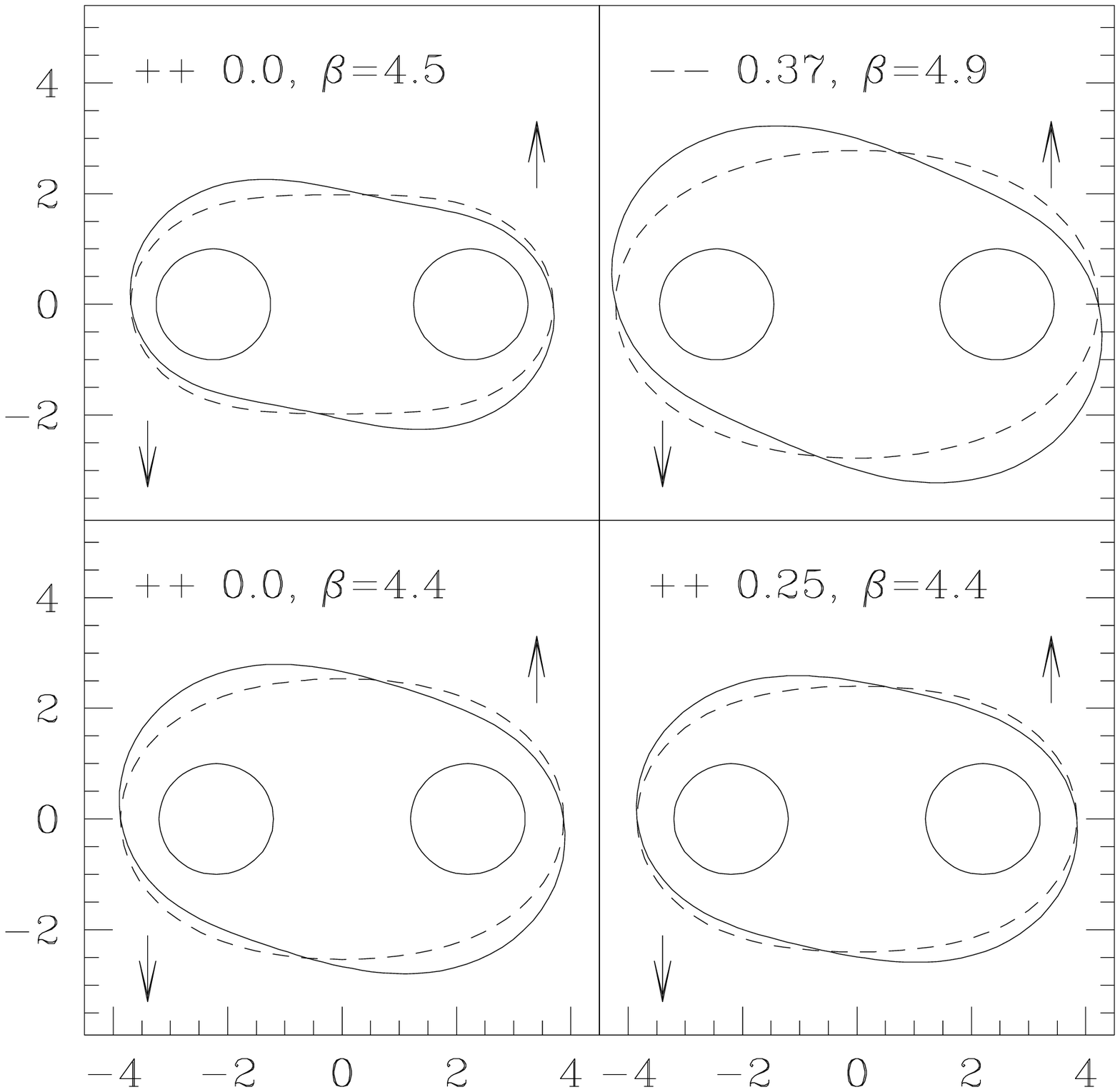}}
\CAP{
Shapes of the common apparent horizons
for different spin configurations
}{\label{fig:AHshapes}Shapes of the common apparent horizons
for different spin configurations. Circles denote the throats of the
holes. The solid lines are cuts in the plane of orbital motion (arrows
indicating the direction of motion), the dashed lines represent cuts
normal to the plane of motion.  }
\end{figure}

The search for the onset of common apparent horizons also provides
the actual surfaces. In Fig.~\ref{fig:AHshapes} some apparent
horizon surfaces just inside the formation of a common apparent
horizon are plotted.  The circles represent the throats of the
holes. The solid lines represent a cut through the plane of orbital
motion of the holes, arrows indicating the direction of linear
momentum of the holes.  The dashed lines are cuts through the plane
perpendicular to the plane of motion and parallel to the spins of the
holes.  We find that the apparent horizons lag behind the orbital
motion, with the amount of lag being larger for counter-rotating than
for co-rotating holes.

\section{Discussion}
\label{sec:Discussion}

We found that the effective potential contours at very small separation
{\em increase} with decreasing separation. This is in contrast to the
usual shape of the effective potential for a Schwarzschild or a Kerr
black hole, which tends to $-\infty$ at sufficiently small
separations.

This behavior can be interpreted in the light of the common
apparent horizon searches. The common apparent horizon that was
found to form at a small separation of the holes might influence the
observed effective potential as follows: The event horizon must
lie outside the apparent horizon.  Therefore a common event
horizon must form before a common apparent horizon forms.  To
accomplish this the event horizons around the individual holes must
grow towards this common event horizon.  Thus, even before
formation of a common event horizon, the individual event horizons
will no longer be close to the individual apparent horizons and the
areas of the event horizons of the individual holes must be larger
than the areas of their apparent horizons.  Therefore, equations
(\ref{eqn:M}) and (\ref{eqn:M_ir}) will {\em under}-estimate the mass
of the holes. We denote this underestimate by $\Delta M$.  Consider
the effect this underestimate of $M$ has on the binding energy.  
The numerator of (\ref{eqn:Eb}) will be {\em over}-estimated by a 
relative amount of
\begin{equation}\label{eqn:Eb2}
\frac{2\Delta M}{|\Eadm-2M|}
=\frac{4}{|E_b/\mu|}\frac{\Delta M}{M}
\gg \frac{\Delta M}{M}.
\end{equation}
At the same time, the denominator of (\ref{eqn:Eb}) and the
denominator of the scaled angular momentum (\ref{eqn:J_root}) change
too, leading to an underestimate of the binding energy $E_b/\mu$.
However, the relative changes of these denominators are only of the
order of $\Delta M/M$, so that the overestimate from
Eqn.~(\ref{eqn:Eb2}) dominates. It might
well be that this overestimate is so large that it counter-balances
the decreasing effective potential that one might expect in analogy to
Schwarzschild or Kerr black holes.

This idea leads to the following picture to explain the observed
effective potential curves: At large separation of the holes, the
masses of the holes and the effective potential are reliable and we
see an effective potential that looks similar to a Schwarzschild black
hole. Consider, for example, the $\pps0.0$ sequence: For $J$ slightly
above its ISCO value we see the (outer) minimum of the stable
quasi-circular orbit and a maximum corresponding to an unstable
circular orbit. As $J$ increases, the stable circular orbit moves
outwards and the unstable one moves inwards. Once the maximum
corresponding to the unstable orbit moves too far in, the $\Delta M/M$
contamination of the effective potential ``eats up'' the maximum and
it disappears.

Now we turn on spin.  We found that a common apparent horizon forms at
approximately the same proper separation, independent of the spin of
the holes.  It seems reasonable that the $\Delta M/M$ error is also
weakly dependent on the spin, and also the separation of the holes,
where $\Delta M/M$ becomes significant.  For the \mms\ sequences the
ISCO moves to larger separations.  Thus the maximum in the effective
potential (the unstable orbit) will survive for a larger range of
separations and angular momenta $J$.  Conversely, for the \pps\
sequences, the ISCO moves inwards, closer to the separation where
$\Delta M/M$ becomes significant. The maximum in $E_b/\mu$ is lost
almost immediately, and in the extreme limit of $S/M^2>0.17$, it does
not show up at all.

This scenario is sufficient to capture the complete behavior of the
effective potential as a function of $J$ and spin.  What does this
picture imply for the validity of our ISCO results from
Table~\ref{tab:ISCO}?  We expect that $\Delta M/M$ decays rapidly with
increasing separation, so the ISCO data for the non-rotating sequence
$\pps0.0$ as well as for the \mms\ sequences should be sound.
However, because $\Delta M$ changes the characteristic behavior for
the \pps\ configurations even for $S/M^2<0.17$, the \pps\ sequences will
be affected. Let us consider how these changes affect our estimates
of circular orbits.

\begin{figure}
\centerline{\epsfxsize=3.5in\epsffile{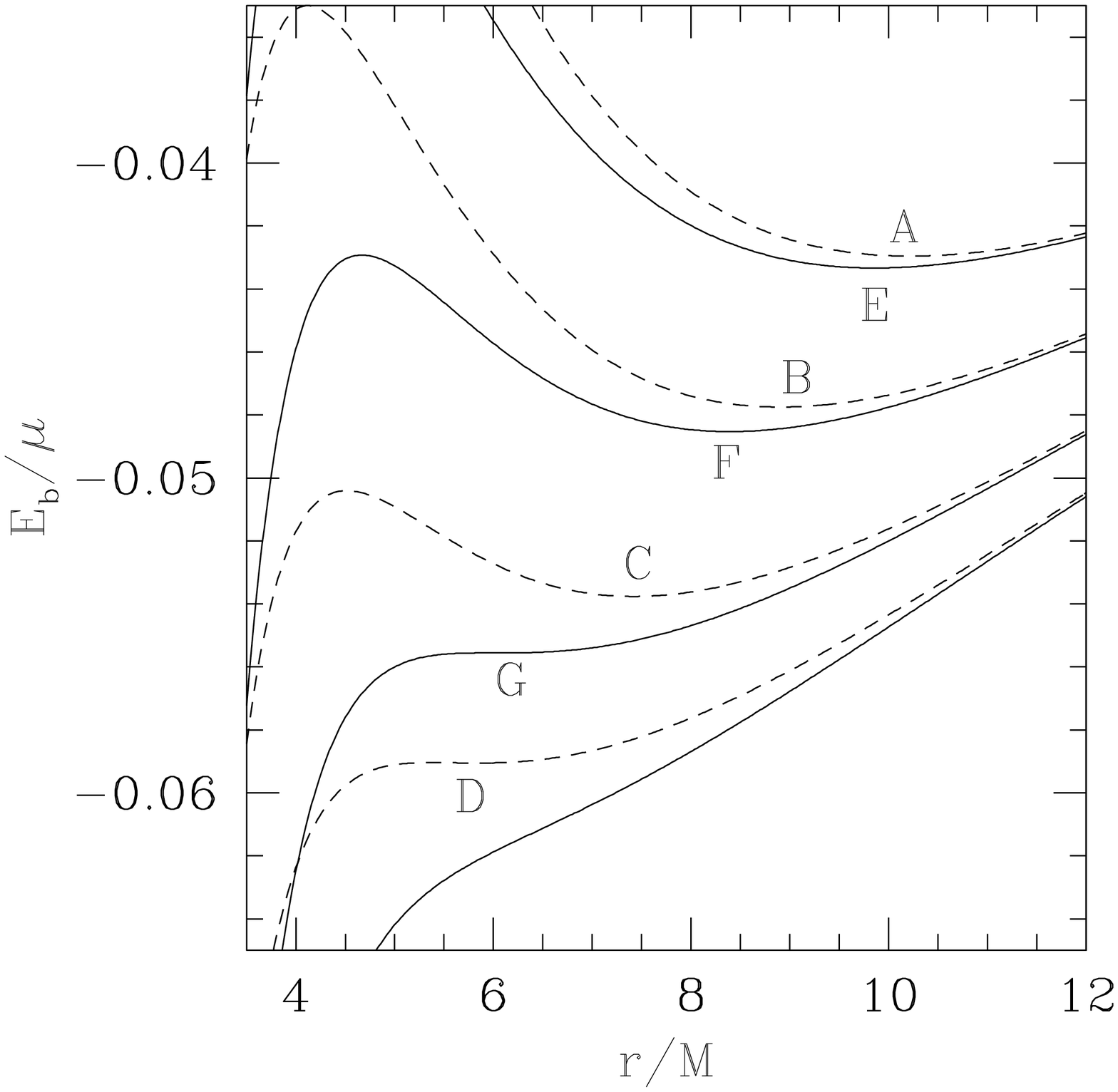}}
\CAP{
Illustration of the effects of a systematic
underestimation of $E_b/\mu$
}{\label{fig:toy}Illustration of the effects of a systematic
underestimation of $E_b/\mu$. The dashed lines represent the observed
effective potential contours for some values of $J$. The points A, B,
and C correspond to circular orbits.  The ISCO is at D.  Assuming that
the true binding energy is smaller, with the deviation increasing as
the separation decreases, yields true effective potential contours
similar to the solid lines. The true circular orbits are at E and F
and the true ISCO is at G.  We find that the minima of the true
contours will lie at smaller separation (for the same $J$).  The
angular frequency is given by $\Omega=dE_b/dJ$. Using the points A and
B, we see that the observed $dE_b$ is smaller than the true one, so we
under-estimate $m\Omega$.  For fixed $J$, true circular orbits will
occur at smaller separation, but the true ISCO will appear at larger
$J$ than we have observed.  These effects counteract each other,
making it impossible to predict their effect on the true ISCO.  }
\end{figure}

Figure~\ref{fig:toy} illustrates the effect of the $\Delta M/M$
contamination on the effective potential contours. As we noted above,
the $\Delta M/M$ contaminations of the binding energy overestimates
the binding energy of an effective potential contour.  Since this
error increases as the separation decreases, our estimates for the
separation at a given value of angular momentum are also too high, and
our estimates of the orbital angular velocity $m\Omega$ are too
low. Unfortunately, we cannot determine whether our estimates for the
location of the ISCO are too high or too low.  While our estimates for
the separation of a given orbit are too high, we see that the true
ISCO will occur at a larger value of the total angular momentum than
we estimate.  These effects oppose each other.

The angular momentum at the ISCO, $J/\mu m$, increases with spin for the
\pps\ configurations. It is interesting to examine whether the final
black hole resulting from a merger of such a spinning binary black
hole can violate the Kerr limit on spin of a black hole. From
(\ref{eqn:M}) we find 
\begin{equation}
M_{ir}^2=\frac{M^2}{2}\left(1+\sqrt{1-\frac{S^2}{M^4}}\right).
\end{equation}

By the area theorem, the final irreducible mass must satisfy
$M_{ir,f}^2\ge2M_{ir}^2$, where equal mass holes were assumed. The
final angular momentum cannot exceed the angular momentum at the ISCO,
$J_f\le J$. With these two constraints and by virtue of the
Christoudoulou formula (\ref{eqn:M}), we find

\begin{equation}
\frac{M_f^2}{M_{ir,f}^2}\le 1+\frac{\left(J/\mu m\right)^2}{4\left(
1+\sqrt{1-\left(S/M^2\right)^2}\;\right)^2}
\end{equation}

A Kerr black hole has always $M^2/M_{ir}^2\le 2$ with equality in the
extreme Kerr limit. With data from Table~\ref{tab:ISCO} we find for
the $\mms0.50$ sequence $M_f^2/M_{ir,f}^2\le 1.43$ and for the $\pps0.17$
sequence $M_f^2/M_{ir,f}^2\le 1.61$. These values correspond to spin
parameters of $J/M_f^2\le 0.92$ and $J/M_f^2\le 0.97$, respectively.
Hence the merged black hole might be close to the Kerr limit, but will
not violate it.

\subsubsection{\pms\ Sequences and conformal flatness}

The $(\mbox{spin})^4$ effect illustrated in Fig.~\ref{fig:spin4}
suggests that the assumption of conformal flatness might lead to
inaccurate results.  This is particularly important for analysis of
gravitational waves.  As seen in Fig.~\ref{fig:spin4}, for spinning
holes with $S/M^2\sim 0.50$ the assumption of conformal flatness
results in an unphysical gravitational wave content of the order
of $\sim 2\cdot 10^{-3}\mu\sim 5\cdot 10^{-4}m$. This is less than
$0.1$ percent of the total mass and a few percent of the binding
energy $E_b$.  If the gravitational energy radiated away is less than
1\% of the total mass, then the gravitational wave content due to an
unsuitable initial data slice is a significant contamination.

\section{Conclusion}
\label{sec:conclusion}

In this work, we have constructed sequences of quasi-circular orbits
for equal-sized, spinning black holes.  At large separations, the
results we have obtained match well with $(\mbox{post})^2$-Newtonian
expansions, although there is a clear contamination of the data because of
the assumption of conformal flatness. The main results of this paper,
displayed in Table~\ref{tab:ISCO} and Fig.~\ref{fig:ISCO}, reveal the
behavior of the ISCO for the cases where the spins of the holes are
either both co-rotating (\pps) or counter-rotating (\mms) with respect
to the orbital motion. For co-rotation, the ISCO moves inwards with
increasing spin and the orbital angular frequency increases. For
counter-rotation the ISCO moves outward and the orbital angular
frequency decreases.  In fact, we find that the orbital angular
frequency changes by almost a factor of 2 between the $\mms0.50$
sequence and the $\pps0.08$ sequence. We have noted a systematic error
in our results that has its origins in an underestimation of the mass
of each black hole when they are close together.  For the ISCO, this
implies that our results are most accurate (ignoring the errors due to
conformal flatness) when the holes have large counter-rotating
spins, and the error increases as we move to configurations with large
co-rotating spins.  In fact, the error becomes so large in the \pps\
sequences that our method cannot locate the ISCO when
$S/M^2\gtrsim0.17$.

Our results clearly show the need to give up the simplifying
assumption of conformal flatness if we are to construct
astrophysically realistic black hole initial data.  This is certainly
not a new realization, but this is the first time that the effects of
the conformal flatness assumption have been seen so clearly in the
context of black hole binaries.  Work toward more astrophysically
realistic initial data has begun \cite{Matzner-Huq-Shoemaker:1999}. This
improvement in the initial data is needed for all separations. It
remains to be seen what impact this improvement will have on the
process of locating quasi-circular orbits when the holes are close
together. It is likely that the systematic underestimate of the mass
will still be significant.  If so, an improved method for locating
quasi-circular orbits and the ISCO will be useful.

\section*{Acknowledgements}

We thank Larry Kidder and Mark Scheel for helpful discussions.
This work was supported in part by NSF grants PHY-9800737 and
PHY-9900672 and NASA Grant NAG5-7264 to Cornell University, and NSF
grant PHY-9988581 to Wake Forest University. Computations were
performed on the IBM SP2 at Cornell Theory Center, and on the Wake
Forest University Department of Physics IBM SP2 with support from an
IBM SUR grant.

\section{Appendix: Common apparent horizons}
\label{sec:AHfinder}
Here we provide details of the apparent horizon (AH) finder.  We use
the AH finder described in \cite{Baumgarte-Cook-etal:1996}.  The AH surface
is expanded in spherical harmonics up to some order $L$.  The
apparent horizon, as a marginally outer trapped surface, has
everywhere vanishing expansion and is located by minimizing the square
of the expansion over the surface.  We use convergence with
increasing expansion order $L$ to diagnose the formation of a
common AH. Therefore high expansion orders $L$ are needed as
well as reliable convergence of the minimization routine to the true
minimum of the square of the expansion.

The Powell minimization used in \cite{Baumgarte-Cook-etal:1996} is too slow
for high-order expansions. We replaced it by a DFP method with finite
difference approximations of the Jacobian \cite{NumericalRecipes}. For the
modest expansion order $L=6$, DFP is already ten times faster than
Powell's method.

Furthermore, we take advantage of the symmetries of the AH
surface. The holes are located along the $\hat z$-axis at $z=\pm
\beta/2$. Their linear momenta point in the $\pm \hat x$-direction and
the spins are directed along the $\pm \hat y$-axis. It is
straightforward to show that these choices imply that the AH surface
is invariant under reflection at the $xz$-plane, $y\to -y$. This
symmetry constrains the coefficients $A_{lm}$ of the expansion in
spherical harmonics to be real. Moreover, for the \pps\ and \mms\ 
configurations with equal sized holes and equal spin magnitudes, the
configuration is symmetric under rotation by $180^\circ$ around the
$\hat y$-axis, this is $(x,y,z)\to (-x, y, -z)$. Both symmetries
together force $A_{lm}=0$ for odd $l$ and $A_{lm}$ to be real for even
$l$.  Hence the number of free parameters in the minimization routine
can be reduced by almost a factor of four.

To prevent convergence to spurious local minima, it is vital that the
function that is minimized be as smooth as possible.  Therefore we use
second order spline interpolation to provide the required data for the
AH finder.  Compared to bicubic interpolation, the
spline interpolation somewhat decreased the number of iterations
needed in the minimization routine, but more importantly it
significantly reduced the probability of getting stuck in a local
minimum.  In addition, many rays were used to reduce the anisotropies
introduced by the discrete position of the rays. Finally, we
distribute the rays {\em non}-uniformly in solid angle. The reason for
this is simple: The common AH surface will be very oblate along
the $\hat z$-axis, since it must encompass the two throats located
along the $\hat z$-axis. The polar regions of the AH surface are close
to the throats and the conformal factor changes rapidly. These regions
are particular important, but the standard distribution uniform in
$\cos\theta$ places relatively few rays in the polar regions.
Therefore we implemented a procedure that distributes the rays in
proportion to an arbitrary ray-density function $f(\theta)$. A
uniform distribution of rays is represented by
$f(\theta)=\mbox{const.}$, whereas we used $f(\theta)=1+\cos^2\theta$,
resulting in a doubled density of rays close to the poles.

With the improved AH finder, we performed extensive tests with various
numbers of rays. As a rule of thumb, about ten times more rays as free
minimization parameters are necessary to ensure reliable convergence
to the true minimum of the square-sum of the expansion.

We used expansions up to order $L=16$ and up to 64x48 rays (64 in
$\theta$ direction, 48 in $\phi$).  We perform a set of AH searches,
starting at $L=2$ and increasing $L$ by $2$ between searches. The
result of the previous search is used as the initial guess for the
next higher expansion order.  Such a set of expansions from $L=2$ to
$L=16$ takes typically about 2 hours CPU time on a RS6000 processor.

A disadvantage of an AH finder based on a minimization routine is
that the minimization routine will {\em always} find a minimum.  It
does not matter whether there actually is a ``true'' apparent horizon,
or whether there is only a surface with small but non-zero
expansion. And even for a true AH, the result of the minimization will
be non-zero because of the finite grid resolution in the underlying
elliptic solver and finite expansion order in spherical
harmonics. Therefore we need a method to discern a ``true'' AH from a
mere minimum in the square of the expansion.

For a true AH, the square of the expansion is exactly zero, therefore
we expect that the residual of the minimization tends to zero as the
resolution of the elliptic solver and the expansion order $L$ are
increased. With increasing $L$, the error in the approximation of the
surface by spherical harmonics should decrease {\em exponentially}. On
the other hand, for a mere minimum in the expansion, we expect that
the residual of the minimization tends towards a {\em non-zero} limit
as the resolution of the elliptic solver and the expansion order $L$
is increased. We use this signature to discern the formation of a
common apparent horizon.

\begin{figure}
\centerline{\epsfxsize=3.5in\epsffile{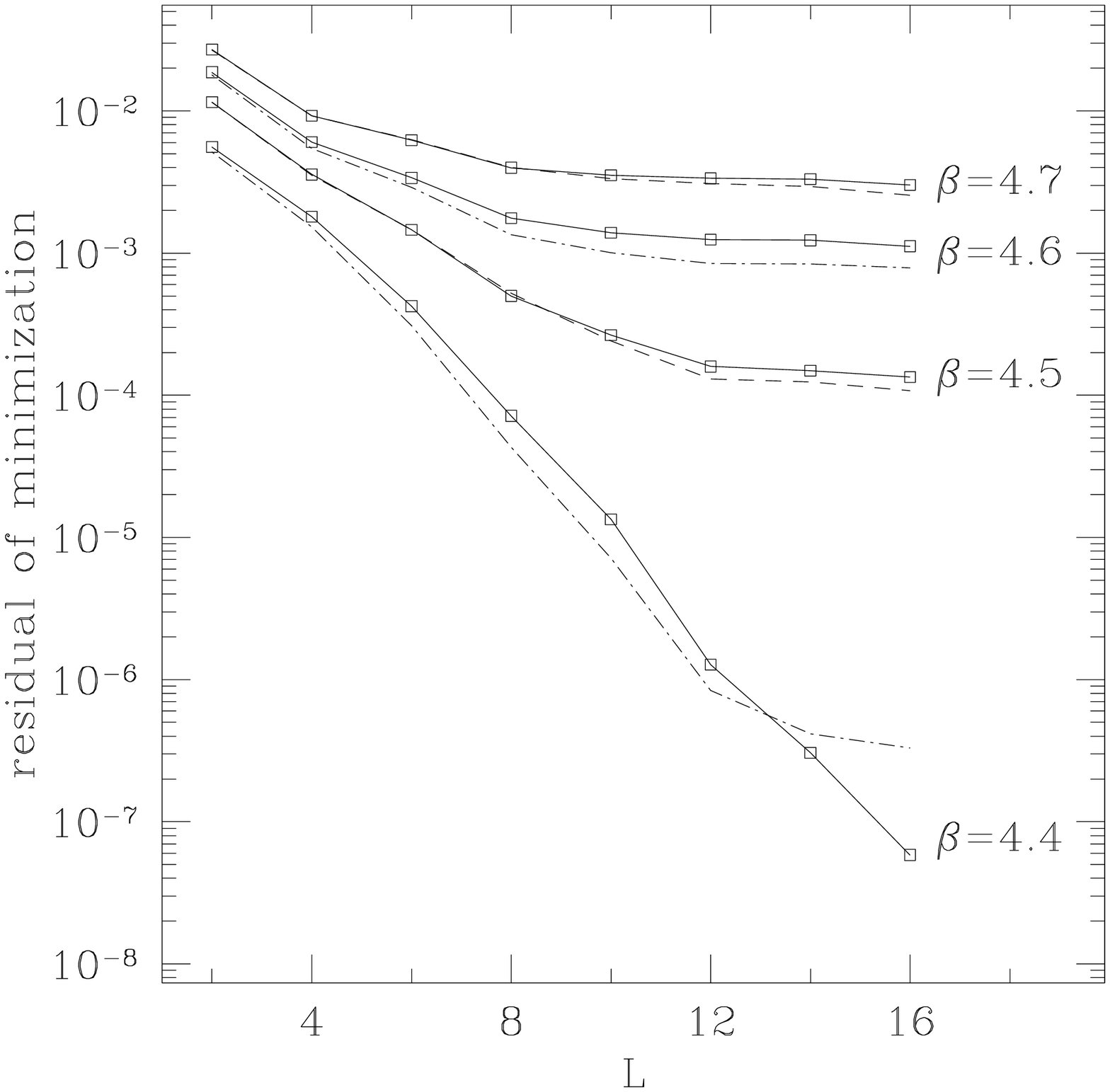}}
\CAP{Residual of the minimization in the apparent horizon finder
 as a function of expansion order L }{\label{fig:AHconv}Residual of
 the minimization in the apparent horizon finder as a function of
 expansion order L. The number of rays used was $N_\theta=64$,
 $N_\phi=48$.  The different solid lines represent different
 separations of the holes along an effective potential contour with
 $J/\mu m=3.29$ on the $\pps0.25$ sequence.  The dashed lines are the
 results of minimizations with $N_\theta=48, N_\phi=32$. The
 dotted-dashed lines show examples of minimizations at lower grid
 resolution and $N_\theta=64, N_\phi=48$.}
\end{figure}

Figure~\ref{fig:AHconv} shows the residual of the minimization for
various values of $L$ and different separations $\beta$.  The solid
lines represent configurations at different separations of the
holes. They are labeled by the background separation of the holes,
$\beta$.  Each solid line represents a set of minimizations with
varying expansion order $2\le L\le 16$ on the {\em same} initial data
set. At large separations, $\beta\ge 4.5$, the residual of the
minimization becomes independent of $L$ for large $L$. At small
separation, $\beta=4.4$, the residual decreases exponentially through
all computed expansion orders up to $L=16$ -- a common AH has
formed.

Neither reducing the number of rays, nor decreasing the resolution of
the Hamiltonian solver changes the convergence behavior
significantly. This is illustrated by some examples in
Fig.~\ref{fig:AHconv}. We conclude that for this particular example a
common AH first forms between $\beta=4.4$ and $\beta=4.5$.

Expansions to high order in $L$ are essential for discerning the
formation of a common AH.  If one had Fig.~\ref{fig:AHconv}
only up to expansions up to $L=8$, it would be impossible to decide
where the common AH first forms. One would probably conclude
that the common AH forms at larger separations than it actually
does.

















\chapter{Quasi-circular orbits in the test-mass limit}
\label{chapter:Testmass}

\section{Introduction}

For the test-mass limit, in which the second black hole (or particle)
has much smaller mass $M_2$ than the first one (with mass $M_1$), many
analytical or semi-analytical results are available.  For example, a
point-particle in circular orbits around a Schwarzschild black hole
satisfies~\cite{Misner-Thorne-Wheeler:1973,Wald:1984}
\begin{align}
\label{eq:Testmass-L}
\frac{L}{\mu m}&= \frac{r/m}{\sqrt{r/m-3}},\\
\label{eq:Testmass-Eb}
\frac{E_b}{\mu}&=\frac{r/m-2}{\sqrt{(r/m)^2-3r/m}}-1,\\
\label{eq:Testmass-Omega}
m\Omega&=\left(\frac{r}{m}\right)^{-3/2}.
\end{align}
Here, $m=M_1+M_2$ and $\mu=M_1M_2/m$ are total mass and reduced mass,
respectively (or, to leading order, mass of the big hole and the small
hole, respectively), $r$ is the areal radius, and $\Omega$ denotes the
orbital angular frequency as seen by an observer at infinity.


Pushing numerical relativity calculations toward this limit is
interesting, because one can compare against analytic results, and
because many approximations involved in constructing black holes in
quasi-circular orbits become exact.  Thus, more stringent tests of the
remaining approximations emerge.

However, the test-mass limit is also computationally more challenging,
as the required resolution is set by the size of the smaller object,
which we are trying to shrink as far as possible.  This
is one reason why fully numerical initial data calculations so far
usually assume equal mass binaries (see e.g. \cite{Cook:1994,
Brandt-Bruegmann:1997, Baumgarte:2000, Pfeiffer-Teukolsky-Cook:2000,
Marronetti-Matzner:2000,
Gourgoulhon-Grandclement-etal:2001,Gourgoulhon-Grandclement-Bonazzola:2001a,
Grandclement-Gourgoulhon-Bonazzola:2001b}).  The restriction to equal
mass reduces also the size of the parameter space under consideration.

The ideas on which these numerical calculations are based do {\em not}
usually require equal mass objects; for example, the effective
potential method \cite{Cook:1994, Pfeiffer-Teukolsky-Cook:2000} which
we apply in chapter \ref{chapter:Spin} is completely general: It can
be used to construct quasi-circular orbits for any choice of masses
and spins. 
The multi-domain elliptic solver developed in chapter
\ref{chapter:Code} can handle very different length scales as was
demonstrated in Figure~\ref{fig:Convergence-MassRatio}.  It is
also sufficiently fast to allow calculations approaching the test-mass
limit.  Consequently, I will now employ this solver to examine
quasi-circular orbits for non-spinning black holes of {\em unequal}
mass.  

\section{Implementation \& Results}

The effective potential method is outlined in section
\ref{sec:Spin:Introduction} and explained in detail in
reference~\cite{Cook:1994}.  Application to non-equal mass binaries is
straightforward.  The root-finder and the remaining infrastructure
developed in chapter \ref{chapter:Spin} can handle non-equal mass
holes anyway (this was necessary for treating one co- and one
counter-rotating hole), and were reused.  
Simply the elliptic solver had to be upgraded from Cook's
finite-difference code \cite{Cook-Choptuik-etal:1993} to the
pseudo-spectral code.  In order to ensure a uniform accuracy of the
pseudo-spectral code, solves of the Hamiltonian constraint were
performed for mass-ratios between unity and $\sim 40$, and for
separations $6\lesssim \beta\lesssim 100$, to ensure that the number
of collocation points in the various subdomains is well-balanced (it
turned out that only the number of radial collocation points in the
spherical shell surrounding the smaller excised sphere had to be
adjusted).

\begin{figure}[tb]
\centerline{\includegraphics[scale=0.4]{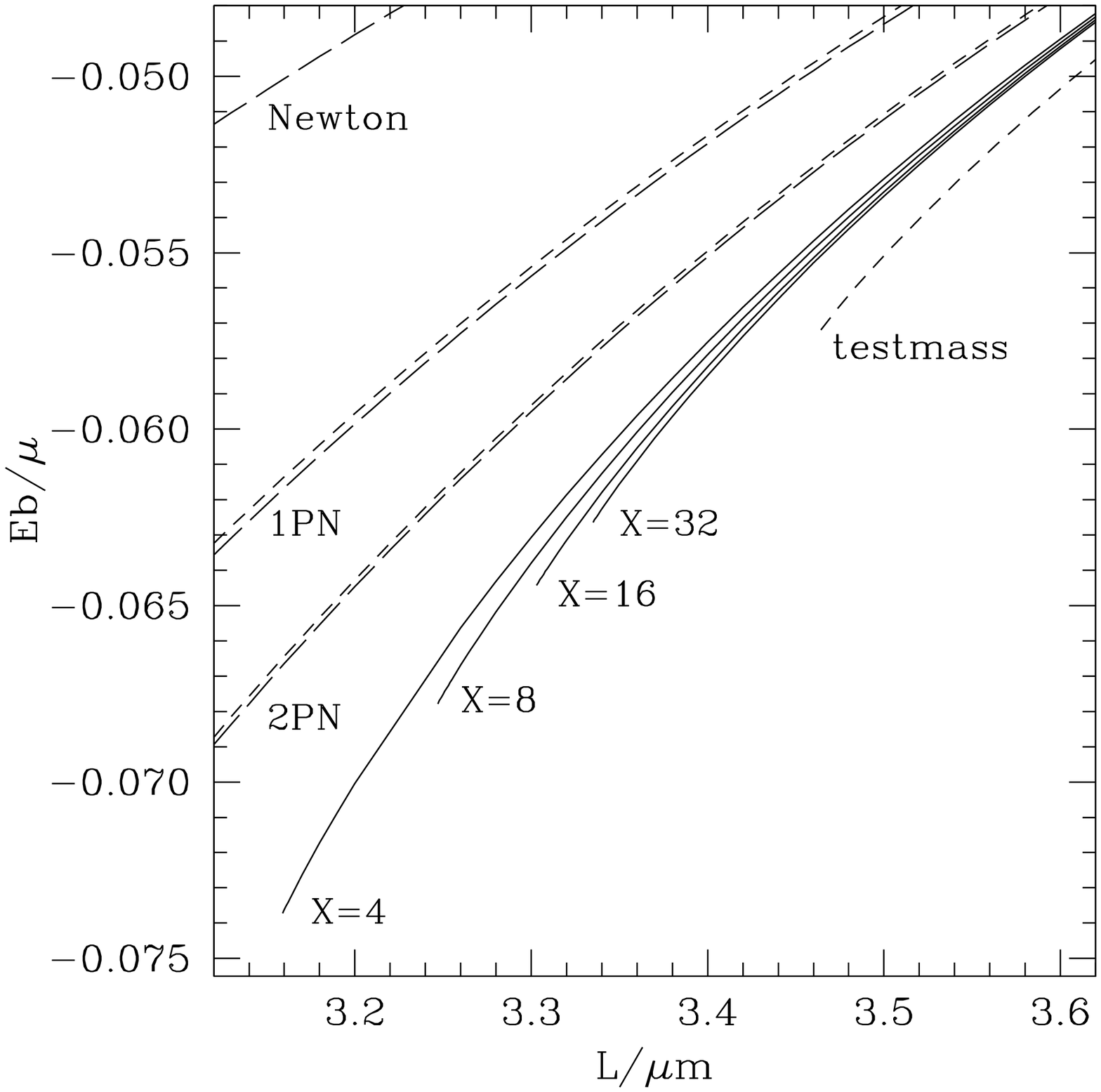}}
\CAP{Sequences of quasi-circular orbits for different mass-ratios
obtained with inversion symmetric Bowen-York initial data (small
$L/\mu m$)} {\label{fig:Testmass1}Sequences of quasi-circular orbits
for different mass-ratios obtained with inversion symmetric Bowen-York
initial data.  The sequences terminate at the location of the ISCO.
Also shown is the sequence of circular orbits for a test-mass orbiting
a black hole, as well as (post-)Newtonian results.  For the latter,
the long-dashed lines correspond to $X=1$, whereas the short-dashed
lines represent the test-mass limit, $X\to\infty$.}
\end{figure}

We traced out sequences of quasi-circular orbits for mass-ratios
$X=M_1/M_2$ up to $32$.  Figure~\ref{fig:Testmass1} presents sequences
for $X=4,8,16,32$.  As the mass-ratio increases, the sequences move
closer to the curve denoting the point-mass result
(cf. Eqs.~(\ref{eq:Testmass-L}) and~(\ref{eq:Testmass-Eb})), as we
expect.  The figure includes post-Newtonian results taken from
\cite{Cook:1994}:
\begin{gather}
\frac{E_b}{\mu}
=-\frac{1}{2}\left(\frac{L}{\mu m}\right)^{-2}
\left[1+\frac{9+\eta}{4}\left(\frac{L}{\mu m}\right)^{-2}
\!\!
+\frac{81-7\eta+\eta^2}{8}\left(\frac{L}{\mu m}\right)^{-4}+\ldots\right],\\
\frac{E_b}{\mu}=-\frac{1}{2}(m\Omega)^{2/3}
\left[1-\frac{9+\eta}{12}(m\Omega)^{2/3}
-\left(\frac{27}{8}-\frac{19\eta}{8}+\frac{\eta^2}{24}\right)
(m\Omega)^{4/3}+\ldots\right],\\
\left(\frac{L}{\mu m}\right)^2\!\!=
(m\Omega)^{-2/3}\!\left[1+\frac{9+\eta}{3}(m\Omega)^{2/3}
+\left(9-\frac{17\eta}{4}+\frac{\eta^2}{9}\right)(m\Omega)^{4/3}+\ldots\right].
\end{gather}
Here, $\eta=\mu/m$, so that equal mass binaries correspond to
$\eta=1/4$, and the test-mass limit is $\eta=0$.

\begin{figure}
\centerline{\includegraphics[scale=0.4]{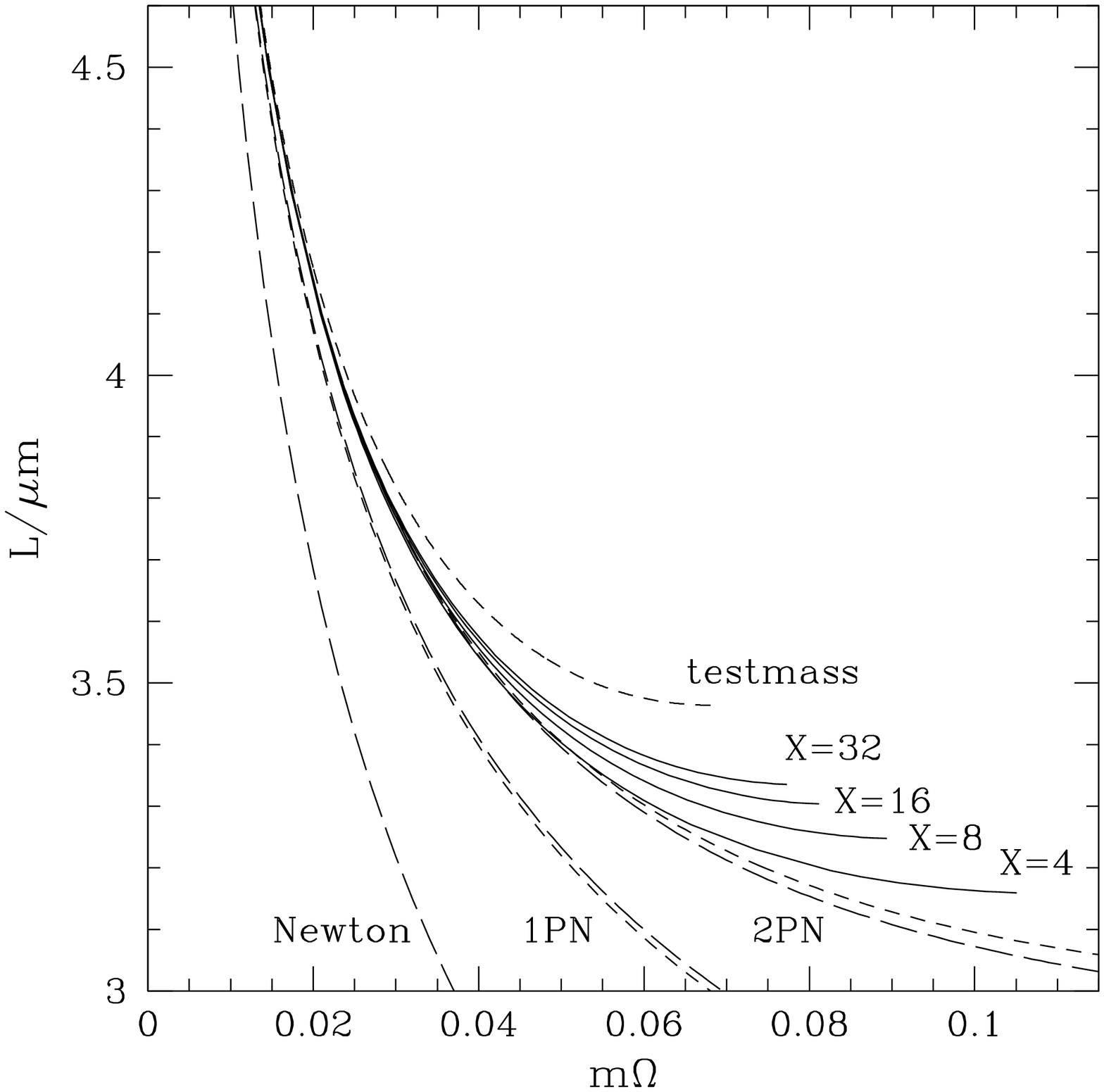} }
\CAP{Sequences of quasi-circular orbits for different mass-ratios
obtained with inversion symmetric Bowen-York initial
data.}{\label{fig:Testmass2}Sequences of quasi-circular orbits for
different mass-ratios obtained with inversion symmetric Bowen-York
initial data.  Symbols as in Figure~\ref{fig:Testmass1}.}
\end{figure}

Figure~\ref{fig:Testmass2} shows the angular momentum $L/\mu m$ 
as a function of orbital angular frequency [cf. Eq.~(\ref{eqn:Omega})]
\begin{equation}
m\Omega=\frac{dE_b/\mu}{dL/\mu m}.
\end{equation}
For this Figure, $m\Omega$ was computed by simple first order finite
differencing between neighboring points along each sequence.  This is
sufficiently accurate for plotting but leads to fairly inaccurate ISCO
locations in this plot.

\begin{figure}[tb]
\centerline{\includegraphics[scale=0.4]{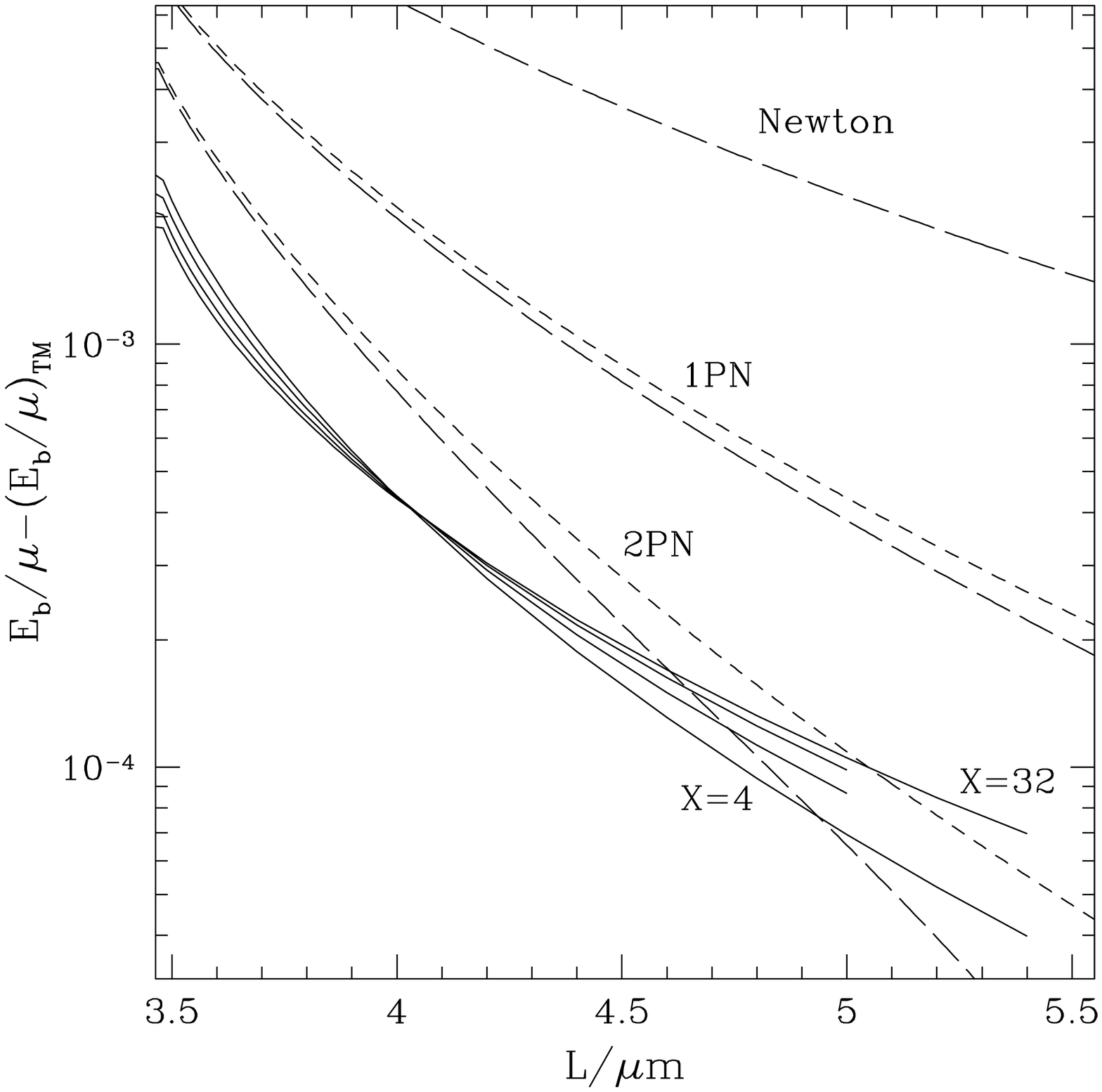}}
\CAP{Sequences of quasi-circular orbits for different mass-ratios
obtained with inversion symmetric Bowen-York initial data (large
$L/\mu m$)} {\label{fig:Testmass3}Sequences of quasi-circular orbits
for different mass-ratios obtained with inversion symmetric Bowen-York
initial data.  Plotted is the difference of $E_b/\mu$ along the
sequence to its value in the test-mass limit.  Symbols as in
Figure~\ref{fig:Testmass1}.  The computed sequences terminate at some
large $(L/\mu m)$ simply because no further data was computed; in
reality they should continue.}
\end{figure}

Figure~\ref{fig:Testmass3} plots the binding energy $E_b/\mu$ along
the sequences for large separation.  To facilitate plotting, the
binding energy in the test-mass limit was {\em subtracted} from each
curve.  The computed sequences, labelled $X=4$ to $X=32$ reverse their
relative ordering around $L/\mu m\approx 4$: At small angular momenta
(i.e. at small separation), $|E_b/\mu|$ increases with increasing
mass-ratio $X$.  At large angular momenta, however, $|E_b/\mu|$
decreases with increasing $X$.  This behavior matches the second
post-Newtonian results, although the crossover occurs at a different
separation.

\begin{figure}
{\includegraphics[scale=0.34]{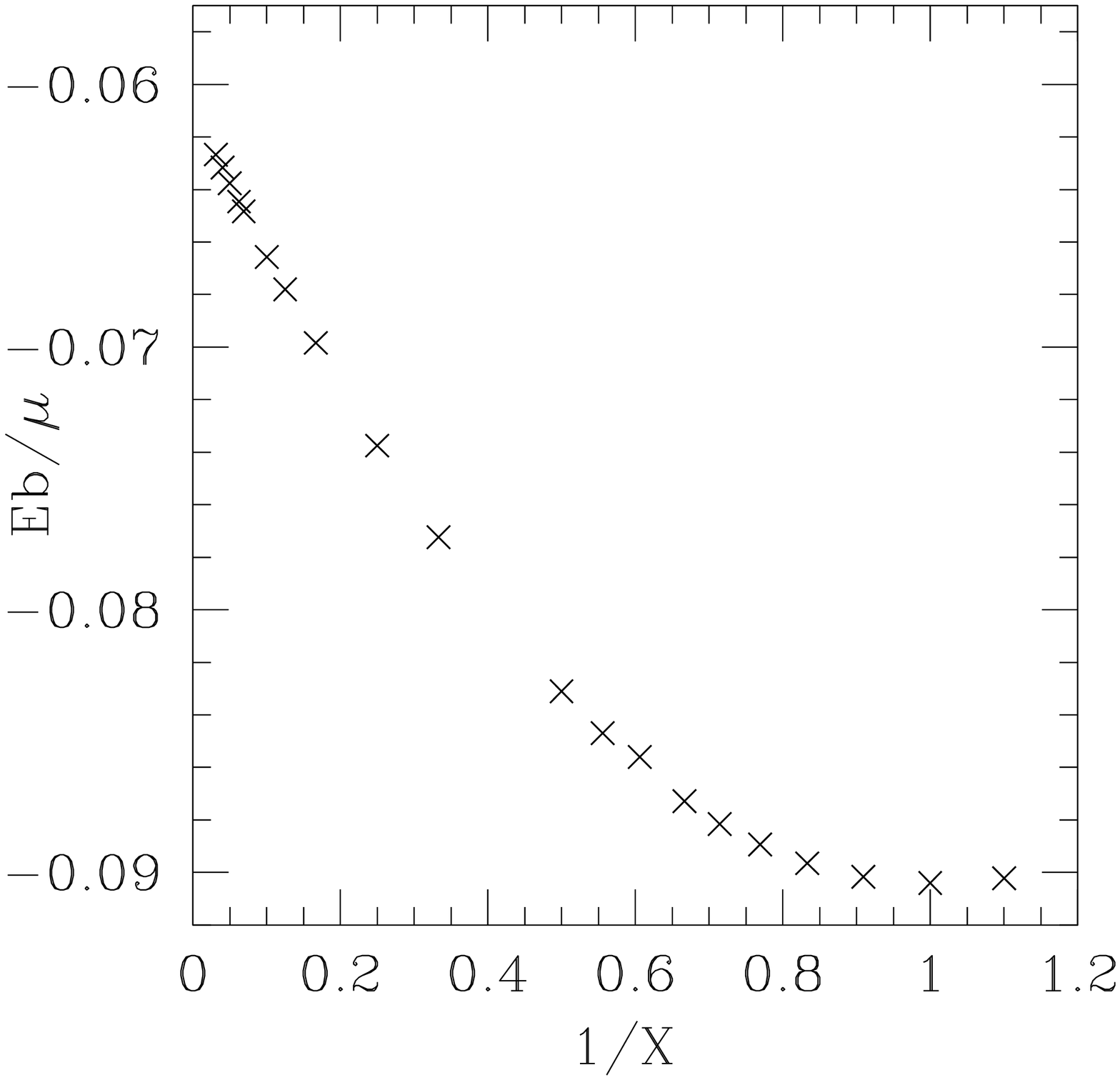}}
\hfill{\includegraphics[scale=0.34]{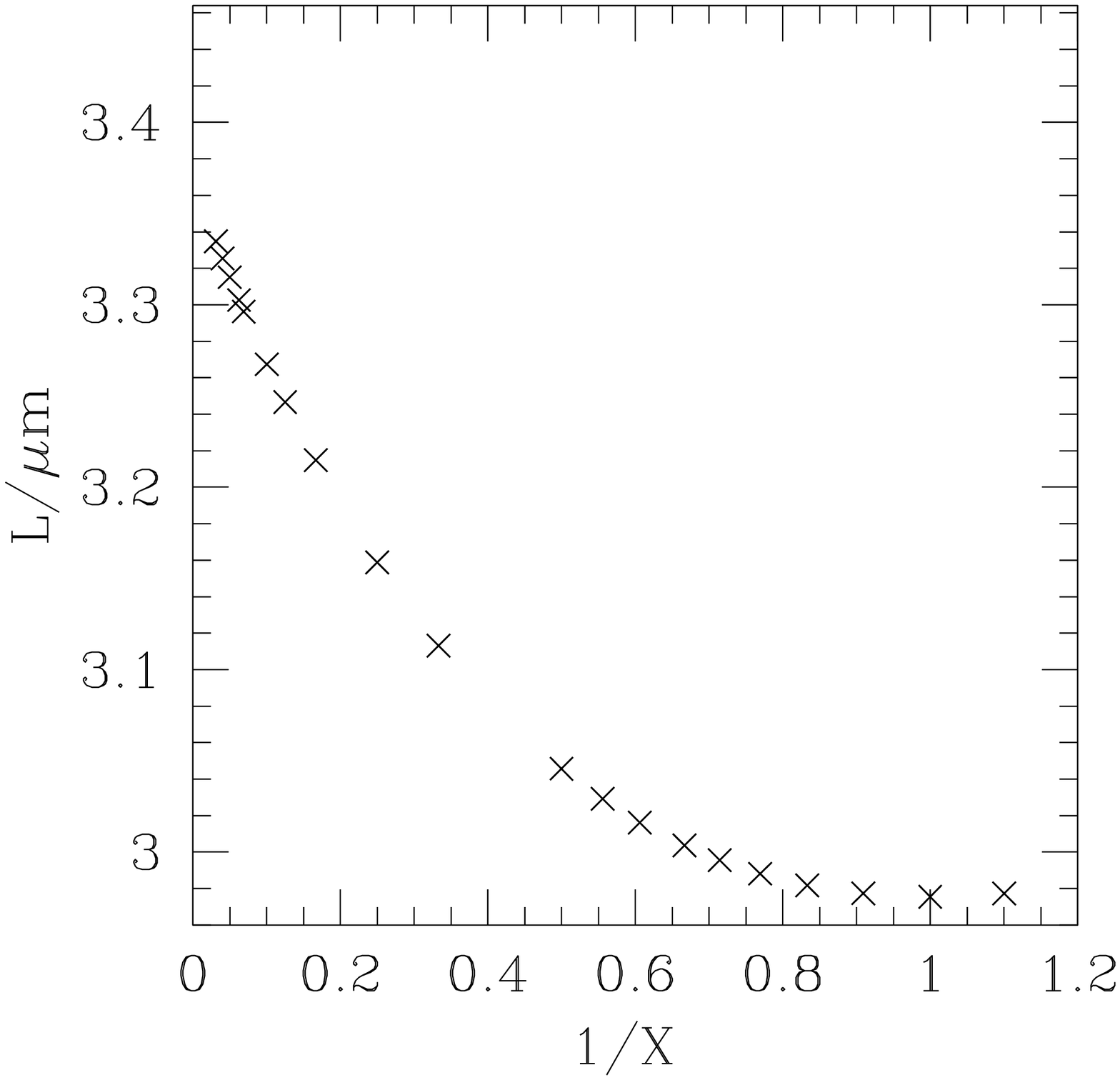}}

{\includegraphics[scale=0.34]{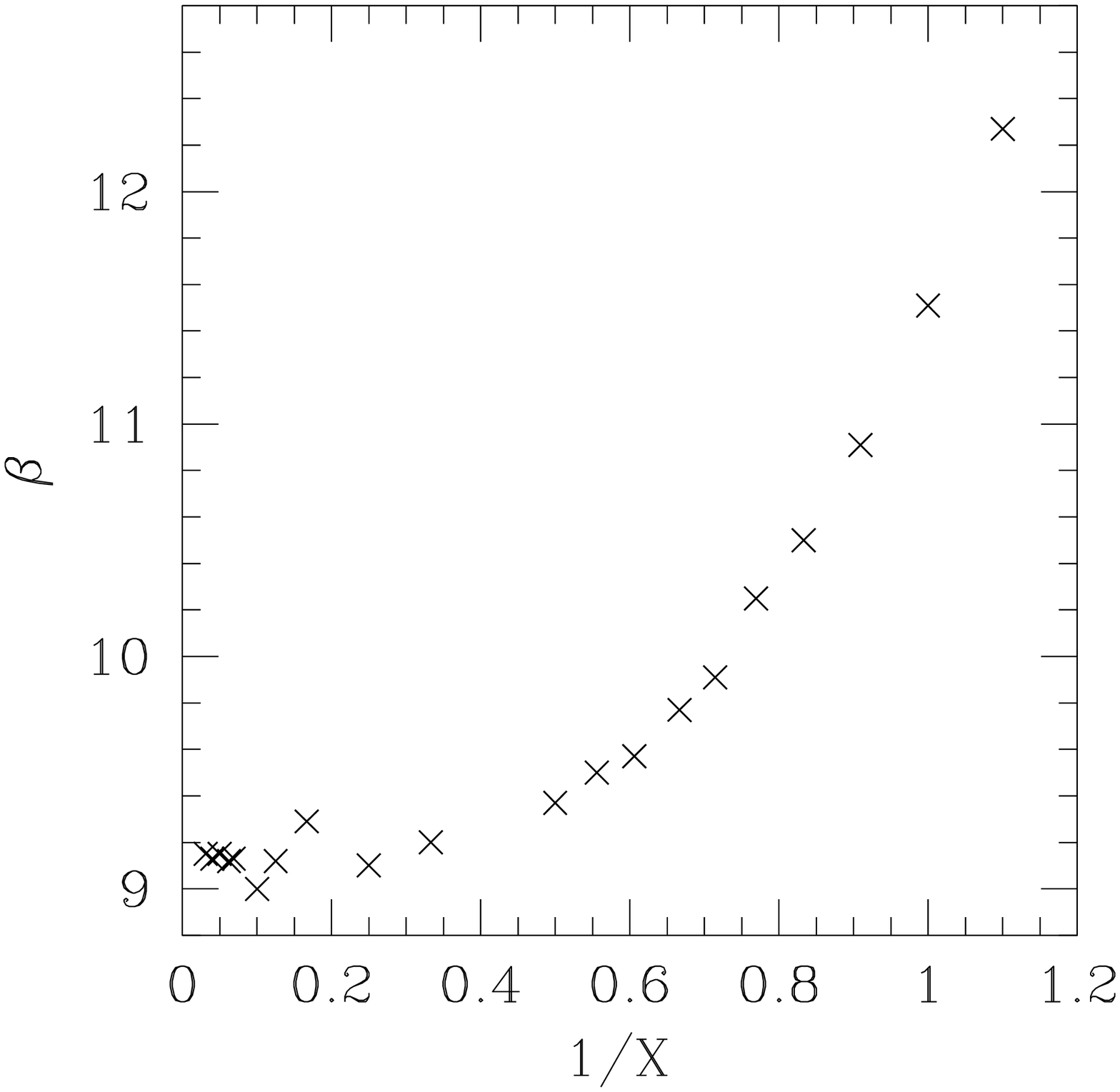}}
\hfill{\includegraphics[scale=0.34]{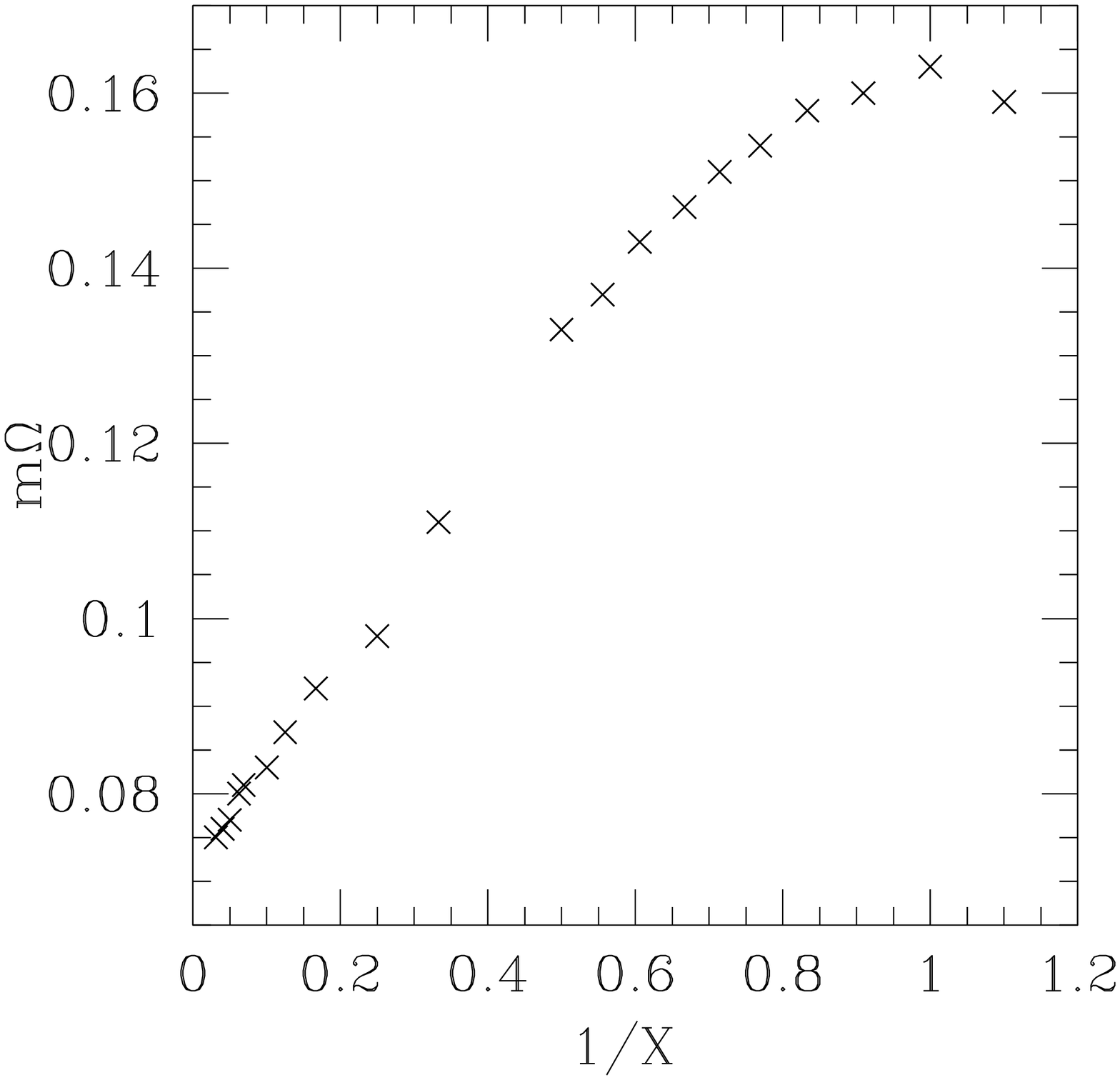}}
\CAP{ISCO's for different mass-ratios obtained with inversion
symmetric Bowen-York initial data.}{\label{fig:Testmass-ISCO}ISCO's
for different mass-ratios obtained with inversion symmetric Bowen-York
initial data.  Plotted are the values of several quantities at ISCO
vs.  the mass-ratio.}
\end{figure}

In Figures~\ref{fig:Testmass1} and~\ref{fig:Testmass2}, the sequences
of quasi-circular orbits terminate simply at the smallest
quasi-circular orbit that was found, which is outside the ISCO by
definition.  The location of the ISCO is determined with greater
accuracy with the following procedure: 

For several quasi-circular orbits close to the tentative ISCO, labeled
by ``$a$'', we recorded the parameters characterizing the orbit:
$\beta_a$, $(L/\mu m)_a$, $(E_b/\mu)_a$.  The Appendix shows that $L$
behaves {\em quadratically} in separation $\beta$ close to ISCO, the
ISCO coinciding with the {\em minimum} of the parabola.  Consequently,
we fit a parabola through the points
\begin{equation}
\Big(\beta_a,\;\; (L/\mu m)_a\Big);
\end{equation}
its minimum gives $\beta_\ISCO$ and $(L/\mu m)_\ISCO$.  Further, we
 fit a parabola through the points
\begin{equation}
\Big((L/\mu m)_a,\;\; (E_b/\mu)_a\Big).
\end{equation}
Evaluation of this parabola at $(L/\mu m)_\ISCO$ yields
$(E_b/\mu)_\ISCO$, evaluation of the {\em derivative} at $(L/\mu
m)_\ISCO$ yields $(m\Omega)_\ISCO$.  We ensure that the
ISCO--parameters are insensitive to the number of points used in the
quadratic fit; usually, five or six points were used.  

The results are presented in Figure~\ref{fig:Testmass-ISCO}.  The data
point at $X=1$ has been computed before, both in \cite{Cook:1994}, and
in chapter \ref{chapter:Spin}.  Close to $X=1$, the physical
parameters $L/\mu m$, $E_b/\mu$ and $m\Omega$ behave quadratically in
$X$, as they must due to the $X\to 1/X$ symmetry (exchange of the two
black holes).  We confirmed that the physical ISCO-parameters for
$X=1.1$ and $X=1/1.1$ are identical, as they should (of course, the
{\em unscaled} background separation $\beta$ differs).  $\beta_\ISCO$
and $(m\Omega)_\ISCO$ are less accurate, as can already be seen by the
noise in Figure~\ref{fig:Testmass-ISCO}.  The values of $\beta$ for
each quasi-circular orbit are obtained as the abscissas of a
minimization, which is inherently inaccurate, whereas $m\Omega$ is
obtained by differentiation of numerical data.

For $1/X\lesssim 0.2$, the ISCO parameters appear to depend linearly
on $1/X$.  We have reached the linear regime so that we can
confidently extrapolate the results to the test-mass limit, $1/X\to
0$.  Table~\ref{tab:testmass} contains the numerical values at ISCO,
as well as an extrapolation to $X=\infty$.  Analytic values for the
test-mass limit are obtained from
Eqs.~(\ref{eq:Testmass-L})--(\ref{eq:Testmass-Omega}) with $r=6m$.  In
this limit, $\beta$ approximates the isotropic radius so that
$\beta_{\ISCO}=\frac{1}{2}\left(5+\sqrt{24}\right)m\approx 9.899$ (the
mass $m\approx 2$ because the large throat has radius $1$ and not
$1/2$.).

\begin{table}
\begin{centering}
\CAP{ISCO parameters for different mass-ratios obtained with the
effective potential method.}{\label{tab:testmass}ISCO parameters for
different mass-ratios obtained with the effective potential method,
``$\infty$'' denoting extrapolation to the test-mass limit. }
\centerline{
\singlespacingplus
\begin{tabular}{|cccc|ccc|}\hline
$X$   &$L/\mu m$& $E_b/\mu$ &$m\Omega$&$\beta$ &$\alpha$&$\bar P$ \\\hline
0.909 &  2.9772 & -0.09022 & 0.159  & 12.27 &  0.8847 & 1.867   \\
1.0   &  2.9755 & -0.09040 & 0.163  & 11.51 &  1.0000 & 1.77   \\
1.1   &  2.9774 & -0.09017 & 0.160  & 10.91 &  1.1301 & 1.639   \\
1.2   &  2.9818 & -0.08965 & 0.158  & 10.50 &  1.2622 & 1.512   \\
1.3   &  2.9881 & -0.08893 & 0.154  & 10.25 &  1.38 & 1.395   \\
1.4   &  2.9955 & -0.08815 & 0.151  &  9.91 &  1.531 & 1.302   \\
1.5   &  3.0036 & -0.08728 & 0.147  &  9.77 &  1.666 & 1.204   \\
1.65  &  3.0162 & -0.08560 & 0.143  &  9.57 &  1.869 & 1.081   \\
1.8   &  3.0292 & -0.0847 & 0.137  &  9.5 &  2.070 & 0.975   \\
2.0   &  3.0457 & -0.08310 & 0.133  &  9.37 &  2.338 & 0.855   \\
3.0   &  3.1132 & -0.07724 & 0.111  &  9.2 &  3.675 & 0.538   \\
4.0  &  3.1588  & -0.07375  & 0.098  &  9.1  &  5.00   & 0.394   \\
6.0   &  3.2149 & -0.06984 & 0.092 &  9.29 &  7.63 & 0.251   \\
8.0   &  3.2467 & -0.06780 & 0.087 &  9.12 & 10.32 & 0.189  \\
10    &  3.2680 & -0.06657 & 0.083 &  9 & 12.96 & 0.152  \\
14.5  &  3.2963 & -0.06483 & 0.081 &  9.13 & 18.94 & 0.1029  \\
16    &  3.3026 & -0.06446 & 0.080 &  9.12 & 20.94 & 0.0933  \\
20    &  3.3151 & -0.06376 & 0.077  & 9.15 & 26.21 & 0.0774 \\
25    &  3.3255 & -0.06317 & 0.076 &  9.13 & 32.9 & 0.059  \\
32    &  3.3348 & -0.06267 & 0.075 &  9.15 & 42.1 & 0.046  \\
\hline
$\infty$& 3.370  & -0.0607    & 0.071  &  9.15  &  & \\\hline 
analytic & 3.463 & -0.0572 &  0.0680 &  9.899 &&\\\hline
\end{tabular}
}\end{centering}
\end{table}

\section{Discussion}

Extrapolation $X\to\infty$ in Table~\ref{tab:testmass} shows that in
this limit, the ISCO parameters computed with the effective potential
method differ by a few per cent from the analytical results; the
computed ISCOs are too tightly bound. 

In the test-mass limit, many assumptions of the effective potential
method become exact: The radiation reaction time-scale grows with the
mass-ratio $X$ so that the adiabatic approximation becomes
increasingly good.  As the inspiral time-scale increases and the
configuration moves closer to stationarity, the deviation between
apparent horizons and event horizons should diminish, and the apparent
horizon masses should increasingly well represent the ``true'' mass
of each black hole.  Moreover, the underlying idea of minimizing
$E_b/\mu$ is equivalent to the method used to find circular orbits
analytically, as the discussion just before and after
Eq.~(\ref{eqn:Schwarzschild}) shows.

The remaining assumptions include conformal flatness, inversion
symmetry, and the Bowen-York solution of the momentum constraint.
Regarding conformal flatness, the space time will be dominated by the
large black hole (which is at rest for infinite mass-ratio) so that we
should recover the Schwarzschild metric in isotropic coordinates.
Close to the small hole, however, the assumption of conformal flatness
might still break down.  I did not analyze or estimate possible errors
associated with conformal flatness.

The major uncertain approximation remains the Bowen-York solution for
the extrinsic curvature.  This is just a {\em convenient} solution of
the momentum constraint, as it is analytical, but there is no reason
for it to be the {\em correct} solution.  In chapter
\ref{chapter:Comparing}, we show that initial data sets computed with
the extrinsic curvature formulation are very sensitive to the choice
of background extrinsic curvature, hence it is likely that the present
results are also sensitive to the choice of extrinsic curvature.

How would a ``wrong'' choice for the extrinsic curvature manifest
itself?  Cook~\cite{Cook:1990} examines single black holes constructed
using the Bowen-York solution and inversion symmetry.  For a black
hole with linear momentum, he finds that the ADM-energy of the
spacetime {\em exceeds} the apparent horizon mass times the
Lorentz-factor.  The difference increases with increasing linear
momentum of the hole, and can be attributed to gravitational radiation
outside the hole, which might fall into the hole, or escape to
infinity.  It seems reasonable that the binary black hole solutions
behave similarly, this is, that the ADM-energy is contaminated by
``gravitational radiation'' outside the holes, such that the
contamination increases with increasing linear momentum $\bar
P$.  Along a single effective potential contour, the momentum
parameter $\bar P$ {\em increases} as the separation decreases: the
observed values of $E_{ADM}$ and $E_b/\mu$ will be systematically too
large for smaller separations.  This is exactly the situation we
discussed in Figure~\ref{fig:toy} on page~\pageref{fig:toy}, and
offers an explanation why the numerical calculation yields $(L/\mu
m)_\ISCO$ smaller than the analytical result.

If the contamination is indeed induced by the Bowen-York extrinsic
curvature, then it will {\em not} disappear in the test-mass limit.
In this limit, the large black hole is simply at rest; unphysical
gravitational radiation will be associated with the small hole, which
moves with a significant fraction of the speed of light.  For given
velocity, the gravitational wave contamination introduced by the small
hole is proportional to its mass $M_2$.  The binding energy $E_b$ is
also proportional to $M_2$, so that $M_2$ scales out of the effective
potential $E_b/\mu$.

This discussion shows that the notion that the Bowen-York extrinsic
curvature contaminates the initial data sets is consistent
with the numerical results.  The same effect might also contribute to
the disappearance of the ISCO for spinning equal mass black holes in
chapter \ref{chapter:Spin}.  In that chapter, we attributed the
failure of the effective potential method to a break down of the
apparent horizon mass, but given the current result and the sensitive
dependence of the initial data sets on the extrinsic curvature
(cf. chapter \ref{chapter:Comparing}), it seems plausible that the
Bowen-York extrinsic curvature plays also a role for the spinning
black holes in chapter~\ref{chapter:Spin}.

The ISCO for equal mass nonspinning black holes has also been computed
with post-Newtonian methods~\cite{Buonanno-Damour:1999,
Damour-Jaranowski-Schaefer:2000b,Blanchet:2002,Grandclement-Gourgoulhon-Bonazzola:2001b,Damour-Gourgoulhon-Grandclement:2002}
and with the helical Killing vector
approximation~\cite{Gourgoulhon-Grandclement-Bonazzola:2001a,Grandclement-Gourgoulhon-Bonazzola:2001b,Damour-Gourgoulhon-Grandclement:2002}.
Both these methods find consistently the ISCO to be {\em less} tightly
bound than the effective potential method with Bowen-York extrinsic
curvature~\cite{Cook:1994,
Pfeiffer-Teukolsky-Cook:2000,Baumgarte:2000}.  As the effective
potential method results in too tightly bound ISCOs for the
test-mass limit, it seems likely that its results for equal mass
binary black holes are too tightly bound, too, thus explaining at least
part of the discrepancy between the different ISCO results.

The results of this chapter point again toward the importance of the
choice of extrinsic curvature, here in the form of the Bowen-York extrinsic
curvature.

\section[Appendix: $L$ depends quadratically on $\beta$ close to ISCO]
{Appendix: \boldmath$L$ depends quadratically on $\beta$ close to ISCO}

Let $x$ be some measure of the separation of the black holes, e. g.,
$\beta$ or the proper separation $\ell/m$.  Denote $\delta
x=x-x_\ISCO$, $\delta L=(L/\mu m)-(L/\mu m)_\ISCO$, and expand the
effective potential as a power series in separation around ISCO:
\begin{equation}
E_b/\mu=a(\delta L) \delta x^3 + b(\delta L) \delta x^2 
+ c(\delta L) \delta x + d(\delta L).
\end{equation}
The coefficients $a(\delta L),\ldots d(\delta L)$ depend on the
angular momentum $\delta L$.  The function $d(\delta L)$ merely shifts
the energy and is irrelevant for determining the sequence of
quasi-circular orbits in the $(\delta x, \delta L)$-plane.  By
definition of ISCO,
\begin{align}
\left.\dnachd{E_b/\mu}{\delta x}\right|_{\delta x=0, \delta L=0}&=0
\qquad\Rightarrow\qquad c(0)=0,\\
\left.\dnachd{^2E_b/\mu}{\delta x^2}\right|_{\delta x=0,\delta L=0}&=0
\qquad\Rightarrow\qquad  b(0)=0,\\
\left.\dnachd{^3E_b/\mu}{\delta x^3}\right|_{\delta x=0,\delta L=0}&>0
\qquad\Rightarrow\qquad a(0)\equiv a_0>0.
\end{align}
Thus, the lowest order behavior of the coefficients is
\begin{equation}
a(\delta L)\approx a_0,\qquad b(\delta L)\approx b_1\delta L, 
\qquad c(\delta L)\approx c_1\delta L,
\end{equation}
with constants $a_0, b_1$ and $c_1$.  For $\delta L>0$, two extrema
exist, therefore the determinant of the quadratic equation
\begin{equation}\label{eq:quadratic}
\left.\dnachd{E_b/\mu}{\delta x}\right|_{\delta L}=3a_0\,\delta
x^2+2b_1\delta L\,\delta x+c_1\delta L=0
\end{equation}
must be positive,
\begin{equation}
4 (b_1 \delta L)^2+ 12 a_0 c_1 \delta L = 12 a_0 c_1\delta L +{\cal
O}(\delta L^2)>0,
\end{equation}
so that $c_1<0$.  Solving Eq.~(\ref{eq:quadratic}) for $\delta L$, we
find that, along the sequence,
\begin{equation}
\delta L=-\frac{3a_0\,\delta x^2}{2b_1\delta x+c_1}
=-\frac{3a_0}{c_1}\delta x^2+{\cal O}(\delta x^3),
\end{equation}
so that $(L/\mu m)$ depends quadratically on the separation and is
extremal at $x_\ISCO$.  Because $a_0\!>\!0$ and $c_1\!<\!0$, $L/\mu m$ has a
minimum at $x_\ISCO$.  We have also established that the energy can be
expanded close to ISCO as
\begin{equation}
 E_b/\mu=a_0 \delta x^3 + b_1 \delta L\,\delta x^2 - c_1 \delta L\,
 \delta x + d(\delta L),
\end{equation}
with $a_0\!>\!0$ and $c_1\!<\!0$.  The sign of $b_1$ is undetermined.




\renewcommand{\thefootnote}{\fnsymbol{footnote}}

\chapter[Comparing initial-data sets for binary black holes]
{Comparing initial-data sets for binary black holes\footnote[1]{H. P. Pfeiffer, G. B. Cook, and S. A. Teukolsky, Pys. Rev. D {\bf 66}, 024047 (2002).}}
\label{chapter:Comparing}

\renewcommand{\thefootnote}{\arabic{footnote}}

\section{Introduction}

Numerical evolutions of black holes have been improved slowly but
steadily over the last few years and now first attempts are being made
to extract physical information from these evolutions. Most notably
one wants to predict the gravitational radiation emitted during black
hole coalescence \cite{Baker-Bruegmann-etal:2001,Alcubierre-Benger-etal:2001,
Baker-Campanelli-etal:2002}.

The quality of the initial data will be crucial to the success of the
predictions of the gravitational wave forms. Unphysical gravitational
radiation present in the initial data will contribute to the
gravitational waves computed in an evolution and might overwhelm the
true gravitational wave signature of the physical process under
consideration.  Therefore an important question is how to control the
gravitational wave content of initial-data sets, and how to specify
{\em astrophysically} relevant initial data with the appropriate
gravitational wave content, for e.g.\ two black holes orbiting each
other.  Unfortunately, assessing and controlling the gravitational wave
content of initial-data sets is not well understood at all.

The mere {\em construction} of an initial-data set alone is
fairly involved, since every initial-data set must satisfy a rather
complicated set of four partial differential equations, the so-called
constraint equations of general relativity.  The question of how to
solve these equations, and how to specify initial data representing
binary black holes in particular, has received considerable attention.

We consider in this paper three different approaches that transform
the constraint equations into elliptic equations: The
{\em conformal transverse-traceless (TT) decomposition}\cite{York:1979},
the {\em physical TT decomposition} \cite{ Murchadha-York:1974b,
  Murchadha-York:1974c, Murchadha-York:1976} and the {\em conformal
  thin sandwich decomposition}\cite{York:1999}.  These decompositions
split the variables on the initial-data surface into various pieces in
such a way that the constraint equations determine some of the pieces while
not restricting the others.  After these freely
specifiable pieces are chosen, the constraint equations are solved and the
results are combined with the freely specifiable pieces to yield a
valid initial-data set.

Any reasonable choice for the freely specifiable pieces will lead to a
valid initial-data set. Furthermore, any one of these decompositions
can generate any desired initial-data set, given the {\em correct}
choices of the freely specifiable pieces.  However, it is not clear
{\em what} choices of freely specifiable pieces lead to initial-data
sets with the desired properties.

The decompositions we consider here lead to four coupled nonlinear
elliptic partial differential equations. Since such equations are difficult to
solve, the early approach to constructing initial data was pragmatic:
One used the conformal TT decomposition with additional restrictions
on the freely specifiable pieces, most notably conformal flatness and
maximal slicing. These assumptions decouple the constraints
and allow for analytical solutions to the momentum
constraints, the so-called {\em Bowen-York extrinsic
  curvature}\cite{Bowen:1979,Bowen-York:1980,Kulkarni-Shepley-York:1983}. 
All that remains is to solve a single elliptic
equation, the Hamiltonian constraint.  This approach has been used in
several variations\cite{Thornburg:1987,Cook-Choptuik-etal:1993,
  Brandt-Bruegmann:1997}.

However, these numerical simplifications come at a cost.
The freely specifiable pieces have been restricted to a small subset of
all possible choices.  One therefore can generate only a subset of
all possible initial-data sets, one that might not contain the desired
astrophysically relevant initial-data sets.

Over the last few years there have been additional developments:
Post-Newtonian results have indicated that binary black hole metrics
are not conformally flat\cite{Rieth:1997,
Damour-Jaranowski-Schaefer:2000b}.
With certain restrictions on the slicing, it has also been shown that a
single stationary spinning black hole cannot be represented with a
conformally flat spatial metric
\cite{Monroe:1976,Garat-Price:2000}.  In \cite{Pfeiffer-Teukolsky-Cook:2000},
it was shown that conformally flat initial data sets for spinning
binary black holes contain an unphysical contamination.  Moreover,
computations in spherical symmetry\cite{Lousto-Price:1998} indicated
that initial-data sets depend strongly on the choice of the extrinsic
curvature and that the use of the Bowen-York extrinsic curvature might
be problematic.

Therefore it is necessary to move beyond conformally flat initial data
and to explore different choices for the extrinsic curvature.  Matzner
et al\cite{Matzner-Huq-Shoemaker:1999} proposed a non-flat conformal
metric based on the superposition of two Kerr-Schild metrics; a
solution based on this proposal was obtained in
\cite{Marronetti-Matzner:2000}. This work demonstrated the existence
of solutions to the 3D set of equations, but did not examine the data
sets in any detail.
Refs.~\cite{Gourgoulhon-Grandclement-Bonazzola:2001a,Grandclement-Gourgoulhon-Bonazzola:2001b} obtained solutions to a
similar set of equations during the computation of quasi-circular
orbits of binary black holes.  However, these works assumed conformal
flatness.

In this paper we present a code capable of solving the three
above-mentioned decompositions of the constraint equations for
arbitrary choices of the freely specifiable pieces.  This code is
based on spectral methods which have been used successfully for several
astrophysical problems (see
e.g. \cite{Bonazzola-Gourgoulhon-etal:1993,
Bonazzola-Gourgoulhon-Marck:1998,
Kidder-Scheel-etal:2000,
Gourgoulhon-Grandclement-etal:2001, 
Kidder-Scheel-Teukolsky:2001, 
Ansorg-Kleinwaechter-Meinel:2002,
Grandclement-Gourgoulhon-Bonazzola:2001b}).  Our code is described in
detail in a separate paper\cite{Pfeiffer-Kidder-etal:2003}.

 We compute solutions of the different decompositions for the non-flat
conformal metric proposed in Ref.~\cite{Matzner-Huq-Shoemaker:1999}.
Each decomposition has certain choices for the freely specifiable
pieces and boundary conditions that seem ``natural'' and which we use
in our solutions.  We compare the computed initial-data sets with each
other and with the ``standard'' conformally-flat solution using the
Bowen-York extrinsic curvature. Our major results confirm that
\begin{enumerate}
\item\label{res:1}the different decompositions generate different
physical initial-data sets for seemingly similar choices for the
freely specifiable pieces.
\item the choice of extrinsic curvature is critical.
\end{enumerate}
The first result is certainly not unexpected, but each of these
factors can cause relative differences of several per cent in 
gauge-invariant quantities like the ADM-energy.

We also find that the conformal TT/physical TT decompositions generate
initial-data sets with ADM-energies $2-3$\% higher than data sets of
the conformal thin sandwich decomposition.  We demonstrate that this
higher ADM-energy is related to the choice of the freely specifiable part
of the extrinsic curvature.  In addition, we find that the solutions
depend significantly on the boundary conditions used.

The paper is organized as follows. In the next section we describe the
three decompositions. Section \ref{sec:FreePieces} explains how we
choose the freely specifiable data within each decomposition.  In
section \ref{sec:Implementation-Comparing} we describe and test our elliptic
solver.  Section \ref{sec:Results-Comparing} presents our results, which we
discuss in section \ref{sec:Discussion-Comparing}.

\section{Decompositions of Einstein's equations and the constraint equations}

\subsection{3+1 Decomposition}

In this paper we use the standard 3+1 decomposition of Einstein's
equations. We foliate the spacetime with $t=\mbox{const}$
hypersurfaces and write the four-dimensional metric as
\begin{equation}
  ^{(4)}ds^2=-N^2\,dt^2+\gamma_{ij}(dx^i+N^idt)(dx^j+N^jdt),
\end{equation}
where $\gamma_{ij}$ represents the induced 3-metric on the
hypersurfaces, and $N$ and $N^i$ represent the lapse function and the
shift vector, respectively.  We define the extrinsic curvature $K_{ij}$
on the slice by 
\begin{equation}\label{eq:K}
{\bf K}=-\frac{1}{2}\,{\perp}\:{\cal L}_n{^{(4)}{\bf g}}
\end{equation}
where ${}^{(4)}{\bf g}$ is the space-time metric, $n$ the unit
normal to the hypersurface, and $\perp$ denotes the projection operator
into the $t=\mbox{const}$ slice.
Einstein's
equations divide into constraint equations, which constrain the data
$(\gamma_{ij}, K^{ij})$ on each hypersurface, and into evolution
equations, which determine how the data $(\gamma_{ij}, K^{ij})$ evolve
from one hypersurface to the next.
The constraint equations are 
\begin{align}
  \label{eq:HamiltonianConstraintPhysical}
  R+K^2-K_{ij}K^{ij}&=16\pi G\rho\\
  \label{eq:MomentumConstraint}
  \nabla_j\left(K^{ij}-\gamma^{ij}K\right)&=8\pi G j^i.
\end{align}
Eq.~(\ref{eq:HamiltonianConstraintPhysical}) is called the {\em
  Hamiltonian constraint}, and Eq.~(\ref{eq:MomentumConstraint}) is
referred to as the {\em momentum constraint}.  $K=\gamma_{ij}K^{ij}$
is the trace of the extrinsic curvature, $\nabla$ and $R$ denote the
three dimensional covariant derivative operator and the Ricci scalar
compatible with $\gamma_{ij}$. $\rho$ and $j^i$ are the energy and
momentum density, respectively. Both vanish for the vacuum spacetimes
considered here.

The evolution equation for $\gamma_{ij}$ is
\begin{equation}
  \label{eq:Evolution-gamma}
  \partial_t\gamma_{ij}=-2N K_{ij}+\nabla_iN_j+\nabla_jN_i,
\end{equation}
which follows from Eq.~(\ref{eq:K}).  There is a similar albeit
longer equation for $\partial_tK_{ij}$ which we will not need in this
paper. The choices of $N$ and $N^i$ are arbitrary. One can in
principle use any lapse and shift in the evolution off the
initial-data surface, although some choices of lapse and shift are better
suited to numerical implementation than others.

Later in this paper we will often refer to the trace-free piece of 
Eq.~(\ref{eq:Evolution-gamma}).  Denote the tracefree piece of a tensor
by $\TF(.)$, and define $\gamma\equiv\det \gamma_{ij}$.  From Eq.~(\ref{eq:Evolution-gamma}) and the fact that
$\delta\ln\gamma=\gamma^{kl}\delta\gamma_{kl}$, it follows that
\begin{equation}\label{eq:Evolution-gamma-TF}
\TF(\partial_t\gamma_{ij})
=\gamma^{1/3}\partial_t\left(\gamma^{-1/3}\gamma_{ij}\right)
=-2N A_{ij}+(\ComparingLong N)_{ij}.
\end{equation}
Here $A_{ij}=K_{ij}-\frac{1}{3}\gamma_{ij}K$ denotes the trace-free
extrinsic curvature, and
\begin{equation}
  \label{eq:DefinitionLong}
  (\ComparingLong N)^{ij}\equiv 
  \nabla^iN^j+\nabla^jN^i-\frac{2}{3}\gamma^{ij}\nabla_kN^k.
\end{equation}
$\ComparingLong$ always acts on a vector, so the 'N' in $(\ComparingLong N)^{ij}$ denotes
the shift vector $N^i$ and not the lapse $N$. 

\subsection{Decomposition of the constraint equations}
\label{sec:Decomposition}

Equations~(\ref{eq:HamiltonianConstraintPhysical}) and
(\ref{eq:MomentumConstraint}) constrain four degrees of freedom of the
12 quantities $(\gamma_{ij}, K^{ij})$. However, it is not immediately
clear which pieces of $\gamma_{ij}$ and $K^{ij}$ are constrained and
which pieces can be chosen at will.  Several decompositions have been
developed to divide the 12 degrees of freedom into freely specifiable
and constrained pieces.  We will now review some properties of
the three decompositions we consider in this paper.

All three decompositions follow the York-Lichnerowicz approach and
use a conformal transformation on the
physical 3-metric $\gamma_{ij}$,
\begin{equation}
  \label{eq:ConformalMetric}
  \gamma_{ij}=\psi^4\tilde\gamma_{ij}.
\end{equation}
$\psi$ is called the {\em conformal factor}, $\tilde\gamma_{ij}$ the
{\em background metric} or {\em conformal metric}.  We will denote all
conformal quantities with a tilde. In particular, $\tilde\nabla$ is
the covariant derivative operator associated with $\tilde\gamma_{ij}$,
and $\tilde R_{ij}$ and $\tilde R$ are the Ricci tensor and Ricci
scalar of $\tilde\gamma_{ij}$.

The extrinsic curvature is split into its trace and trace-free part,
\begin{equation}
  \label{eq:SplitK_Aij}
  K^{ij}=A^{ij}+\frac{1}{3}\gamma^{ij}K.
\end{equation}
The three decompositions of the constraint equations we discuss in
this paper differ in how $A^{ij}$ is decomposed.  For each
decomposition, we discuss next the relevant equations, and describe
how we choose the quantities one has to specify before solving the
equations.  We use the conventions of \cite{Cook:2000}.

\subsubsection{Conformal TT Decomposition}

In this decomposition one first conformally transforms the traceless
extrinsic curvature,
\begin{equation}
  \label{eq:ConfTT-AijConfAij}
  A^{ij}=\psi^{-10}\tilde A^{ij},
\end{equation}
and then applies a TT decomposition with respect to the background
metric $\tilde\gamma_{ij}$:
\begin{equation}
\label{eq:ConfTT-Aij}
\tilde A^{ij}=\tilde A_{TT}^{ij}+(\tilde\ComparingLong X)^{ij}.
\end{equation}
The operator $\tilde\ComparingLong$ is defined by Eq.~(\ref{eq:DefinitionLong}) but
using the conformal metric $\tilde\gamma_{ij}$ and derivatives
associated with $\tilde\gamma_{ij}$. $\tilde A^{ij}_{TT}$ is
transverse with respect to the conformal metric, $\tilde\nabla_j\tilde
A^{ij}_{TT}=0$, and is traceless. 

Substituting Eqs.~(\ref{eq:ConfTT-AijConfAij}) and (\ref{eq:ConfTT-Aij})
into the momentum constraint (\ref{eq:MomentumConstraint}), one finds
that it reduces to an elliptic equation for $X^i$, whereas $\tilde
A^{ij}_{TT}$ is unconstrained. 

In order to specify the transverse-traceless tensor $\tilde
A^{ij}_{TT}$ one usually has to {\em construct} it from a general
symmetric trace-free tensor $\tilde M^{ij}$ by subtracting the
longitudinal piece.  As described in \cite{Cook:2000} one can
incorporate the construction of $\tilde A^{ij}_{TT}$ from $\tilde
M^{ij}$ into the momentum constraint, arriving at the following
equations:

\begin{gather}
\label{eq:ConfTT-1-Comparing}
    \tilde\nabla^2\psi-\frac{1}{8}\psi\tilde R-\frac{1}{12}\psi^5K^2
    +\frac{1}{8}\psi^{-7}\tilde A_{ij}\tilde A^{ij}=-2\pi G\psi^5\rho,\\
\label{eq:ConfTT-2-Comparing}
    \tildeLapLong V^i-\frac{2}{3}\psi^6\tilde\nabla^iK
    +\tilde\nabla_j\tilde M^{ij}=8\pi G\psi^{10}j^i,
\end{gather}
where $\tilde A^{ij}$ and the operator $\tildeLapLong$ are defined by
\begin{equation}
\label{eq:ConfTT-Aij_tilde}
  \tilde A^{ij}=(\tilde\ComparingLong V)^{ij}+\tilde M^{ij}
\end{equation}
and
\begin{equation}
  \label{eq:DefinitionLapLong}
  \tildeLapLong V^i\equiv\tilde\nabla_j(\tilde\ComparingLong V)^{ij}.
\end{equation}

After solving these equations for $\psi$ and $V^i$, one obtains the
physical metric $\gamma_{ij}$ from (\ref{eq:ConformalMetric}) and the
extrinsic curvature from
\begin{equation}
\label{eq:ConfTT-3}
    K^{ij}=\psi^{-10}\tilde A^{ij}+\frac{1}{3}\psi^{-4}\tilde\gamma^{ij}K. 
\end{equation}
We will refer to Eqs.~(\ref{eq:ConfTT-1-Comparing}) and (\ref{eq:ConfTT-2-Comparing})
together with (\ref{eq:ConfTT-Aij_tilde}), (\ref{eq:ConfTT-3}) and
(\ref{eq:ConformalMetric}) as the {\em conformal TT equations}.  In
these equations we are free to specify the background metric
$\tilde\gamma_{ij}$, the trace of the extrinsic curvature $K$, and a
symmetric traceless tensor $\tilde M^{ij}$.  The solution $V^i$ will
contain a contribution that removes the longitudinal piece from
$\tilde M^{ij}$ and the piece that solves the momentum constraint if
$\tilde M^{ij}$ were transverse-traceless.

This decomposition has been the most important in the past, since if
one chooses a constant $K$ and if one considers vacuum spacetimes then
the momentum constraint (\ref{eq:ConfTT-2-Comparing}) decouples from the
Hamiltonian constraint (\ref{eq:ConfTT-1-Comparing}). Moreover, if one assumes
conformal flatness and $\tilde M^{ij}=0$, it is possible to write down
analytic solutions to Eq.~(\ref{eq:ConfTT-2-Comparing}), the so-called
Bowen-York extrinsic curvature.  In that case one has to deal with
only one elliptic equation for $\psi$.  The Bowen-York extrinsic
curvature can represent multiple black holes with arbitrary momenta
and spins.  One can fix boundary conditions for $\psi$ by requiring
that the initial-data slice be inversion symmetric at both
throats\cite{Misner:1963,Cook:1991}.  In that case one has to modify the
extrinsic curvature using a method of images.  We will
include initial-data sets obtained with this approach below, where we
will refer to them as {\em inversion symmetric} initial data.

Reasonable choices for the freely specifiable pieces $\tilde\gamma_{ij}$, $K$,
$\tilde M^{ij}$ will lead to an initial-data set $(\gamma_{ij},
K^{ij})$ that satisfies the constraint equations. How should we choose
all these functions in order to obtain a desired physical
configuration, say a binary black hole with given linear momenta and
spins for the individual holes?  We can gain insight into this question
by considering how the conformal TT decompositions can recover a known
solution.

Suppose we have a known solution $(\gamma_{0\,ij}, K^{ij}_0)$
of the constraint equations. Denote the trace and trace-free parts of
this extrinsic curvature by $K_0$ and $A^{ij}_0$, respectively.  If we
set 
\begin{equation}
\tilde\gamma_{ij}=\gamma_{0\,ij},\quad K=K_0,
\quad\tilde M^{ij}=A^{ij}_0
\end{equation}
then 
\begin{equation}
\psi=1,\quad V^i=0
\end{equation}
trivially solve Eqs.~(\ref{eq:ConfTT-1-Comparing}-\ref{eq:ConfTT-2-Comparing}). Note
that we have to set $\tilde M^{ij}$ equal to the trace-free part of the
extrinsic curvature.

Now suppose we have a guess for a metric and an extrinsic curvature,
which ---most likely--- will not satisfy the constraint equations
(\ref{eq:HamiltonianConstraintPhysical}) and
(\ref{eq:MomentumConstraint}). Set $\tilde\gamma_{ij}$ to the guess
for the metric, and set $K$ and $\tilde M^{ij}$ to the trace and
trace-free piece of the guess of the extrinsic curvature.  By solving
the conformal TT equations we can compute $(\gamma_{ij}, K^{ij})$ that
satisfy the constraint equations.  If our initial guess is ``close''
to a true solution, we will have $\psi\approx 1$ and $V^i\approx 0$,
so that $\gamma_{ij}$ and $K^{ij}$ will be close to our initial guess.

Thus one can guess a metric and extrinsic curvature as well as
possible and then solve the conformal TT equations to obtain corrected
quantities that satisfy the constraint equations.

An artifact of the conformal TT decomposition is that one has no
direct handle on the transverse traceless piece with respect to the
{\em physical} metric.  For any vector $X^i$,
\begin{equation}\label{eq:ConformalLong}
  (\ComparingLong X)^{ij}=\psi^{-4}(\tilde\ComparingLong X)^{ij}.
\end{equation}
Thus, Eqs.~(\ref{eq:ConfTT-AijConfAij}) and (\ref{eq:ConfTT-Aij}) imply
\begin{equation}\label{eq:ConfTT-Decomposition-psi}
  A^{ij}=\psi^{-10}\tilde A^{ij}_{TT}+\psi^{-6}(\ComparingLong X)^{ij}.
\end{equation}
For any symmetric traceless tensor $S^{ij}$
\begin{equation}
  \label{eq:DecompositionSymmetricTrace-Free}
  \nabla_jS^{ij}=\psi^{-10}\tilde\nabla_j\left(\psi^{10}S^{ij}\right).
\end{equation}
Therefore the first term on the right hand side of
Eq.~(\ref{eq:ConfTT-Decomposition-psi}) is transverse-traceless with
respect to the physical metric, 
\begin{equation}
  \nabla_j\left(\psi^{-10}\tilde A^{ij}_{TT}\right)=0.
\end{equation}
However, the second term on the right hand side of
Eq.~(\ref{eq:ConfTT-Decomposition-psi}) is conformally weighted.
Therefore, Eq.~(\ref{eq:ConfTT-Decomposition-psi}) does not represent the
usual TT decomposition.

\subsubsection{Physical TT Decomposition}

In this case one decomposes the physical traceless extrinsic
curvature directly:
\begin{equation}
  \label{eq:PhysTT-Aij}
  A^{ij}=A^{ij}_{TT}+(\ComparingLong X)^{ij}.
\end{equation}
As above in the conformal TT decomposition, the momentum constraint
becomes an elliptic equation for $X^i$. We can again incorporate the
construction of the symmetric transverse traceless tensor
$A^{ij}_{TT}$ from a general symmetric tensor $\tilde M^{ij}$ into the
momentum constraint. Then one obtains the {\em physical TT equations}:

\begin{gather}
\label{eq:PhysTT-1}
    \tilde\nabla^2\psi-\frac{1}{8}\psi\tilde R-\frac{1}{12}\psi^5K^2
    +\frac{1}{8}\psi^{5}\tilde A_{ij}\tilde A^{ij}=-2\pi G\psi^5\rho,\\
\label{eq:PhysTT-2}
    \tildeLapLong V^i+6(\tilde\ComparingLong V)^{ij}\tilde\nabla_j\ln\psi
    -\frac{2}{3}\tilde\nabla^iK
    +\psi^{-6}\tilde\nabla_j\tilde M^{ij}=8\pi G\psi^4j^i,
\end{gather}
where $\tilde A^{ij}$ is defined by
\begin{equation}
    \tilde A^{ij}=(\tilde\ComparingLong V)^{ij}+\psi^{-6}\tilde M^{ij}.
\end{equation}
When we have solved (\ref{eq:PhysTT-1}) and (\ref{eq:PhysTT-2}) for
$\psi$ and $V^i$, the physical metric is given by
(\ref{eq:ConformalMetric}), and the extrinsic curvature is
\begin{equation}
\label{eq:PhysTT-3}
K^{ij}=\psi^{-4}\left(\tilde A^{ij}+\frac{1}{3}\tilde\gamma^{ij}K\right).
\end{equation}

We are free to specify the background metric $\tilde\gamma_{ij}$, the
trace of the extrinsic curvature $K$, and a symmetric traceless tensor
$\tilde M^{ij}$.  As with the conformal TT equations, the solution
$V^i$ will contain a contribution that removes the longitudinal piece
from $\tilde M^{ij}$ and a piece that solves the momentum constraint
if $\tilde M^{ij}$ were transverse-traceless.

These equations can be used in the same way as the conformal TT
equations.  Guess a metric and extrinsic curvature, set
$\tilde\gamma_{ij}$ to the guess for the metric, and $K$ and $\tilde
M^{ij}$ to the trace and trace-free pieces of the guess for the
extrinsic curvature. Then solve the physical TT equations to obtain a
corrected metric $\gamma_{ij}$ and a corrected extrinsic curvature
$K^{ij}$ that satisfy the constraint equations.

The transverse traceless piece of $K^{ij}$ (with respect to
$\gamma_{ij}$) will be the transverse traceless piece of
$\psi^{-10}\tilde M^{ij}$.  One can also easily rewrite the physical
TT equations such that $\psi^{-10}\tilde M^{ij}$ can be freely chosen
instead of $\tilde M^{ij}$.  So, in this decomposition we can directly
control the TT piece of the physical extrinsic curvature.  We have
chosen to follow \cite{Cook:2000} since it seems somewhat more natural
to specify two conformal quantities, $\tilde\gamma_{ij}$ and $\tilde
M^{ij}$ than to specify one conformal and one physical quantity.

\subsubsection{Conformal thin sandwich decomposition}

The conformal and physical TT decompositions rely on a tensor
splitting to decompose the trace-free part of the extrinsic curvature.
In contrast, the conformal thin sandwich decomposition simply defines
 $A^{ij}$ by Eq.~(\ref{eq:ConfTT-AijConfAij}) and the decomposition
\begin{equation}
\label{eq:SandwichTT-3}
  \tilde A^{ij}\equiv\frac{1}{2\tilde\alpha}
\left((\tilde\ComparingLong \beta)^{ij}-\tilde u^{ij}\right),
\end{equation}
where $\tilde u^{ij}$ is symmetric and tracefree.
Eq.~(\ref{eq:SandwichTT-3}) is motivated by
Eq.~(\ref{eq:Evolution-gamma-TF}): If one evolves an initial-data set
with $A^{ij}$ of the form (\ref{eq:SandwichTT-3}) using as lapse and
shift
\begin{equation}\label{eq:SandwichTT-LapseShift}
\begin{aligned}
  N&=\psi^6\tilde\alpha,\\
  N^i&=\beta^i,
\end{aligned}
\end{equation}
then 
\begin{equation}\label{eq:TF-sandwich}
\TF(\partial_t\gamma_{ij})=\psi^4\tilde u_{ij}.
\end{equation}
Therefore, the decomposition (\ref{eq:SandwichTT-3}) is closely
related to the kinematical quantities in an evolution. Although
$\tilde\alpha$ and $\beta^i$ are introduced in the context of initial
data, one usually refers to them as the ``conformal lapse'' and ``shift''.
While the form of Eq.~(\ref{eq:SandwichTT-3}) is similar in form to
the conformal and physical TT decompositions, there are differences.
In particular, $\tilde{u}^{ij}$ is {\em not} divergenceless.

Within the {\em conformal thin sandwich decomposition}, the constraint
equations take the form:
\begin{gather}
\label{eq:SandwichTT-1}
    \tilde\nabla^2\psi-\frac{1}{8}\psi\tilde R-\frac{1}{12}\psi^5K^2
    +\frac{1}{8}\psi^{-7}\tilde A_{ij}\tilde A^{ij}=-2\pi G\psi^5\rho\\
\tildeLapLong\beta^i-(\tilde\ComparingLong\beta)^{ij}\tilde\nabla_j\ln\tilde\alpha
-\frac{4}{3}\tilde\alpha\psi^6\tilde\nabla^iK
\qquad\qquad\qquad\qquad\nonumber\\
\label{eq:SandwichTT-2}
\qquad\qquad\qquad\qquad -\tilde\alpha\tilde\nabla_j\Big(\frac{1}{\tilde\alpha}\tilde u^{ij}\Big)
=16\pi G\tilde\alpha\psi^{10}j^i
\end{gather}
Having solved Eqs.~(\ref{eq:SandwichTT-1}) and (\ref{eq:SandwichTT-2}) 
for $\psi$ and the vector $\beta^i$, one obtains the physical metric
from (\ref{eq:ConformalMetric}) and the extrinsic curvature from
\begin{equation}
    K^{ij}=\psi^{-10}\tilde A^{ij}+\frac{1}{3}\psi^{-4}\tilde\gamma^{ij}K.
\end{equation}
In this decomposition we are free to specify a conformal metric
$\tilde\gamma_{ij}$, the trace of the extrinsic curvature $K$, a
symmetric trace-free tensor $\tilde u^{ij}$ and a function $\tilde\alpha$.

It seems that the conformal thin sandwich decomposition contains
additional degrees of freedom in the form of the function
$\tilde\alpha$ and three additional unconstrained components of
$\tilde{u}^{ij}$. This is not the case.  The longitudinal piece of
$\tilde u^{ij}$ corresponds to the gauge choice of the actual shift
vector used in an evolution.  Thus $\tilde{u}^{ij}$ really only
contributes two degrees of freedom, just like $\tilde{M}^{ij}$ in the
conformal and physical TT decompositions.  Furthermore, we can reach
any {\em reasonable} physical solution $(\gamma_{ij}, K^{ij})$ with
any {\em reasonable} choice of $\tilde\alpha$; each choice of
$\tilde\alpha$ simply defines a new decomposition.  A forthcoming
article by York\cite{Pfeiffer-York:2003} will elaborate on these ideas.  Note
that for $\tilde \alpha=1/2$ we recover the conformal TT
decomposition.

Let us now turn to the question of how one should pick the freely
specifiable data in the conformal thin sandwich approach. 
We motivate our prescription again by considering how to recover a
known spacetime: Assume we are given a full four-dimensional spacetime
with 3+1 quantities $\gamma_{0\,ij}$, $K_0^{ij}$, $N_0^i$ and $N_0$.
Further assume the spacetime is stationary and the slicing is such that
$\partial_t\gamma_{ij}=\partial_tK_{ij}=0$.  An example for such a
situation is a Kerr black hole in Kerr-Schild or
Boyer-Lindquist coordinates.

Using $\partial_t\gamma_{0\,ij}=0$ in
Eq.~(\ref{eq:Evolution-gamma-TF}) yields a relation for the trace-free
extrinsic curvature
\begin{equation}
  A_0^{ij}=\frac{1}{2N_0}(\ComparingLong N_0)^{ij}.
\end{equation}
This is a decomposition of the form (\ref{eq:SandwichTT-3}) with
$\tilde u^{ij}=0$. Therefore, if we choose the freely specifiable data
for the conformal thin sandwich equations as 
\begin{equation}
\begin{aligned}
\tilde\gamma_{ij}&=\gamma_{0\,ij},&
\tilde\alpha&=N_0,\\
K&=K_0,&\tilde u^{ij}&=0,
\end{aligned}
\end{equation}
and if we use appropriate boundary conditions, then the solution of
the conformal thin sandwich equations will be $\psi=1$ and $\beta^i=N_0^i$.
As part of the solution, we obtain the shift vector needed for an
evolution to produce $\TF(\partial_t\gamma_{ij})=0$.
Not needing a guess for the trace-free extrinsic curvature, and having
the solution $\beta^i$ automatically provide an initial shift for
evolution, make the conformal thin sandwich equations very attractive.

In order to generate initial-data slices that permit an evolution with
zero time derivative of the conformal metric --- a highly desirable
feature for quasi-equilibrium data, or for a situation with holes
momentarily at rest --- one can proceed as follows: Set
$\tilde\gamma_{ij}$ and $K$ to the guess for the metric and trace of
extrinsic curvature, respectively. Set $\tilde\alpha$ to the
lapse function that one is going to use in the evolution, and set
$\tilde u^{ij}=0$. If these guesses are good, the conformal factor
$\psi$ will be close to 1, and $N=\psi^6\tilde\alpha$ as well as
$N^i=\beta^i$ give us the actual lapse function and shift vector to use in
the evolution.

\section{Choices for the freely specifiable data}
\label{sec:FreePieces}

\subsection{Kerr-Schild coordinates}
\label{sec:Choices:Kerr-Schild}

We base our choice for the freely specifiable data on a superposition of two
Kerr black holes in Kerr-Schild coordinates. In this section we
describe the Kerr-Schild solution and collect necessary
equations. We also describe how we compute the 3-metric, lapse, shift
and extrinsic curvature for a boosted black hole with arbitrary spin.

A Kerr-Schild metric is given by
\begin{equation}\label{eq:KerrSchild}
  g_{\mu\nu}=\eta_{\mu\nu}+2Hl_\mu l_\nu,
\end{equation}
where $\eta_{\mu\nu}$ is the Minkowski metric, and $l_\mu$ is a
null-vector with respect to both the full metric and the Minkowski
metric: $g^{\mu\nu}l_\mu l_\nu=\eta^{\mu\nu}l_\mu l_\nu=0$.  The
3-metric, lapse and shift are
\begin{align}
  \label{eq:KerrSchild-gamma}
  \gamma_{ij}&=\delta_{ij}+2Hl_il_j,\\
  N&=(1+2Hl^tl^t)^{-1/2},\\
\label{eq:KerrSchild-beta}
  N^i&=-\frac{2Hl^tl^i}{1+2Hl^tl^t}.
\end{align}

For a black hole at rest at the origin with mass $M$ and angular
momentum $M\vec a$, one has
\begin{align}
  H&=\frac{Mr^3}{r^4+(\vec a\cdot\vec x)^2},\\
  l^{\mbox{\footnotesize rest}}_\mu&=(1, \vec l_{\mbox{\footnotesize rest}}),\\
  \vec l_{\mbox{\footnotesize rest}}\;\,&=\frac{r\vec x-\vec a\times\vec x+(\vec a\cdot\vec x)\vec a/r}
              {r^2+a^2},
\end{align}
with
\begin{equation}
  r^2=\frac{\vec x^2-\vec a^2}{2}
       +\left(\frac{(\vec x^2- \vec a^2)^2}{4}
       +(\vec a\cdot\vec x)^2\right)^{1/2}.
\end{equation}
For a nonrotating black hole with $\vec a=0$, $H$ has a pole at the
origin, whereas for rotating black holes, $r$ has a ring singularity.
We will therefore have to excise from the computational domain a
region close to the center of the Kerr-Schild black hole.

Under a boost, a Kerr-Schild coordinate system transforms into a
Kerr-Schild coordinate system. Applying a Lorentz transformation with
boost velocity $v^i$ to $l^{\mbox{\footnotesize rest}}_\mu$, we obtain
the null-vector $l_\mu$ of the boosted Kerr-Schild coordinate system.
Eqs.~(\ref{eq:KerrSchild-gamma}-\ref{eq:KerrSchild-beta}) give then
the boosted 3-metric, lapse, and shift.  Since all time-dependence is
in the uniform motion, evolution with lapse $N$ and shift $N^i$ yields
$\partial_t\gamma_{ij}=-v^k\partial_k\gamma_{ij}$, and from
Eq.~(\ref{eq:Evolution-gamma}) one can compute the extrinsic curvature
\begin{equation}\label{eq:KerrSchild-Kij}
  K_{ij}=\frac{1}{2N}\left(v^k\partial_k\gamma_{ij}
         +\nabla_{\!i} N_j+\nabla_{\!j} N_i\right).
\end{equation}

If this initial-data set is evolved with the shift $N^i$, the black
hole will move through the coordinate space with velocity $v^i$.
However, if the evolution uses the shift vector $N^i+v^i$, the
coordinates will move with the black hole, and the hole will be at
rest in coordinate space.  The spacetime is nonetheless different from
a Kerr black hole at rest. The ADM-momentum will be $P^i_{ADM}=\gamma
M v^i$, where $M$ is the rest-mass of the hole and $\gamma=(1-\vec
v^2)^{-1/2}$.

\subsection{Freely specifiable pieces}

We want to generate initial data for a spacetime containing two black
holes with masses $M_{\!A,B}$, velocities $\vec v_{\!A,B}$ and spins
$M_{\!A}\vec a_A$ and $M_{\!B}\vec a_B$.

We follow the proposal of Matzner et al
\cite{Matzner-Huq-Shoemaker:1999, Marronetti-Matzner:2000} and base
our choices for the freely specifiable choices on two Kerr-Schild
coordinate systems describing two individual black holes.  The first
black hole with label A has an associated Kerr-Schild coordinate
system with metric
\begin{equation}
  \label{eq:KerrSchild-HoleA-gamma}
\gamma_{A\,ij}=\delta_{ij}+2H_{\!A}\,l_{A\,i}\,l_{A\,j},
\end{equation}
and with an extrinsic curvature $K_{\!A\,ij}$, a lapse $N_{\!A}$ and a
shift $N^i_A$. The trace of the extrinsic curvature is $K_A$. All
these quantities can be computed as described in the previous section,
\ref{sec:Choices:Kerr-Schild}. The second black hole has a similar set
of associated quantities which are labeled with the letter B.

For all three decompositions, we need to choose a conformal metric and
the trace of the extrinsic curvature. We choose
\begin{gather}
  \label{eq:BinaryKerrSchild-gamma}
  \tilde\gamma_{ij}=\delta_{ij}+2H_{\!A}\,l_{A\,i}\,l_{A\,j}
  +2H_{\!B}\,l_{B\,i}\,l_{B\,j}\\
K=K_{\!A}+K_B\label{eq:BinaryKerrSchild-K}
\end{gather}
The metric is singular at the center of each hole. Therefore we have
to excise spheres around the center of each hole from the
computational domain.  We now specify for each decomposition the
remaining freely specifiable pieces and boundary conditions.

\subsubsection{Conformal TT and physical TT decompositions}

For the conformal TT and physical TT decompositions we will be solving
for a correction to our guesses. 
As guess for the trace-free extrinsic curvature, we use a superposition
\begin{equation}
  \label{eq:BinaryKerrSchild-Mij}
  \tilde M^{ij}=\left(K^{(i}_{A\,k}+K^{(i}_{B\,k}
    -\frac{1}{3}\delta^{(i}_k(K_A+K_B)\right)\tilde\gamma^{j)k}.
\end{equation}
$\tilde M^{ij}$ is symmetric and trace-free with respect to the
conformal metric, $\tilde\gamma_{ij}\tilde M^{ij}=0$.  Solving for a
correction only, we expect that $\psi\approx 1$ and $V^i\approx 0$,
hence we use Dirichlet boundary conditions
\begin{equation}
  \label{eq:BC-ConfTT-PhysTT}
  \psi=1, \qquad V^i=0.
\end{equation}

\subsubsection{Conformal thin sandwich}

For conformal thin sandwich, we restrict the discussion to either two
black holes at rest, or in a quasi-circular orbit in corotating
coordinates.  In these cases, one expects small or even vanishing
time-derivatives, $\partial_t\gamma_{ij}\approx 0$, and so
Eq.~(\ref{eq:TF-sandwich}) yields the simple choice
\begin{equation}\label{eq:CTS-uij}
\tilde u^{ij}=0.
\end{equation}

The conformal 3-metric and the trace of the extrinsic curvature are
still given by Eqs.~(\ref{eq:BinaryKerrSchild-gamma}) and
(\ref{eq:BinaryKerrSchild-K}). Orbiting black holes in a corotating
frame will not move in coordinate space, therefore we do not boost the
individual Kerr-Schild metrics in this decomposition: $v^i_{A/B}=0$.
The lapse functions $N_{A/B}$ and the shifts $N^i_{A/B}$ are also for
unboosted Kerr-Schild black holes.
   
We use Dirichlet boundary conditions:
\begin{subequations}\label{eq:BC-sandwich}
\begin{align}
\psi&=1                            &&\mbox{all boundaries}\\
\label{eq:Sandwich-BC2}
\beta^i&=N_{\!A}^i                 &&\mbox{sphere inside hole A}\\
\beta^i&=N_B^i                     &&\mbox{sphere inside hole B}\\
\label{eq:Sandwich-BC4}
\beta^i&=\vec\Omega\times\vec r    &&\mbox{outer boundary}
\end{align}
\end{subequations}

Eq.~(\ref{eq:Sandwich-BC4}) ensures that we are in a corotating
reference frame; the cross-product is performed in flat space, and
$\vec\Omega=0$ corresponds to two black holes at rest.  Close to the
holes we force the shift to be the shift of a single black hole in the
hope that this choice will produce a hole that is at rest in
coordinate space.

For the conformal lapse we use
\begin{equation}\label{eq:Sandwich-N1}
  \tilde\alpha=N_{\!A}+N_B-1
\end{equation}
or
\begin{equation}\label{eq:Sandwich-N2}
  \tilde \alpha=N_{\!A}\;N_B.
\end{equation}
The first choice of $\tilde\alpha$ follows the philosophy of adding
quantities of each individual hole. However, $\tilde\alpha$ of
Eq.~(\ref{eq:Sandwich-N1}) becomes negative sufficiently close to the
center of each hole and is therefore a bad choice if the excised
spheres are small. The choice (\ref{eq:Sandwich-N2}) does not change
sign and has at large distances the same behavior (same $1/r$ term) as
(\ref{eq:Sandwich-N1}).

\section{Numerical Implementation}
\label{sec:Implementation-Comparing}

We implemented an elliptic solver that can solve all three
decompositions we described above in complete generality.  The solver
uses domain decomposition and can handle nontrivial topologies. It is
based on pseudospectral collocation, that is, it expresses the
solution in each subdomain as an expansion in basis functions.  This
elliptic solver is described in detail in a separate paper
\cite{Pfeiffer-Kidder-etal:2003}.

\begin{figure}
\centerline{{\includegraphics[scale=0.35,angle=90]{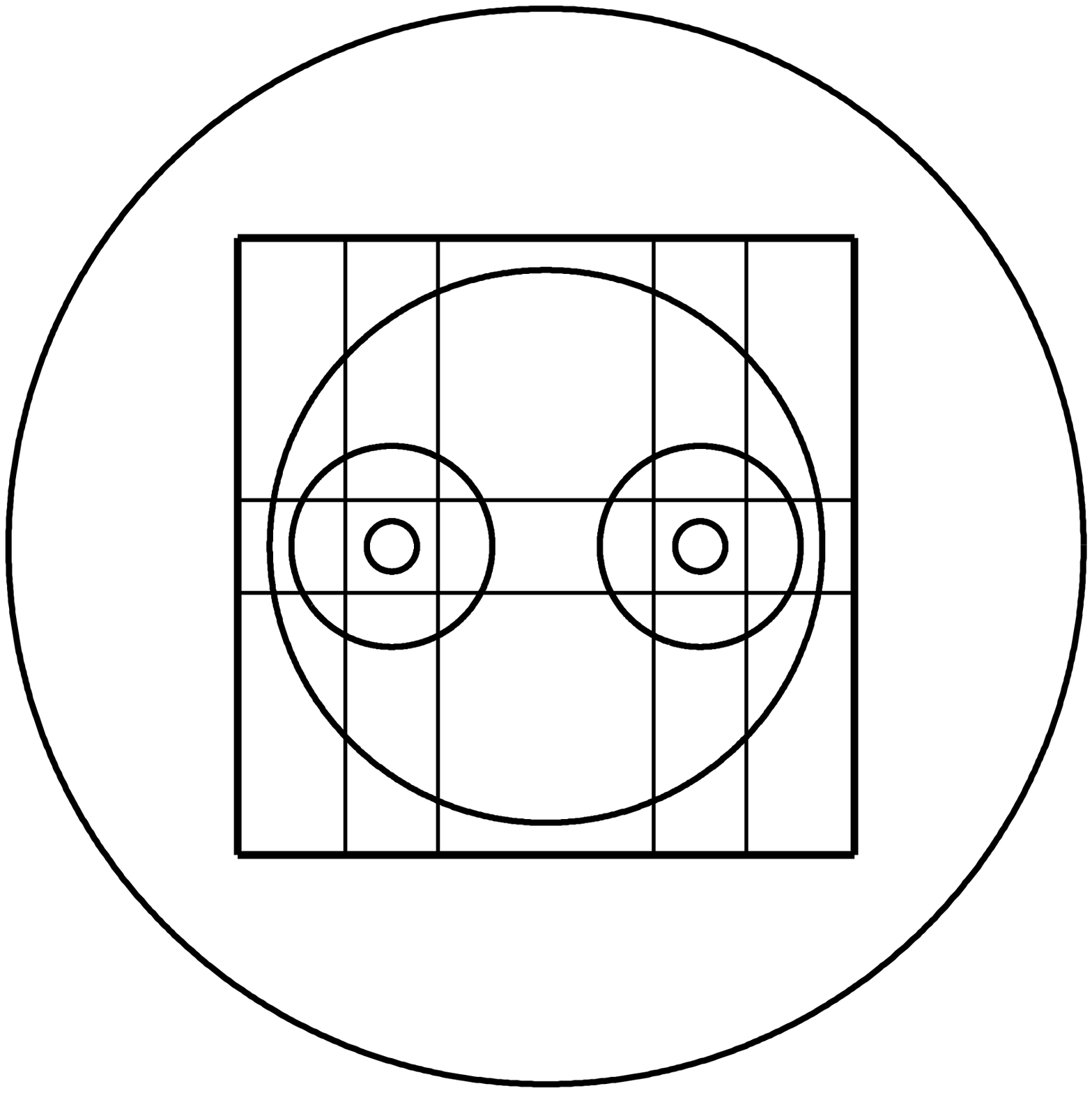}}}
\CAP{Domain decomposition for binary black
hole}{\label{fig:Domains}Structure of domains. Spherical shells around
each excised sphere are surrounded by 43 rectangular blocks and
another spherical shell.  The rectangular blocks touch each other and
overlap with all three spherical shells.}
\end{figure}

From the computational domain we excise two spheres containing the
singularities of the Kerr-Schild metric close to the center of each
hole.  Around each of the excised spheres we place a spherical shell.
These shells are patched together with $5\times 3\times 3=45$
rectangular blocks, with the two blocks at the location of the spheres
removed. Around these 43 blocks, another spherical shell is placed
that extends far out, typically to an outer radius of $10^7M$.  In the
rectangular blocks, we expand in Chebyshev polynomials, while in the
spheres we use Chebyshev polynomials radially and spherical harmonics
for the angular variables.  This setup is depicted in Fig.~\ref{fig:Domains}.

The domain decomposition in Fig.~\ref{fig:Domains} is fairly
complicated. Even if the shells were made as large as possible, they do
not cover the full computational domain when the excised spheres are
close together.  Thus additional subdomains are needed in any case.
Choosing the 43 cubes as depicted allows for relatively small inner
shells and for a relatively large inner radius of the outer shell.
Thus each shell covers a region of the computational domain in
which the angular variations of the solution are fairly low, allowing
for comparatively few angular basis-functions.

The code can handle a general conformal metric. In principle, the user
needs to specify only $\tilde\gamma_{ij}$. Then the code computes
$\tilde\gamma^{ij}$, and ---using numerical derivatives--- the Christoffel
symbols, Ricci tensor and Riemann scalar.  For the special case of the
Kerr-Schild metric of a single black hole and the superposition of two
Kerr-Schild metrics, Eq.~(\ref{eq:BinaryKerrSchild-gamma}), we compute
first derivatives analytically and use numerical derivatives only to
compute the Riemann tensor.

The solver implements Eqs.~(\ref{eq:ConfTT-1-Comparing}) and (\ref{eq:ConfTT-2-Comparing})
for the conformal TT decomposition, Eqs.~(\ref{eq:PhysTT-1}) and
(\ref{eq:PhysTT-2}) for the physical TT decomposition, and
Eqs.~(\ref{eq:SandwichTT-1}) and (\ref{eq:SandwichTT-2}) for the
conformal thin sandwich decomposition.

After solving for $(\psi, V^i)$ [conformal TT and physical TT], or
$(\psi, \beta^i)$ [thin sandwich] we compute the physical metric
$\gamma_{ij}$ and the physical extrinsic curvature $K^{ij}$ of the
solution.  Utilizing these physical quantities $(\gamma_{ij},
K^{ij})$, we implement several analysis tools.  We evaluate the
constraints in the form of
Eq.~(\ref{eq:HamiltonianConstraintPhysical}) and
(\ref{eq:MomentumConstraint}), we compute ADM-quantities and we search
for apparent horizons. Note that these analysis tools are completely
independent of the particular decomposition; they rely only on
$\gamma_{ij}$ and $K^{ij}$.

Next we present tests ensuring that the various systems of equations
are solved correctly.  We also include tests of the
analysis tools showing that we can indeed compute constraints,
ADM-quantities and apparent horizons with good accuracy.

\subsection{Testing the conformal TT and physical TT decompositions}
\label{sec:Comparing:TestingTT}

We can test the solver by conformally distorting a known solution of
the constraint equations.  Given a solution to the constraint
equations $(\gamma_{0\,ij}, K^{ij}_0)$ pick functions
\begin{gather}
  \Psi>0,\qquad  W^i
\end{gather}
and set
\begin{align}
  \label{eq:Test-gamma}
  \tilde\gamma_{ij}&=\Psi^{-4}\gamma_{0\,ij},\\
  \label{eq:Test-K}
  K\;&=K_0,
\end{align}
and 
\begin{align}\label{eq:Test-ConfTT-M}
\tilde M^{ij}&=\Psi^{10}\left(K_0^{ij}-\frac{1}{3}\gamma_0^{ij}K_0\right)
                 -\Psi^4(\ComparingLong_0W)^{ij}
\intertext{for conformal TT or}
\label{eq:Test-PhysTT-M}
\tilde M^{ij}&=\Psi^{10}\left(K_0^{ij}-\frac{1}{3}\gamma_0^{ij}K_0
                 -(\ComparingLong_0W)^{ij}\right)
\end{align}
for physical TT.
With these freely specifiable pieces and appropriate boundary
conditions, a solution of the conformal TT equations
(\ref{eq:ConfTT-1-Comparing}), (\ref{eq:ConfTT-2-Comparing}) or the physical TT equations
(\ref{eq:PhysTT-1}), (\ref{eq:PhysTT-2}) will be
\begin{align}
  \psi&=\Psi\\
  V^i&=W^i.
\end{align}
From Eq.~(\ref{eq:ConformalMetric}) we recover our initial metric
$\gamma_{0\,ij}$, and from Eq.~(\ref{eq:ConfTT-3}) [conformal TT] or
Eq.~(\ref{eq:PhysTT-3}) [physical TT] we recover the extrinsic
curvature $K^{ij}_0$.

In our tests we used the particular choices
\begin{gather}
\label{eq:Psi}
  \Psi=1+\frac{8(r-2)}{36+x^2+0.9y^2+1.3(z-1)^2}\\
\label{eq:W^i}
  W^i=\frac{50(r-2)}{(6^4+r^4)} (-y, x, 1).
\end{gather}

\begin{figure}
\centerline{\includegraphics[scale=0.35]{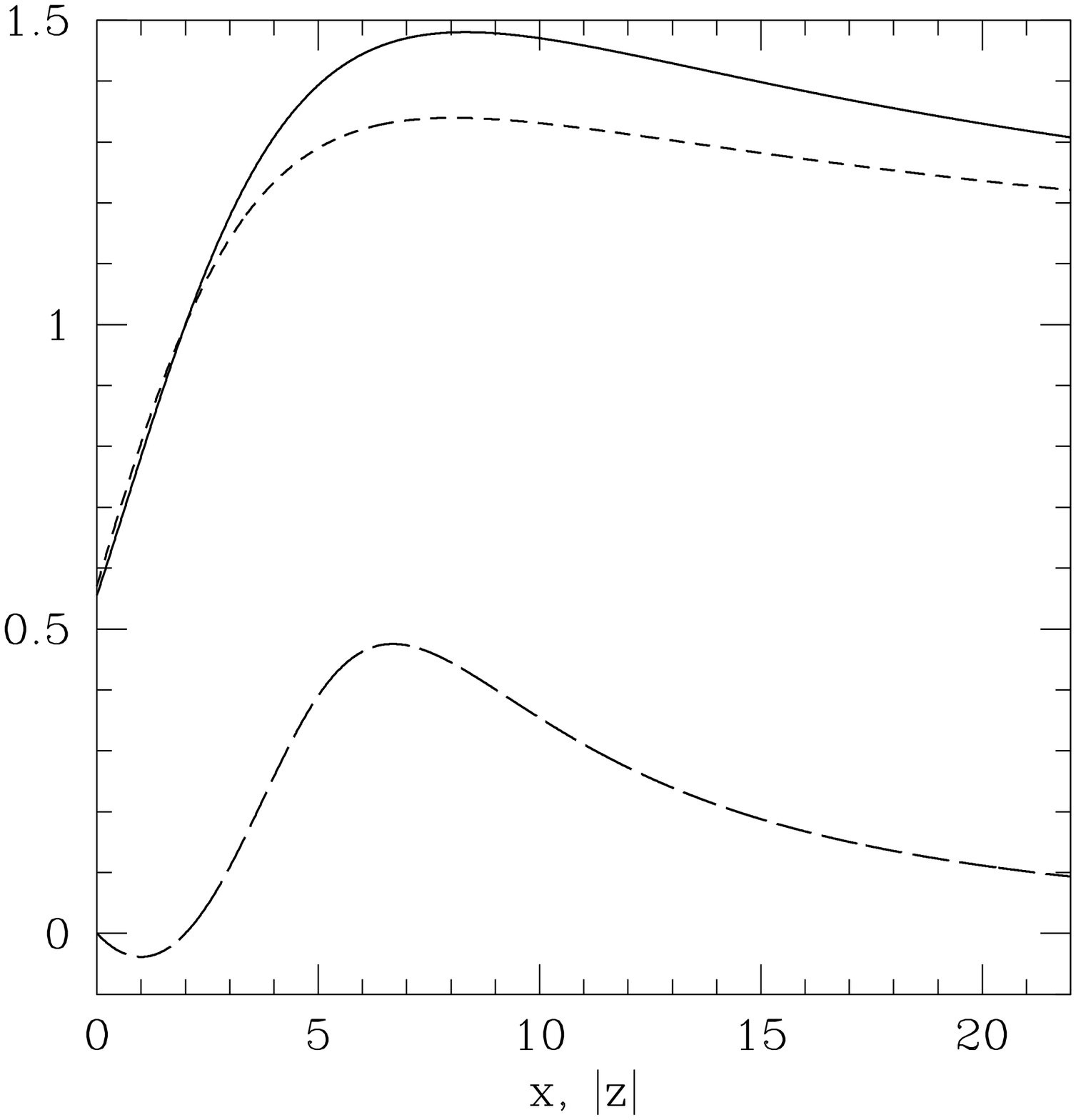}}
\CAP{Plot of the functions $\Psi$ and $W^i$ from Eqs.~(\ref{eq:Psi}) 
  and (\ref{eq:W^i})}{\label{fig:Testing-PlotDistortion}Plot of the functions $\Psi$ and $W^i$ from Eqs.~(\ref{eq:Psi}) 
  and (\ref{eq:W^i}).  The solid line depicts $\Psi$ along the
  positive $x$-axis, the short dashed line depicts $\Psi$ along the
  negative $z$-axis. The long dashed line is a plot of $W^y$ along the
  positive $x$-axis.}
\end{figure}

These functions are plotted in Fig.~\ref{fig:Testing-PlotDistortion}.
$\Psi$ varies between 0.8 and 1.5, $W^i$ varies between $\pm 0.5$, and
both take their maximum values around distance $\sim 7$ from the
center of the hole.  We used for $(\tilde\gamma_{0\,ij}, K^{ij}_0)$ a
single, boosted, spinning black hole in Kerr-Schild coordinates.

\begin{figure}
  \centerline{\includegraphics[scale=0.35]{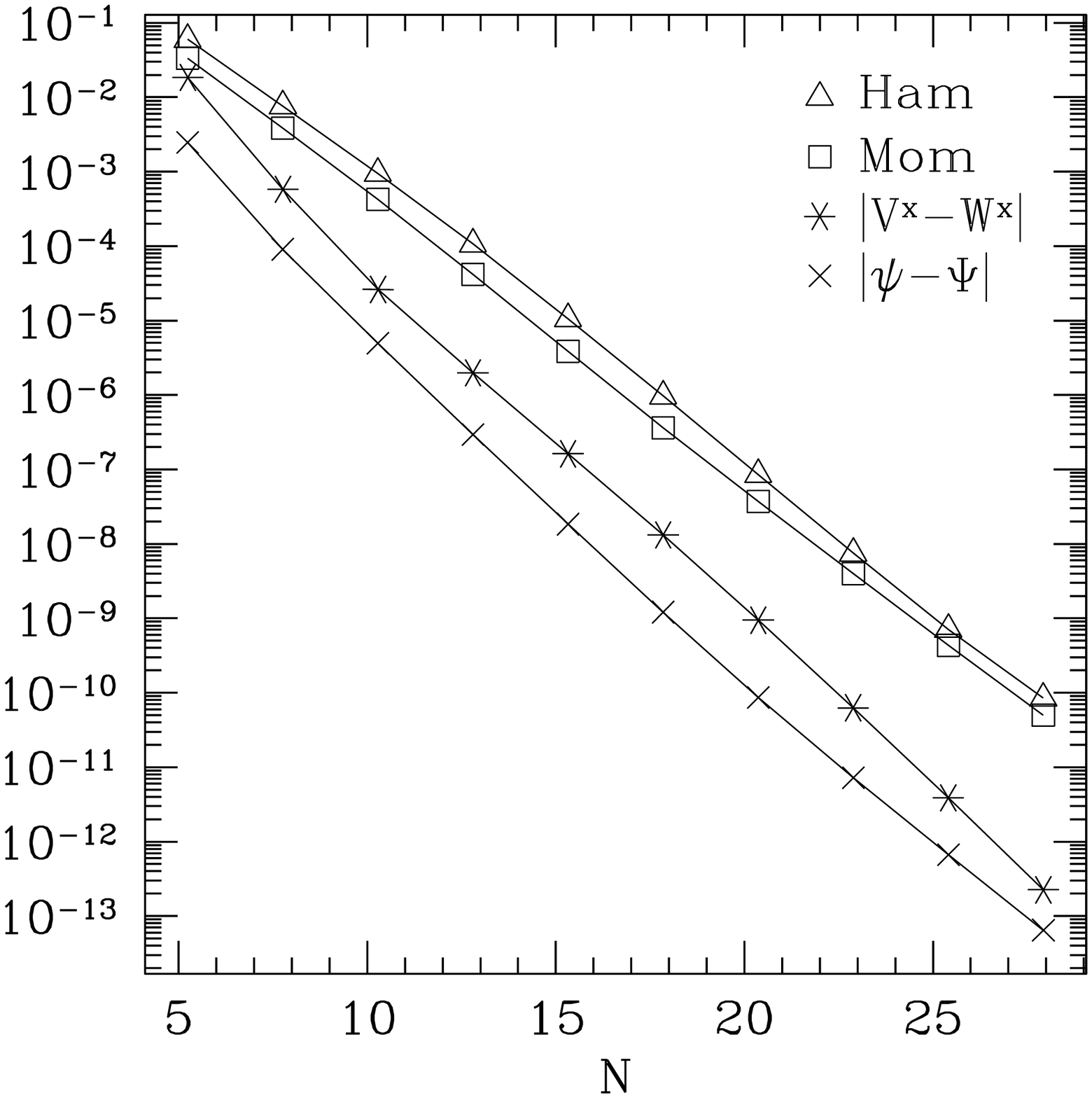}}
\CAP{Testing the solver for the conformal TT
decomposition}{\label{fig:Testing-ConfTT}Testing the solver for the
conformal TT decomposition.
Eqs.~(\ref{eq:ConfTT-1-Comparing}-\ref{eq:ConfTT-2-Comparing}) with
freely specifiable data given by
Eqs.~(\ref{eq:Test-gamma}-\ref{eq:Test-ConfTT-M}) are solved in a
single spherical shell with $1.5M<r<10M$. $N$ is the cube root of the
total number of unknowns.  Plotted are the L2-norms of $\psi-\Psi$,
$V^x-W^x$, and the residuals of Hamiltonian and momentum constraints,
Eqs.~(\ref{eq:HamiltonianConstraintPhysical}) and
(\ref{eq:MomentumConstraint}).  }
\end{figure}

Figure~\ref{fig:Testing-ConfTT} shows results of testing the conformal
TT decomposition on a single spherical shell. The numerical solution
$(\psi, V^i)$ converges to the analytic solutions $(\Psi, W^i)$
exponentially with the number of basis functions as expected for a
properly constructed spectral method. Moreover, the reconstructed
metric and extrinsic curvature satisfy the constraints. 

\begin{figure}[bt]
\centerline{\includegraphics[scale=0.35]{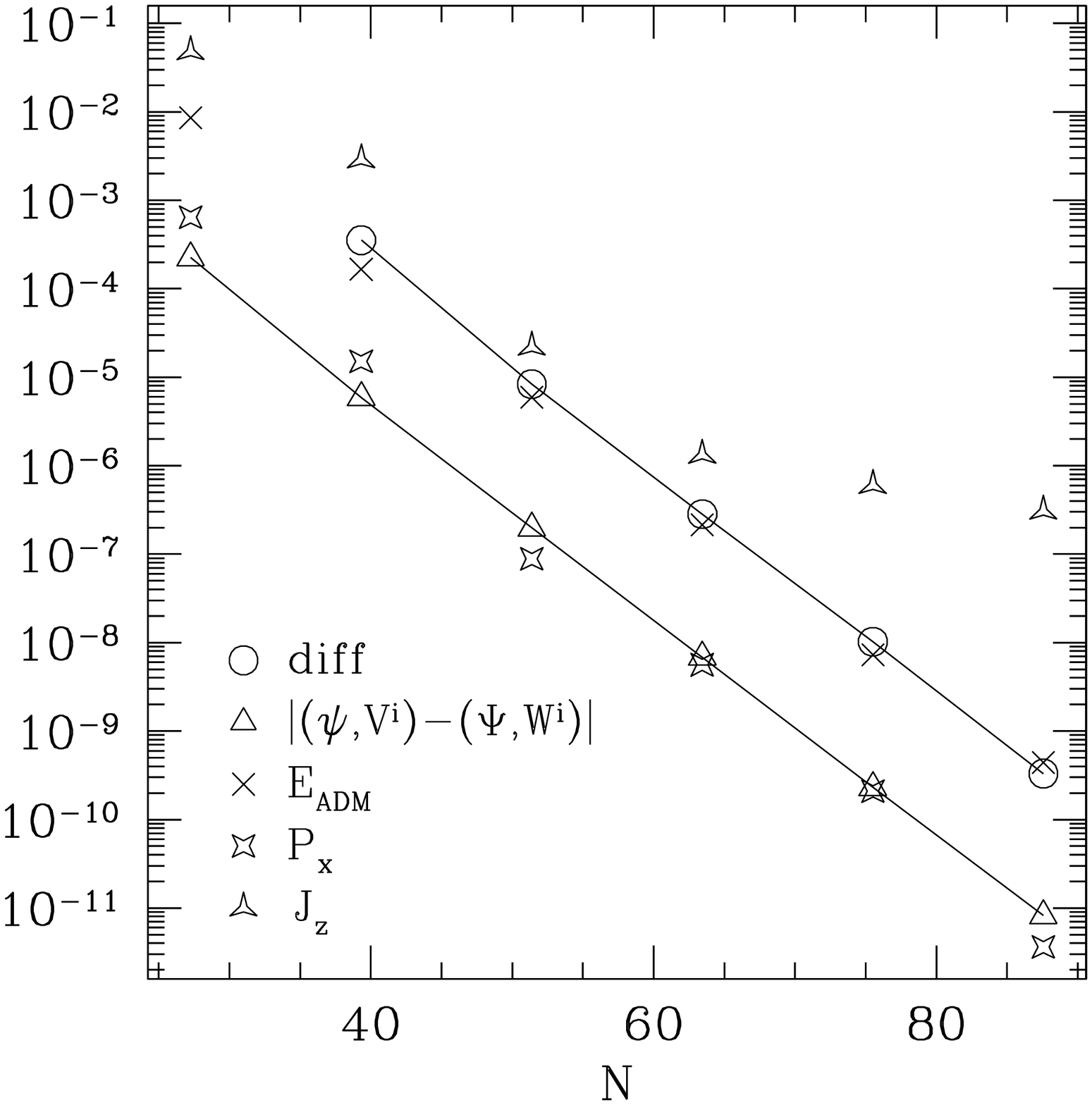}}
\CAP{Testing the solver for the physical TT decomposition with domain
decomposition}{\label{fig:Testing-PhysTT}Physical TT decomposition
with domain decomposition.  Eqs.~(\ref{eq:PhysTT-1}),
(\ref{eq:PhysTT-2}) with freely specifiable data given by
Eq.~(\ref{eq:Test-gamma}, \ref{eq:Test-K}, \ref{eq:Test-PhysTT-M}) are
solved in multiple domains (one inner spherical shell, 26 rectangular
blocks, one outer spherical shell).  $N$ is the cube root of the total
number of grid-points.  {\em diff} denotes the L2-norm of the
difference of the solution and the solution at next lower resolution.
Triangles denote the L2-norm of the difference to the analytic
solution.  The remaining symbols denote the errors of numerically
extracted ADM-quantities. }
\end{figure}
 
Now we test the solver for the physical TT decomposition, and
demonstrate that we can correctly deal with multiple domains.  In this
example the computational domain is covered by an inner spherical
shell extending for $1.5M\le r\le 10M$.  This shell is surrounded by
26 rectangular blocks that overlap with the shell and extend out to
$x,y,z=\pm 25M$.  Finally another spherical shell covers the region
$20M\le r\le 10^6M$.  As can be seen in Fig.~\ref{fig:Testing-PhysTT},
the solution converges again exponentially.

For realistic cases we do not know the analytic solution and therefore
need a measure of the error.  Our major tool will be the change in
results between different resolution. In particular we consider the
$L_2$ norm of the point-wise differences of the solution at some resolution
and at the next lower resolution. This diagnostic is labeled by circles
in Fig.~\ref{fig:Testing-PhysTT}.  Since the solution converges
exponentially, these circles essentially give the error of the {\em
  lower} of the two resolutions.

In addition to testing the equations, this example tests domain
decomposition and the integration routines for the ADM quantities.
The ADM quantities are computed by the standard integrals at infinity
in Cartesian coordinates,
\begin{align}\label{eq:EADM}
E_{ADM}&=\frac{1}{16\pi}\int_{\infty} 
  \left(\gamma_{ij,j}-\gamma_{jj,i}\right)\,d^2S_i,\\
\label{eq:JADM}
J_{(\xi)}&=\frac{1}{8\pi}\int_{\infty}
  \left(K^{ij}-\gamma^{ij}K\right)\xi_j\,d^2S_i.
\end{align}
For the $x$-component of the linear ADM-momentum, $\xi=\hat e_x$ in
Eq.~(\ref{eq:JADM}).  The choice $\xi=x\hat e_y-y\hat e_x$ yields the
$z$-component of the ADM-like angular momentum as defined by York
\cite{York:1979}.  Since the space is asymptotically flat
there is no distinction between upper and lower indices in
Eqs.~(\ref{eq:EADM}) and (\ref{eq:JADM}). Note that
Eq.~(\ref{eq:EADM}) reduces to the familiar monopole term
\begin{equation}
-\frac{1}{2\pi}\int_{\infty}\partial_r\psi\,dA
\end{equation}
{\em only} for quasi-isotropic coordinates.
Our outer domain is large, but since
it does not extend to infinity, we extrapolate $r\to\infty$.

For a Kerr black hole with mass $M$ and spin $\vec a$, that is boosted
to velocity $\vec v$, the ADM-quantities will be
\begin{align}
  E_{ADM}&=\gamma M,\\
  \vec P_{ADM}&=\gamma M\vec v,\\
\label{eq:J-boostedBH}
  \vec J_{ADM}&=\left[\gamma\vec a-(\gamma-1)\frac{(\vec a\,\vec v)\,\vec v}{\vec  v\,^2}\right]M,
\end{align}
where $\gamma=(1-\vec v\,^2)^{-1/2}$ denotes the Lorentz factor.
Eq.~(\ref{eq:J-boostedBH}) reflects the fact that under a boost, the
component of the angular momentum perpendicular to the boost-direction
is multiplied by $\gamma$.

The example in Fig.~\ref{fig:Testing-PhysTT} uses 
$\vec v=(0.2, 0.3, 0.4)$, and $\vec a=(-1/4, 1/4, 1/6)M$.  The
evaluation of the angular momentum $J_z$ seems to be less accurate
since our current procedure to extrapolate to infinity magnifies
roundoff. We plan to improve this in a future version of the code.
Until then we seem to be limited to an accuracy of $\sim 10^{-6}$.

\subsection{Testing conformal thin sandwich equations}
\label{sec:Comparing:TestingCTS}

The previous decompositions could be tested with a conformally
distorted known solution.  In order to test the conformal thin
sandwich decomposition we need to find an analytic decomposition of
the form (\ref{eq:SandwichTT-3}).  To do this, we
start with a stationary solution of Einstein's equations and boost it with
uniform velocity $v^i$.  Denote the metric, extrinsic curvature, lapse
and shift of this boosted spacetime by $\tilde\gamma_{0\,ij}$,
$K^{ij}_0=A^{ij}_0 +\frac{1}{3}\gamma^{ij}_0K_0$, $N_0$ and $N^i_0$,
respectively. Since we boosted the static solution, we will {\em not}
find $\partial_t\gamma_{ij}=0$ if we evolve it with the shift $N^i_0$.
However, all time-dependence of this spacetime is due to the uniform
motion, so in the comoving reference frame specified by the
shift $N^i_0+v^i$, we will find $\partial_t\gamma_{ij}=0$. In this
case, Eq.~(\ref{eq:Evolution-gamma-TF}) yields 
\begin{equation}\label{eq:Aij-boost}
A^{ij}_0=\frac{1}{2N_0}(\ComparingLong(N_0+v))^{ij}.
\end{equation}
If we choose $\tilde \alpha=N_0$ and
$\tilde u^{ij}=0$, the thin sandwich equations (\ref{eq:SandwichTT-1})
and (\ref{eq:SandwichTT-2}) will thus be solved by $\psi=1$ and
$\beta^i=N^i_0+v^i$.  Similar to the conformal TT and physical TT
decomposition above, we can also conformally distort the metric
$\gamma_{0\,ij}$.  Furthermore, we can consider nonvanishing $\tilde
u^{ij}$. We arrive at the following method to test the solver for the
conformal thin sandwich equations:

Given a boosted version of a stationary solution with shift $N_0^i$,
lapse $N_0$, 3-metric $\gamma_{0\,ij}$, trace of extrinsic curvature
$K_0$, and boost-velocity $v^i$. Pick any functions
\begin{gather}
\Psi>0\\
W^i
\end{gather}
and set
\begin{align}
\label{eq:Test-Sandwich-gamma}
\tilde\gamma_{ij}&=\Psi^{-4}\gamma_{0\,ij}\\
K\;&=K_0\\
\tilde\alpha\;\,&=\Psi^{-6}N_0\\
\label{eq:Test-Sandwich-u}
\tilde u^{ij}&=\Psi^4(\ComparingLong_0 W)^{ij}
\end{align}
Then a solution to the thin sandwich equations
(\ref{eq:SandwichTT-1}-\ref{eq:SandwichTT-2}) will be
\begin{align}
\psi&=\Psi\label{eq:psi-static-boosted}\\
\beta^i&=N_0^i+v^i+W^i\label{eq:beta-static-boosted}
\end{align}
assuming boundary conditions respecting this solution.

\begin{figure}
\centerline{\includegraphics[scale=0.35]{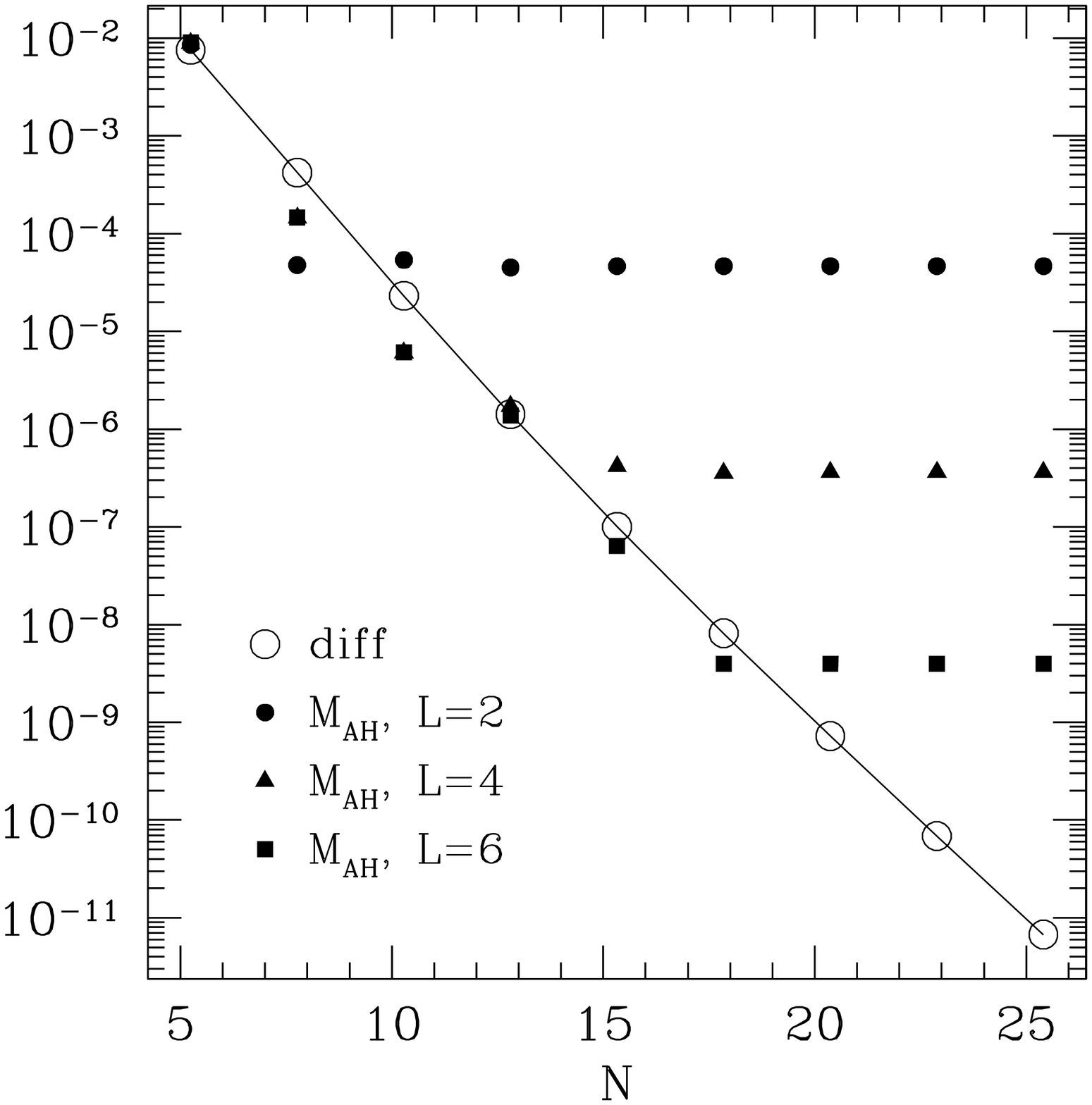}}
\CAP{Testing thin sandwich decomposition with apparent horizon
searches}{\label{fig:Testing-Sandwich}Testing thin sandwich
decomposition with apparent horizon searches.
Equations~(\ref{eq:SandwichTT-1}) and (\ref{eq:SandwichTT-2}) with
freely specifiable data given by
Eqs.~(\ref{eq:Test-Sandwich-gamma})--(\ref{eq:Test-Sandwich-u}) are
solved in a single spherical shell.  $N$ and {\em diff} as in
Fig.~\ref{fig:Testing-PhysTT}.  Apparent horizon searches with
different surface expansion order $L$ were performed, and the errors
of the apparent horizon mass $M_{AH}$ are plotted. }
\end{figure}

Figure~\ref{fig:Testing-Sandwich} shows results of this test for a
single spherical shell and a Kerr black hole with $\vec v=(0.2,
-0.3, 0.1)$, $\vec a=(0.4, 0.3, 0.1)M$. The solution converges to the
expected analytical result exponentially.  In addition, apparent
horizon searches were performed. For the numerically found apparent
horizons, the apparent horizon area $A_{AH}$ as well as the apparent
horizon mass
\begin{equation}
  \label{eq:M_AH}
  M_{AH}=\sqrt{\frac{A_{AH}}{16\pi}}
\end{equation}
were computed. The figure compares $M_{AH}$ to the expected value
\begin{equation}
M\left(\frac{1}{2}+\frac{1}{2}\sqrt{1-\frac{\vec a^2}{M^2}}\;\right)^{1/2}.
\end{equation}
As described in \cite{Baumgarte-Cook-etal:1996,
  Pfeiffer-Teukolsky-Cook:2000}, the apparent horizon finder expands
the apparent horizon surface in spherical harmonics up to a fixed
order $L$.  For fixed $L$, the error in the apparent horizon mass is
dominated by discretization error of the elliptic solver at low
resolution $N$.  As $N$ is increased, the discretization error of the
elliptic solver falls below the error due to finite $L$.  Then the
error in $M_{AH}$ becomes independent of $N$. Since the expansion in
spherical harmonics is {\em spectral}, the achievable resolution
increases exponentially with $L$.  Note that for exponential
convergence it is necessary to position the rays in the apparent
horizon finder at the abscissas of Gauss-Legendre integration.

\subsection{Convergence of binary black hole solutions}

\begin{figure}
  \centerline{\includegraphics[scale=0.35]{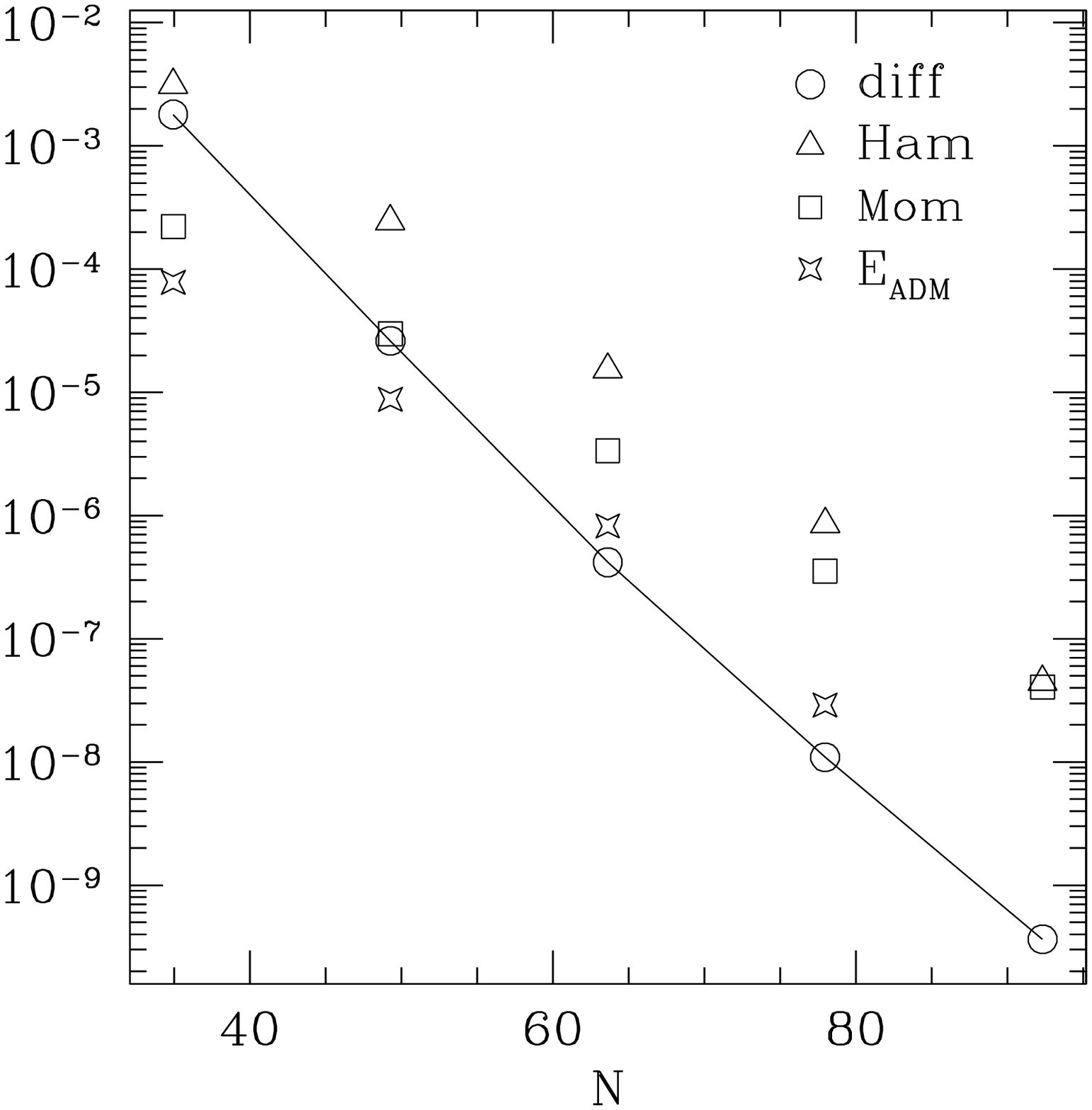}}
  \CAP{Convergence of binary black hole solution with conformal TT
  decomposition}{\label{fig:Convergence-09F}Binary black hole with
  conformal TT decomposition. The residuals of several quantities are
  plotted as a function of the cube root of the total number of grid
  points. {\em diff} as in Fig.~\ref{fig:Testing-PhysTT}, {\em Ham}
  and {\em Mom} are the residuals of Hamiltonian and momentum
  constraints.  $E_{ADM}$ denotes the difference between ADM-energy at
  resolution $N$ and ADM-energy at highest resolution.}
\end{figure}

Figure~\ref{fig:Convergence-09F} present the convergence of the solver
in the binary black hole case. In this particular example, the
conformal TT equations were solved for two black holes at rest with
coordinate separation of $10M$.  The computational domain is structured
as in Fig.~\ref{fig:Domains}.  The excised spheres have radius
$r_{exc}=2M$, the inner spherical shells extend to radius $4M$. The
rectangular blocks cover space up to $x,y,z=\pm 25M$, and the outer
spherical shell extending from inner radius $20M$ to an outer radius
of $R=10^7M$.

We do not use fall-off boundary conditions at the outer boundary; we
simply set $\psi=1$ there.  This limits the computations presented in
this paper to an accuracy of order $1/R\sim 10^{-7}$. 
Figure~\ref{fig:Convergence-09F} shows that even for the next to
highest resolution ($N\approx 80$) the solution will be limited by the
outer boundary condition.  All results presented in the following
section are obtained at resolutions around $N\approx 80$.  If the need
arises to obtain solutions with higher accuracy, one can easily change
to a fall-off boundary condition, or just move the outer boundary
further out.

\section{Results}
\label{sec:Results-Comparing}

The purpose of this paper is to compare the initial-data sets
generated by different decompositions using simple choices for
the freely specifiable pieces in each decomposition.  We solve 

$\bullet$ {\bf ConfTT:} Conformal TT equations (\ref{eq:ConfTT-1-Comparing})
and (\ref{eq:ConfTT-2-Comparing}) with freely specifiable pieces and boundary
conditions given by Eqs.~(\ref{eq:BinaryKerrSchild-gamma}),
(\ref{eq:BinaryKerrSchild-K}), (\ref{eq:BinaryKerrSchild-Mij})
and (\ref{eq:BC-ConfTT-PhysTT}).
  
$\bullet$ {\bf PhysTT:} Physical TT equations (\ref{eq:PhysTT-1})
and (\ref{eq:PhysTT-2}) with same freely specifiable pieces and
boundary conditions as ConfTT.
  
$\bullet$ {\bf CTS:} Conformal thin sandwich
equations (\ref{eq:SandwichTT-1}) and (\ref{eq:SandwichTT-2}) with
freely specifiable pieces and boundary conditions given by
Eqs.~(\ref{eq:BinaryKerrSchild-gamma}), (\ref{eq:BinaryKerrSchild-K}),
(\ref{eq:CTS-uij}) and (\ref{eq:BC-sandwich}). The lapse
$\tilde\alpha$ is given by either Eq.~(\ref{eq:Sandwich-N1}), or by
Eq.~(\ref{eq:Sandwich-N2}).
  
We will apply the terms ``ConfTT'', ``PhysTT'' and ``CTS'' only to
these particular choices of decomposition, freely specifiable pieces
and boundary conditions.  When referring to different freely
specifiable pieces, or a decomposition in general, we will not use
these shortcuts.  If we need to distinguish between the two choices of
$\tilde\alpha$ in CTS, we will use ``CTS-add'' for the additive lapse
Eq.~(\ref{eq:Sandwich-N1}) and ``CTS-mult''for the multiplicative
lapse Eq.~(\ref{eq:Sandwich-N2}).
Below in section \ref{Comparing:sec:mConfTT} we will also introduce as a forth
term ``mConfTT''.

\subsection{Binary black hole at rest}
\label{sec:ResultsBHatRest}

We first examine the simplest possible configuration: Two black holes
at rest with equal mass, zero spin, and with some fixed proper
separation between the apparent horizons of the holes.  We solve
\begin{itemize}
\item ConfTT
\item PhysTT
\item CTS (with both choices of $\tilde\alpha$).  
\end{itemize}
In the comparisons, we also include inversion symmetric conformally
flat initial data obtained with the conformal-imaging formalism.

We excised spheres with radius $r_{exc}=2M$, which is close to the
event horizon for an individual Eddington-Finkelstein black hole.
This results in the boundary conditions being imposed close to, but
within the apparent horizons of the black holes.  
The centers of the excised spheres have
a coordinate separation of $d=10M$.

We now discuss the solutions.  The conformal factor $\psi$ is very
close to $1$ for each of the three decompositions. It deviates from
$1$ by less than 0.02, indicating that a conformal metric based of a
superposition of Kerr-Schild metrics does not deviate far from the
constraint surface.

\begin{figure}[bt]
\centerline{\includegraphics[scale=.42]{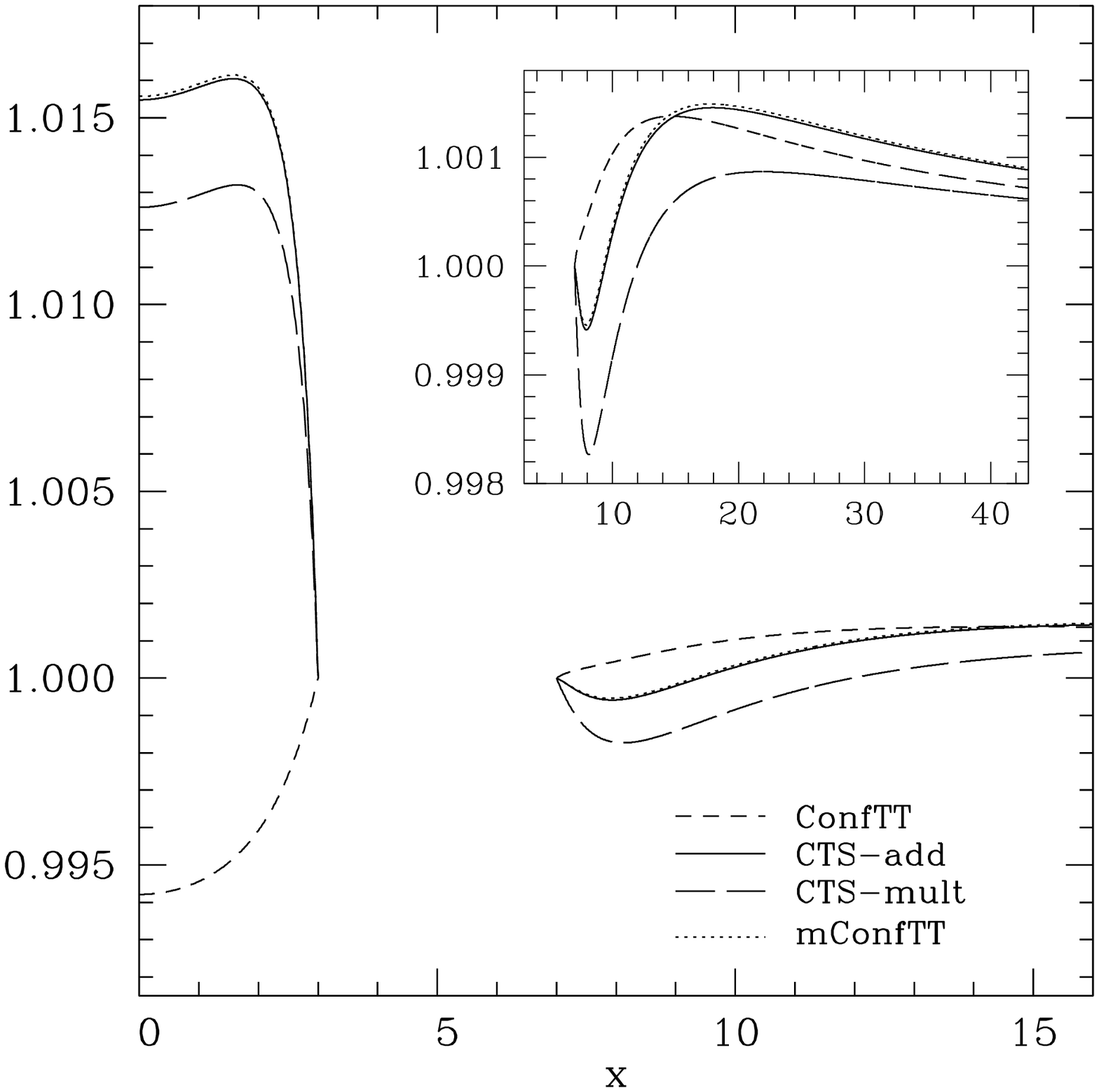}}
\CAP{The conformal factor $\psi$ along the axis connecting the holes
for several decompositions}{\label{fig:Compare-cuts}The conformal
factor $\psi$ along the axis connecting the holes for several
decompositions. $x$ measures the distance from the center of mass, so
that the excised sphere is located between $3<x<7$.  mConfTT is
explained below in section \ref{Comparing:sec:mConfTT}. The solution of PhysTT
is not plotted since it is within the line thickness of ConfTT.  The
insert shows an enlargement for large $x$.  }
\end{figure}

\begin{figure}
\centerline{
\includegraphics[scale=0.34]{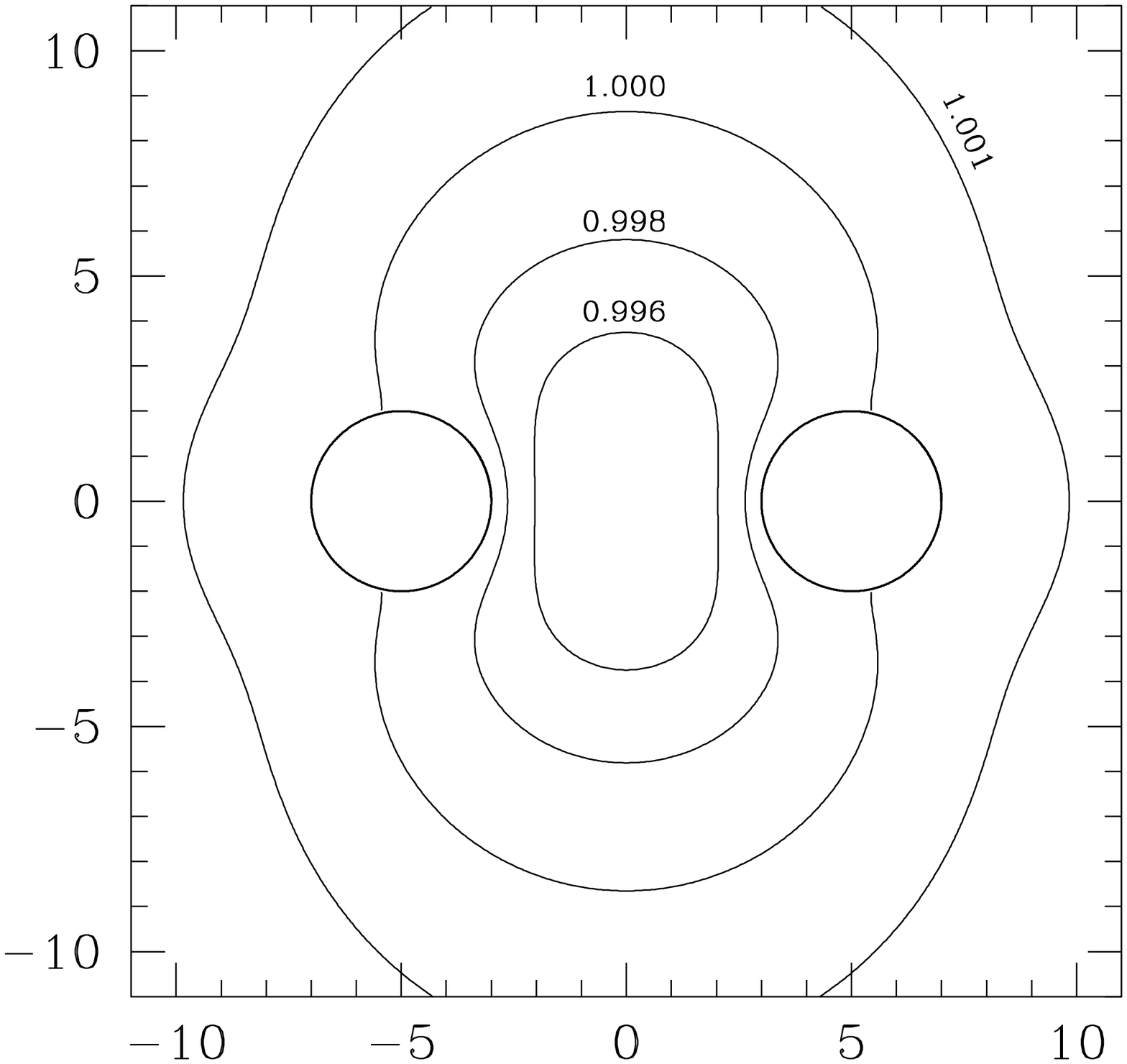}
\includegraphics[scale=0.34]{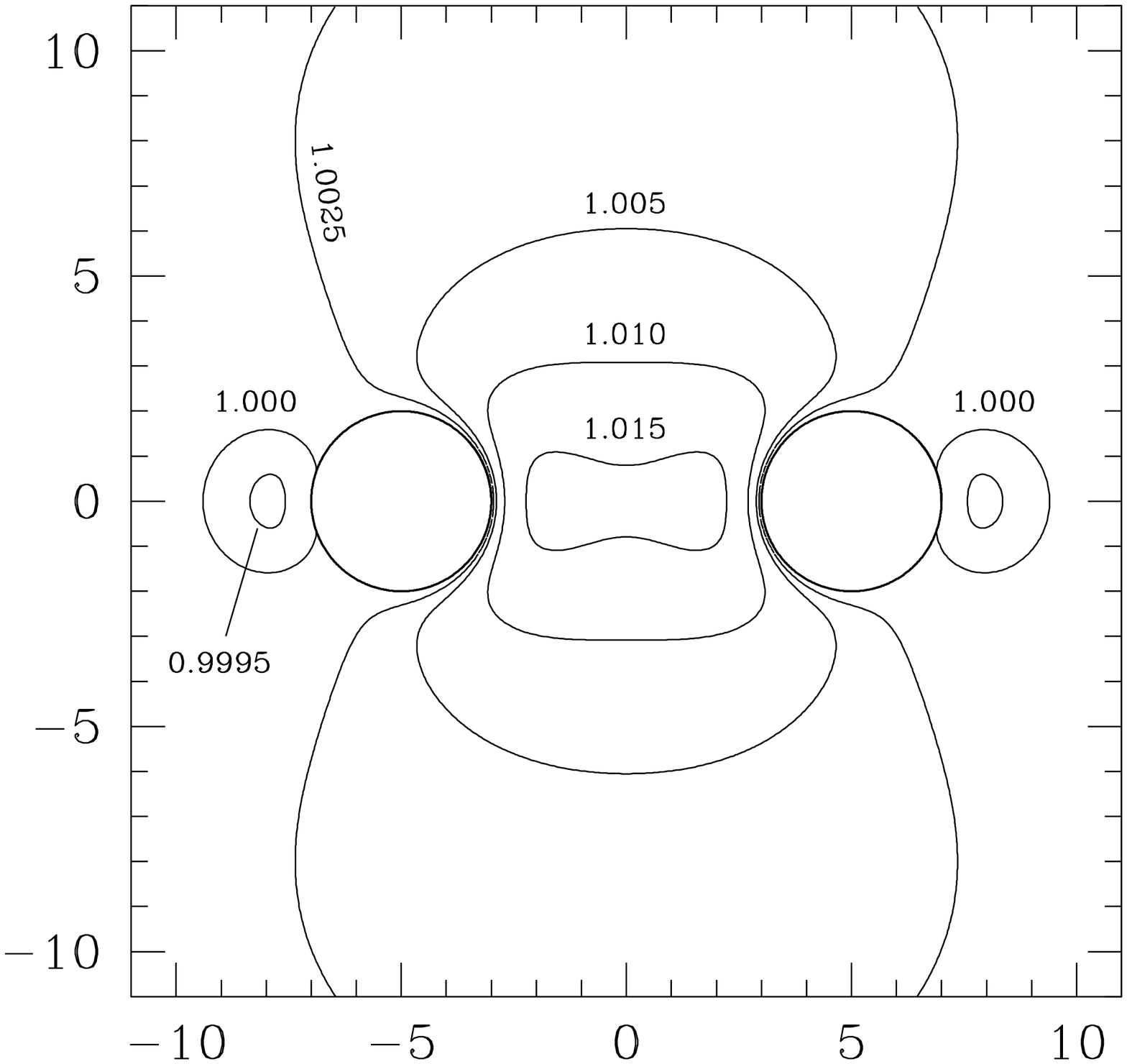}
} \CAP{Black holes at rest: Contour plots of the conformal factor
$\psi$ for ConfTT and CTS-add}{\label{fig:Contours}Black holes at
rest: Contour plots of the conformal factor $\psi$ for ConfTT (left)
and CTS-add (right). The circles denote the excised spheres of radius
$2$. }
\end{figure}

Figure~\ref{fig:Compare-cuts} presents a plot of the conformal factor
along the axis through the centers of the holes. One sees that $\psi$
is close to $1$; however, between the holes ConfTT and CTS force
$\psi$ in {\em opposite} directions. For CTS, $\psi>1$ between the
holes, for ConfTT, $\psi<1$!  The contour plots in Fig.~\ref{fig:Contours} 
also show this striking difference between the
decompositions.

\begin{sidewaystable}
\CAP{Solutions of different decompositions for two black holes at
  rest}{\label{tab:EADM}Solutions of different decompositions for two
  black holes at rest.  Ham and Mom are the rms residuals of the
  Hamiltonian and momentum constraint, $\ell$ is the proper separation
  between the apparent horizons. {\em mConfTT} represents the modified
  conformal TT decomposition which is explained in section
  \ref{Comparing:sec:mConfTT}.  {\em inv. symm.} represents a conformally flat,
  time symmetric and inversion symmetric solution of the Hamiltonian
  constraint.  }
\centerline{\begin{tabular}{ccccccccccc}
       & Ham & Mom & $E_{ADM}$ & $A_{AH}$ & $M_{AH}$ & $\ell$ & $\ell/m$ & $E_{ADM}/m$ & $E_{MPRC}/E_{ADM}$ & $E_b/\mu$ 
\rule[-.65em]{0.em}{1.3em}\\\hline
ConfTT & $9\times 10^{-7}$ & $4\times 10^{-7}$ & 2.06486 & 57.7369 & 1.07175 & 8.062 & 3.761 & 0.9633 & 0.2660 & -0.1467\\
PhysTT & $9\times 10^{-7}$ & $3\times 10^{-7}$ & 2.06490 & 57.7389 & 1.07176 & 8.062 & 3.761 & 0.9633 & 0.2660 & -0.1467\\
CTS-add  & $2\times 10^{-6}$ & $4\times 10^{-7}$ & 2.08121 & 62.3116 & 1.11340 & 8.039 & 3.610 & 0.9346 & 0.2434 & -0.2615\\ \hline
CTS-mult & $2\times 10^{-6}$ & $5\times 10^{-7}$ & 2.05851 & 60.8113 & 1.09991 & 8.080 & 3.672 & 0.9358 & 0.2444 & -0.2569\\
mConfTT & $3\times 10^{-6}$ & $1\times 10^{-6}$ & 2.0827 & 62.404 & 1.1142 &&& 0.9346 & 0.2434 & -0.2617\\ \hline
inv. symm. & - & -       & 4.36387 & 284.851 & 2.38053 & 17.731 & 3.724 & 0.9166 & 0.2285 & -0.3337
\end{tabular}}
\end{sidewaystable}

The result of PhysTT was found to be almost identical with ConfTT.
This is reasonable, since these two decompositions differ only in that
in one case the TT decomposition is with respect to the conformal
metric, and in the other case the TT decomposition it is with respect
to the physical metric. Since $\psi\approx 1$, the conformal metric is
almost identical to the physical metric, and only minor differences
arise. In the following we will often use ConfTT/PhysTT when referring
to both data sets.

We performed apparent horizon searches for these cases. For all
data sets, the apparent horizon is outside the sphere with radius
$2M$, that is outside the coordinate location for the apparent horizon
in a single hole spacetime.
For ConfTT/PhysTT, the radius of the apparent horizon surface is
$\approx 2.05M$, for CTS it is $\approx 2.15M$.  We computed
the apparent horizon area $A_{AH}$, the apparent horizon mass
\begin{equation}
  \label{eq:MAH}
  M_{AH}=\sqrt{\frac{A_{AH}}{16\pi}}
\end{equation}
of either hole, and the combined mass of both holes, 
\begin{equation}\label{eq:m}
  m=2M_{AH}.
\end{equation}
There is no rigorous definition of the mass of an individual black
hole in a binary black hole spacetime, and Eq.~(\ref{eq:MAH})
represents the true mass on an individual black hole
only in the limit of wide separation of the black holes.
A hard upper limit on the possible gravitational radiation
emitted to infinity during the coalescence process of a binary
will be
\begin{equation}\label{eq:MPRC}
  E_{MPRC}=E_{ADM}-\sqrt{\frac{2A_{AH}}{16\pi}},
\end{equation}
where $2A_{AH}$ is the combined apparent horizon area of both holes.
Thus, $E_{MPRC}$ represents the maximum possible radiation content
(MPRC) of the initial data.  This, of course, makes the unlikely
assumption that the binary radiates away all of its angular momentum.

We also compute the proper separation~$\ell$ between the apparent
horizon surfaces along the straight line connecting the centers of the excised
spheres.  In order to compare different data sets we consider the
dimensionless quantities $\ell/E_{ADM}$, $E_{ADM}/m$ and
$E_{MPRC}/E_{ADM}$.  We will also use $E_b/\mu$ which will be defined
shortly.

Table~\ref{tab:EADM} summarizes these quantities for all three
decompositions.  It also includes results for inversion symmetric
initial data, which for black holes at rest reduces to the Misner
data\cite{Misner:1963}\footnote{Although the Misner solution can be
  obtained analytically, we found it more convenient to solve the
  Hamiltonian constraint numerically.  The configuration in 
  Table~\ref{tab:EADM} corresponds to a separation $\beta=12$ in terms of
  \cite{Cook:1991}.}.  The results in Table~\ref{tab:EADM} are
intended to represent nearly the {\em same} physical configuration.

From Table~\ref{tab:EADM}, one finds that the black holes have roughly
the same dimensionless proper separation.  However, the scaled
ADM-energy $E_{ADM}/m$ differs by as much as 4.7\% between the
different data sets.  $E_{MPRC}/E_{ADM}$, which does not depend on any
notion of individual black hole masses at all, differs by 16\% between
the different data sets.

The inversion symmetric data has lowest $E_{ADM}/m$ and
$E_{MPRC}/E_{ADM}$, CTS has somewhat larger values, and ConfTT/PhysTT
lead to the biggest values for $E_{ADM}/m$ and $E_{MPRC}/E_{ADM}$.
This indicates that, relative to the sizes of the black holes,
ConfTT/PhysTT and CTS probably contain some excess energy.

A slightly different argument uses the binding energy which is defined
as
\begin{equation}
  \frac{E_b}{\mu}\equiv\frac{E_{ADM}-2M_{AH}}{\mu},
\end{equation}
where $\mu=M_{AH}/2$ is the reduced mass.  Two Newtonian point masses
at rest satisfy
\begin{equation}\label{eq:Eb-Newton}
  \frac{E_b}\mu = -\frac{1}{\ell/m}.
\end{equation}
From Table~\ref{tab:EADM} we see that for ConfTT/PhysTT, $|E_b/\mu| >
(l/m)^{-1}$, and for CTS, $|E_b/\mu|\approx (l/m)^{-1}$. Since gravity
in general relativity is typically {\em stronger} than Newtonian
gravity, we find again that CTS and ConfTT/PhysTT contain too much
energy relative to the black hole masses, ConfTT/PhysTT having even
more than CTS.

The proper separation between the apparent horizons $\ell/m$ is about
4\% smaller for CTS than for ConfTT/PhysTT. By
Eq.~(\ref{eq:Eb-Newton}) this should lead to a relative difference in binding
energy of the same order of magnitude.  Since $E_b/\mu$ differs by
almost a factor of two between the different decompositions, the
differences in $\ell/m$ play only a minor role.

\subsection{Configurations with angular momentum}
\label{sec:resultOrbiting}

Now we consider configurations which are approximating two black holes
in orbit around each other. The conformal metric is still a
superposition of two Kerr Schild metrics. The black holes are located
along the $x$-axis with a coordinate separation $d$.  For
ConfTT/PhysTT, we boost the individual holes to some velocity $\pm
v\hat e_y$ along the $y$-axis. For CTS we go to a co-rotating frame
with an angular frequency $\vec{\Omega}=\Omega\hat e_z$.  Thus, for
each decomposition we have a two parameter family of solutions, the
parameters being $(d,v)$ for ConfTT and PhysTT, and
$(d,\Omega)$ for CTS.

By symmetry, these configuration will have an ADM angular momentum
parallel to the $z$-axis which we denote by $J$. In order to compare
solutions among each other, and against the conformally flat inversion
symmetric data sets, we adjust the parameters $(d,v)$ and
$(d,\Omega)$, such that each initial-data set has angular momentum
$J/\mu m=2.976$ and a proper separation between the apparent horizons
of $l/m=4.880$. In Ref.~\cite{Cook:1994}, these values were determined to be
the angular momentum and proper separation of a binary black hole at
the innermost stable circular orbit.

\begin{sidewaystable}
\CAP{Initial-data sets generated by different decompositions for
binary black holes with orbital angular momentum}{\label{tab:EADM2}Initial-data sets generated by
different decompositions for binary black holes with the same angular
momentum $J/\mu m$ and separation $\ell/m$.  The mConfTT dataset is
explained in section \ref{Comparing:sec:mConfTT}. It should be compared to
CTS-add.  }
\centerline{\begin{tabular}{ccccccccc}
 & parameters & $M_{AH}$ & $E_{ADM}$ & $J/\mu m$ & $\ell/m$ & $E_{ADM}/m$ & $E_{MPRC}/E_{ADM}$ & $E_b/\mu$
\rule[-.65em]{0.em}{1.3em}\\\hline
ConfTT & $d=11.899,v=0.26865$ & 1.06368 & 2.12035 & 2.9759 & 4.879 & 0.9967 & 0.2906 & -0.0132\\
PhysTT & $d=11.899, v=0.26865$ & 1.06369 & 2.12037 & 2.9757 & 4.879 & 0.9967 & 0.2906 & -0.0132\\
CTS-add  & $d=11.860, \Omega=0.0415$ & 1.07542 & 2.10391 & 2.9789 & 4.884 & 0.9782 & 0.2771 &-0.0873\\
CTS-mult & $d=11.750, \Omega=0.0421$ & 1.06528 & 2.08436 & 2.9776 & 4.880 & 0.9783 & 0.2772 & -0.0867\\
\hline
mConfTT & $d=11.860, \Omega=0.0415$ & 1.0758 & 2.1061 & 3.011 & 4.883 & 0.979 & 0.278 & -0.085\\
inv. symm.\footnote{Data taken from \cite{Cook:1994}} & & & & 2.976 & 4.880 & 0.9774 & 0.2766 & -0.09030
\end{tabular}}
\end{sidewaystable}

Table~\ref{tab:EADM2} lists the parameters corresponding to this
situation as well as results for each initial-data set\footnote{
  Because of the Lorentz contraction, the apparent horizons for
  ConfTT/PhysTT intersect the sphere with radius $2$. In order to have
  the full apparent horizon inside the computational domain, the
  radius of the excised spheres was reduced to 1.9 for these data
  sets.}.  As with the configuration with black holes at rest, we find
again that ConfTT/PhysTT and CTS lead to different ADM-energies.
Now, $E_{ADM}/m$ and $E_{MPRC}/E_{ADM}$ differ by $0.02$ and $0.013$,
respectively, between CTS and ConfTT/PhysTT.  However, in contrast to
the cases where the black holes at rest, now CTS and the inversion symmetric data set
have very similar values for $E_{ADM}/m$ and $E_{MPRC}/E_{ADM}$.

\subsection{Reconciling conformal TT and thin sandwich}
\label{Comparing:sec:mConfTT}

We now investigate further the difference between ConfTT/PhysTT and
CTS.  Since the resulting initial-data sets for PhysTT and ConfTT are
very similar, we restrict our discussion to ConfTT.

\subsubsection{Motivation}

The construction of binary black hole data for the ConfTT/PhysTT
cases produces an extrinsic curvature that almost certainly
contains a significant TT component.  It would be interesting to know
how significant this component is to the value of the various physical
parameters we are comparing.  Ideally, we would like to completely
eliminate the TT component and see what effect this has on the resulting
data sets.  Unfortunately, this is a difficult, if not impossible,
task.

The TT component of a symmetric tensor $M^{ij}$ is defined as
\begin{equation}
        M^{ij}_{TT} \equiv M^{ij} - (\ComparingLong Y)^{ij},
\end{equation}
where the vector $Y^i$ is obtained by solving an elliptic equation
of the form
\begin{equation}\label{eq:Long_eqn}
        \LapLong Y^i = \nabla_jM^{ij}.
\end{equation}
The problem resides in the fact that the meaning of the TT component
depends of the boundary conditions used in solving (\ref{eq:Long_eqn}).

For the ConfTT/PhysTT cases we are actually solving for a vector $V^i$
that is a linear combination of two components, one that solves an
equation of the form of (\ref{eq:Long_eqn}) to obtain the TT component
of $\tilde{M}^{ij}$ and one that solves the momentum constraint.  But
by imposing inner-boundary conditions of $V^i=0$, we don't specify the
boundary conditions on either part independently.  Nor is it clear
what these boundary conditions should be.  Since we cannot explicitly
construct the TT component of the extrinsic curvature, we cannot
eliminate it.  Although it is not ideal, there is an alternative we
can consider that does provide some insight into the importance of
the initial choice of $\tilde{M}^{ij}$.

\subsubsection{Black holes at rest}
\label{sec:mConfTT:Rest}

Consider the following numerical experiment for two black holes at
rest: Given $\tilde M^{ij}$ from Eq.~(\ref{eq:BinaryKerrSchild-Mij}),
make a transverse traceless decomposition by setting
\begin{equation}\label{eq:mConfTT}
  2N \tilde M^{ij}=\tilde M_{TT}^{ij}+(\tilde\ComparingLong Y)^{ij}
\end{equation}
where $\tilde\nabla_j \tilde M^{ij}_{TT}=0$ and $N=N_A+N_B-1$.  
Notice that we are decomposing $2N\tilde{M}^{ij}$, not $\tilde{M}^{ij}$.
Taking
the divergence of Eq.~(\ref{eq:mConfTT}) one finds 
\begin{equation}\label{eq:mConfTT-X}
  \tildeLapLong Y^i=\tilde\nabla_j\left(2N\tilde M^{ij}\right).
\end{equation}
The decomposition chosen in Eq.~(\ref{eq:mConfTT}) is motivated by the
conformal thin sandwich decomposition. With this decomposition we can,
in fact, use the shift vector $N^i$ to fix boundary conditions on
$Y^i$, just as it was used to fix the boundary conditions in
Eqs.~(\ref{eq:Sandwich-BC2}---\ref{eq:Sandwich-BC4}).  For the black
holes at rest in this case, we have $\Omega=0$. After solving
Eq.~(\ref{eq:mConfTT-X}) for $Y^i$, we can construct a new 
conformal extrinsic curvature by
\begin{equation}\label{eq:mConfTT-Mij}
  {\tilde{M}{'}}^{\;ij}=\frac{1}{2N}(\tilde\ComparingLong Y)^{ij}
\end{equation}
which is similar to what would result if we could eliminate $\tilde
M^{ij}_{TT}$ from $\tilde M^{ij}$. Using $\tilde M^{\prime ij}$ in
place of $\tilde M^{ij}$, we can again solve the conformal TT
equations.  The result of this modified conformal TT decomposition
{\bf ``mConfTT''} is striking: Figure~\ref{fig:Compare-cuts} shows
that mConfTT generates a conformal factor $\psi$ that is very similar
to $\psi$ of CTS. mConfTT is also included in Table~\ref{tab:EADM}
where it can be seen that the quantities $E_{ADM}/m$ and
$E_{MPRC}/E_{ADM}$ differ only slightly between mConfTT and CTS.

The fact that modification of the extrinsic curvature changes the
ADM-energy by such a large amount underlines the importance of a
careful choice for the extrinsic curvature $\tilde M^{ij}$ in
ConfTT/PhysTT.  The extremely good agreement between CTS and mConfTT
is probably caused by our procedure to determine $\tilde M{'}^{ij}$.
We force the extrinsic curvature of mConfTT into the form
Eq.~(\ref{eq:mConfTT-Mij}). This is precisely the form of the
extrinsic curvature in CTS, Eq.~(\ref{eq:SandwichTT-3}), even using
the same function $N$ and the same boundary conditions on the vectors
$Y^i$ and $\beta^i$.

\subsubsection{Black holes with angular momentum}

We now apply the modified conformal TT decomposition to the orbiting
configurations of section~\ref{sec:resultOrbiting}.  In the corotating
frame, the black holes are at rest, and we start with
$\tilde\gamma_{ij}$ and $\tilde M^{ij}$ of two black holes {\em at
  rest} with coordinate separation $d=11.860$. We now solve
Eq.~(\ref{eq:mConfTT-X}) with
\begin{equation}
  N=N_A+N_B-1
\end{equation}
and corotating boundary conditions on $Y^i$ [cf.\
Eqs.~(\ref{eq:Sandwich-BC2})--(\ref{eq:Sandwich-BC4})]:
\begin{subequations}
\begin{align}
Y^i&=N_{\!A}^i                 &&\mbox{sphere inside hole A,}\\
Y^i&=N_B^i                     &&\mbox{sphere inside hole B,}\\
Y^i&=\vec\Omega\times\vec r    &&\mbox{outer boundary.}
\end{align}
\end{subequations}
$N_{A/B}$ and $N_{A/B}^i$ are lapse and shift of individual Kerr-Schild
black holes at rest.  $\tilde M{'}^{ij}$ is again constructed by
Eq.~(\ref{eq:mConfTT-Mij}) and used in solving the conformal TT equations.

Results from this procedure are included in Table~\ref{tab:EADM2}.
Again, mConfTT generates results very close to CTS.  $E_{ADM}/m$ 
changes by 1.8\% of the total mass between ConfTT and mConfTT, again
highlighting the importance of the extrinsic curvature.

\subsection{Dependence on the size of the excised spheres}

The framework presented in this paper requires the excision of the
singularities at the centers of each hole\footnote{Marronetti and
  Matzner\cite{Marronetti-Matzner:2000} effectively excised the
  centers, too, by using ``blending functions''.}.  So far we have used
$r_{exc}=2M$ or $r_{exc}=1.9M$ in order to impose boundary conditions
close to the apparent horizons, but different choices can be made.
Indeed, one might expect that the boundary conditions
(\ref{eq:BC-ConfTT-PhysTT}) and (\ref{eq:BC-sandwich}) become ``better''
farther inside the apparent horizon, where the metric and extrinsic
curvature of that black hole dominate the superposed metric
$\tilde\gamma_{ij}$ and superposed extrinsic curvature $\tilde
M^{ij}$.

In order to test this assumption, we solve the constraint equations
for two black holes at rest for different radii $r_{exc}$. We find
that for all three decompositions, the data sets depend strongly on
the radius of the excised spheres.

\begin{figure}
  \centering 
 \includegraphics[scale=0.38]{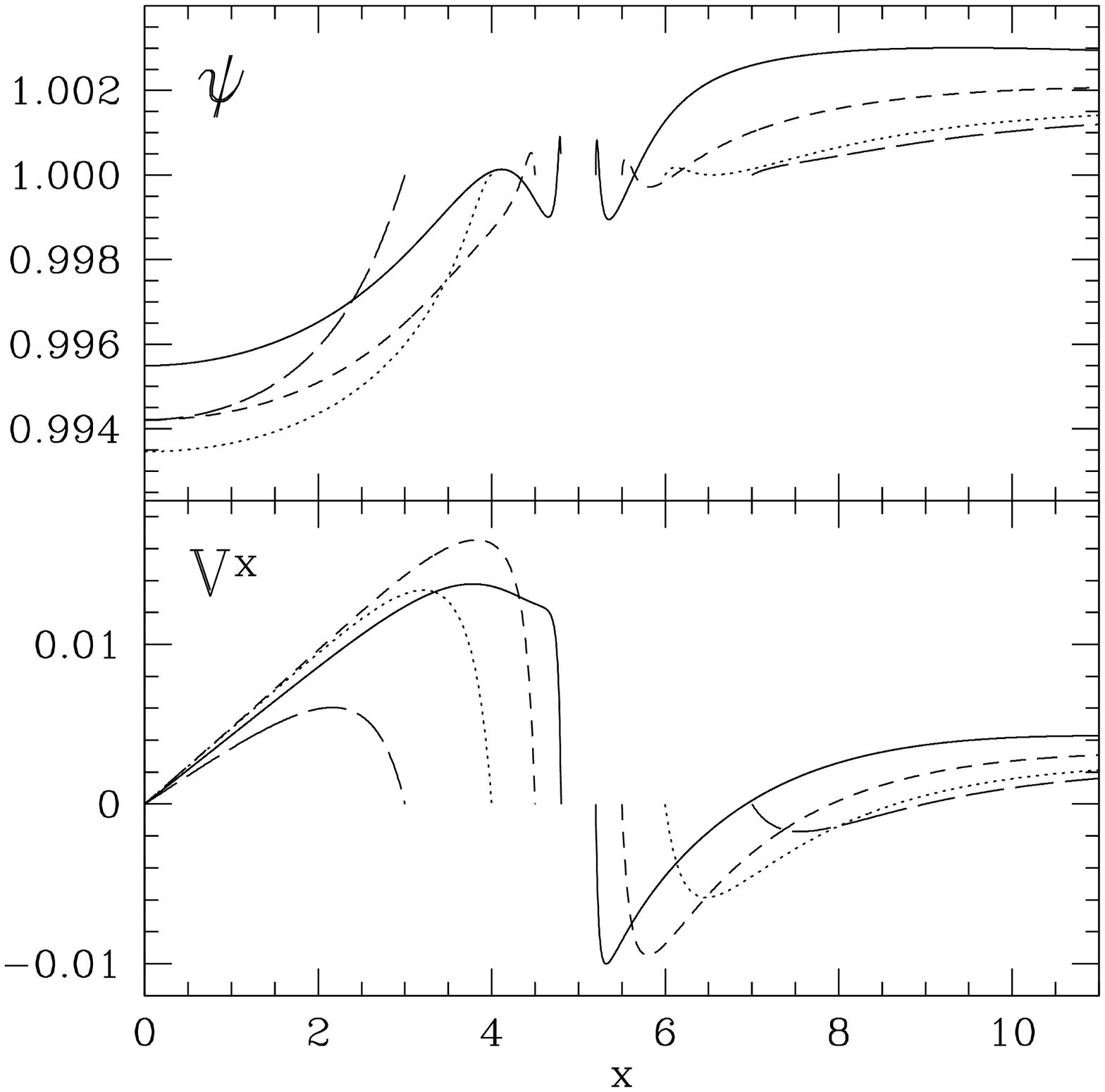} \CAP{Plots of
 $\psi$ and $V^x$ along the positive $x$-axis for ConfTT for different
 radii $r_{exc}=2M, M, 0.5M, 0.2M$}{Plots of $\psi$ and $V^x$ along
 the positive $x$-axis for ConfTT for different radii $r_{exc}=2M, M,
 0.5M, 0.2M$. The excised spheres are centered on the $x$-axis at
 $x=\pm 5$.  The position where a line terminates gives $r_{exc}$ for
 that line.}
    \label{fig:ConfTT-cuts}
 \end{figure}

\begin{figure}
  \centerline{\includegraphics[scale=0.38]{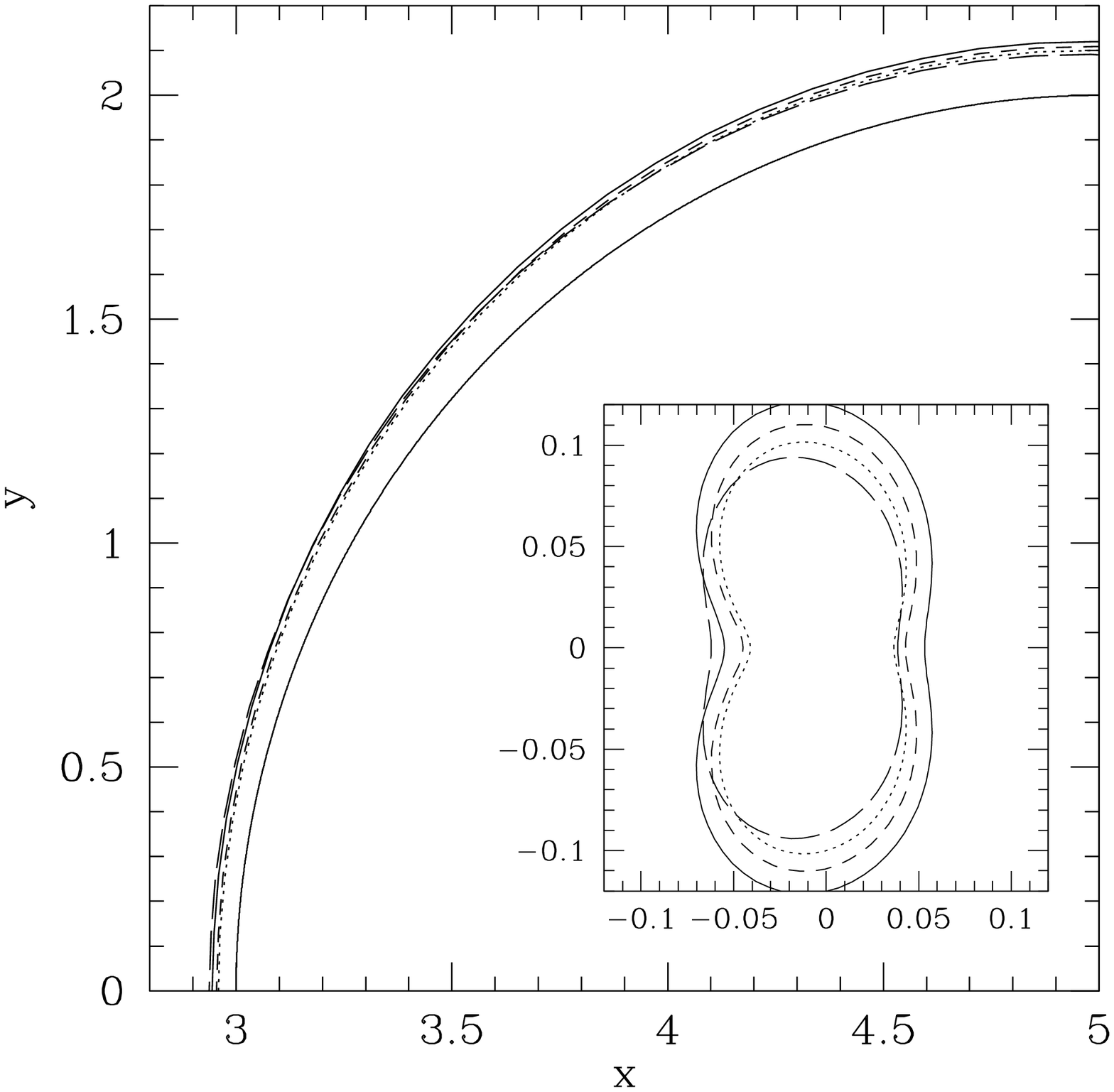}}
  \CAP{Apparent horizons for ConfTT with different radii of excised
  spheres}{\label{fig:ConfTT-AH}Apparent horizons for ConfTT with
  different radii of excised spheres. Results shown are for
  $r_{exc}=2M$ (long dashed line), $M$ (dotted line), $0.5M$ (short
  dashed line) and $0.2M$ (outer solid line). The inner solid line is
  a circle with radius 2.  The insert shows a parametric plot of
  $r(\phi)-2$, which emphasizes the differences between the different
  apparent horizons.}
\end{figure}

\begin{table}
\CAP{Solutions of ConfTT for different radii of the excised spheres,
$r_{exc}$}{\label{tab:TT}Solutions of ConfTT for different radii of
the excised spheres, $r_{exc}$.  The results for PhysTT are nearly
identical.  }
\centerline{\begin{tabular}{cccccc}
$r_{exc}$ & $E_{ADM}$ & $A_{AH}$ & $\ell$ & $E_{ADM}/m$ & $\ell/E_{ADM}$
\rule[-.65em]{0.em}{1.3em}\\\hline
\multicolumn{6}{c}{ Conformal TT }\\
2.0 &    2.0649 &    57.737 &   8.062 &   0.9633 &   3.904 \\ 
1.0 &    2.0682 &    57.825 &   8.101 &   0.9641 &   3.917 \\ 
0.5 &    2.0808 &    58.520 &   8.101 &   0.9642 &   3.893 \\ 
0.2 &    2.0978 &    59.514 &   8.093 &   0.9640 &   3.858 \\ 
0.1 &    2.1064 &    60.025 &   8.089 &   0.9638 &   3.840 \\ 
\end{tabular}}
\end{table}

Figure~\ref{fig:ConfTT-cuts} presents plots of the conformal factor
$\psi$ and $V^x$ for ConfTT with different $r_{exc}$.  There is no
clear sign of convergence of $\psi$ as $r_{exc}\to 0$.  For
$r_{exc}=0.2M$, the conformal factor $\psi$ even oscillates close to
the excised sphere.  Table~\ref{tab:TT} displays various
quantities for the ConfTT decomposition for
different $r_{exc}$.
As $r_{exc}$ varies
between $2.0M$ and $0.1M$, the ADM-energy varies between $2.065$ and
$2.106$, whereas the apparent horizon area changes by nearly 4\%.  The
apparent horizons move around somewhat as $r_{exc}$ changes. 
Figure~\ref{fig:ConfTT-AH} shows the location of the apparent horizons for
different $r_{exc}$.

\begin{figure}[bt]
  \centering
  \includegraphics[scale=0.38]{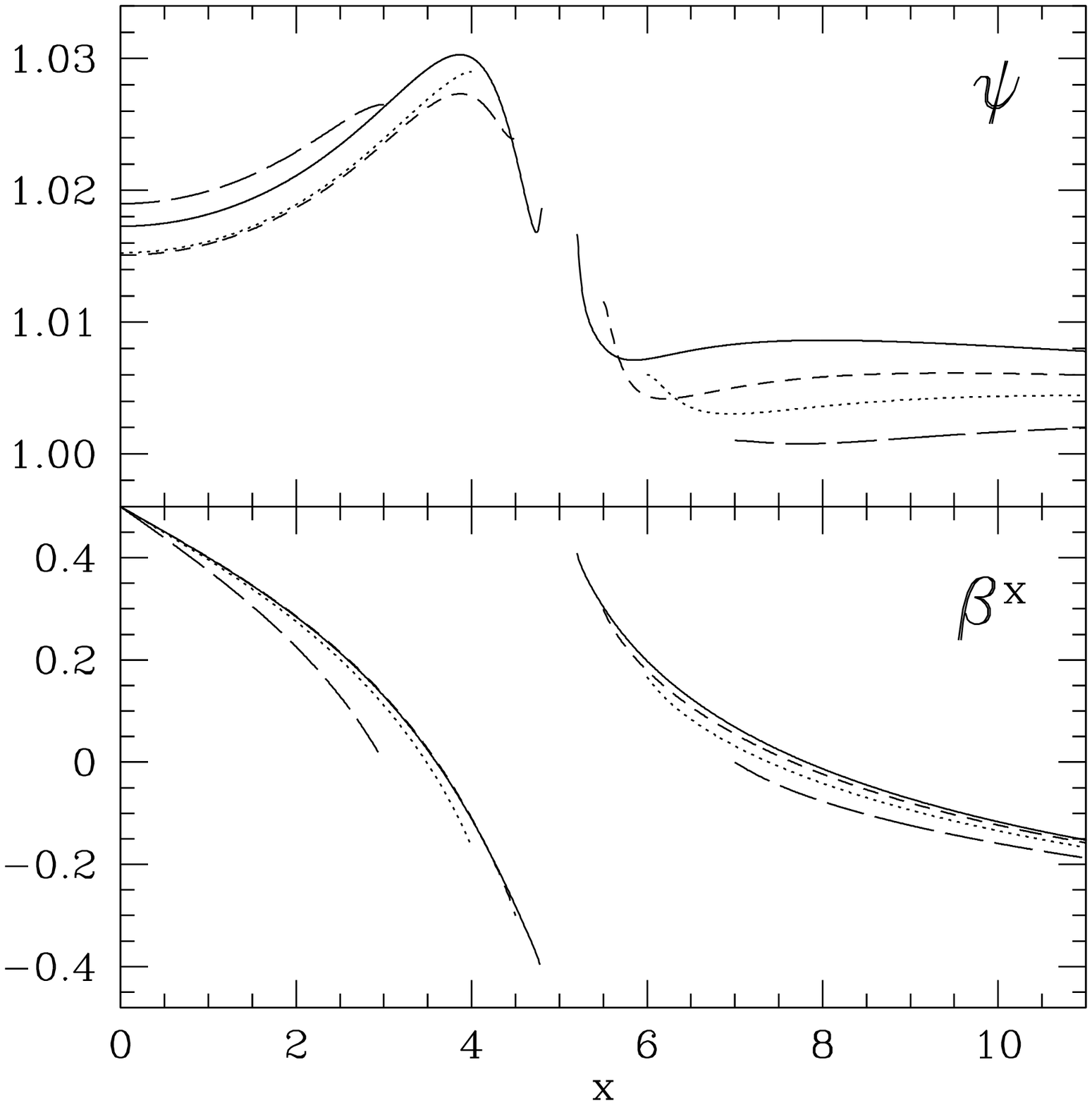} \CAP{Cuts
  through $\psi$ and $\beta^x$ for CTS-mult for different radii
  $r_{exc}$}{\label{fig:Sandwich-cuts-N1N2}Cuts through $\psi$ and
  $\beta^x$ for CTS-mult for different radii $r_{exc}$. Here $\tilde
  \alpha=N_{\!A}N_B$ and the boundary condition on $\psi$ at the
  excised spheres is $d\psi/dr=0$.  The curves for $\beta^x$ are
  shifted up by 0.5 for $x<5$, and are shifted down by 0.5 for $x>5$
  to allow for better plotting.  $d\psi/dr$ approaches zero at the
  inner boundary on scales too small to be seen in this figure.  }
\end{figure}

\begin{figure}
 \centerline{\includegraphics[scale=0.38]{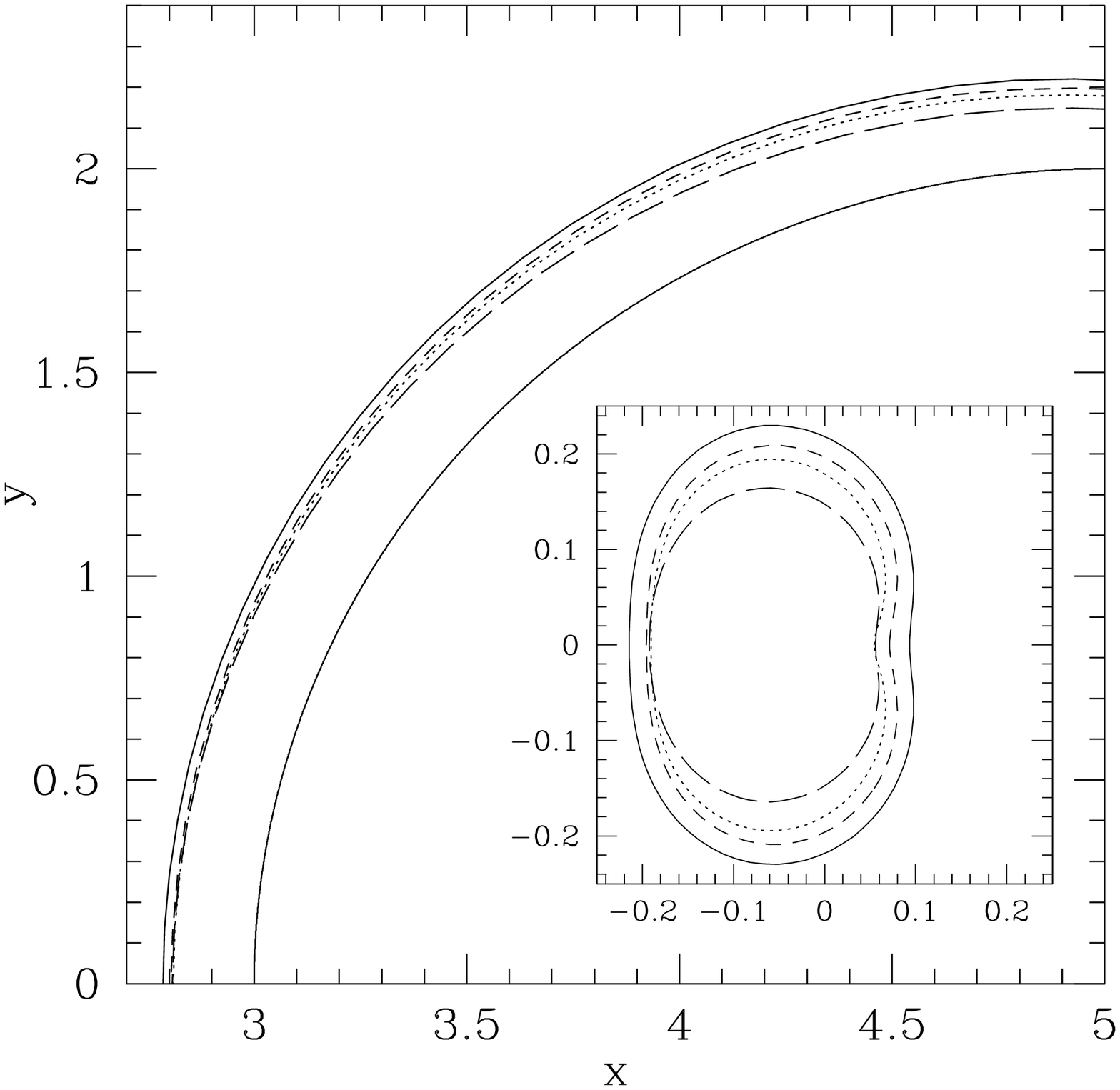}}
 \CAP{Apparent horizons for CTS with $\tilde\alpha=N_A\,N_B$ and inner
 boundary condition $d\psi/dr=0$}{\label{fig:Sandwich-AH-2}Apparent
 horizons for CTS with $\tilde\alpha=N_A\,N_B$ and inner boundary
 condition $d\psi/dr=0$.  The different curves belong to different
 $r_{exc}$ as explained in Fig.~\ref{fig:ConfTT-AH} }
\end{figure}

For CTS-add (with $\tilde\alpha=N_A+N_B-1$), the initial-data sets
seem to diverge as $r_{exc}$ is decreased.  This has to be expected,
since this choice for $\tilde\alpha$ changes sign if the excised
spheres become sufficiently small.  Changing to $\tilde\alpha=N_AN_B$
so that the lapse does not change sign reduces this divergent behavior.  Von
Neumann boundary conditions on $\psi$ at the excised spheres,
\begin{equation}
  \frac{\partial\psi}{\partial r}=0,
\end{equation}
lead to an increase in $A_{AH}$ especially for large excised spheres.
This combination of lapse $\tilde\alpha$ and boundary conditions
exhibits the smallest variations in $E_{ADM}/m$; cuts through $\psi$,
$\beta^x$ and through the apparent horizons are shown in
Figs.~\ref{fig:Sandwich-cuts-N1N2} and \ref{fig:Sandwich-AH-2}.  From
the three examined combinations of lapse and boundary conditions, the
one shown behaves best, but there is still no convincing sign of
convergence.

Table~\ref{tab:CTS} presents ADM-energies and apparent
horizon areas and masses for CTS with different $r_{exc}$ and
different choices of lapse and boundary condition.  From the unscaled
ADM-energy $E_{ADM}$ it is apparent that $\tilde\alpha=N_A+N_B-1$
diverges most strongly.  Note that between all choices of lapse,
boundary conditions and $r_{exc}$, the unscaled quantities $E_{ADM}$,
$M_{AH}$, and $\ell$ exhibit a much broader variation than the scaled
quantities $E_{ADM}/m$ and $\ell/E_{ADM}$.

\begin{table}
\CAP{Solutions of CTS as a function of radius of excised spheres,
  $r_{exc}$.  Different choices of the lapse $\tilde\alpha$ and
  boundary conditions for $\psi$ at the excised spheres are
  explored.}{\label{tab:CTS}Solutions of CTS as a function of radius
  of excised spheres, $r_{exc}$.  Different choices of the lapse
  $\tilde\alpha$ and boundary conditions for $\psi$ at the excised
  spheres are explored.  }
\centerline{\begin{tabular}{ccrccc}
$r_{exc}$ & $E_{ADM}$ & $A_{AH}\;$& $\ell$ & $E_{ADM}/m$ & $\ell/E_{ADM}$ 
\rule[-.5em]{0.em}{1.5em}\\\hline
\multicolumn{6}{c}{ $\tilde\alpha=N_A+N_B-1,\;\; \psi=1$ }\\
2.0 &    2.0812 &    62.312 &   8.039 &   0.9346 &   3.863 \\ 
1.0 &    2.1846 &    68.279 &   8.000 &   0.9372 &   3.662 \\ 
0.5 &    2.3085 &    76.253 &   7.925 &   0.9371 &   3.433 \\ 
0.2 &    2.5463 &    93.534 &   7.750 &   0.9333 &   3.044 \\ 
0.1 &    2.8543 &   118.834 &   7.489 &   0.9282 &   2.624 \\ 
\hline
\multicolumn{6}{c}{ $\tilde\alpha=N_AN_B,\;\; \psi=1$ }\\
2.0 &    2.0585 &    60.811 &   8.080 &   0.9358 &   3.925 \\ 
1.0 &    2.1216 &    64.080 &   8.044 &   0.9395 &   3.792 \\ 
0.5 &    2.1696 &    66.790 &   8.017 &   0.9411 &   3.695 \\ 
0.2 &    2.2120 &    69.456 &   7.991 &   0.9409 &   3.613 \\ 
0.1 &    2.2326 &    70.809 &   7.978 &   0.9405 &   3.573 \\ 
\hline
\multicolumn{6}{c}{ $\tilde\alpha=N_AN_B,\;\; \partial\psi/\partial r=0$ }\\
2.0 &    2.1110 &    64.229 &   8.085 &   0.9337 &   3.830 \\ 
1.0 &    2.1533 &    66.128 &   8.030 &   0.9387 &   3.729 \\ 
0.5 &    2.1794 &    67.427 &   8.011 &   0.9409 &   3.676 \\ 
0.2 &    2.2136 &    69.559 &   7.990 &   0.9409 &   3.609 \\ 
0.1 &    2.2330 &    70.836 &   7.978 &   0.9405 &   3.573 \\ 
\end{tabular}}
\end{table}

\section{Discussion}
\label{sec:Discussion-Comparing}

Our results clearly show that different decompositions lead to
different initial-data sets, even when seemingly similar choices for
the freely specifiable pieces are used.  From Tables~\ref{tab:EADM}
and \ref{tab:EADM2}, one sees that $E_{ADM}/m$ changes by as much as
0.029 between ConfTT/PhysTT and CTS.  The difference between
ConfTT/PhysTT and the inversion symmetric data is even larger, 0.047.
These numbers seem to be small; however, current evolutions of binary
data usually find the total energy emitted in gravitational radiation
$E_{GW}/m$ to be between 0.01 and 0.03
\cite{Baker-Bruegmann-etal:2001,Alcubierre-Benger-etal:2001,Brandt-Correll-etal:2000},
which is the same order of magnitude as the changes in $E_{ADM}/m$ we
find.  This means that, in principle, most of the energy radiated in
these simulations could originate from ``spurious'' energy in the
system and not from the dynamics of the binary system we are
interested in.

These findings highlight the fact that current binary black hole
initial data sets are inadequate for the task of accurately describing
realistic binary systems.  We see that the choices of the conformal
3-geometry $\tilde\gamma_{ij}$ and the freely specifiable portions of
the extrinsic curvature, embedded in $\tilde{M}^{ij}$, influence the
content of the initial data at a significant level.  Furthermore, the
results suggest that small changes in the free data associated with
the extrinsic curvature are more significant than small changes in the
choice of $\tilde\gamma_{ij}$.\footnote{Following submission of this
paper, a preprint by Damour et
al.\cite{Damour-Gourgoulhon-Grandclement:2002} has appeared
that lends support to our idea that the extrinsic curvature plays a
key role in constructing quasi-equilibrium binary black hole initial
data.}  This assertion is supported by the fact that $E_{ADM}/m$
is consistently larger for the ConfTT solutions than for the CTS
solutions but the two approaches can be made to produce quite
consistent results by using the modified extrinsic curvature of the
mConfTT method.  All of these decompositions use the same non-flat
conformal metric, but differ in the extrinsic curvature. On the other
hand, results for the conformally-flat inversion-symmetric data agree
rather well with the results from the CTS method when we consider
orbiting black holes.  For black holes at rest, CTS differs from the
inversion symmetric data, which seems to contradict our conclusion.
However, this difference is likely due to the time-symmetric nature of
the inversion symmetric data, which is especially adapted to the
time-symmetry of the particular configuration of ``two black holes at
rest''.

Improved binary black hole initial data will require choices for the
freely specifiable data that are physically motivated, rather than
chosen for computational convenience.  The same is true for the
boundary conditions used in solving the constraints.  The boundary
conditions used in this paper carry the implicit assumptions that the
approximate metric and extrinsic curvature are correct at the
excision boundaries and that the value of the single-hole Kerr-Schild
shift at the excision boundary is correct in a multi-hole situation.
This is clearly not true, but we might hope that the impact of the
error in this choice would diminish as we decrease the radius of the
excision boundary.  However, our results presented in
Tables~\ref{tab:TT} and \ref{tab:CTS} do not support this conjecture.
Examining the change in $E_{ADM}/m$ as we vary $r_{exc}$ shows only a
small change, but more importantly, it shows no sign of converging as
we decrease $r_{exc}$.  The effects of changing $r_{exc}$ are much
more significant for $\ell/E_{ADM}$, changing its value by as much as
10\% in the case of CTS-mult for the range of values considered.
Furthermore, as with the energy, we see no sign of convergence in
$\ell/E_{ADM}$ as $r_{exc}$ decreases.  Interestingly, although the
solutions show no sign of convergence as we shrink the excision
radius, we do find that the dimensionless quantities $E_{ADM}/m$ and
$\ell/E_{ADM}$ do become independent of the choice of the
inner-boundary condition as $r_{exc}$ decreases.  This can be seen in
comparing the result in Table~\ref{tab:CTS} for the cases using
$\psi=1$ and $\partial\psi/\partial r=0$ as inner-boundary conditions.
Additional tests, not reported in this paper, further support this
assertion.

\section{Conclusion}

Using a new elliptic solver capable of solving the initial-value
problem of general relativity for any of three different
decompositions and any choice for the freely specifiable data, we have
examined data sets representing binary black hole spacetimes.  We find
that the choices for the freely specifiable data currently in use are
inadequate for the task of simulating the gravitational radiation
produced in astrophysically realistic situations.  In particular,
we studied the results of using a superposition of two Kerr-Schild
black holes to fix the freely specifiable data and compared them
to the results obtained from conformally flat initial data.

Although the new Kerr-Schild based data provide a valuable point of
comparison, it is not clear that the data produced are significantly
superior to previous conformally-flat data.  What is clear is that the
choice of the freely specifiable data will be very important in
constructing astrophysically realistic binary black hole initial data.
Progress will require that these data, {\em and} the boundary
conditions needed to solve the constraints, must be chosen based on
physical grounds rather than computational convenience.

How can better initial data be achieved and how can the quality of
initial data be measured?  We believe that the conformal thin sandwich
decomposition will be especially useful.  Genuine radiative degrees of
freedom cannot {\em in principle} be recognized on a single time
slice. The conformal thin sandwich method uses in effect two nearby
surfaces, giving it a potential advantage over other methods.  Also,
it avoids much of the uncertainty related to specifying a conformal
extrinsic curvature.
Moreover, the conformal thin sandwich approach is especially well
suited for the most interesting configurations, a black-hole binary in
a quasi-equilibrium orbit.  In this case time derivatives of all
quantities are small and the choice $\tilde u^{ij}=0$ is physically
motivated.  One should exploit the condition of quasi-equilibrium as
fully as possible, i.e. one should use the conformal thin sandwich
approach together with the constant $K$ equation, $\partial_tK=0$. The
latter yields another elliptic equation for the lapse which removes
the arbitrariness inherent in choosing a conformal lapse
$\tilde\alpha$.  One will also need more physical boundary conditions.
Work in this direction was begun
in\cite{Gourgoulhon-Grandclement-Bonazzola:2001a,
Grandclement-Gourgoulhon-Bonazzola:2001b} and refined in
\cite{Cook:2002}.

Ultimately, the gravitational wave content of an initial-data set can
be determined only by long term evolutions. One must compute an
initial-data set representing a binary black hole in quasi-circular
orbit and evolve it. Then one must repeat this process with an 
initial-data set representing the {\em same} binary black hole, say, one
orbital period earlier, and evolve that data set, too.  If both
evolutions lead to the same gravitational waves (modulo time offset)
then one can be confident that the gravitational radiation is indeed
astrophysically realistic.  This approach has recently been used for
the first time in conjunction with conformally flat puncture data,
where it proved remarkably successful
\cite{Baker-Campanelli-etal:2002}.

\section*{Acknowledgements} 

We thank Lawrence Kidder, Mark Scheel, and James York for helpful
discussions.  This work was supported in part by NSF grants
PHY-9800737 and PHY-9900672 to Cornell University, and by NSF grant
PHY-9988581 to Wake Forest University.  Computations were performed on
the IBM SP2 of the Department of Physics, Wake Forest University, with
support from an IBM SUR grant.




\chapter{Quasi-equilibrium initial data}
\label{chapter:QE}

In this chapter we explore another route to initial data for binary
black holes in circular orbits.  The basic idea is that if a black
hole binary were in an exact circular orbit, then in a {\em
corotating} frame, the configuration would be {\em time-inde\-pend\-ent}.
Thus, for an exact circular orbit,
\begin{equation}\label{eq:exactcircularorbits}
\partial_t\g_{ij}=0,\qquad
\partial_t\K_{ij}=0
\end{equation}
in a corotating coordinate system.  Of course, in reality, a binary
system slowly spirals in due to gravitational wave emission, and
therefore all 12 equations (\ref{eq:exactcircularorbits}) cannot be
satisfied\footnote{Unless the emitted gravitational radiation is
balanced with an equal amount of {\em ingoing}
radiation~\cite{Whelan-Beetle-etal:2002}.}.  However, one can still
try to enforce as many of these equations as possible.  Moreover, the
radiation reaction time-scale increases very rapidly with separation,
so that at large separations it should be possible to satisfy
Eq.~(\ref{eq:exactcircularorbits}) to a very good approximation.

The quasi-equilibrium idea was first used by Wilson \&
Mathews~\cite{Wilson-Mathews:1989, Wilson-Mathews:1995} and later by
many others to construct binary neutron stars.  Later, it was applied
by Gourgoulhon, Grandcl{\'e}ment \&
Bonazzola~\cite{Gourgoulhon-Grandclement-Bonazzola:2001a,
Grandclement-Gourgoulhon-Bonazzola:2001b} to construct black hole
binaries in circular orbits.  Gourgoulhon et al's work is slightly
deficient in that their initial data sets violate the
constraints~\cite{Cook:2002}, although at a very small level.  Also,
their boundary conditions require the lapse to vanish on the apparent
horizons of the black holes.  Evolutions with black hole excision,
however, need initial data that extends smoothly {\em through} the
horizon.

The conformal thin sandwich formalism very clearly displays {\em
which} of the time-derivatives in (\ref{eq:exactcircularorbits}) are
freely specifiable, namely $\partial_t\cg_{ij}=\cu_{ij}$ and
$\partial_t\trK$.  It therefore provides a unified and general
framework for quasi-equilibrium initial data.  Accordingly, we use the
conformal thin sandwich equations with
\begin{equation}\label{eq:QE-zero}
\tilde u_{ij}=0,\qquad\partial_t\trK=0,
\end{equation}
throughout this chapter.  Since we only enforce
Eq.~(\ref{eq:QE-zero}), the quasi-equilibrium method has a
built-in consistency check: For the ``correct'' quasi-equilibrium
solution, {\em all} time-derivatives in
Eq.~(\ref{eq:exactcircularorbits}) should be small.

In section \ref{sec:QE:StationarySpacetimes} we consider the
consequences of~(\ref{eq:QE-zero}) for stationary spacetimes.  This
will lead us to the question of {\em boundary conditions} in
section~\ref{sec:QE:QE-BC}.  After some implementation details
(section \ref{sec:QE:Implementation}), we present solutions for single
black hole spacetimes in section~\ref{sec:QE:SingleBlackHoles}, and
binary black hole spacetimes in section~\ref{sec:QE:BinaryBlackHoles}.
We close with a brief discussion in section \ref{sec:QE:Discussion}.
The appendices contain details of the calculations, tests, and remarks
on how to prepare these initial data sets for an evolution.

\section{Stationary spacetimes}
\label{sec:QE:StationarySpacetimes}

Let $\Sigma$ be a spacelike hypersurface with future pointing unit
normal $n^\mu$ through a spacetime with Killing vector $l$ such that
$-n\cdot l$ is strictly positive.  Denote by $\g^0_{ij}$,
$\K^{ij}_0$, and $\trK_0=\g_{ij}^0\K^{ij}_0$ the induced metric,
extrinsic curvature and mean curvature of $\Sigma$.  Furthermore, drag
the coordinates along the Killing vector, i.e. 
\begin{equation}\label{eq:Killing-projection}
\begin{aligned}
  \beta_0 & =\; \perp l ,\\
  \N_0 &= - n\cdot l,
\end{aligned}
\end{equation}
where $\perp$ is the projection operator into $\Sigma$.  The gauge
(\ref{eq:Killing-projection}) evolves along the Killing
vector, so the 3+1-quantities will be time-independent:
$\partial_t\g_{ij}=\partial_t\K_{ij}=0$.

We now set up this hypersurface for a solution of the conformal thin
sandwich equations (\ref{eq:Mom3}), (\ref{eq:Ham3}) and
(\ref{eq:dtK3}).  If we choose as free data~(\ref{eq:CTS-freedata2})
\begin{equation}
\label{eq:Stationary-free-data}
\cg_{ij}=\Psi^{-4}\g^0_{ij},\qquad\tilde u_{ij}=0,\qquad\trK=\trK_0,
\qquad\partial_t\trK=0,
\end{equation}
(where $\Psi$ is some uniformly positive function), then by construction, 
\begin{equation}
\label{eq:Stationary-solution}
\CF=\Psi,\qquad\beta^i=\beta^i_0,\qquad\cN=\Psi^{-6}\N_0,
\end{equation}
will solve the conformal thin sandwich equations, given boundary
conditions compatible with Eq.~(\ref{eq:Stationary-solution}).  We point
out that in Eq.~(\ref{eq:Stationary-free-data}) we specify only the conformal
equivalence class of $\g_{ij}$ and the mean curvature of $\Sigma$; the
determinant of $\g_{ij}$ and the tracefree extrinsic curvature are
recovered by the conformal thin sandwich equations.

Assuming that the conformal thin sandwich equations have a unique
solution, we can reverse these considerations: 

{\em If the free data $\cg_{ij}$,
and $\trK$ as well as the boundary conditions are compatible with a
time-independent solution of Einstein's equations, then the choice
$\cu^{ij}=0$, $\partial_t\trK=0$ will generate complete initial data
on $\Sigma$, as well as lapse and shift such that the evolution is
completely time independent.  The Killing vector of the spacetime is
then given by $l=\N n+\beta$.}

This result is very useful in two ways: First, the only non-trivial
free data is $\cg_{ij}$ and $\trK$.  In particular, one need not
provide a guess for the trace-free extrinsic curvature $\A^{ij}$, on
which solutions of the extrinsic curvature decomposition depend very
sensitively (see chapter \ref{chapter:Comparing}).  In addition, any
spherically symmetric slice can be written in conformally flat
coordinates, so that $\cg_{ij}=f_{ij}$ is a completely general choice
for spherical symmetry.

Second, one is generally interested in {\em evolving} the initial
data, which requires lapse and shift.  We see that the conformal thin
sandwich equations not only yield lapse and shift, but, for stationary
spacetimes, yield the {\em optimal} lapse and shift.

Of course, the present discussion depends on {\em boundary
conditions}.  The conformal thin sandwich equations generally will
have a solution even if the boundary conditions are not compatible
with (\ref{eq:Killing-projection}).  In such a case, however, the
time-vector of the evolution will not coincide with the Killing
vector, and the evolution will not be time-independent.  Thus we are led to
consider boundary conditions next.

\section{Quasi-equilibrium boundary conditions}
\label{sec:QE:QE-BC}

The idea of quasi-equilibrium is also very useful for deriving
boundary conditions at interior boundaries of the computational domain
which surround the singularities inside each black hole.  Here, I
present the proposal of Cook~\cite{Cook:2002} in my own words, with
some improvements and simplifications which have been found since
publication of~\cite{Cook:2002} by Cook and by myself.

Our goal is to construct initial data for either a single black hole,
or a binary black hole in quasi-circular orbit.  For a binary, the
orbital angular frequency is denoted by $\Omega$.  For a single black
hole one should replace $\Omega$ by zero in what follows.  Thus we
seek initial data $(\g_{ij}, \K^{ij})$ on a hypersurface $\Sigma$ as
well as lapse $\N$ and shift $\beta^i$ so that the configuration is
initially as time-independent as possible.

The initial data should contain one or two black holes which are
``time-inde\-pen\-dent.''  Therefore, we require that $\Sigma$
contains one or two {\em apparent horizons},\footnote{Technically, a
``marginally outer trapped surface'' (MOTS) -- the apparent horizon is
defined as the outermost MOTS, and we do not rigorously show that no
other MOTS exists outside $\S$.} denoted by $\S$, and
that these apparent horizons have constant area as one moves to the
neighboring hypersurfaces.  The situation is illustrated in
Figure~\ref{fig:InnerBdry}, which also introduces several vectors on
$\S$: $n^\mu$ denotes as usual the future pointing unit normal to
$\Sigma$, $s^\mu$ is the unit-normal to $\S$ {\em within} $\Sigma$,
and
\begin{equation}
k^\mu=s^\mu+n^\mu
\end{equation}
denotes the outgoing null normal to $\S$ with a convenient
normalization.  Furthermore, the induced metric $h_{ij}$ on $\S$ is
\begin{equation}
h_{ij}=\g_{ij}-s_is_j.
\end{equation}

\begin{figure}
\centerline{\input{InnerBdry.pictex}} \CAP{Inner boundary for
quasi-equilibrium boundary conditions}{\label{fig:InnerBdry}Inner
boundary for quasi-equilibrium boundary conditions.  The 2-surface
$\S$ has the induced metric $h_{ij}$. }
\end{figure}

For the apparent horizon $\S$, by definition, the expansion of $k^\mu$
vanishes:
\begin{equation}\label{eq:theta=0}
\theta=0\qquad\mbox{on $\S$.}
\end{equation}
Therefore, if $\S$ is moved along $k^\mu$, initially, the surface area
of $\S$ will not change.  Of course, so far there is no guarantee that
the apparent horizon moves along $k^\mu$.

We consider Raychaudhuri's equation
\begin{equation}\label{eq:Raychaudhuri}
\frac{d\theta}{d\lambda}=-\frac{1}{2}\theta^2-\sigma_{\mu\nu}\sigma^{\mu\nu}
+\omega_{\mu\nu}\omega^{\mu\nu}-R_{\mu\nu}k^\mu k^\nu,
\end{equation}
where $\sigma_{\mu\nu}$, and $\omega_{\mu\nu}$ denote shear and twist
of $k^\mu$, respectively, and $\lambda$ is the affine parameter along
the null-geodesics tangent to $k^\mu$ such that $k=d/d\lambda$.  Since
$\theta=0$ (on $\S$), $R_{\mu\nu}=0$ in vacuum and $\omega_{\mu\nu}=0$
(because $k^\mu$ is surface-forming by construction),
Eq.~(\ref{eq:Raychaudhuri}) simplifies to
\begin{equation}\label{eq:Raychaudhuri2}
\frac{d\theta}{d\lambda}=-\sigma_{ij}\sigma^{ij}\qquad\mbox{on $\S$.}
\end{equation}
Therefore, the expansion will stay zero along $k^\mu$ (initially), if
\begin{equation}\label{eq:shear=0}
\sigma_{ij}=0\qquad\mbox{on $\S$.}
\end{equation}
The apparent horizon is defined by $\theta=0$; Eq.~(\ref{eq:shear=0})
thus ensures that the apparent horizon moves along $k^\mu$ and that
its area is (initially) constant during an evolution.

We remark that the conditions on $\S$ we have derived so far are in
close contact to the {\em isolated horizon framework} (see,
e.g. \cite{Ashtekar-Beetle-etal:2000}, or
\cite{Dreyer-Krishnan-etal:2003} for an easily accessible
introduction).  This framework tries to capture the concept of a black
hole that is not changing for a certain period of time in a spacetime,
which is otherwise allowed to change.  Within this framework, one can
then define {\em locally} mass and angular momentum of the hole.  One
of the defining properties of an isolated horizon is that it can be
foliated by null-geodesics with zero expansion, which is very similar
to our requirement that $\theta=0$ and $d\theta/d\lambda=0$ on $\S$.

So far we have been concerned with coordinate independent conditions
on the spacetime; we now consider coordinate choices which make the
configuration as time-independent as possible in the specific
coordinate system we are constructing.

It will be convenient to know where the apparent horizon $\S$ is
located in $\Sigma$.  We achieve this by simply {\em choosing} the
surface $\S$ in our coordinate system, and then {\em demanding} that
$\S$ is an apparent horizon.  In Appendix \ref{QE:appendix} we show
that from Eq.~(\ref{eq:theta=0}) it follows that
\begin{equation}\label{eq:BC-AH}
\tilde s^k\tilde\nabla_k\ln\CF
=-\frac{1}{4}\left(\tilde h^{ij}\tilde\nabla_i\tilde s_j-\CF^2J\right)
\qquad\quad\mbox{on $\S$}.
\end{equation}
Here, $\tilde s^k$ and $\tilde h_{ij}$ are the conformal unit-normal
to $\S$ and the conformal induced metric of $\S$,
\begin{equation}
s_i=\CF^2\tilde s_i,\qquad
h_{ij}=\CF^4\tilde h_{ij},
\end{equation}
and $J=h_{ij}K^{ij}$.  Equation~(\ref{eq:BC-AH}) has already been
expressed in conformal quantities which are known before the conformal
thin sandwich equations are solved.

Equations (\ref{eq:Raychaudhuri2}) and (\ref{eq:shear=0}) ensure that
the apparent horizon is a null surface; we now require that the set of
{\em coordinate} points that coincide with $\S$ in the initial
hypersurface $\Sigma$ coincides with a null surface during the
evolution (initially).  This ensures that the apparent horizons will
not move through coordinate space in the evolution (at least
initially).  We decompose the shift $\beta^\mu$ into pieces
perpendicular and tangential to $\S$,
\begin{equation}\label{eq:beta-split}
\beta^\mu=\beta_\perp s^\mu+\beta_\parallel^\mu
\qquad\quad\mbox{on $\S$,}
\end{equation}
with $\beta_\perp=s_\mu\beta^\mu$ and $s_\mu\beta_\parallel^\mu=0$.
The time-vector then splits into
\begin{equation}\label{eq:t-split}
t=Nn+\beta_\perp s+\beta_\parallel=\zeta+\beta_\parallel
\qquad\quad\mbox{on $\S$,}
\end{equation}
where $\zeta^\mu\equiv Nn^\mu+\beta_\perp s^\mu$ describes the radial
motion of the boundary whereas $\beta_\parallel^\mu$ shifts
coordinates {\em tangential} to $\S$.  In order for the inner boundary to
coincide with a null surface, $\zeta^\mu$ must be null,
\begin{equation}
\zeta^2 = -N^2+\beta^2_\perp=0
\qquad\quad\mbox{on $\S$,}
\end{equation}
so that
\begin{equation}\label{eq:BC-shift-perp}
\beta_\perp=N
\qquad\quad\mbox{on $\S$,}
\end{equation}
(which also implies that $\zeta^\mu$ is proportional to $k^\mu$).
This condition relates the radial part of the shift-vector to the
lapse.  The overall scaling of $\zeta^\mu$ is not fixed by the
requirement of $\zeta^\mu$ being null.  

Having the results of the last two pages, we analyze the condition
(\ref{eq:shear=0}).  We show in Appendix
\ref{QE:appendix} that $\sigma_{ij}=0$ together with
(\ref{eq:beta-split}) and (\ref{eq:BC-shift-perp}) implies that
\begin{equation}\label{eq:BC-shift-parallel}
(\tilde\Long_\S\beta_\parallel)^{ij}=0.
\end{equation}
The operator on the left-hand side of~(\ref{eq:BC-shift-parallel}) is
the two-dimensional longitudinal operator (the {\em conformal Killing
operator}), given by Eq.~(\ref{eq:L}), acting in $\S$.  The solutions
of (\ref{eq:BC-shift-parallel}) are therefore conformal Killing
vectors of $\S$ (or on each connected component of $\S$, for a binary
black hole).  For the special case of a sphere in conformal flatness,
we note that a rotation centered on the sphere satisfies
Eq.~(\ref{eq:BC-shift-parallel}) for any choice of rotation-angle and
orientation of the axis of rotation.

Cook~\cite{Cook:2002, Cook:2003} suggests to use the freedom in
$\beta^i_\parallel$ to construct {\em spinning} black holes.  The
choice $\beta^i_\parallel=0$ corresponds to corotating black holes,
whereas any other solution of (\ref{eq:BC-shift-perp}) would impart an
additional rotation on the hole.  In the special case of conformal
flatness with excised spheres, for example, the choice
\begin{equation}
\beta^i_{\parallel}=\Omega\left(\left.\dnachd{}{\phi}\right._{\!\!H}\right)^i
\end{equation}
with $\partial/\partial\phi_H$ the rotational conformal Killing vector
centered on the excised sphere with rotation axis parallel to the
orbital angular momentum, should generate irrotational black holes.

Looking back we see that Eqs.~(\ref{eq:BC-AH}), (\ref{eq:beta-split})
and (\ref{eq:BC-shift-perp}) have the character of boundary conditions
on $\CF$ and $\beta^i$, which are four of the five unknowns of the
conformal thin sandwich equations (the fifth one is the lapse $\cN$).
Therefore, we can guarantee that these equations are satisfied by
using them as boundary conditions in the elliptic solve of the
conformal thin sandwich equations\footnote{The apparent horizon
equation was first used as a boundary condition
in~\cite{Thornburg:1987}.}.

At the {\em outer} boundary, the boundary conditions are dictated by
asymptotic flatness:
\begin{align}
\label{eq:BC-psi-infty}
\CF&\to 1\qquad\mbox{as $r\to\infty$},\\ 
\label{eq:BC-N-infty0}
\N&\to
1\qquad\mbox{as $r\to\infty$},
\end{align}
both in an inertial frame and in the rotating frame.
In the inertial frame, furthermore, $\beta^i\to 0$.  In the corotating
frame, therefore
\begin{equation}\label{eq:BC-shift-infty}
\beta^i\to\Omega\left(\dnachd{}{\phi}\right)^i\qquad\mbox{as $r\to\infty$},
\end{equation}
where $\partial/\partial\phi$ represents a rotation in the
asymptotically flat region around the orbital axis (for a single black
hole, $\Omega=0$).

We can summarize the {\em quasi-equilibrium method} as follows:
\begin{enumerate}
\item Pick a conformal metric $\cg_{ij}$, mean curvature $\trK$, and a
two-surface $\S$ consisting of one or two topologically spherical
components.
\item Solve the conformal thin sandwich equations with $\tilde
u_{ij}=\dot\trK=0$ in the exterior of $\S$ with boundary
conditions
\begin{itemize}
\item on $\S$ given by Eqs.~(\ref{eq:BC-AH}), (\ref{eq:beta-split})
and (\ref{eq:BC-shift-perp}), and $\beta^i_\parallel$ satisfying
(\ref{eq:BC-shift-parallel}),
\item at infinity given by Eqs.~(\ref{eq:BC-psi-infty}), (\ref{eq:BC-N-infty0}) and (\ref{eq:BC-shift-infty}).
\end{itemize}

\end{enumerate}

We have not yet specified a boundary condition on the lapse on $\S$.
Cook~\cite{Cook:2002} proposed to require that along the inner
boundary, the expansion of the {\em ingoing} null geodesics be
constant in time.  This translates to a very complicated condition,
namely
\begin{equation}\label{eq:BC-lapse}
\BC\equiv J\tilde s^i\partial_i\N+\CF^2(J^2-JK+\tilde{\cal D})\N=0,
\end{equation}
with 
\begin{equation}\label{eq:calD}
\tilde{\cal D}\equiv 
\CF^{-4}\left(-\frac{1}{2}\;^{(2)}\tilde R+2\tilde D^2\ln\CF
              +\tilde h^{ij}(\tilde D_i-J_i)(\tilde D_j-J_j)
\right),
\end{equation}
where $J_i=h_{ik}s_lK^{kl}$.  (Note that $\tilde{\cal D}$ involves
second two-dimensional covariant derivatives within the surface $\S$).

In spherical symmetry, however, I show that (\ref{eq:BC-lapse}) is
degenerate with the quasi-equilibrium boundary conditions on $\CF$ and
$\beta^i$, and cannot be used.  Furthermore, for non-spherical
symmetric situations, I present evidence in
section~\ref{sec:QE:LapseBC} that Eq.~(\ref{eq:BC-lapse}) is ill-posed
due to a term that vanishes identically in spherical symmetry.

Consequently, I do not believe that Eq.~(\ref{eq:BC-lapse}) represents
a usable boundary condition.  Instead I will use Dirichlet or von
Neumann boundary conditions on the lapse.  But although
Eq.~(\ref{eq:BC-lapse}) cannot be used as a boundary condition,
$\BC=0$ must nonetheless hold for {\em time-independent} slicings;
therefore, we will use the residual $\BC$ as a diagnostic of the
time-independence of computed initial data sets.

\section{Implementation details}
\label{sec:QE:Implementation}

In the code, we always excise {\em spheres}.  For a binary, the
centers of the spheres are located on the $x$-axis at $x=\pm s/2$.
We also choose $\vec\Omega$ parallel to the $z$-axis,
$\vec\Omega=\Omega\hat e_z$.

For equal mass black holes in a configuration that is symmetric under
exchange of the two holes, the center of the corotating frame must
coincide with the origin.  For {\em unequal} mass black holes the
center will no longer be halfway between the holes and therefore not
at the origin.  Moreover, the $+y$ direction is different from the
$-y$ direction, as the holes moving in either direction differ in
mass.  Thus it seems possible that the center of the corotating frame
is not on the $x$-axis, although we would expect it to be close to it.
Accordingly we center the corotating frame at $\vec R=(R_x, R_y,0)$ in
Cartesian coordinates.

The shift boundary condition Eq.~(\ref{eq:BC-shift-infty}) reads in
Cartesian coordinates
\begin{equation}
\beta^i\to \big(\vec\Omega\times(\vec r-\vec R)\big)^i=\xi^i+(\Omega
R_y,-\Omega R_x,0),
\end{equation}
where
\begin{equation}
\xi^i\equiv (\vec\Omega\times\vec r)^i=(-\Omega\, y, \Omega\, x, 0)
\end{equation}
denotes the divergent part of the shift, and $\vec r=(x,y,z).$

The outer sphere of the domain decomposition extends to very large
outer radius and expands functions in inverse powers of $r$ (via the
inverse mapping of Eq.~(\ref{eq:Mappings}) on
p.~\pageref{eq:Mappings}). The shift is proportional to $r$ for large
radii, and therefore cannot be represented as an expansion in
$1/r$.  Instead of solving for $\beta^i$ directly, we write
\begin{equation}
\beta^i=B^i+\xi^i
\end{equation}
and solve for $B^i$ (which is finite as $r\to\infty$).

With these changes, the conformal thin sandwich equations 
are
\begin{subequations}\label{eq:QE}
\begin{align}
\label{eq:QE-Mom}
\cderiv_j\left(\frac{1}{2\cN}(\cLong B)^{ij}\right)
+\cderiv_j\left(\frac{1}{2\cN}(\cLong\xi)^{ij}\right)
-\frac{2}{3}\CF^6\cderiv^i\trK
&=0,\\
\label{eq:QE-Ham}
\cderiv^2\CF-\frac{1}{8}\CF\cR
-\frac{1}{12}\CF^5\trK^2+\frac{1}{8}\CF^{-7}\cA_{ij}\cA^{ij}
&=0,\\
\nonumber
\cderiv^2(\cN\CF^7)-(\cN\CF^7)\bigg[\frac{1}{8}\cR+\frac{5}{12}\CF^4\trK^2
+\frac{7}{8}\CF^{-8}\cA_{ij}\cA^{ij}\bigg]\quad&\\
\label{eq:QE-dtK}
-\CF^5B^k\partial_k\trK
-\CF^5\xi^k\partial_k\trK&=0,
\end{align}
where 
\begin{equation}
\cA^{ij}=(2\cN)^{-1}\Big((\cLong B)^{ij}+(\cLong\xi)^{ij}\Big).
\end{equation}
The quasi-equilibrium boundary conditions are
\begin{align}
\label{eq:BC-AH2}
\tilde s^k\tilde\nabla_k\ln\CF
&=-\frac{1}{4}\left(\tilde h^{ij}\tilde\nabla_i\tilde s_j-\CF^2J\right)
&&\mbox{on $\S$},\\
\label{eq:BC-shift}
B^i&=N\CF^{-2}\tilde s^i+\beta^i_\parallel-\xi^i&&\mbox{on $\S$},\\
\label{eq:BC-CF-infty}
\CF&\to 1&&\mbox{as $r\to\infty$},\\
B^i&\to (\Omega R_y, -\Omega R_x,0)&&\mbox{as $r\to\infty$},\\
\label{eq:BC-N-infty}
\N&\to 1&&\mbox{as $r\to\infty$},
\end{align}
\end{subequations}
with $\beta^i_\parallel$ satisfying
Eq.~(\ref{eq:BC-shift-parallel}).  Equation~(\ref{eq:QE-dtK}) is coded
as an equation for $\cN\CF^7=\N\CF$, so that, in the code, we impose
boundary conditions on $\N\CF$.  We will loosely refer to these
boundary conditions as ``lapse boundary conditions,'' nonetheless.

We finally note that one can absorb the $(\cLong \xi)^{ij}$-terms in a
redefinition of $\tilde u^{ij}$: If one uses $\tilde
u^{ij}=-(\cLong\xi)^{ij}$, then Eqs.~(\ref{eq:QE-Mom}) and
(\ref{eq:QE-Ham}) for $B^i$ take the same form as the original thin
sandwich equations (\ref{eq:Mom3}) and (\ref{eq:Ham3}) for $\beta^i$;
$\xi^i$ only appears in the advection term in (\ref{eq:QE-dtK}) and in
the boundary conditions.  Also, in the important case of conformal
flatness, $(\cLong\xi)^{ij}=0$.  We report on some code tests in
Appendix \ref{sec:QE:CodeTest}, and proceed now to compute initial
data sets.
                         
\section{Single black hole solutions}
\label{sec:QE:SingleBlackHoles}

We will first apply the quasi-equilibrium ideas to a spherically
symmetric spacetime with a single black hole.  This demonstrates the
practicability of the formalism, and tests the code.

\subsection{Eddington-Finkelstein coordinates}

The Eddington-Finkelstein coordinate system is well suited for
numerical relativity, as it is {\em horizon penetrating} and
comparatively simple.  We collect here some relevant formulae and a
few useful numbers.  The standard Eddington-Finkelstein coordinate
system is given by
\begin{subequations}\label{eq:EF}
\begin{align}
\label{eq:gij-EF}   \g^{EF}_{ij}&=\delta_{ij}+\frac{2M}{r}n^in^j, \\
\label{eq:lapse-EF} \N_{EF}&=\left(1+\frac{2M}{r}\right)^{-1/2},\\
\label{eq:shift-EF}
\beta^i_{EF}&=\left(1+\frac{2M}{r}\right)^{-1}\frac{2M}{r}n^i,
\end{align}
\end{subequations}
where $n^i=x^i/r$, and $r^2=\delta_{ij}x^ix^j$.  The mean curvature is
\begin{align}
\label{eq:trK-EF}
\trK_{EF}(r)&=\frac{2M}{r^2}\left(1+\frac{2M}{r}\right)^{-3/2}
      \left(1+\frac{3M}{r}\right).
\end{align}
The tracefree piece of the extrinsic curvature is not needed here; it
could be obtained via $\A^{ij}_{EF}=(2\N_{EF})^{-1}(\Long\beta_{EF})^{ij}$ if necessary.

The 3-metric~(\ref{eq:gij-EF}) in spherical coordinates reads
\begin{equation}\label{eq:gij-EF-2}
^{(3)}ds^2=\left(1+\frac{2M}{r}\right)dr^2+r^2d\Omega^2
\end{equation}
where $d\Omega^2$ is the metric on the sphere.
As with any spherically symmetric metric a suitable radial coordinate
transformation $r\to \hat r(r)$ will transform~(\ref{eq:gij-EF-2}) into
conformally flat form,
\begin{equation}\label{eq:gij-EF-3}
^{(3)}ds^2=\CF^4\Big(d\hat r^2+\hat r^2d\Omega^2\Big).
\end{equation}
Comparing the coefficients of $d\Omega^2$ in (\ref{eq:gij-EF-2}) and
(\ref{eq:gij-EF-3}) yields $r^2=\CF^4\hat r^2$, so that $\CF^2=r/\hat
r$.  Comparison of coefficients along the radial direction gives then
\begin{equation}\label{eq:EF-coordtrafo2}
\left(1+\frac{2M}{r}\right)^{1/2}dr=\frac{r}{\hat r}d\hat r,
\end{equation}
with solution (subject to the boundary condition $\hat r\to r$ as
$r\to\infty$)~\cite{Cook:2002}
\begin{equation}\label{eq:EF-flat-r}
\hat r=\frac{r}{4}\left(1+\sqrt{1+\frac{2M}{r}}\right)^2e^{2-2\sqrt{1+2M/r}}.
\end{equation}
The conformal factor is therefore
\begin{equation}\label{eq:CF-EF-flat}
\CF=\sqrt{\frac{r}{\hat r}}
=\frac{2e^{\sqrt{1+2M/r}-1}}{1+\sqrt{1+2M/r}}.
\end{equation}
The mean curvature $\trK$ behaves as a scalar under such a spatial
coordinate transformation,
\begin{equation}\label{eq:trK-EFCF}
\trK_{EFCF}(\hat r)=\trK_{EF}\big(r(\hat r)\big).
\end{equation}

The coordinate transformation (\ref{eq:EF-flat-r}) maps the radius of
the horizon from $r_{AH}/M=2$ to
\begin{align}\label{eq:r-AH-EF}
\frac{\hat r_{AH}}{M}&=\frac{e^{2-2\sqrt{2}}}{2}\left(1+\sqrt{2}\right)^2
\approx 1.27274, 
\end{align}
whereas, on the horizon,
\begin{align}
\label{eq:psi-EF}
\CF(\hat r_{AH})
&=\frac{2e^{\sqrt{2}-1}}{1+\sqrt{2}}\approx 1.25356,\\
\label{eq:Npsi-EF}
(\N\CF)(\hat r_{AH})
&=\sqrt{\frac{1}{2}}\;\CF(\hat r_{AH})\approx 0.88640.
\end{align}
These three numerical values will be used below.  We confirm in
Appendix~\ref{sec:QE:CodeTest} that the Eddington-Finkelstein metric
indeed satisfies the quasi-equilibrium equations~(\ref{eq:QE}).

\subsection{Solving for spherically symmetric spacetimes}
\label{sec:QE:SphericalSymmetricSolves}

We now employ the quasi-equilibrium method to solve for the standard
Eddington-Finkelstein slice.  We set the conformal metric equal to
Eq.~(\ref{eq:gij-EF}), so that the solution should have $\CF=1$,
choose the mean curvature by Eq.~(\ref{eq:trK-EF}), and set
$\beta^i_\parallel=0$.  Furthermore, we use a Dirichlet boundary
condition on the lapse,
\begin{equation}
\N\CF=\frac{1}{\sqrt{2}}\qquad \mbox{on $\S$},
\end{equation}
which follows from Eq.~(\ref{eq:lapse-EF}) and $\CF=1$.  

\begin{figure}[tb]
\centerline{\includegraphics[scale=0.4]{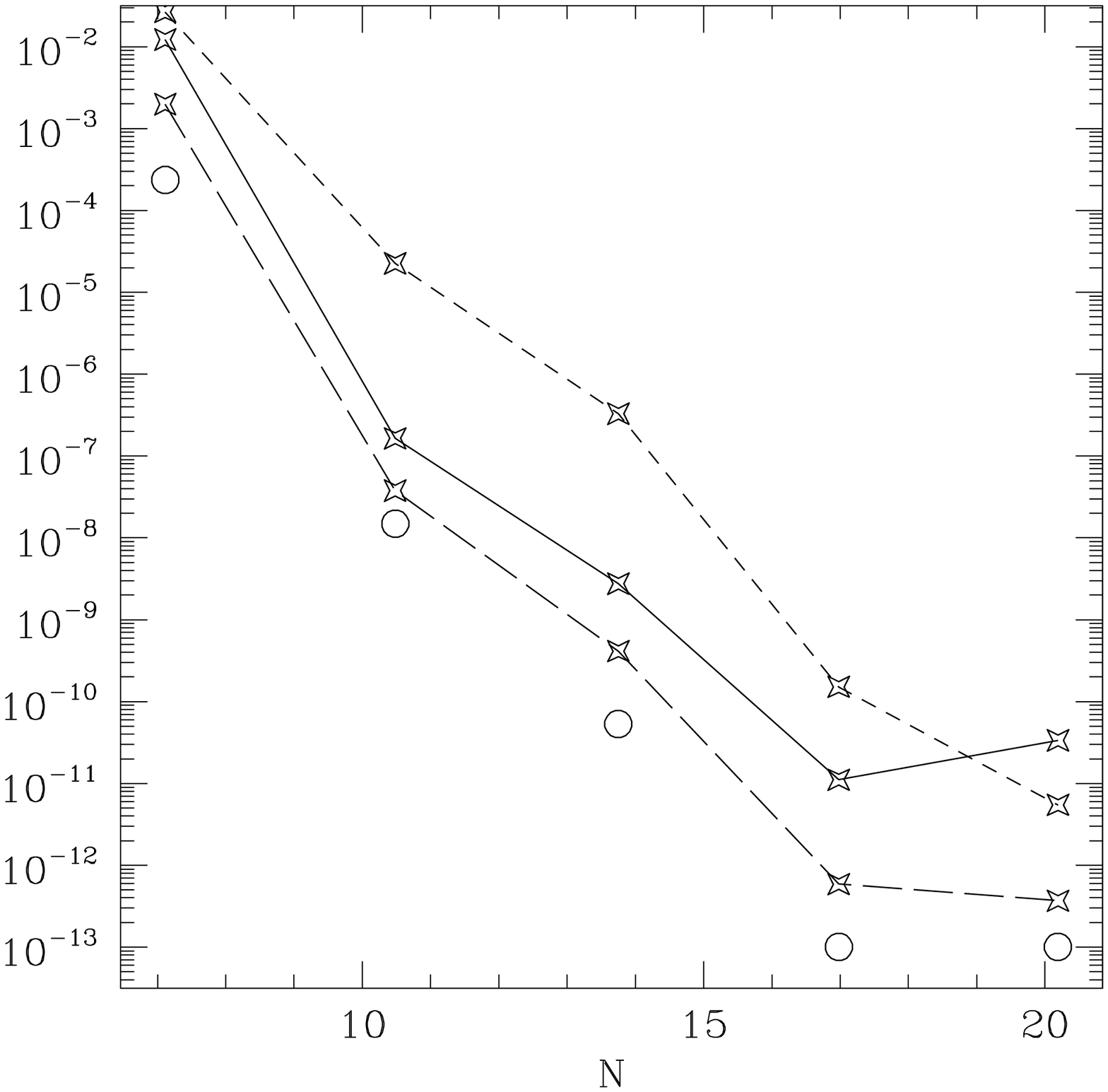}}
\CAP{Solving the quasi-equilibrium equations~(\ref{eq:QE}) on an
Eddington-Finkelstein slice}{\label{fig:SolveEF}Solving the
quasi-equilibrium equations~(\ref{eq:QE}) on an Eddington-Finkelstein
slice.  The lines denote the $L_\infty$ residuals of the Hamiltonian
constraint (solid), $\partial_t\ln\CF$ (long-dashed) and the lapse
condition~(\ref{eq:BC-lapse}) (short-dashed). 
The circles give the deviation of the ADM-energy from the
expected value $1$.}
\end{figure}  

With the most primitive initial guess, $\CF=1$, $\beta^i=0$, and
$N\CF=1$, the Newton-Raphson method inside the elliptic solver does
not converge.  A suitable initial guess can be obtained by performing
an elliptic solve with {\em Dirichlet} boundary condition on the
conformal factor, $\CF=1$ on $\S$, and then using the result of this
solve as initial guess for the ``real'' solve.  Similar convergence
problems of the Newton-Raphson procedure are encountered sometimes for
binary black hole solutions.  In practice, a similar ``preprocessing
solve'' with Dirichlet boundary conditions $\CF=\CF_\S$ resolves this
difficulty (this procedure is fairly insensitive to the exact value of
the constant $\CF_\S$--- choosing it based on the underlying single
black hole solutions is sufficient).  For binary black hole solves,
using the underlying single black hole solutions as initial guess
usually results in fast convergence of the Newton-Raphson method, too.

With a suitable initial guess, the Newton-Raphson method converges
nicely for each resolution.  The solutions converge exponentially with
increasing resolution, as they should (see Figure~\ref{fig:SolveEF}),
and reproduce the standard Eddington-Finkelstein slice.

If we replace the Dirichlet boundary condition on the lapse by the
lapse condition $\BC=0$, Eq.~(\ref{eq:BC-lapse}), then the elliptic
solver does {\em not} converge.  To gain further insight, we perform
nine solves for spherically symmetric single black hole spacetimes.
For all solves, we choose conformal flatness, $\cg_{ij}=\delta_{ij}$,
and vary the lapse boundary condition as well as the mean curvature:
\begin{enumerate}
\item For the lapse, we use two different Dirichlet boundary
  conditions (as indicated in Table~\ref{tab:SingleHoleSolves}) and
  the von Neumann boundary condition $\partial_r(\N\CF)=0$.
\item For $\trK$, we use $\trK=\trK_{EFCF}(\hat r)$
(Eddington-Finkelstein with conformally flat radial coordinate),
$\trK=2/r^2$ and $\trK=0$.
\end{enumerate}
The radius of the excised sphere is $\hat r_{AH}/M\approx 1.2727$ as
given by Eq.~(\ref{eq:r-AH-EF}), so that the first of the nine
solutions should recover the Eddington-Finkelstein solution with
conformally flat radial coordinate.

\begin{sidewaystable}
\begin{centering}
\CAP{Spherically symmetric single black hole
solutions}{\label{tab:SingleHoleSolves}Spherically symmetric single
black hole solutions with $\cg_{ij}\!=$flat and different choices for
the lapse boundary condition and for $\trK$.  Radius of the excised sphere
is $\hat r_{AH}$.}

\centerline{
\begin{tabular}{|cc|cc|ccc|cc|}\hline
lapse BC & $\trK$ & Hamilt. & moment. &
$M_{AH}$ & $E_{ADM}$ & $M_K$ & $L_\infty(\partial_t\ln\CF)$ & 
$L_\infty(\BC)$\\
&&\multicolumn{2}{c|}{(upper limits)} 
&&&& \multicolumn{2}{c|}{(upper limits)}\\\hline
  $(\N\CF)(\hat r_{AH})$  &  $\trK_{EFCF}$
  & $1.3\cdot 10^{-11}$ & $6.1\cdot 10^{-12}$ 
  & $ 1.0000000$ & $1.000000$ & $1.000000$ 
  & $1.0\cdot 10^{-13}$ & $2.5\cdot 10^{-12}$ \\
  $(\N\CF)(\hat r_{AH})$  & $2/r^2$ 
  & $3.3\cdot 10^{-9}$   &  $1.9\cdot 10^{-8}$ 
  & $0.772329$ &   $0.77235$  &  $0.77236$  
  & $2.2\cdot 10^{-10}$  &  $9.3\cdot 10^{-10}$  \\
  $(\N\CF)(\hat r_{AH})$  & $0$ 
  & $1.5\cdot 10^{-10}$  &  $1.5\cdot 10^{-11}$ 
  & $1.2897483$  &  $1.289748$  &  $1.28975$  
  & $2.0\cdot 10^{-12}$  &  $5.7\cdot 10^{-11}$ \\
  $3/4$   &    $\trK_{EFCF}$
  &  $1.9\cdot 10^{-10}$  &  $2.0\cdot 10^{-9}$ 
  &  $1.085413$  &  $1.08541$  &  $1.08541$ 
  & $2.5\cdot 10^{-11}$  &  $1.4\cdot 10^{-10}$ \\
  $3/4$  &  $2/r^2$  
  & $1.0\cdot 10^{-9}$  &  $1.2\cdot 10^{-8}$
  &  $0.8351808$ &  $0.83518$  &  $0.8352$  
  &  $5.1\cdot 10^{-11}$  &  $2.2\cdot 10^{-10}$  \\
  $3/4$  &  $0$
  & $5.3\cdot 10^{-11}$  &  $1.1\cdot 10^{-11}$ 
  & $1.3998627$  &  $1.39986$  &  $1.39986$ 
  & $6.2\cdot 10^{-13}$   &  $1.7\cdot 10^{-11}$ \\
  $\partial_r(\N\CF)=0$ & $\trK_{EFCF}$ 
  &$3.0\cdot 10^{-11}$  &  $5.2\cdot 10^{-11}$  
  & $0.9942475$  &  $0.994247$  &  $0.994247$ 
  & $4.3\cdot 10^{-13}$  &  $6.4\cdot 10^{-12}$ \\
$\partial_r(\N\CF)=0$ & $2/r^2$
  &  $2.0\cdot 10^{-10}$  &  $2.9\cdot 10^{-9}$  
  &  $0.6828056$  & $0.68281$  &  $0.68281$  
  &  $1.7\cdot 10^{-11}$  &  $1.1\cdot 10^{-10}$ \\
$\partial_r(\N\CF)=0$ & $0$
  &  $3.4\cdot 10^{-11}$  &  $1.6\cdot 10^{-11}$
  &  $1.4807928$  & $1.480792$  &  $1.480793$ 
  &  $1.0\cdot 10^{-11}$  &  $2.7\cdot 10^{-11}$ 
\\\hline
  \end{tabular}
}
\end{centering}
\end{sidewaystable}

The runs are summarized in Table~\ref{tab:SingleHoleSolves}.  For each
run we compute the apparent horizon mass
\begin{equation}
M_{AH}=\sqrt{\frac{A_{AH}}{16\pi}},
\end{equation}
the ADM-energy
\begin{equation}\label{eq:E-ADM}
E_{ADM}=-\frac{1}{8\pi}\int_{r=\infty}\dnachd{\CF}{r}d^2A,
\end{equation}
and the Komar mass\footnote{These definitions for $E_{ADM}$ and $M_K$
are valid only in conformal flatness.  More general definitions can be
found in \cite{York:1979} and \cite{Gourgoulhon-Bonazzola:1994}.}
\begin{equation}\label{eq:Komar-Mass}
M_K=\frac{1}{4\pi}\int_{r=\infty} \dnachd{N}{r}d^2A.
\end{equation}
We also evaluate $\partial_t\ln\CF$ and $\BC$.  For single hole
solutions, we can push the accuracy of the pseudo-spectral method
close to round-off, as can be seen by the residuals of the constraints
(these are $L_\infty$-norms; the $L_2$ norms are smaller by roughly
one order of magnitude).  $E_{ADM}$ and $M_K$ are determined by
surface integrals extrapolated to infinite radius, which makes them
less accurate.  Nonetheless, we can compute them to better than
$10^{-6}$.  We also remark that the data sets with $\trK=2/r^2$
display a slower exponential convergence; although more collocation
points are used for these solves, they are still somewhat less
accurate.

The first data set in Table~\ref{tab:SingleHoleSolves} has a mass of
exactly one, as it should, because it {\em is} just
Eddington-Finkelstein with unit-mass with the radial coordinate
transformation (\ref{eq:EF-flat-r}).  It is not surprising
that the time-derivative of the conformal factor vanishes and that
this dataset also satisfies the lapse condition $\BC$,
Eq.~(\ref{eq:BC-lapse}).

Perhaps more surprising is that {\em all} datasets presented in
Table~\ref{tab:SingleHoleSolves} apparently satisfy the lapse
condition (\ref{eq:BC-lapse}).  This explains why the lapse
condition~(\ref{eq:BC-lapse}) fails as a boundary condition together
with the other quasi-equilibrium boundary conditions: {\em Any} value
of the lapse (with the other quasi-equilibrium boundary conditions)
leads to a data set satisfying (\ref{eq:BC-lapse}).
Equation~(\ref{eq:BC-lapse}) is degenerate with the other
quasi-equilibrium boundary conditions and cannot be used to single out
a unique value of the lapse on the horizon.

From Table~\ref{tab:SingleHoleSolves} we see further, that{\em all}
datasets apparently have vanishing $\partial_t\ln\CF$, indicating that
all these data sets might be time-independent.  Moreover, for each
dataset, apparent horizon mass, ADM energy and Komar mass agree up to
the numerical accuracy of those numbers.  $E_{ADM}$ measures the total
energy of the hypersurface, including gravitational waves, whereas
$M_{AH}$ just includes the black hole mass.  As the apparent horizon
mass is a lower bound of the event horizon mass, which in turn must be
smaller than the ADM energy, we find that none of these datasets
contains gravitational radiation.  Finally, $E_{ADM}=M_K$ for
stationary spacetimes \cite{Beig:1978, Gourgoulhon-Bonazzola:1994},
hence equality of these masses once again hints toward a stationary
spacetime.

The fact that the quasi-equilibrium equations in spherical symmetry
seem to generate stationary slices for all our choices of the lapse
boundary condition and the mean curvature can be understood as
follows: The apparent horizon boundary condition guarantees the
existence of a black hole in the spherically symmetric space time; by
Birkhoff's theorem, the generated initial data set must therefore be a
slice through a Schwarzschild black hole.  Schwarzschild is
stationary, and we recall from the discussion about stationary
spacetimes in section~\ref{sec:QE:StationarySpacetimes} that lapse and
shift along the Killing vector solve the conformal thin sandwich
equations {\em if compatible} with the free data.

Any spherically symmetric hypersurface is conformally flat, so that
$\cg_{ij}=\mbox{flat}$ cannot lead to incompatibilities.  The inner
boundary is an apparent horizon (by Eq.~(\ref{eq:BC-AH2})~) and the
time vector $t^\mu$ is radial and null there (because of
Eq.~(\ref{eq:BC-shift}) together with $\beta^i_\parallel=0$),
therefore at the inner boundary, $t^\mu$ is proportional to the
Killing vector.  At the outer boundary, the time vector is identical
to the Killing vector by the outer boundary
conditions~(\ref{eq:BC-CF-infty})--(\ref{eq:BC-N-infty}).  So the only
incompatibility left could be a different constant of proportionality
between $t^\mu$ and the Killing vector on the boundaries.  But for the
data sets contained in Table~\ref{tab:SingleHoleSolves}, apparently no
such inconsistency exists.  Thus it seems likely that for all
(reasonable) mean curvatures $\trK$ and values of the lapse on the
apparent horizon, there exists a spherically symmetric slice through
Schwarzschild.  In each case, the black hole mass will have some
value, based on the radius of the excised sphere, the lapse boundary
condition and $\trK$.

This interpretation is also consistent with the fact that there exists
a one-parameter family of spherically symmetric maximal slices through
Schwarzschild~\cite{Cook:2003,
Estabrook-Wahlquist-etal:1973}\footnote{Estabrook et
al.~\cite{Estabrook-Wahlquist-etal:1973} use $T=CM^2$ instead of $C$.
Equations~(\ref{eq:MaximalSlice}) can be obtained by substituting a
{\em time-independent} $T$ into their more general results.}, given by
\begin{subequations}\label{eq:MaximalSlice}
\begin{align}
\label{eq:MaximalSlice-gij}
^{(3)}ds^2
&=\left(1-\frac{2M}{r}+C^2\frac{M^4}{r^4}\right)^{-1}dr^2+r^2d\Omega^2,\\
\label{eq:MaximalSlice-N}
\N&=\left(1-\frac{2M}{r}+C^2\frac{M^4}{r^4}\right)^{1/2},\\
\label{eq:MaximalSlice-shift}
\beta^r&=\left(1-\frac{2M}{r}+C^2\frac{M^4}{r^4}\right)^{-1/2},
\end{align}
\end{subequations}
where $M$ is the mass of the black hole and $C$ is a dimensionless
constant ($C=0$ yields the standard Schwarzschild slice).
Substituting $r=2M$ into~(\ref{eq:MaximalSlice-N}) we find the value
of the lapse on the horizon to be $C/4$, so that the slicings differ
indeed by the value of the lapse on the horizon.

A suitable radial coordinate transformation $r\!\to\! \hat r(r)$ will make
the metric~(\ref{eq:MaximalSlice-gij}) conformally flat\footnote{This
coordinate transformation is not available in closed form; it must be
obtained by numerical integration~\cite{Cook:2003}.}.  Picking values
for $M$ and $C$ in
Eqs.~(\ref{eq:MaximalSlice-gij})--(\ref{eq:MaximalSlice-shift}) and
transforming to conformally flat coordinates will result in an initial
data set with apparent horizon at $\hat r_{AH}=\hat r(2M)$ and lapse
on the horizon of $N(\hat r_{AH})=C/4$.  Conversely, picking the
radius of the excised sphere $r_{exc}$ (which is an apparent horizon
due to the quasi-equilibrium boundary conditions) and a lapse boundary
condition on the excised sphere will determine some numerical values
for $M$ and $C$ in Eq.~(\ref{eq:MaximalSlice-gij}): $M$ will simply be
the ADM-energy, and $C=4N(r_{exc})$.

Finally, we remark, that Table~\ref{tab:SingleHoleSolves} suggests that
for any value of the mean curvature datasets with $\partial_t\ln\CF=0$
exist.  This implies that we cannot use quasi-equilibrium ideas to
single out a unique $\trK$.  One cannot, for example, use
$\partial_t\ln\CF=0$ to fix $\trK$.  This is perfectly consistent with
the result in section~\ref{sec:IVP:ImplicationsForEvolution} that
using $\partial_t\CF=0$ to fix $\trK$ leads to a non-invertible
differential operator in the momentum constraint.

\subsection{The lapse boundary condition}
\label{sec:QE:LapseBC}

In the last section we have seen that the lapse condition
(\ref{eq:BC-lapse}) is degenerate with the other quasi-equilibrium
boundary conditions in {\em spherical symmetry}.  For binary black
hole configurations which are {\em not} strictly stationary, this
degeneracy might disappear.  In that case, Eq.~(\ref{eq:BC-lapse})
might represent a valid lapse boundary condition which sets one
additional time derivative to zero, bringing the solution even closer
to stationarity.

However, in the limit of large separation, we recover two isolated
black holes, and therefore the boundary conditions must become
degenerate.  Therefore it is likely that even at finite separation the
lapse condition will be nearly degenerate with the remaining
quasi-equilibrium boundary conditions, resulting in a system of
equations which is difficult to solve numerically.

Nonetheless, I have put considerable effort into using the lapse
condition (\ref{eq:BC-lapse}) for solutions representing binary black
hole spacetimes, or non-spherically symmetric single black hole
spacetimes, but {\em without success}.

The rest of this subsection demonstrates one further problematic point
regarding the lapse condition (\ref{eq:BC-lapse}) in non-spherically
symmetric single black hole solutions: We excise a sphere with radius
$\hat r_{AH}\approx 1.2727$, choose conformal flatness
$\cg_{ij}=\delta_{ij}$, and use $\trK_{EFCF}$ as mean curvature
centered at $(x,y,z)=(0.02, 0, 0.01)$ so that the center of $\trK$
does not coincide with the center of the excised sphere.  We use the
quasi-equilibrium boundary conditions
(\ref{eq:BC-AH2})--(\ref{eq:BC-N-infty}), where we set
$\beta^i_\parallel$ to a rotation around the z-axis with angular
frequency $0.1$ to further break spherical symmetry.

With these free data and boundary conditions, and with the lapse-condition 
as boundary condition on $\N\CF$, the linear solver inside the Newton-Raphson
procedure fails to converge.

From spherically symmetric spacetimes, we recall that the lapse
boundary condition was degenerate with the remaining quasi-equilibrium
boundary conditions; the overall mass-scale had to be set.  Similarly,
we now try to fix the value of $\N\CF$ at one point on the surface
of the excised sphere, in the hope that the lapse condition
(\ref{eq:BC-lapse}) will then determine it at all other points.  Since
we solve for the spherical harmonic coefficients
(cf. Eq.~(\ref{eq:ExpansionSphere2}) on
p. \pageref{eq:ExpansionSphere2}), it is difficult to fix $\N\CF$ at
one point in real-space.  Instead, we require that the {\em average}
value of $\N\CF$ on the surface equals $0.866$, which translates to a
condition on the coefficient of $Y_{00}$.  This average-fixing
condition replaces the $Y_{00}$ component of the lapse condition
(\ref{eq:BC-lapse}), i.e. we do not enforce the monopole component
$\BC=0$, but require only that all higher spherical harmonic pieces of
$\BC$ vanish.

This fairly elaborate procedure alone does not yet cure the
convergence problems of the linear solver.  In addition, we need to
consider the lapse boundary condition (\ref{eq:BC-lapse}) in more
detail.  Expanding Eq.~(\ref{eq:BC-lapse}) yields the version of this
equation which is actually implemented in the code:
\begin{equation}\label{eq:LapseBC1}
\begin{aligned}
0=&\left(\frac{\CF}{2\N\CF}\tilde J+\frac{2}{3}K\right)
\tilde s^i\tilde\nabla_i(\N\CF)\\
  &+\frac{3\CF^4}{16\N\CF}\tilde J^2
  -\frac{1}{3}\CF^2(\N\CF)K^2
  +\left(\frac{\CF}{8}\tilde J+\frac{\N\CF}{6}K\right)
    \tilde h^{ij}\tilde\nabla_j\tilde s_i
  +\CF^3\tilde{\cal D}\N,
\end{aligned} 
\end{equation}
where
\begin{equation}\label{eq:LapseBC2}
\begin{aligned}
  \CF^3\tilde{\cal D}\N
=&\frac{1}{\CF^2}\tilde D^2(\N\CF)
  +\frac{\N\CF}{\CF^3}\tilde D^2\CF
  -\frac{\N\CF}{2\CF^2}\;^{(2)}R
  -\frac{2}{\CF^3}\tilde D^i(\N\CF)\tilde D_i\CF\\
  & -\frac{\CF}{2\N\CF}\tilde h^{ij}\tilde J_i\tilde D_j(\N\CF)
  -\frac{1}{2}\tilde h^{ij}\tilde J_i\tilde D_j\CF
  +\frac{\CF^4}{4\N\CF}\tilde h^{ij}\tilde J_i\tilde J_j
  -\frac{\CF}{2}\tilde h^{ij}\tilde D_i\tilde J_j,
\end{aligned}
\end{equation}
and
\begin{align}
  J&=h_{ij}K^{ij}=\frac{1}{2\alpha}\tilde J+\frac{2}{3}K,\\
\tilde J_i&=\tilde h_{ik}\tilde s_l\left((\tilde\Long\beta)^{kl}
  -\tilde u^{kl}\right).
\end{align}
We will refer to the last term of Eq.~(\ref{eq:LapseBC2}),
$-\frac{\CF}{2}\tilde h^{ij}\tilde D_i\tilde J_j$ as ``the divergence
term.''  We multiply this term by a coefficient $\Xi$; $\Xi=1$
recovers Eq.~(\ref{eq:BC-lapse}).  We find, that for $\Xi\neq 1$, we
{\em can} solve the system of equations (albeit the linear solver
converges very slowly).  For $\Xi=1$, we recover the original
formulation where the solver fails.

\begin{figure}[tb]
\centerline{\includegraphics[scale=0.4]{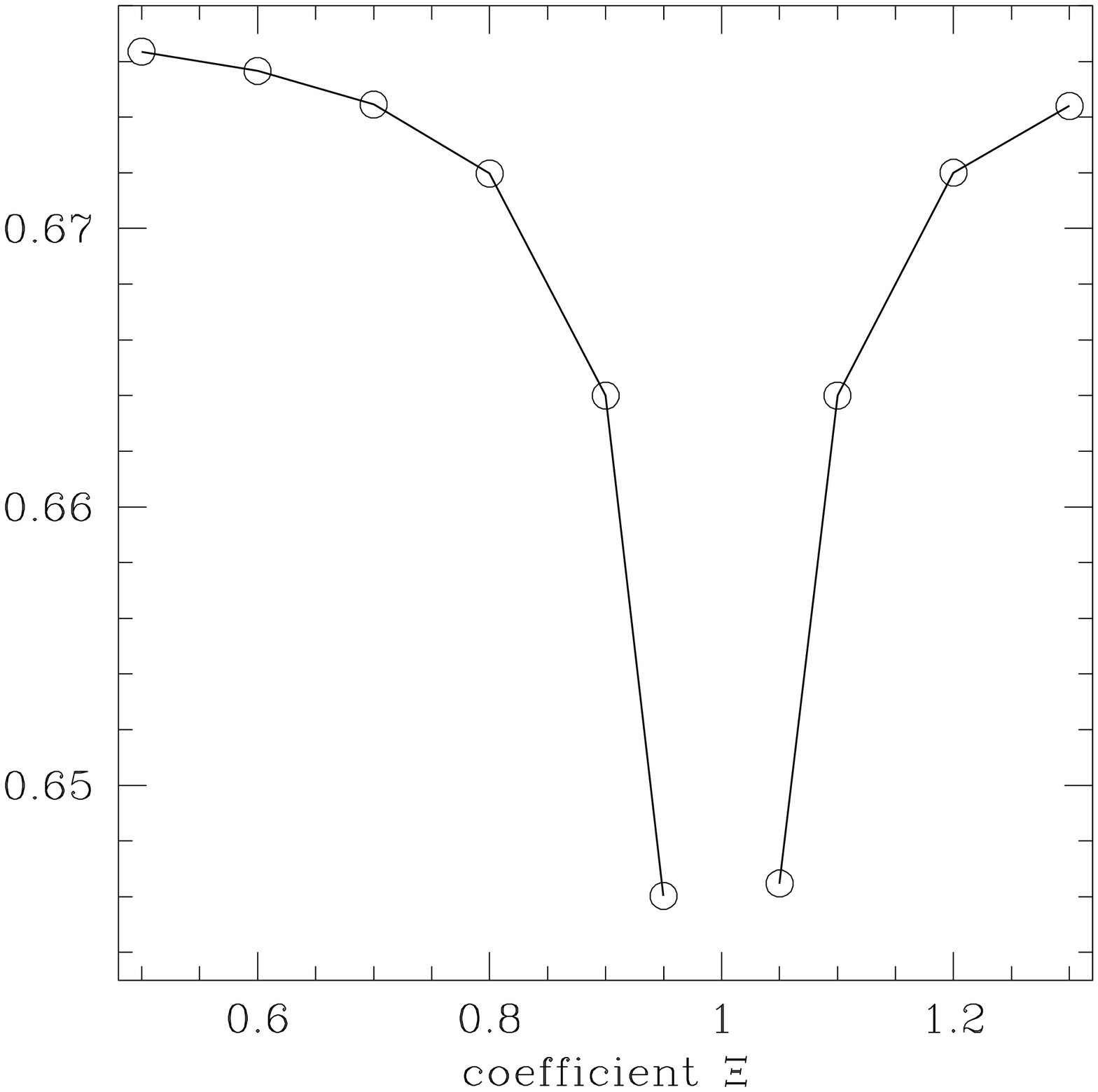}}
\CAP{Minimum of $\N$ as a function of the coefficient $\Xi$ in the
modified lapse boundary condition.}{\label{fig:QE-Lapse-BC}Minimum of
$\N$ as a function of the coefficient $\Xi$ multiplying the last term
in Eq.~(\ref{eq:LapseBC2}).}
\end{figure}

Figure~\ref{fig:QE-Lapse-BC} plots the minimum of $\N$ of the solution
versus the coefficient $\Xi$ in the modified lapse boundary condition.
Every data point in this figure represents a convergent solution (in
the number of basis functions\footnote{For given resolution, the
linear solve inside the Newton-Raphson method converges very slowly.
This might be caused by either insufficient preconditioning of the
lapse condition (which is not preconditioned at all), or by a nearly
degenerate system of equations.}) of the quasi-equilibrium equations
with the modified lapse condition.  Close to $\Xi=1$, the solution
depends very sensitively on the parameter $\Xi$, apparently with a
singularity for $\Xi=\Xi_{crit}\approx 1$.  This observation shows
that the lapse-condition with $\Xi=1$ is probably ill-posed.

We note that the divergence term under consideration vanishes
identically in spherical symmetry.  Moreover, the divergence term
contains {\em second} derivatives of the shift $\beta^i$.  Thus it is
possible that the lapse condition somehow interferes with the {\em
second order} elliptic equation for $\beta^i$.

\section{Binary black hole solutions}
\label{sec:QE:BinaryBlackHoles}

\subsection{Choices for the remaining free data}

We consider initial data sets with two black holes labeled $A$ and
$B$.  The quasi-equilibrium framework requires excision of two
topologically spherical regions, which we choose to be spheres of
radius $r_{A/B}$.  We need further choices for the conformal
metric $\cg_{ij}$, the mean curvature $\trK$, and a lapse boundary
condition.

In the spirit of Matzner et al \cite{Matzner-Huq-Shoemaker:1999,
Marronetti-Matzner:2000} and chapter \ref{chapter:Comparing} of this
thesis, we will continue to base these free data on a superposition of
single black hole quantities.  For convenience, we reproduce here
Eqs.~(\ref{eq:BinaryKerrSchild-gamma})--(\ref{eq:BinaryKerrSchild-Mij}),
which are the choices advocated by Matzner et al:
\begin{align}\label{eq:Matzner-gij}
  \cg_{ij}&=\delta_{ij}+2H_{\!A}\,l_{A\,i}\,l_{A\,j}
  +2H_{\!B}\,l_{B\,i}\,l_{B\,j}\\
\label{eq:Matzner-K}
K&=K_{\!A}+K_B\\
\label{eq:Matzner-Aij}
  \tilde M^{ij}&=\left(K^{(i}_{\!A\;k}+K^{(i}_{B\;k}
    -\frac{1}{3}\delta^{(i}_k(K_{\!A}+K_B)\right)\cg^{j)k}.
\end{align}

We focus here on non-spinning black holes as the first step in
understanding binary black hole data sets within the quasi-equilibrium
framework.  For non-spinning black holes, we have seen in the previous
section that a multitude of single black hole solutions exist that are
{\em conformally flat} (namely all spherically symmetric solutions,
after an appropriate radial coordinate transformation).  Basing the
superposition on conformally flat single black hole slices, the
equivalent of (\ref{eq:Matzner-gij}) is simply
\begin{equation}\label{eq:BBH-gij}
\cg_{ij}=\delta_{ij}.
\end{equation}
For the mean curvature, we will explore two different choices:
\begin{itemize}
\item Eddington-Finkelstein slices with ``conformally flat''
radial coordinate,
\begin{equation}\label{eq:BBH-trK}
\trK(x^i) = \trK_{EFCF\,A}(r_A) + \trK_{EFCF\,B}(r_B).
\end{equation}
The functions $\trK_{EFCF\;A/B}$ are given by Eqs.~(\ref{eq:trK-EFCF})
and (\ref{eq:trK-EF}) with parameters appropriate for hole $A/B$.  The
radii of the excised spheres are given by $\hat r_{AH}$ from
Eq.~(\ref{eq:r-AH-EF}).

\item maximal slices with a ``conformally flat'' radial
coordinate.  Each slice has $\trK_{A/B}=0$, hence trivially 
\begin{equation}\label{eq:BBH-trK2}
\trK=0.
\end{equation}
\end{itemize}

As mentioned already, the quasi-equilibrium approach based on the
conformal thin sandwich equations does not require specification of a
background extrinsic curvature analogous to (\ref{eq:Matzner-Aij}).
The choice (\ref{eq:BBH-gij}) looks certainly much simpler than
(\ref{eq:Matzner-gij}), however, both of these choices are inherently
ambiguous in that both are based on a {\em specific} coordinate system
within the Eddington-Finkelstein slice.

As lapse boundary condition on the excised spheres, we use a von
Neumann condition,
\begin{equation}\label{eq:Neumann-lapse-BC}
\tilde s^k\cderiv_k(\N\CF)=0.
\end{equation}
The reasons for this condition are rather heuristic: With Dirichlet
boundary condition, fairly often steep gradients arise in $\N$ close
to the inner boundary.  Using a von Neumann condition allows the lapse
to adjust itself to reduce these gradients, and to better adjust to
tidal distortion due to the other hole.  We choose
$\partial_t(\N\CF)=0$ rather than $\partial_t\N=0$ to mimic single
black hole solutions, which often have a lapse that {\em increases} as
one moves away from the horizon, for example, Eqs.~(\ref{eq:lapse-EF})
and (\ref{eq:MaximalSlice-N}).

Finally, the quasi-equilibrium framework presented so far does not
yet fix the orbital angular frequency $\Omega$ which appears in the
outer boundary condition~(\ref{eq:BC-shift-infty}).  Gourgoulhon et al
\cite{Gourgoulhon-Grandclement-Bonazzola:2001a,
Grandclement-Gourgoulhon-Bonazzola:2001b} suggest that $\Omega$ be
chosen such that the $1/r$--terms of $\g_{ij}$ and $\N$ behave as for
a Schwarzschild black hole;  in conformal flatness, this implies that
\begin{align}\label{eq:Omega-condition}
\CF\sim 1+\frac{M}{2r}\qquad\mbox{and}\qquad
\N\sim 1-\frac{M}{r}
\end{align}
for the same constant $M$.  Equation~(\ref{eq:Omega-condition}) is
equivalent to requiring that the ADM-energy Eq.~(\ref{eq:E-ADM})
equals the quantity $M_K$ defined in
Eq.~(\ref{eq:Komar-Mass}).\footnote{The Komar-mass is only defined for
stationary spacetimes which have a timelike Killing vector at
infinity, whereas the helical Killing vector is spacelike at infinity.
Hence, for binary black holes in the corotating frame $M_K$ is,
strictly speaking, not the Komar-mass.}
Below, we take the approach of computing several data sets for
different $\Omega$ and we will find that $E_{ADM}=M_K$ indeed singles
out a unique $\Omega$.

\subsection{Eddington-Finkelstein slicings}
\label{sec:QE:BBH:EFslicings}

We solve the quasi-equilibrium equations~(\ref{eq:QE}) with lapse
boundary condition (\ref{eq:Neumann-lapse-BC}) for equal mass black
holes.  According to the discussion in the previous paragraphs, we
choose a {\em flat} conformal metric and give the mean curvature
by~(\ref{eq:BBH-trK}) with $M_A=M_B=1$.  We excise two spheres with
radius $\hat r_{AH}\approx 1.27$ [cf. Eq.~(\ref{eq:r-AH-EF})] with
centers separated by $s=10$.  By symmetry, the rotation axis of the
corotating frame must pass through the origin, $R_x=R_y=0$.

\begin{figure}
\centerline{\includegraphics[scale=0.4]{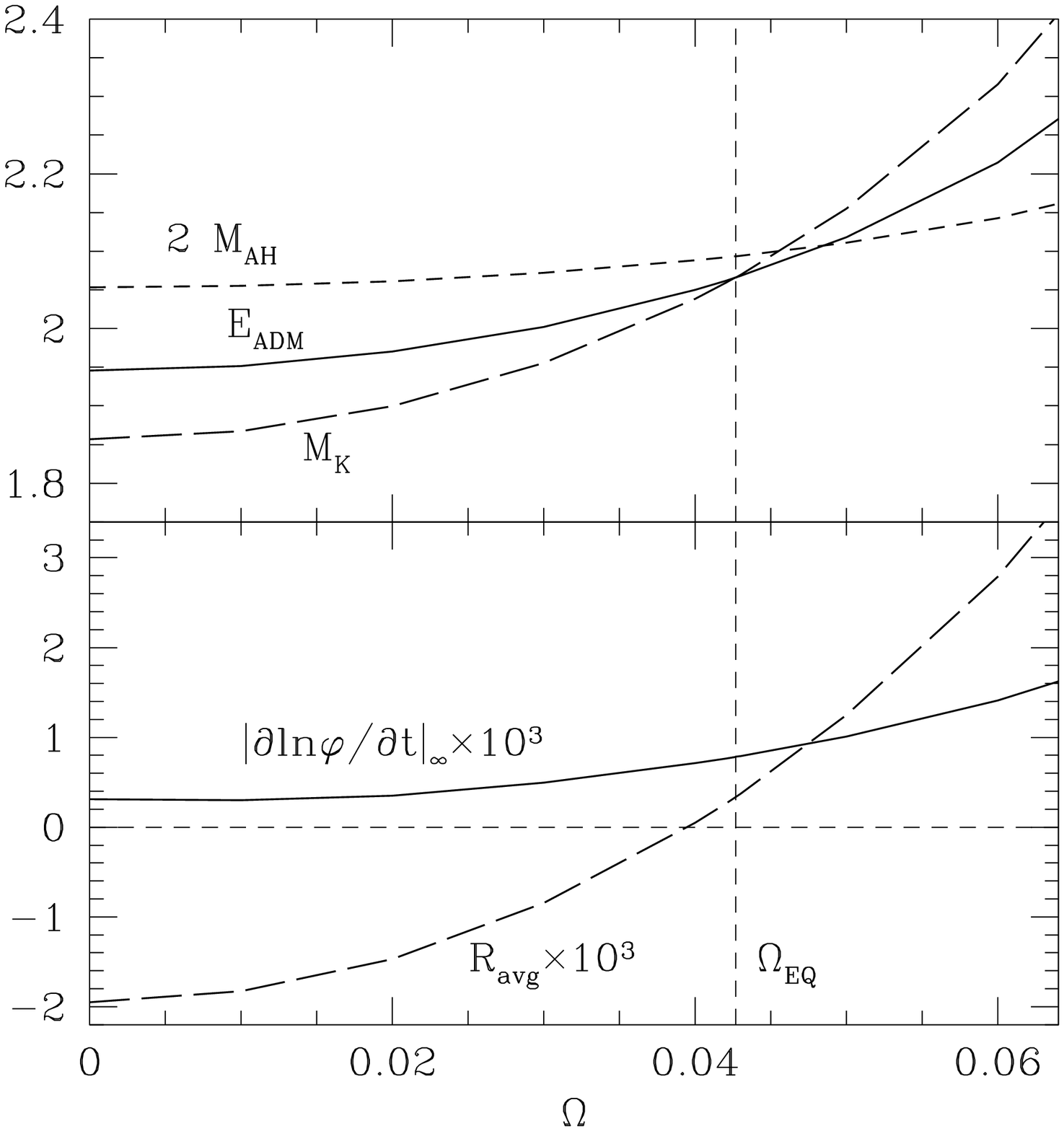}}
\CAP{Quasi-equilibrium binary black hole initial data sets on
Eddington-Finkelstein slices: dependence on
$\Omega$}{\label{fig:EF_vs_Omega}Quasi-equilibrium binary black hole
initial data sets on Eddington-Finkelstein slices.  Plotted are
several quantities as a function of the orbital angular velocity
$\Omega$. $R_{avg}$ denotes the residual of $\BC$ averaged over the
boundary $\S$.}
\end{figure}

We perform several solves with different values for the last free
parameter,~$\Omega$, and compute several diagnostic quantities, in
particular $M_{AH}\!=\!\sqrt{A_{AH}/16\pi}$ (the
apparent horizon mass of one black hole), $E_{ADM}$ and $M_K$, as well as the
time-derivative of the conformal factor, $\partial_t\ln\CF$ and the
violation of the lapse condition, $\BC$.  Figure~\ref{fig:EF_vs_Omega}
presents these quantities as a function of $\Omega$.  We see that
there is indeed a unique $\Omega_{EQ}$ such that $E_{ADM}=M_K$; in
this case $\Omega_{EQ}\approx 0.04266$. 

Turning our attention to the lower panel of
Figure~\ref{fig:EF_vs_Omega}, we see that the maximal value of
$|\partial_t\ln\CF|$ is smaller than $2\cdot 10^{-3}$ for all
considered values of $\Omega$.  This time derivative is smallest for
$\Omega=0$.  Finally, we note that the average value of the lapse
condition $\BC$, Eq.~(\ref{eq:BC-lapse}) passes through zero at
$\Omega$ close to $\Omega_{EQ}$, exhibiting consistency between
Eq.~(\ref{eq:BC-lapse}) and the stationarity condition $E_{ADM}=M_K$.
Of course, an {\em average} value of $\BC$ of zero does not imply that
$\BC\equiv 0$, for example, for $\Omega=0.04$, $\BC$ has an average of
only $\BC_{avg}=5\cdot 10^{-5}$, but a maximum value of
$|\BC|_\infty=8\cdot 10^{-3}$.

\begin{figure}
\includegraphics[scale=0.36]{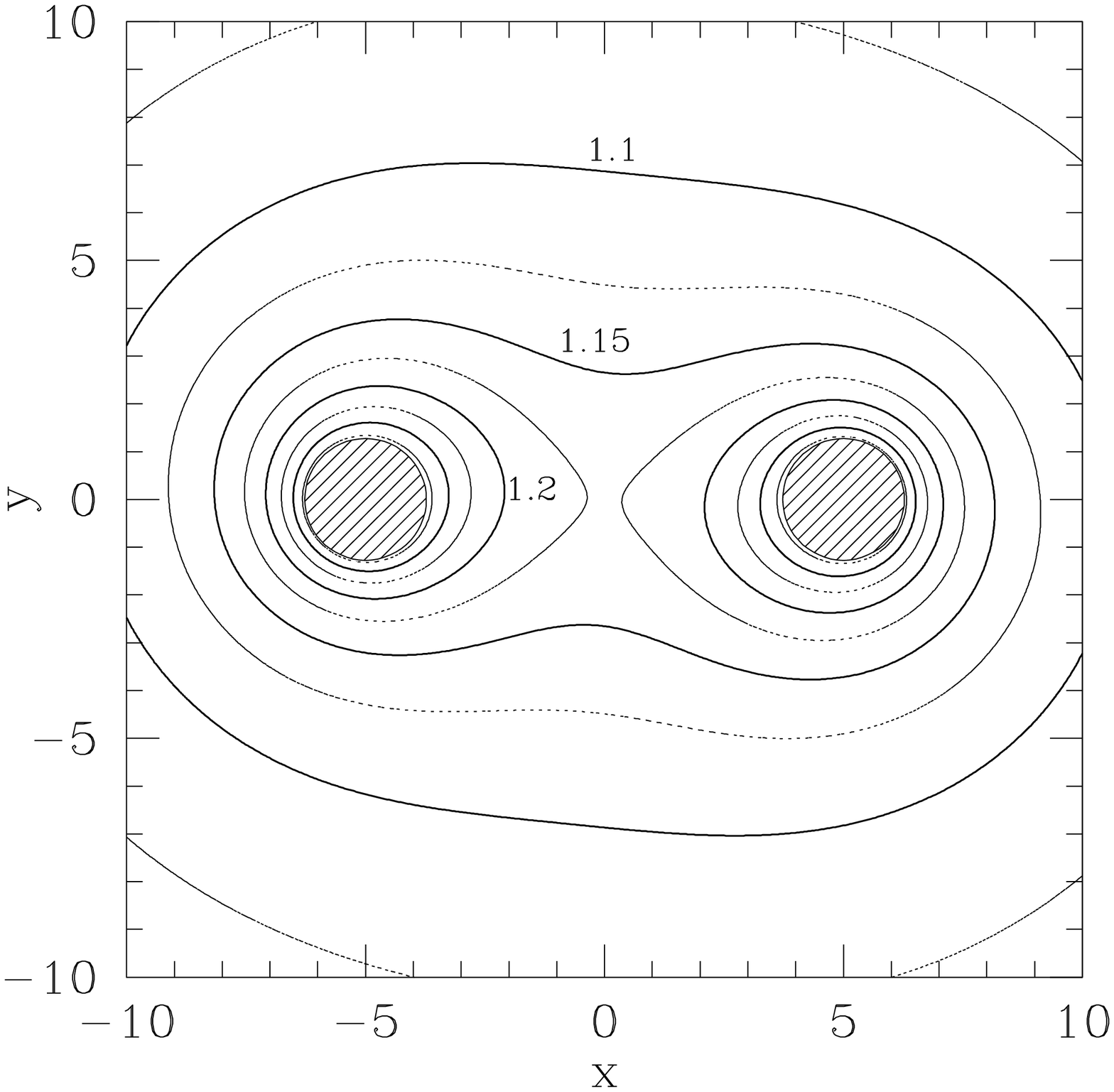}
\includegraphics[scale=0.36]{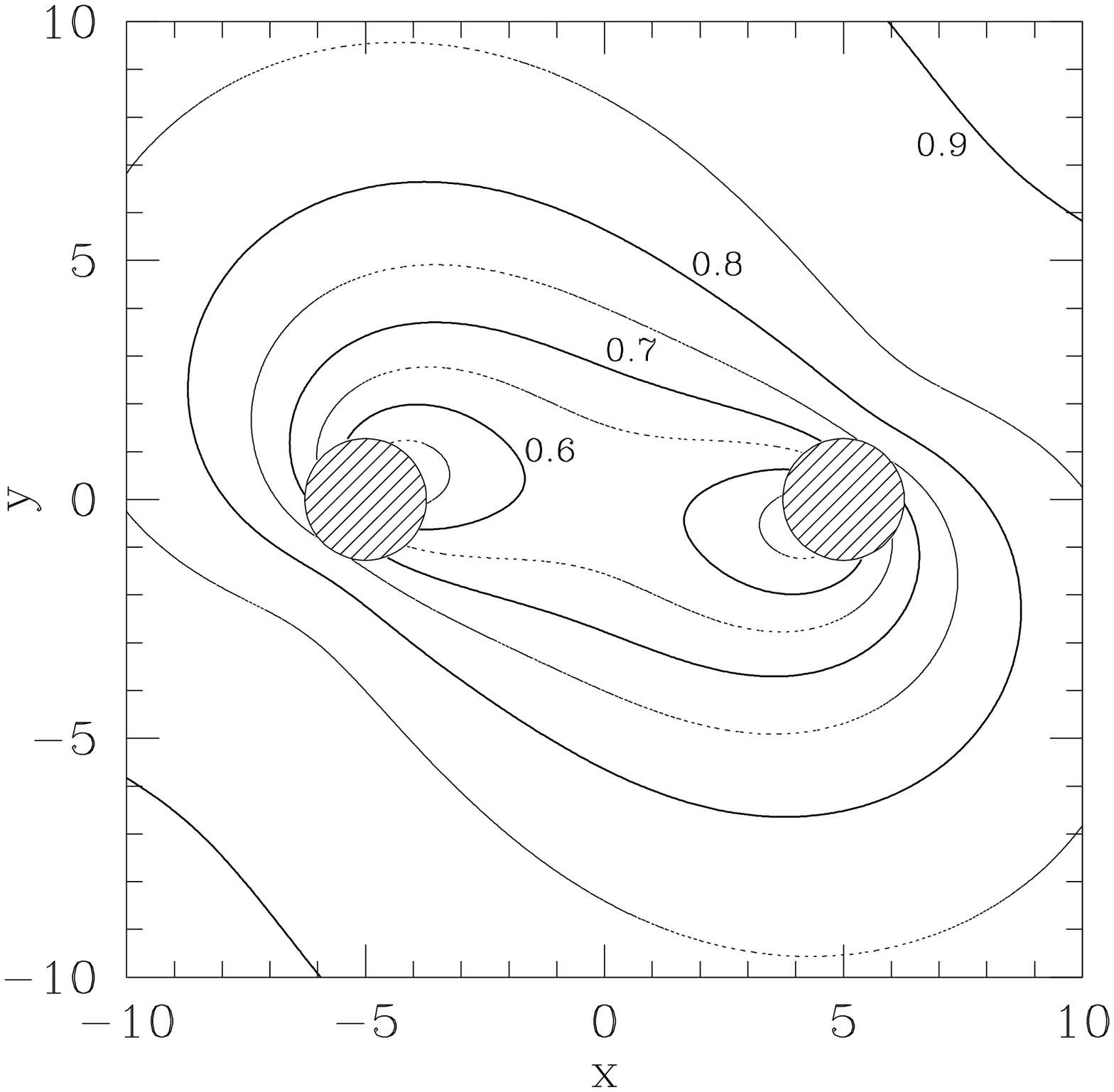}
\includegraphics[scale=0.36]{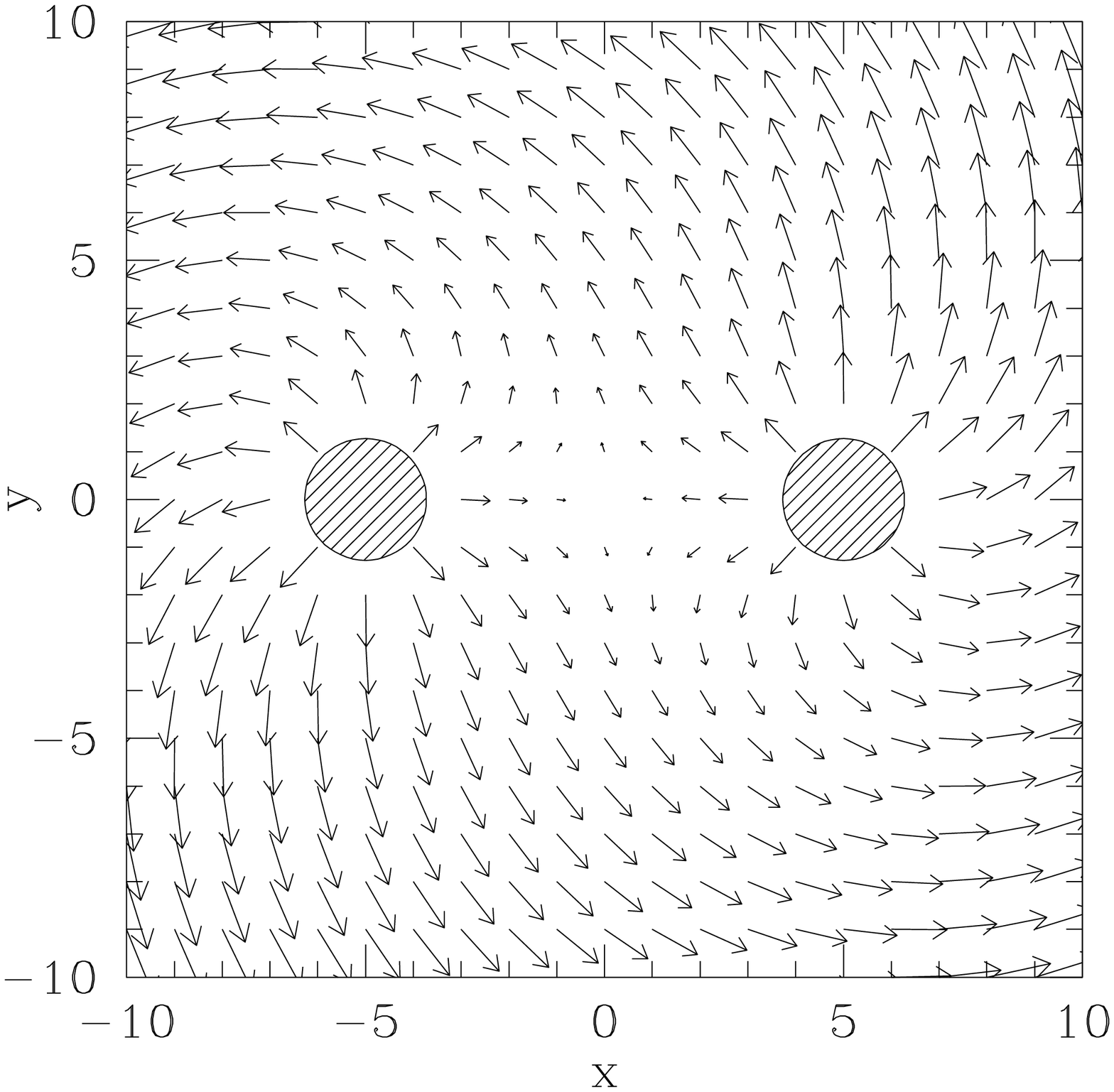}
\includegraphics[scale=0.36]{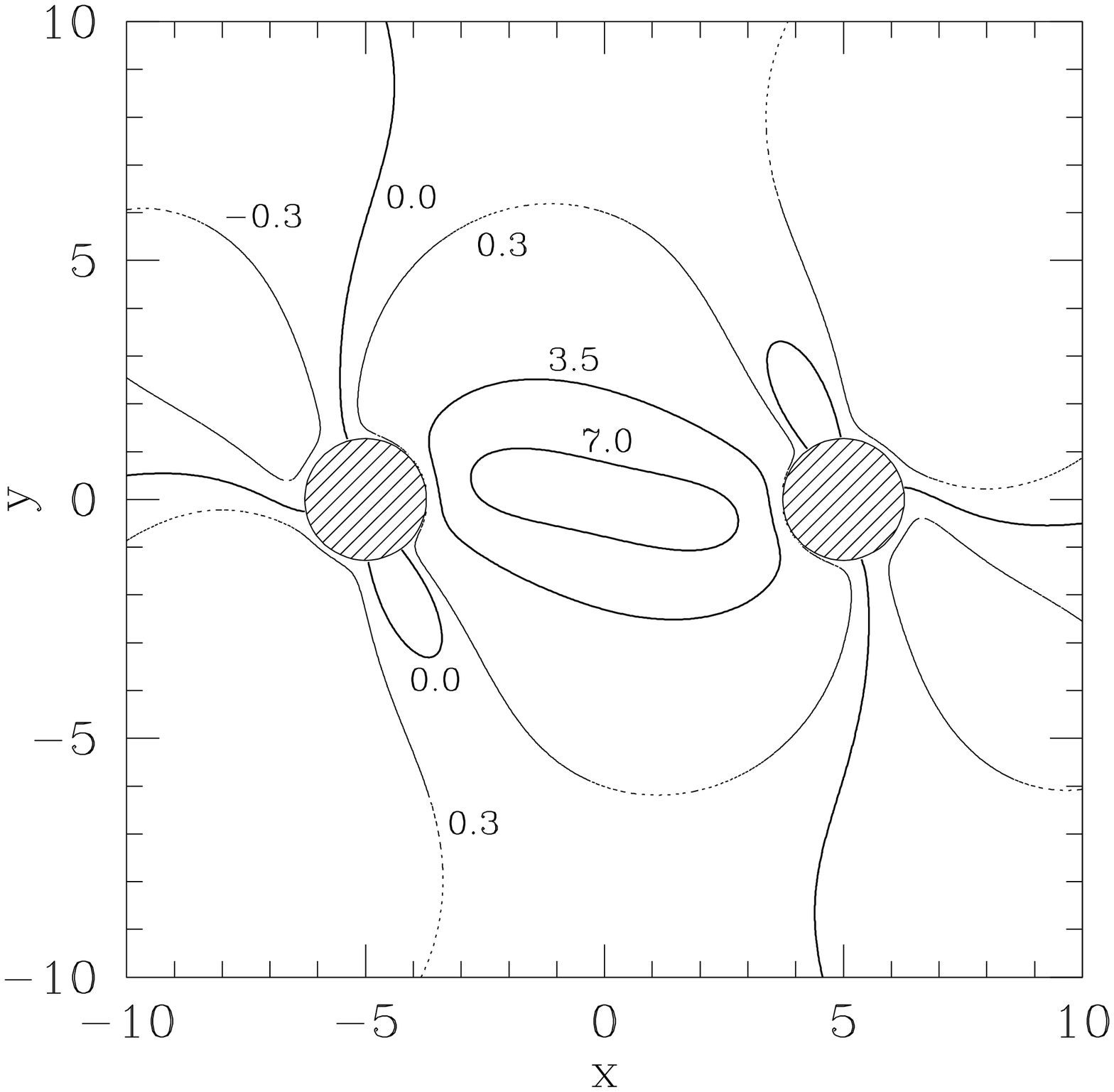}

\CAP{Quasi-equilibrium binary black hole initial data set on an
Eddington-Finkelstein slice}{\label{fig:EF_contours}Quasi-equilibrium
binary black hole initial data set on an {\em Eddington-Finkelstein
slice}.  From top right, counter-clockwise: lapse $\N$, 
conformal factor $\CF$, shift $\beta^i$ (in arbitrary units), and
$\partial_t\ln\CF$ (in units of $10^{-4}$). }
\end{figure}

Figure~\ref{fig:EF_contours} examines the solution with
$\Omega=\Omega_{EQ}$ in more detail.  The conformal factor around each
hole is spherically symmetric to a good approximation, whereas the
lapse is very asymmetric, varying on the horizon between $0.53$ and
$0.75$. Another interesting feature is that $\partial_t\ln\CF$ is more
than ten times bigger {\em between} the holes than anywhere else.

The results in this section confirm that it is feasible to solve the
quasi-equilibrium equations on a slice that is a superposition of two
Eddington-Finkelstein slices.  The condition $E_{ADM}=M_K$ indeed
singles out a unique $\Omega$.

\subsection{Maximal slices}
\label{sec:QE:BBH:MaxSlice}

We now repeat the computations of the last section for maximal slices.
Again, the quasi-equilibrium equations~(\ref{eq:QE}) are solved with
lapse boundary condition (\ref{eq:Neumann-lapse-BC}) for equal mass
black holes.  The conformal metric is {\em flat}, and the mean
curvature is $\trK=0$.

Table~\ref{tab:SingleHoleSolves} includes a single black hole solve
with these free data.  It shows that an excised sphere with radius
$r_{exc}\approx 1.2727$ leads to a black hole mass of $M_{AH}\approx
1.4808$.  Thus, $r_{exc}=1.27274103/1.4807928\approx 0.8594997$ in a
single hole solve will result in a black hole of unit
mass\footnote{One can show analytically, that $\partial_r(\N\CF)=0$
corresponds to $C=\frac{2}{3}\left(\sqrt{13}-1\right)$ in
Eqs.~(\ref{eq:MaximalSlice}).  Numerical integration of the coordinate
transformation to conformal flatness yields the same horizon radius as
we find here \cite{Cook:2003}.}.  For ease of comparing with the
previous section we therefore excise spheres with radius
$r_{exc}=0.8594997$.  Although the excised spheres here and for the
Eddington-Finkelstein slices considered earlier have the same
conformal separation $10$, the resulting initial data sets will
nonetheless have black holes at different {\em physical} separation.
For example, we will find that on maximal slices, the conformal factor
tends to be larger resulting in a larger separation.

\begin{figure}
\centerline{\includegraphics[scale=0.4]{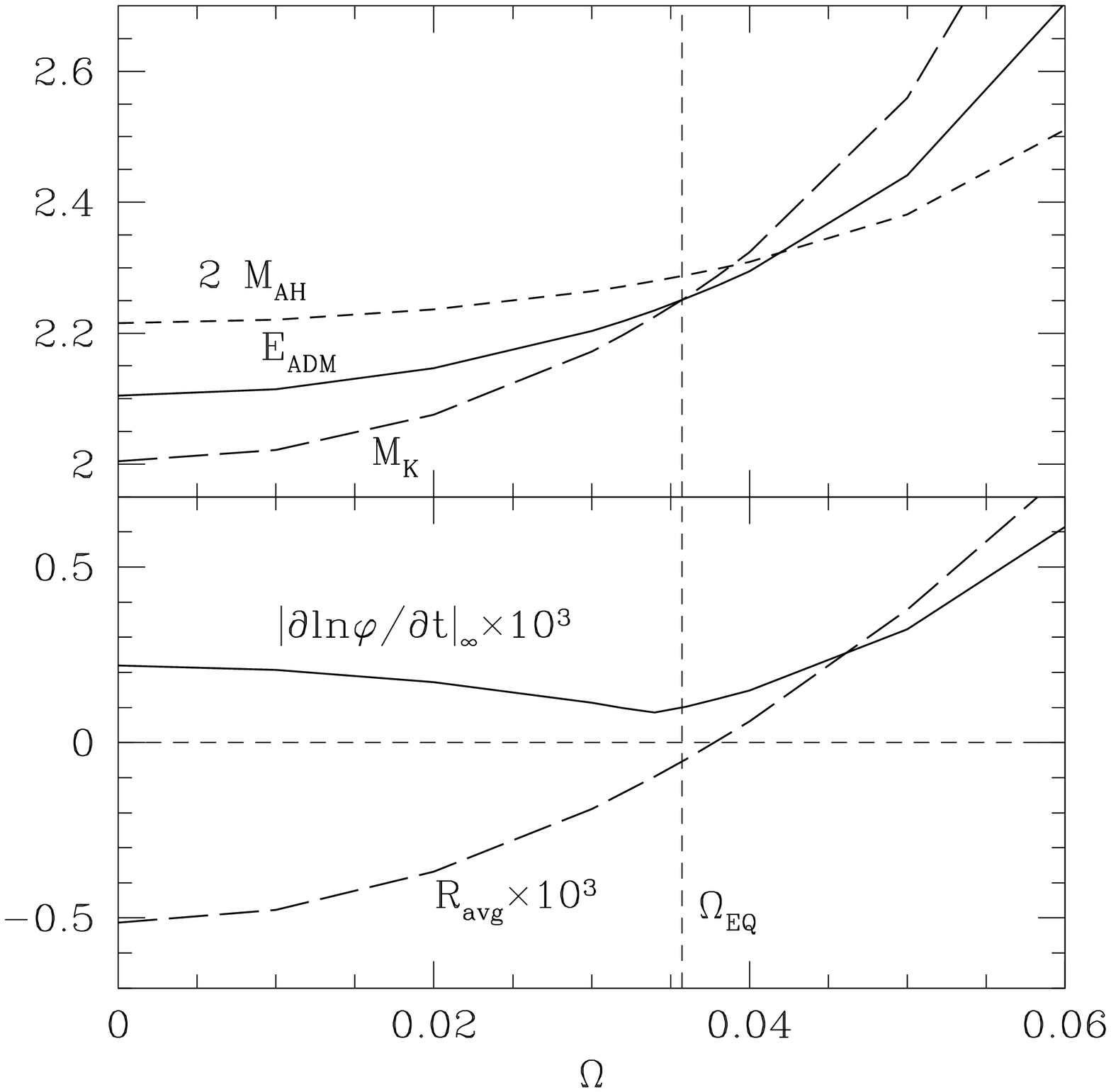}}
\CAP{Quasi-equilibrium binary black hole initial data sets on maximal
slices: Dependence on $\Omega$.}
{\label{fig:MaxSlice_vs_Omega}Quasi-equilibrium binary black hole
initial data sets on {\em maximal slices}.  Plotted are several
quantities as a function of the orbital angular velocity
$\Omega$. $R_{avg}$ denotes the residual of $\BC$ averaged over the
boundary $\S$.  Compare to Figure~\ref{fig:EF_vs_Omega}.}
\end{figure}

Figure~\ref{fig:MaxSlice_vs_Omega} presents several quantities as
functions of $\Omega$.  There is a unique $\Omega_{EQ}\approx
0.03572$, such that $E_{ADM}=M_K$. We also note that the average value
of the lapse condition $\BC$ passes though zero close to
$\Omega_{EQ}$, and $\partial_t\ln\CF$ is closest to zero around
$\Omega_{EQ}$, too.  These findings again strongly indicate that
$E_{ADM}=M_K$ is a reasonable condition to select $\Omega$.

\begin{figure}

\includegraphics[scale=0.36]{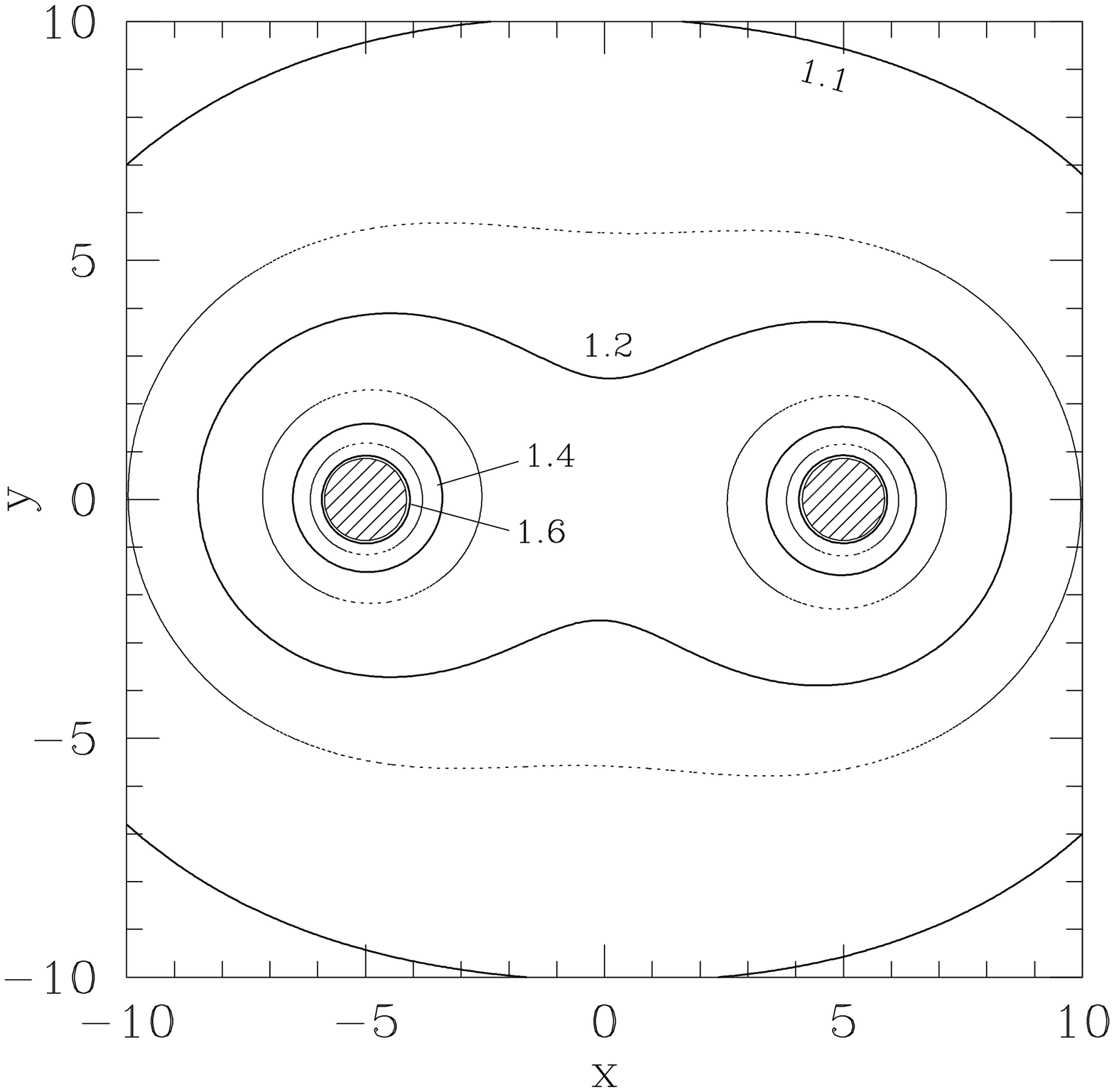}
\includegraphics[scale=0.36]{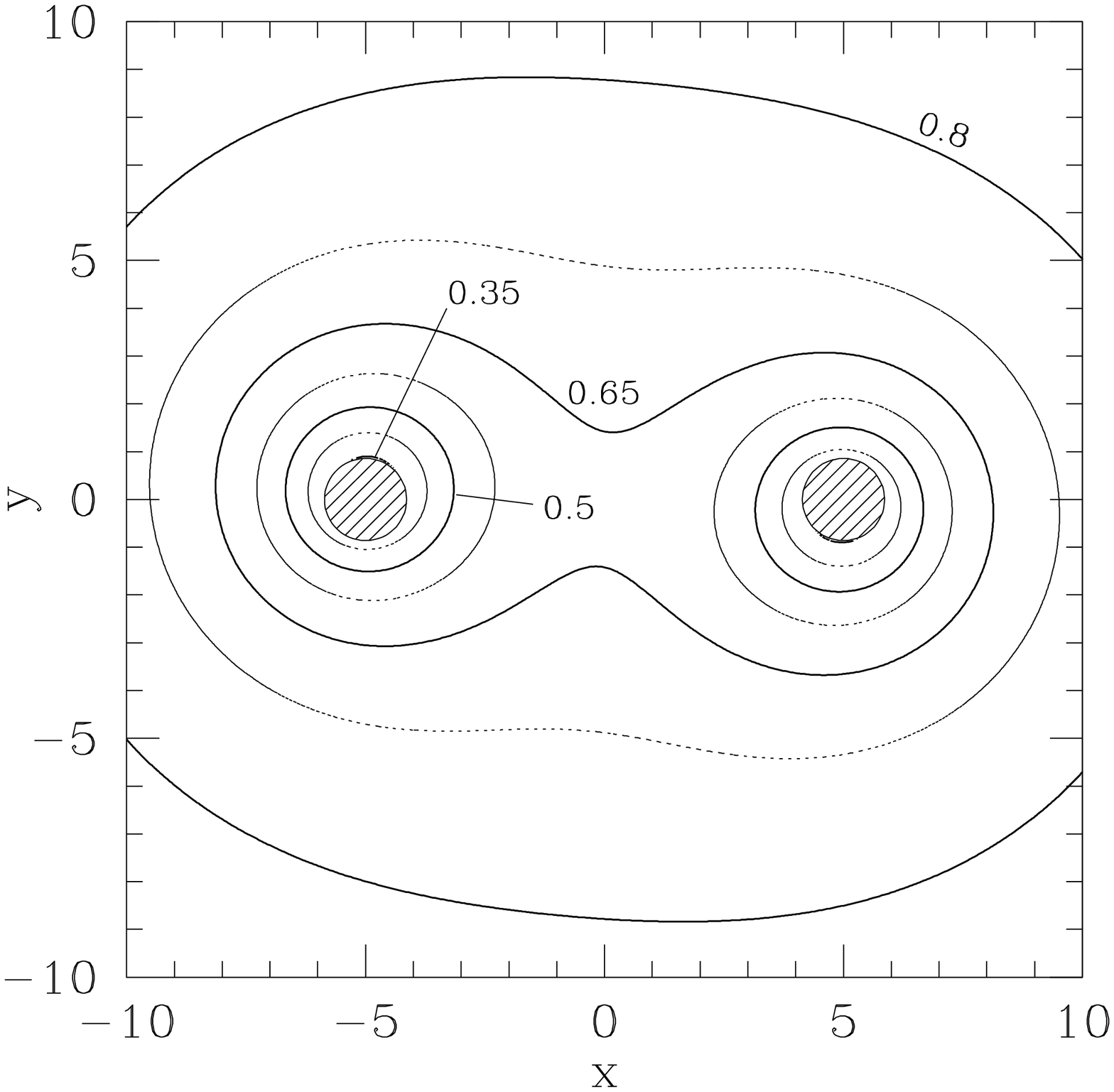}
\includegraphics[scale=0.36]{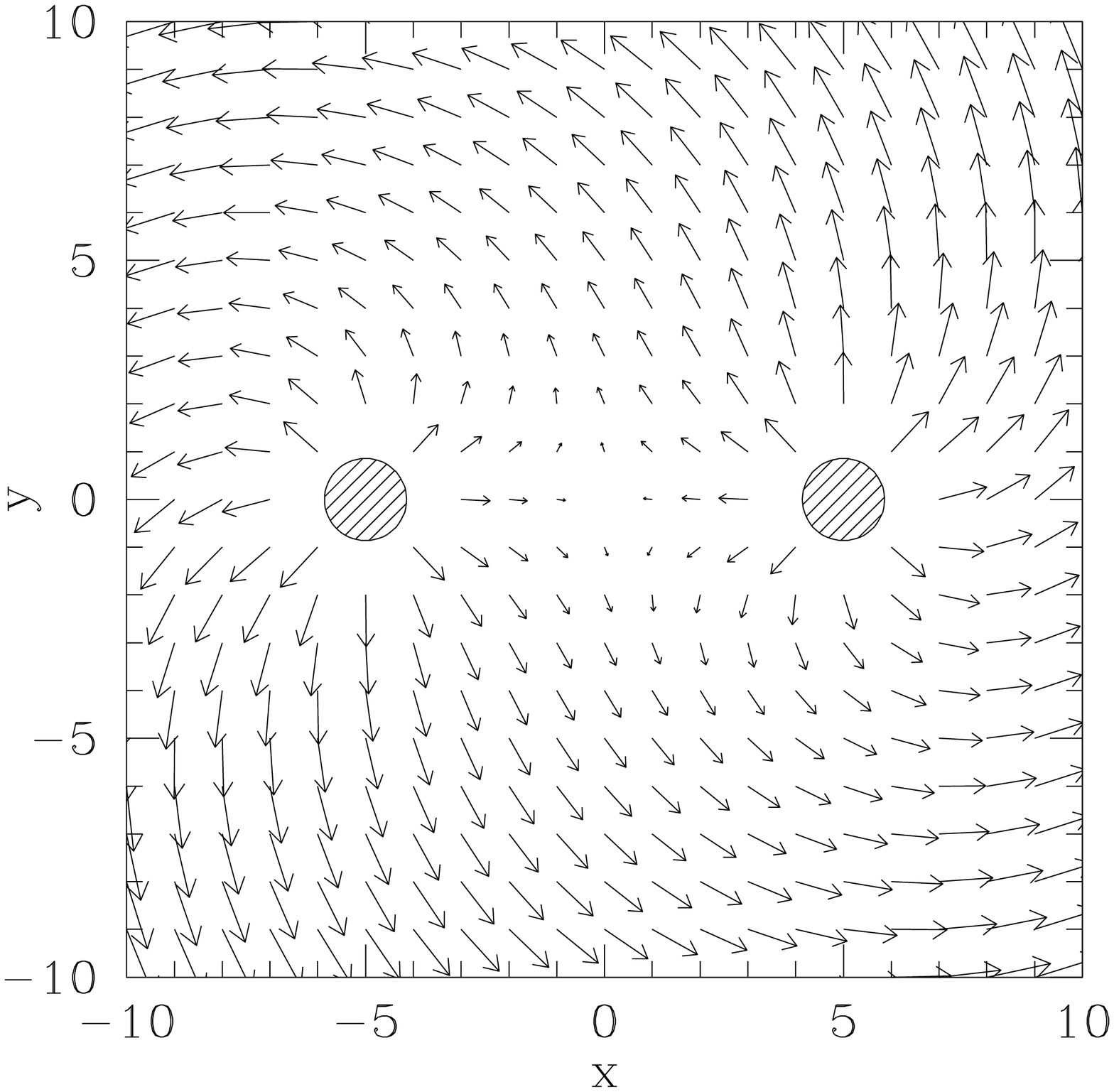}
\includegraphics[scale=0.36]{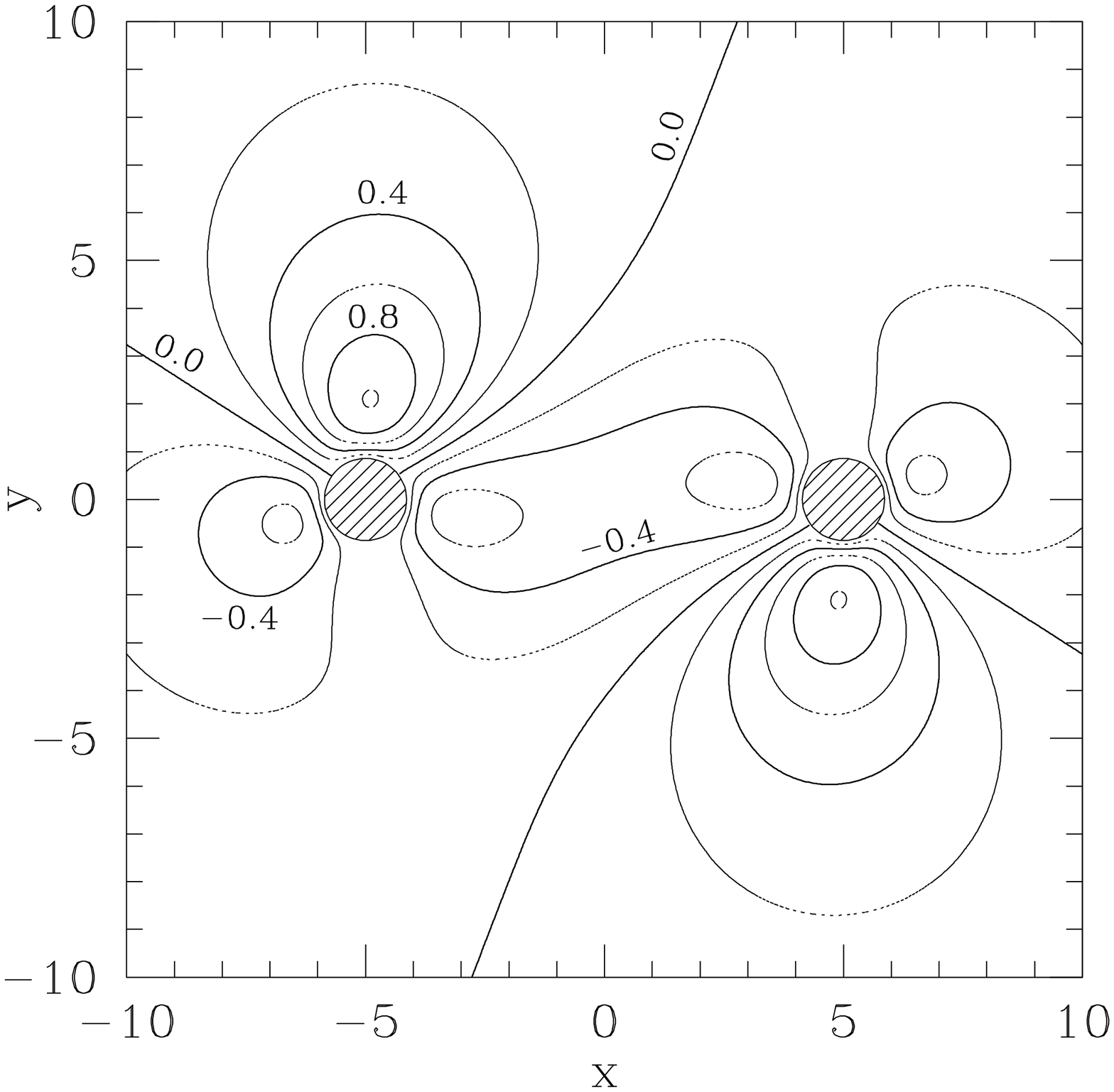}

\CAP{Quasi-equilibrium binary black hole initial data set on an
maximal slice}{\label{fig:MS_contours}Quasi-equilibrium binary black
hole initial data set on a {\em maximal slice}.  From top right,
counter-clockwise: lapse $\N$, conformal factor $\CF$, shift $\beta^i$
(in arbitrary units), and $\partial_t\ln\CF$ (in units of
$10^{-4}$). Compare to Figure~\ref{fig:EF_contours}.}
\end{figure}

Figure~\ref{fig:MS_contours} presents cuts through the data set with
$\Omega=\Omega_{EQ}$.  The lapse $\N$ varies between $0.345$ and
$0.395$ on the horizon, a smaller variation than for the
Eddington-Finkelstein slice.  The time-derivative of the conformal
factor is approximately one order of magnitude smaller than for the
Eddington-Finkelstein slice (Figure~\ref{fig:EF_contours}) and
exhibits more complex behavior: Along the axis connecting the black
holes, $\partial_t\ln\CF$ is negative, whereas in the wake of each
hole, it is positive.  As a function of $\Omega$, the extrema of
$\partial_t\ln\CF$ change in a complicated way: As $\Omega$ increases,
the minimum between the holes and the peaks behind the holes increase
whereas the minima outside the holes on the $x$-axis decrease.  All
five extrema change sign at angular frequencies close to
$\Omega_{EQ}$.

\subsection[Toward the test mass limit]{Toward the test mass limit}

We now attempt to compute initial data sets for unequal mass black
holes.  From the computational domain, we excise two spheres with
radii $r_1$ and $r_2$, centered on the x-axis at location $x=\pm s/2$,
$s$ being the separation between the centers of the excised spheres.
$r_1$ is fixed depending on $\trK$ such that the resulting hole has
approximately unit-mass (e.g. $r_1=\hat r_{AH}$ for $\trK$ based on
Eddington-Finkelstein, cf. Eq.~(\ref{eq:r-AH-EF})).

We place the center of the rotating frame at $\vec R=(R_x, R_y,0)$,
and choose the orbital angular frequency along the $z$-axis,
$\vec\Omega=\Omega\hat e_z$.  As before, $\Omega$ will be determined by
the requirement
\begin{subequations}
\begin{equation}\label{eq:EADM=MK}
E_{ADM}=M_K.
\end{equation}
To fix $R_x$ and $R_y$, we require that the linear ADM momentum vanishes,
\begin{equation}
P_{ADM}^i=0.
\end{equation}
Finally, we choose the radius of the second hole such that the ratio of
the apparent horizon masses equals a specified mass-ratio $X$,
\begin{equation}\label{eq:M1oM2=X}
\frac{M_1}{M_2}=\frac{\sqrt{A_1/16\pi}}{\sqrt{A_2/16\pi}}=X.
\end{equation}
\end{subequations}
(In the present exploratory work, we do not take into account the
black hole spins in defining $M_{1,2}$ for corotating configurations.)
This procedure leads to four-dimensional rootfinding in $\Omega,
r_2/r_1, R_x$, and $R_y$ which is performed with Broyden's
method~\cite{NumericalRecipes}.

We take $s=10$ throughout this section, and perform solves (including
rootfinding to satisfy Eqs.~(\ref{eq:EADM=MK})--(\ref{eq:M1oM2=X}) )
for $X=1, 2, 4, 8, 16$.  We explore four families of solutions, with
two different slicings (Eddington-Finkelstein slicing and maximal
slicing) and corotating black holes as well as irrotational black
holes.  In the latter case,
$\beta^i_\parallel=\Omega(\partial/\partial\phi)^i$ on each excised
sphere, with the axis of rotation centered on the respective hole.

\begin{figure}
\begin{centering}
\includegraphics[scale=0.3]{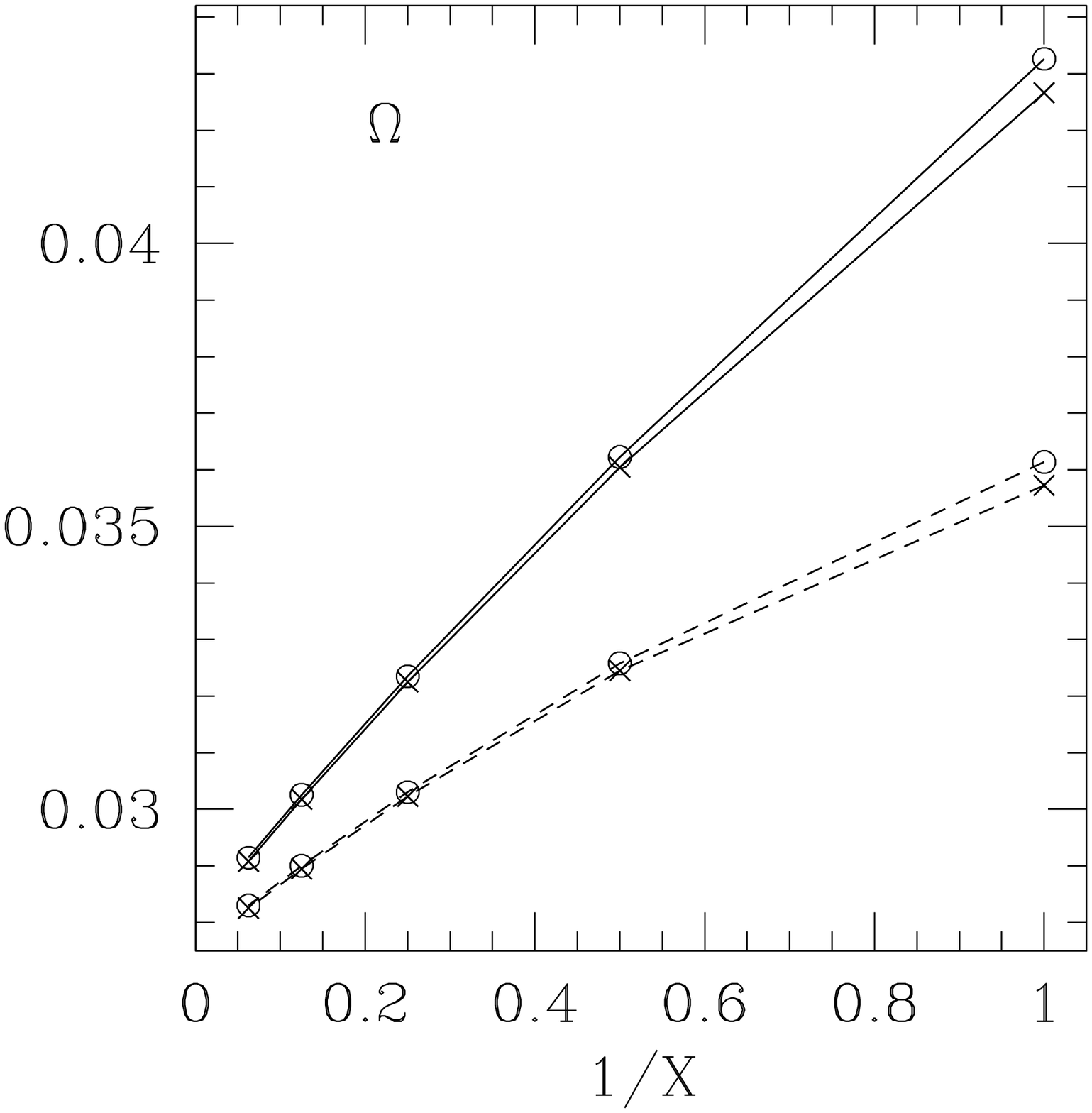}
\includegraphics[scale=0.3]{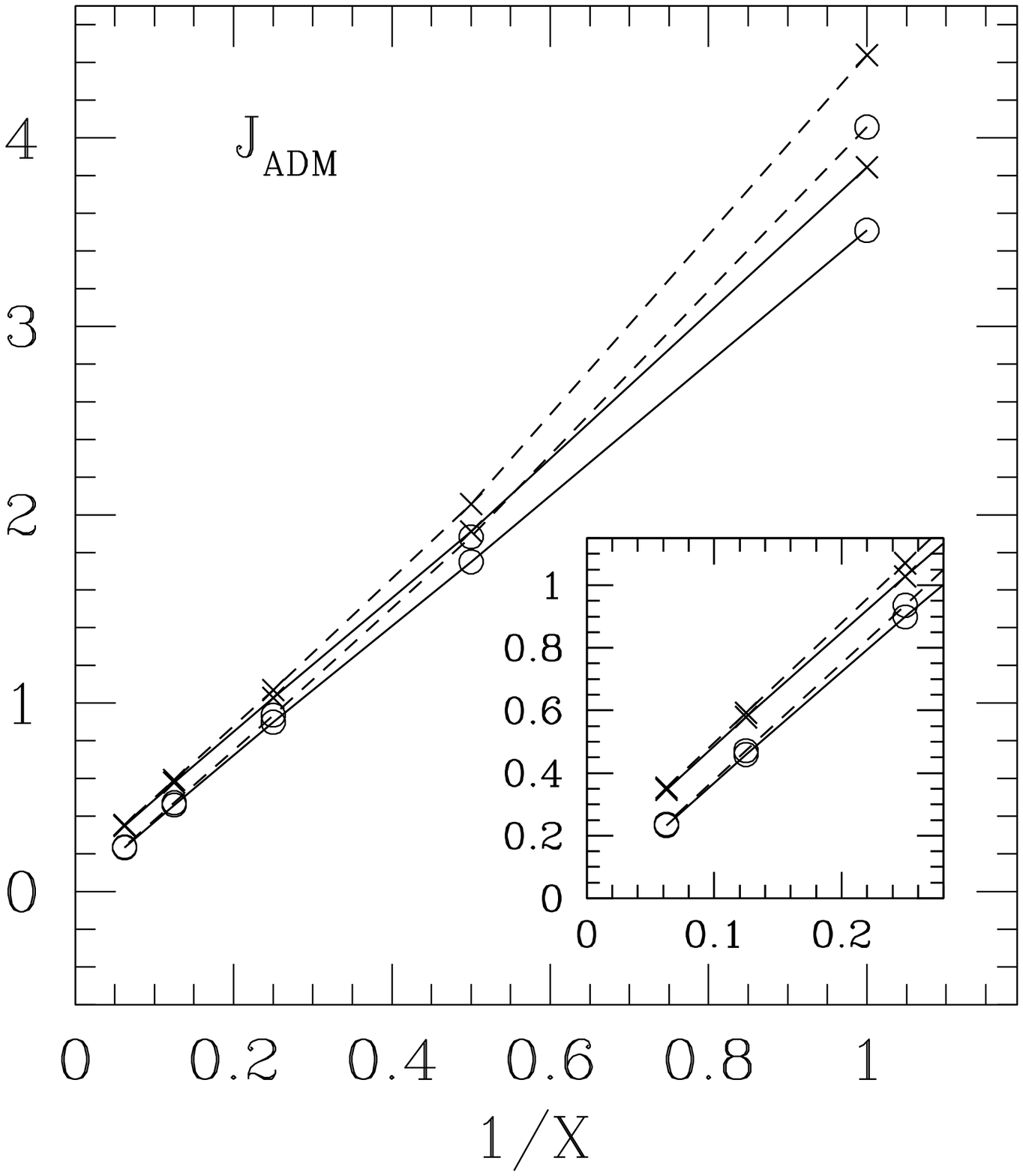}

\includegraphics[scale=0.3]{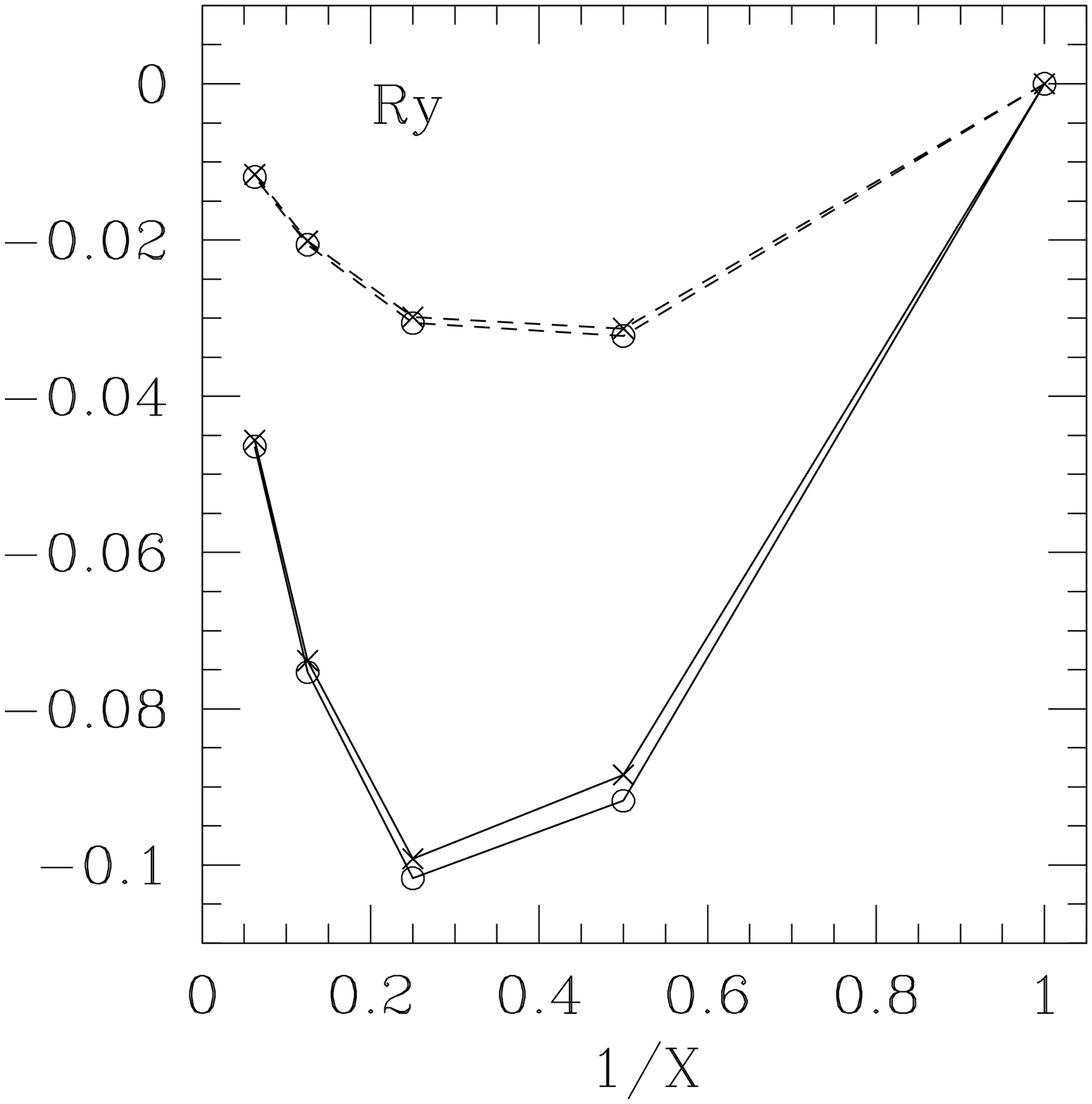}
\includegraphics[scale=0.3]{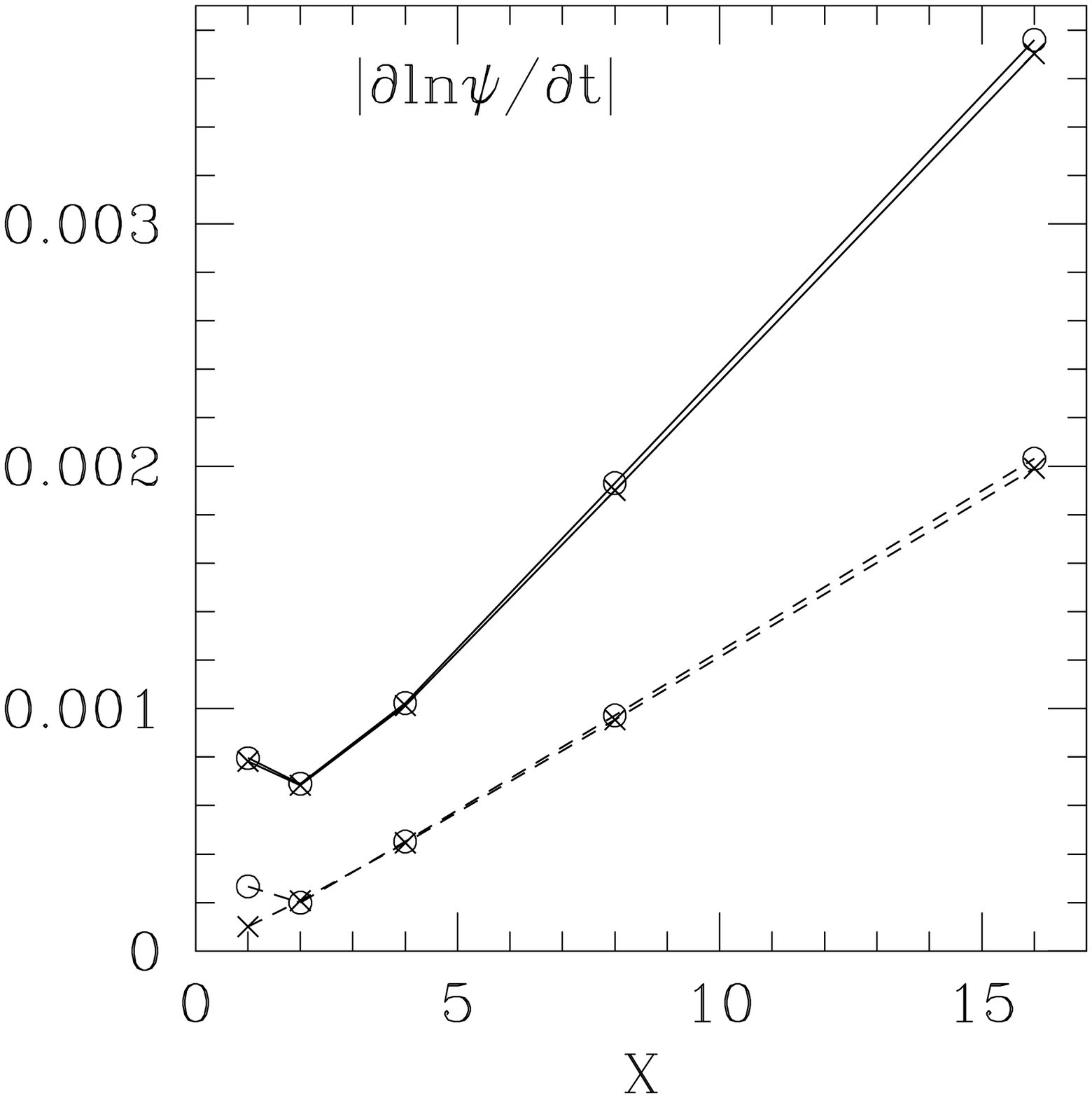}

\includegraphics[scale=0.3]{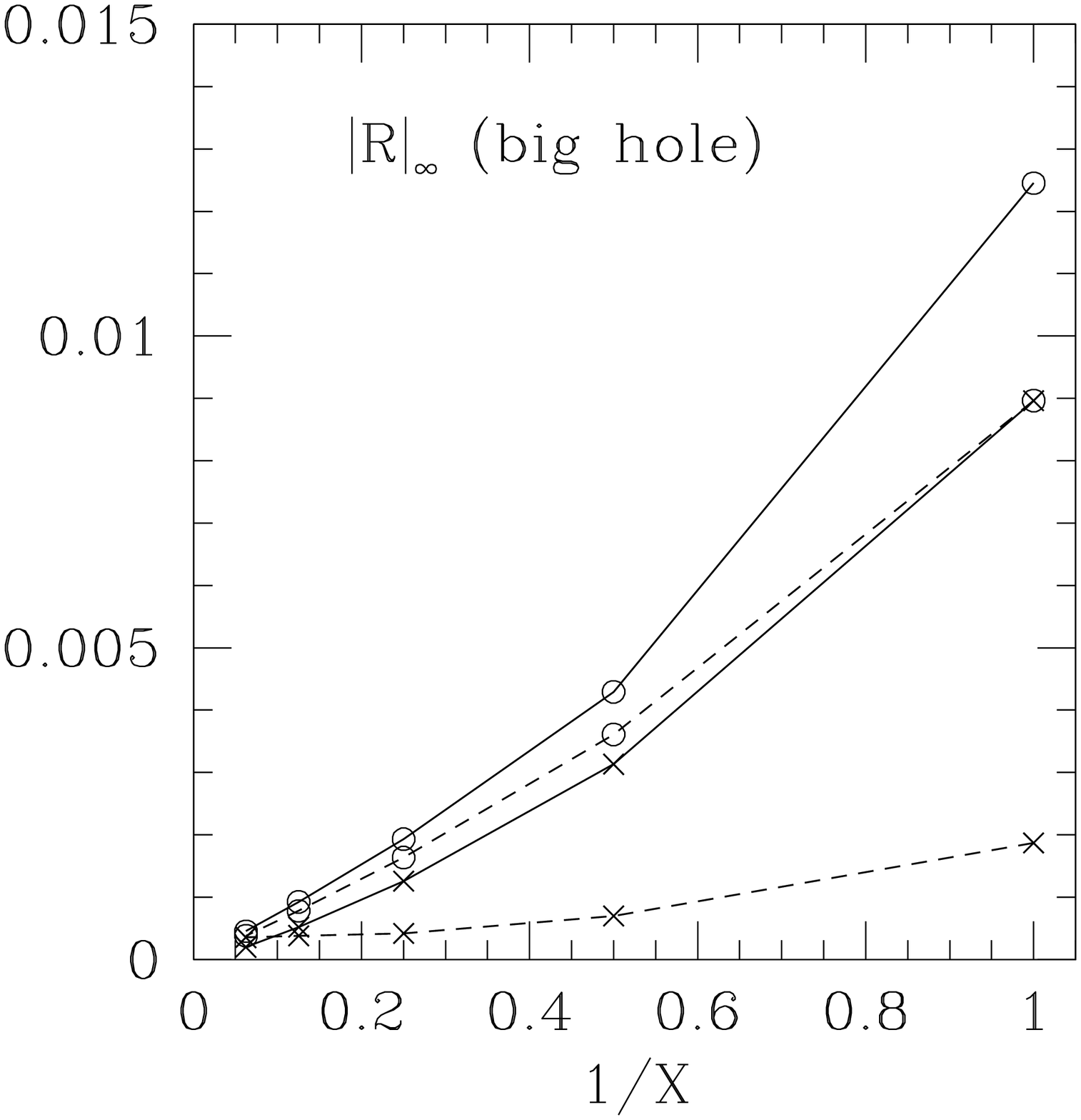}
\includegraphics[scale=0.3]{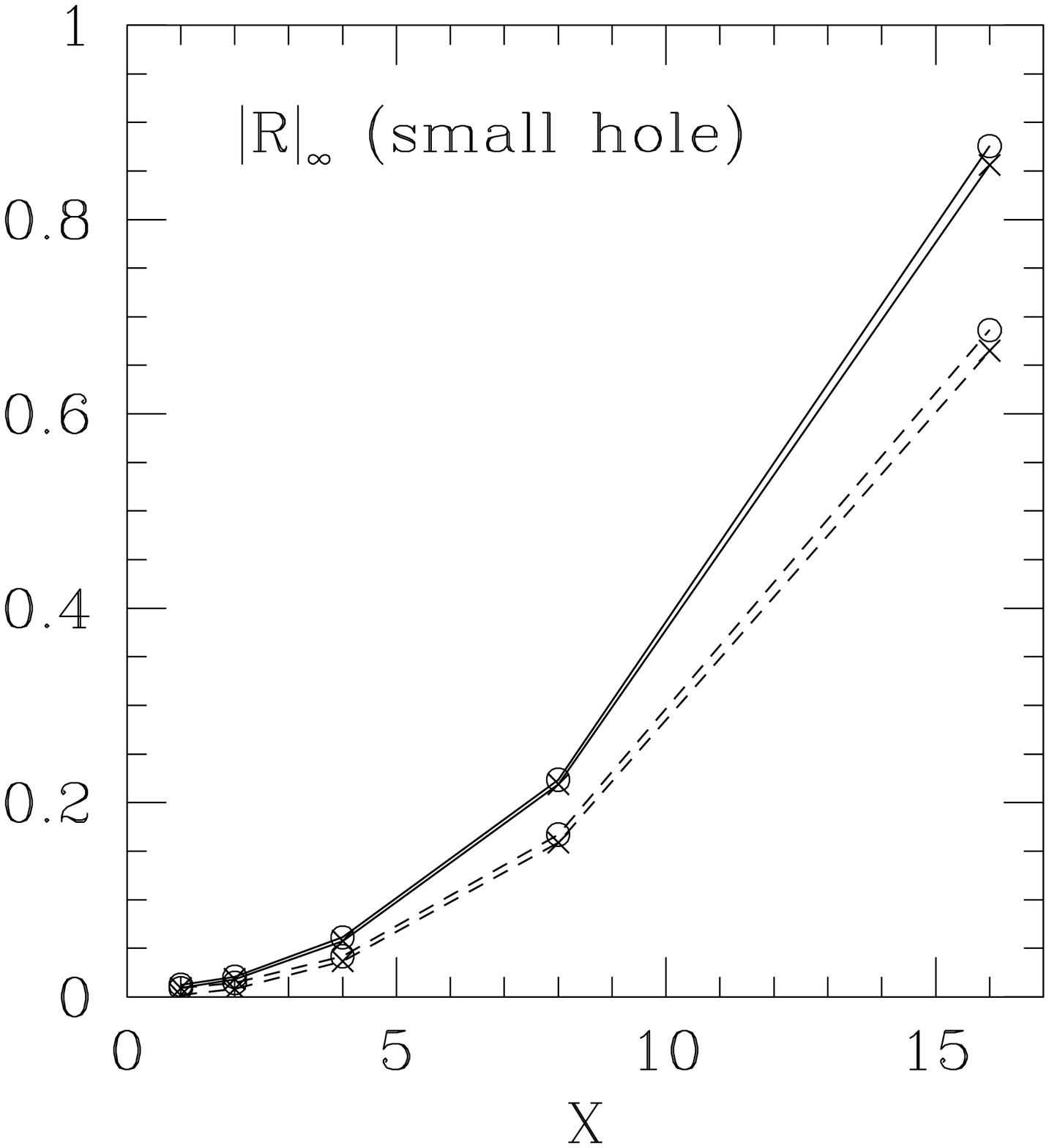}

\end{centering}
\CAP{Quasi-equilibrium method in the test-mass limit: Various
quantities as a function of the
mass-ratio}{\label{fig:QEtestmass}Quasi-equilibrium method in the
test-mass limit: Various quantities as a function of the mass-ratio.
Solid lines denote Eddington-Finkelstein slices, dashed lines maximal
slices.  Crosses stand for corotation, circles for irrotation.}
\end{figure}

Figure~\ref{fig:QEtestmass} presents results as a function of $X$.  We
see that, as $X\to\infty$, the orbital angular frequency $\Omega$
approaches a constant value for each of the four families (the
families will lead to different values in the limit).  The angular
momentum appears to be approximately linear in the smaller mass, as it
should be.  The enlargement of the $J_{ADM}$--plot shows that at
large mass-ratios for {\em irrotational} families, $J_{ADM}$ is
proportional to $1/X$ as expected.  For {\em corotating} families,
$J_{ADM}$ has an additional contribution due to the intrinsic spin of
the large black hole.  We also find that the residual $\BC$ of the
lapse condition on the {\em large} black hole tends to zero as
$X\to\infty$.  These findings are so far consistent with the
expectation that for $X\to\infty$, we should recover a spacetime with
an exact helical Killing vector.

\begin{figure}
\begin{centering}
\includegraphics[scale=0.36]{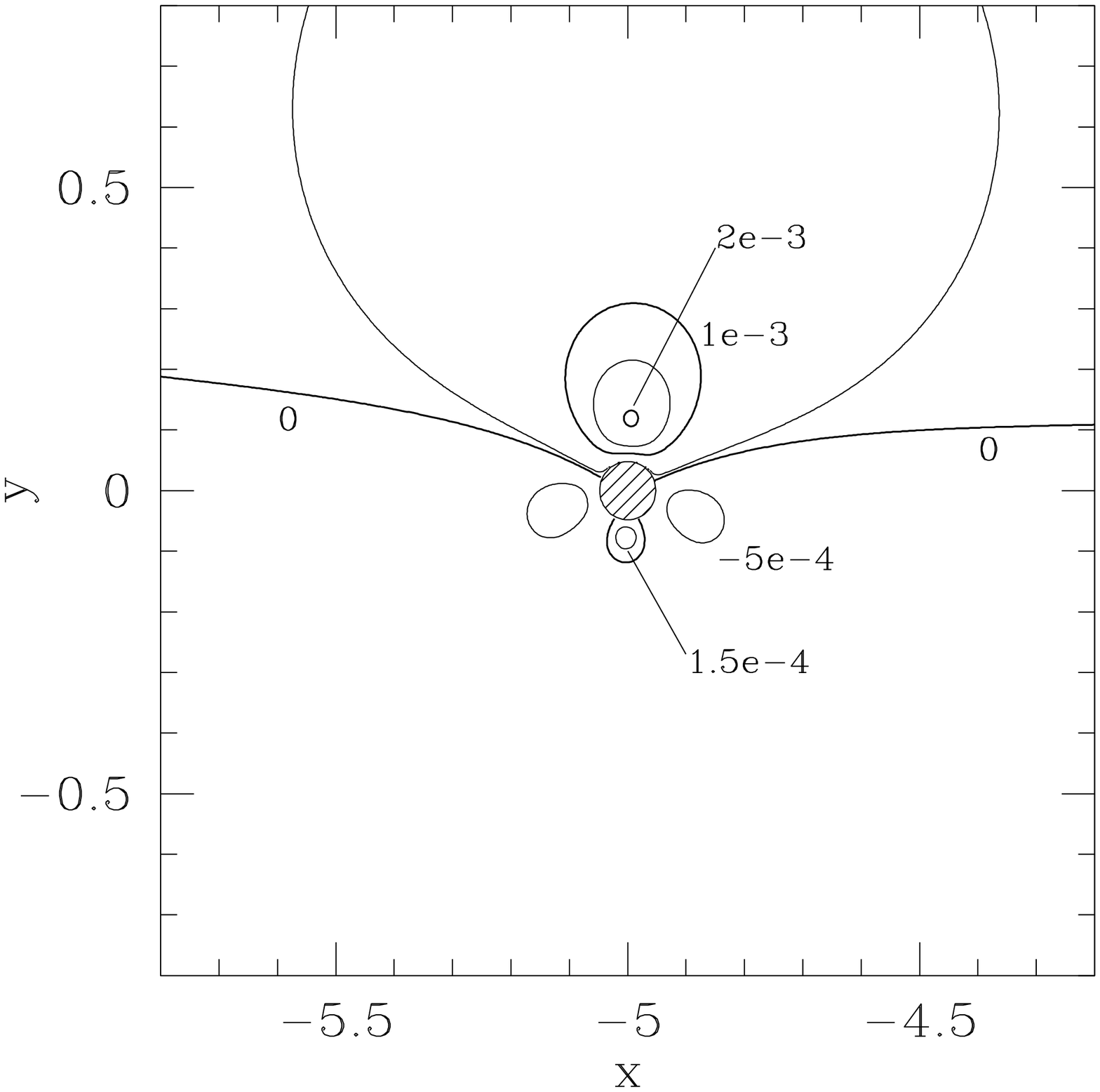}
\includegraphics[scale=0.36]{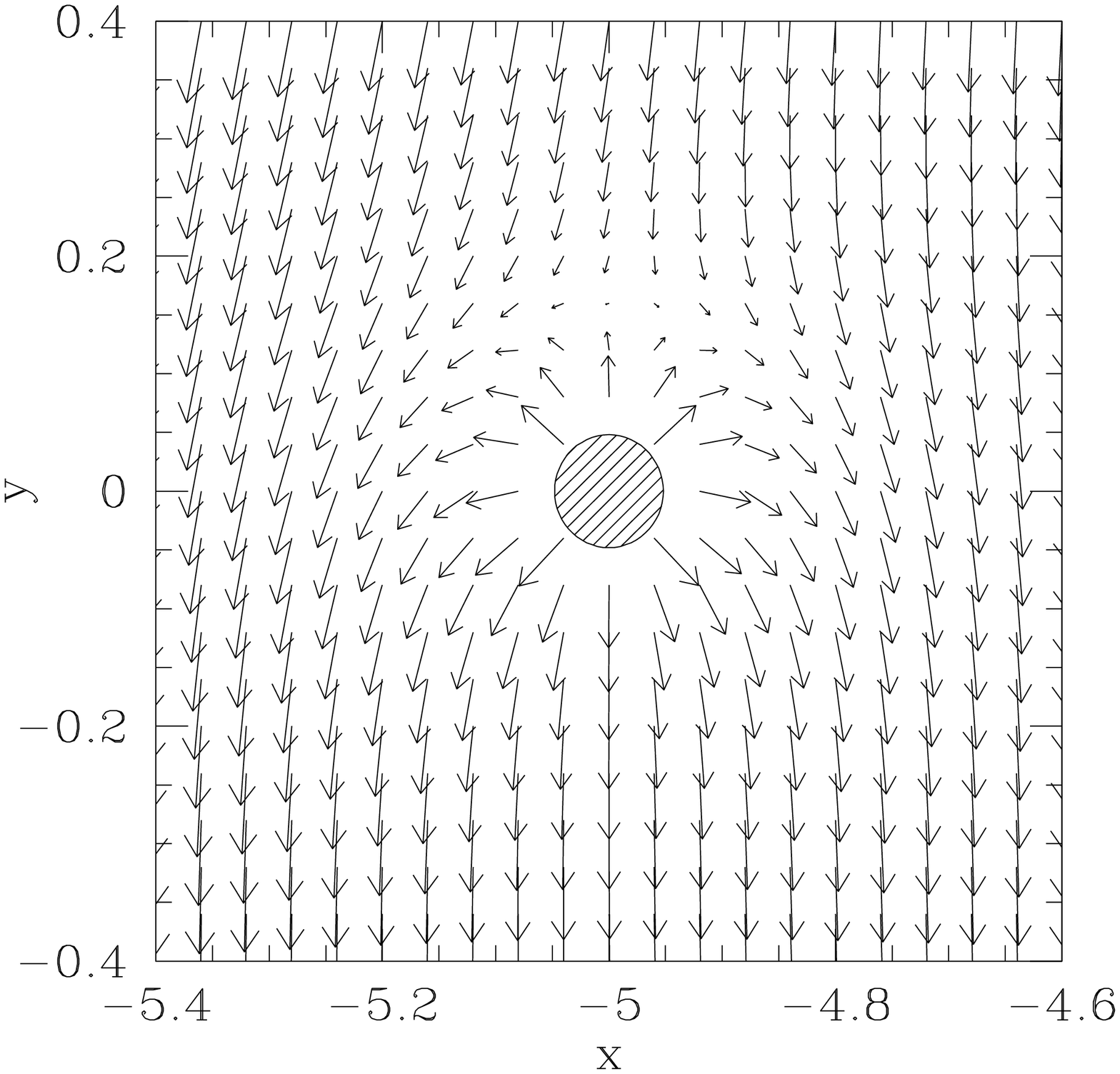}

\includegraphics[scale=0.36]{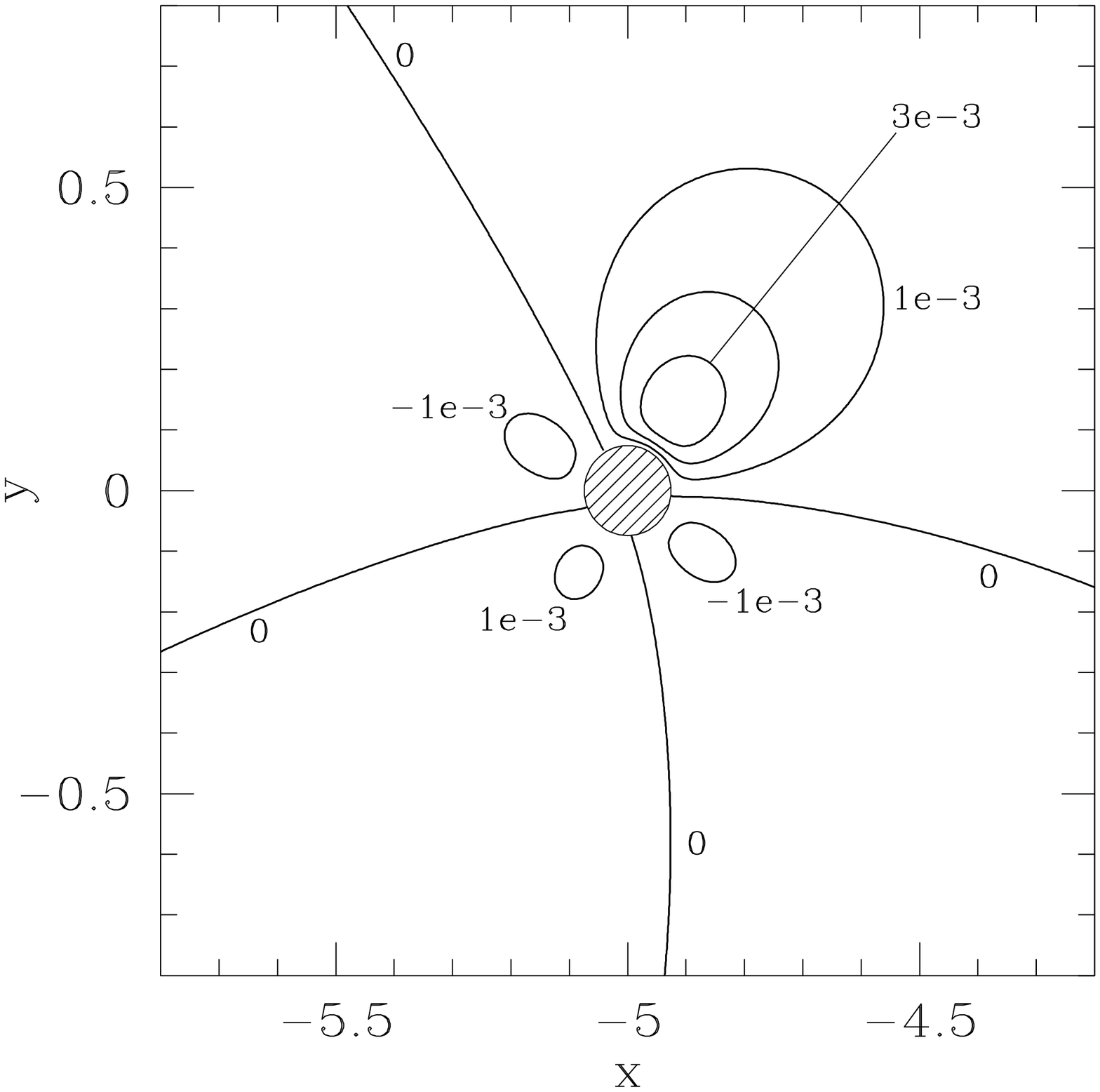}
\includegraphics[scale=0.36]{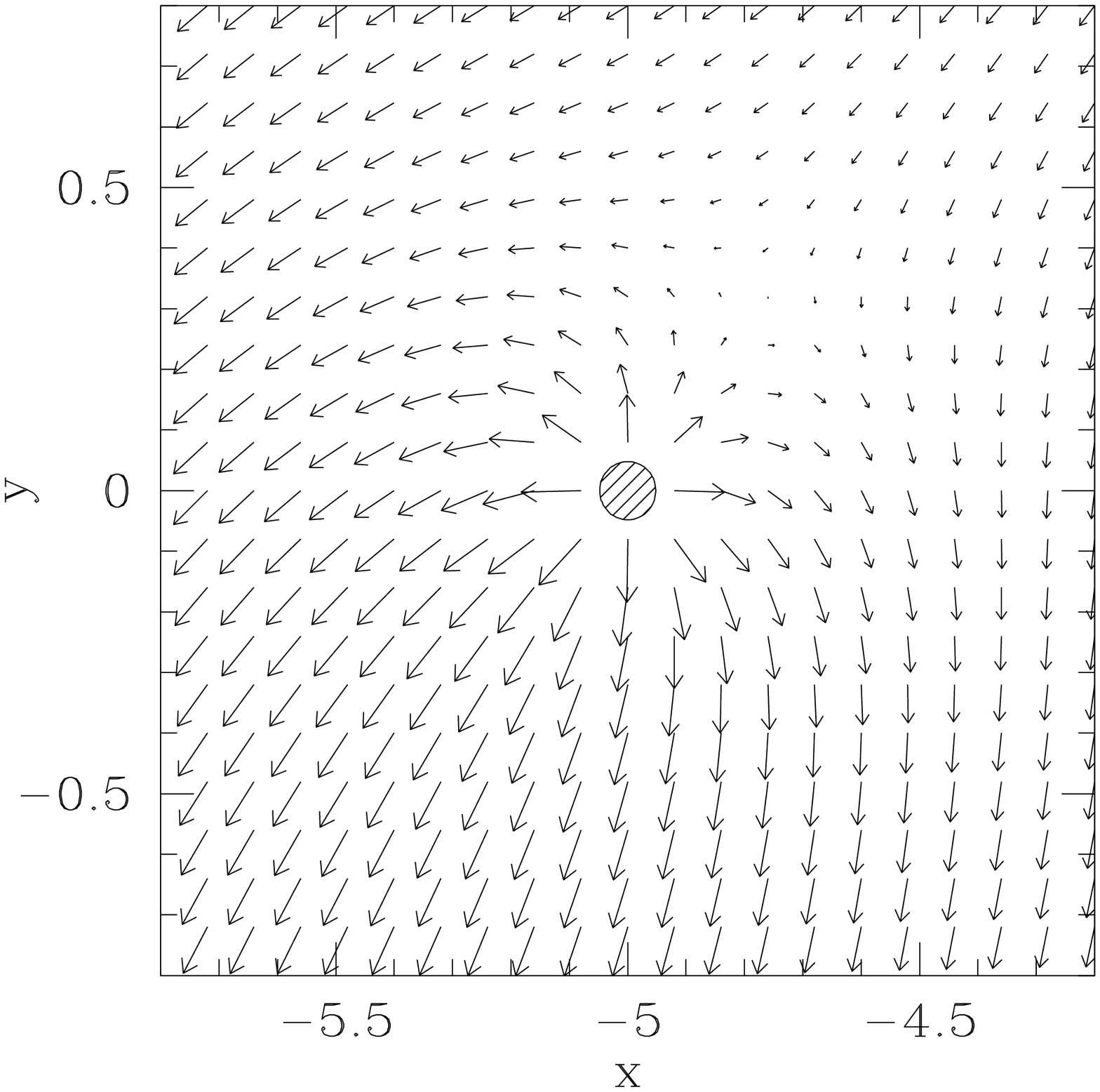}

\end{centering}
\CAP{$\partial_t\ln\CF$ and shift $\beta^i$ for initial data sets with
mass-ratio 16.}{\label{fig:X=16}$\partial_t\ln\CF$ (left plots) and
shift $\beta^i$ (right plots) for the $X\!=\!16$ initial data sets on
a maximal slice (top plots) and Eddington-Finkelstein slice (bottom
plots).  Only a small region in the $xy$ plane around the small black
hole is shown.  The upper right plot has a different scale.}
\end{figure}

However, two more indicators of time-independence fail: The maximum of
$\partial_t\ln\CF$ seems to increase linearly with $X$, and the
residual of $\BC$ on the small black hole appears to diverge even
faster with $X$, probably quadratically.  These results do not depend 
on the choice of slicing, or on the choice between corotation or
irrotation.  The maximum of $\partial_t\ln\CF$ occurs very close to
the small black hole as Figure~\ref{fig:X=16} confirms.

The behavior of $\partial_t\ln\CF$ in the vicinity of the small black
hole seems to be self-similar: Scaled to the radius $r_2$ of the small
black hole, the shape of $\partial_t\ln\CF$ appears to be independent
of $r_2$ for sufficiently large mass-ratio, while the amplitude grows
linearly with $X$.  So far we have not scaled the time-coordinate in
$\partial_t\ln\CF$.  The large black hole has approximately unit mass
in all solutions, so that $X=M_1/M_2\approx 1/M_2$, from which we see
that the conformal factor $\CF$ changes on the time-scale of the small
black hole.

Figure~\ref{fig:X=16} also presents plots of the shift close to the
small black hole.  At the horizon, the shift points radially outward,
consistent with the boundary condition (\ref{eq:BC-shift}), whereas
far away from the small hole, the Keplerian shift of the corotating
frame dominates.  The cross-over between these two regimes depends on
the slicing: For Eddington-Finkelstein slices, the shift of the hole
is larger than for maximal slices, hence the cross-over occurs at a
larger distance.  We also note that the maximum of $\partial_t\ln\CF$ 
seems to be in the vicinity of the stagnation point of the shift. 

The apparent self-similarity of $\partial_t\ln\CF$ for $X\to\infty$,
suggests that the small black hole influences the overall solution
only in a small region around it; one can think of the small black
hole as ``immersed'' in the ``background'' generated from the large
black hole.
If we place a sphere $S$ around the small black hole with radius $r$
satisfying $M_2\ll r\ll s$, then on this surface the solution $(\CF,
\beta^i, \N)$ of the conformal thin sandwich equations will be the
solution of the large hole alone, i.e. it will not depend on $M_2$ (as
long as $M_2\ll r$ holds).

The values $(\CF|_S, \beta^i|_S,N|_S)$ constitute now {\em outer
boundary conditions} for the local solution around the small black
hole, which thus can be interpreted as a quasi-equilibrium solution
for {\em one} black hole.  However, in contrast to the single hole
solutions in section \ref{sec:QE:SingleBlackHoles}, the outer boundary
condition on the shift $\beta^i$ is here {\em not} zero, but the
Keplerian shift of the large black hole.  (The conformal factor
$\CF|_S$ and lapse $\N|_S$ will also slightly differ from $1$,
however, an overall shift in those quantities corresponds merely to an
overall scaling of the spatial coordinates and the time-coordinate,
respectively.)  As the outer boundary is far away ($M_2\ll
r\approx\infty$), the solution scales with the size $M_2$ of the small
hole, explaining why $\partial_t\ln\CF$ is self-similar and changes on
the time-scale of $M_2$.  The lapse condition (\ref{eq:BC-lapse}) is
essentially the time-derivative of the ingoing expansion.  The ingoing
expansion is proportional to $1/M_2$, and the time-derivative scales
as $1/M_2$, too, thus $\BC\propto M_2^{-2}$, as observed in
Figure~\ref{fig:QEtestmass}.

\begin{figure}
\begin{centering}
\includegraphics[scale=0.36]{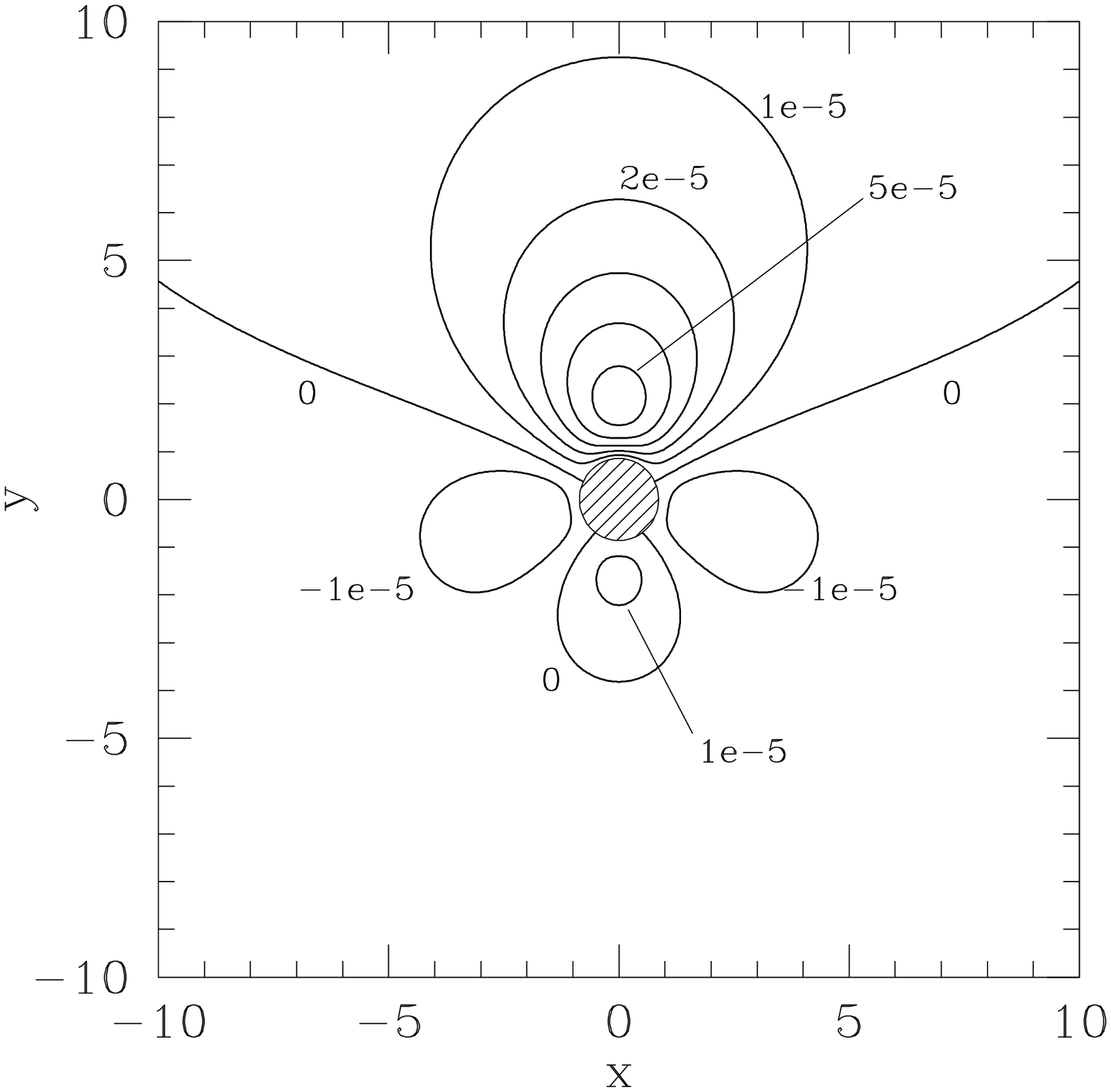}
\includegraphics[scale=0.36]{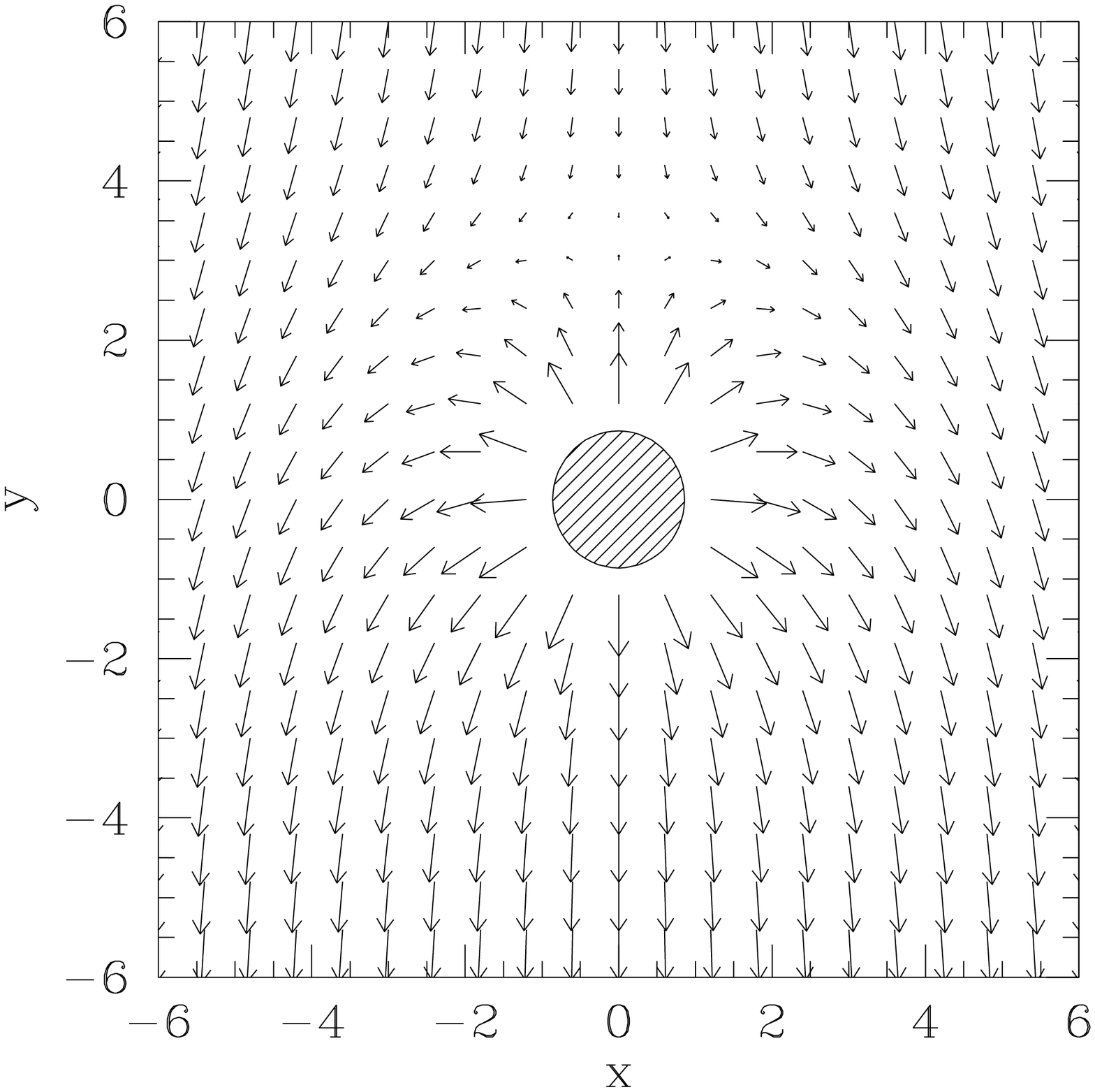}

\end{centering}
\CAP{$\partial_t\ln\CF$ and shift $\beta^i$ for a single boosted black
hole.}  {\label{fig:QEtestmass-SingleHole}$\partial_t\ln\CF$ (left)
and shift $\beta^i$ (right) for a single black hole with outer
boundary conditions (\ref{eq:QEtestmass-outerBC}).  Compare to the
upper panels of Figure~\ref{fig:X=16}.}
\end{figure}

We confirm this picture by performing a single black hole solve that
mimics the situation around the small black hole: We solve the
quasi-equilibrium equations (\ref{eq:QE}) on a maximal slice with
inner boundary condition on the lapse $\partial_r(\N\CF)=0$, and outer
boundary conditions
\begin{equation}
\label{eq:QEtestmass-outerBC}
\CF=1,\qquad \beta^i=(0, -0.22, 0),\qquad \N=1, \qquad\mbox{as $r\to\infty$}.
\end{equation}
The particular value for $\beta^i$ (in Cartesian coordinates) is the
ambient Keplerian shift in the top-right panel of
Figure~\ref{fig:X=16}.  The radius of the excised sphere is taken to
be $0.8594997$, appropriate for unit-mass if the outer shift boundary
condition were $\beta^i\to 0$.  Note that this solve differs from the
spherical symmetric single black hole solves of section
\ref{sec:QE:SingleBlackHoles} only in the outer boundary condition of
the shift.  Figure~\ref{fig:QEtestmass-SingleHole} presents
$\partial_t\ln\CF$ and shift for this solve; it is strikingly similar
to the binary black hole solution in Figure~\ref{fig:X=16}.

In essence, a quasi-equilibrium solve with outer boundary conditions
(\ref{eq:QEtestmass-outerBC}) attempts to generate a
time-independent solution with a constant shift $\beta^i_{\infty}$
toward infinity -- a boosted black hole in a co-moving frame.
These considerations lead to two conclusions:
\begin{enumerate}
\item The free quantities in the quasi-equilibrium approach
($\cg_{ij}$, $\trK$, the lapse-boundary condition on $\S$ and the
shape of $\S$) must be consistent with a stationary boosted black hole
in a co-moving coordinate system.
\item The free quantities we use so far do not satisfy this
requirement (because $\partial_t\ln\CF\neq 0$ in
Figure~\ref{fig:QEtestmass-SingleHole}).
\end{enumerate}

In sections~\ref{sec:Comparing:TestingTT}
and~\ref{sec:Comparing:TestingCTS}, we use Eddington-Finkelstein
initial data sets of boosted black holes in co-moving coordinates.
Those data sets would be appropriate here.  Due to the
Lorentz-contraction, the apparent horizons of those data sets are
ellipsoids, indicating that in a fully consistent quasi-equilibrium
approach, one might have to abandon exactly spherical surfaces
$\S$. Once we abandon spherical surfaces, however, we have to find a
means to pick the shape.  An ellipsoid might be appropriate for the
test-mass limit $M_2\ll M_1$, but for comparable sized black holes
tidal forces might deform $\S$ even further.  In addition, even in the
test-mass limit, one has to fix the eccentricity of the ellipsoid
(i.e. the Lorentz-factor of the boost).

\subsection{Toward the post-Newtonian limit}

For binary black holes at large separations, the post-Newtonian
approximation to general relativity becomes increasingly accurate.
Moreover, the inspiral time-scale set by radiation-reaction forces
increases much more rapidly with separation than the orbital period,
so that the approximation of quasi-stationarity should more accurately
reflect reality. 

We therefore perform calculations at larger separations on a maximal
slice (cf. section~\ref{sec:QE:BBH:MaxSlice}), which was found to have
a smaller $\partial_t\ln\CF$ than the superposed Eddington-Finkelstein
slice.  At separations $s=30$ and $s=100$, we adjust $\Omega$ such
that $E_{ADM}=M_K$, and examine the resulting data sets.  Our results
are summarized in Table~\ref{tab:BBH-QE}.  As this table contains a
wealth of information we will spend quite some time discussing it
(the table contains additional information about the data sets
discussed in sections~\ref{sec:QE:BBH:EFslicings} and
\ref{sec:QE:BBH:MaxSlice} as well).

\begin{table}
\begin{centering}
\CAP{Parameters for equal mass binary black hole solutions constructed
with the quasi-equilibrium method.}{\label{tab:BBH-QE}Parameters for
equal mass binary black hole solutions.  The $s=10$ data sets were
presented in Figures~\ref{fig:EF_contours} and \ref{fig:MS_contours}.
$M$ is given by Eq.~(\ref{eq:M-physical}), $m=2M$ and $\mu=M/2$ are
combined mass of both black holes, and reduced mass, respectively, and
$E_b=E_{ADM}-2M$.  } 
\centerline{
\singlespacingplus
\begin{tabular}{|l|r|rrr@{\hspace*{0.55cm}}|}\hline
%
& EF& \multicolumn{3}{|c|}{Maximal slice}\\\hline
$s$                  &  $10$           &  $10$           & $30$         & $100$       \\
$\Omega$        &  $0.0426633$    &  $0.0357221$   & $0.0080078$  & $0.0013840$ \\
$E_{ADM}$            &  $2.0659$       &  $2.2506$       & $2.06299$    & $2.01751$   \\
$M_{K}$              &  $2.0659$       &  $2.2506$       & $2.06299$    & $2.01751$   \\
$M_{AH}$             &  $1.0466$       &  $1.1436$       & $1.03934$    & $1.011223$   \\
$J_{ADM}$            &  $3.8443$       &  $4.4406$       & $4.75536$    & $7.49276$   \\
$M$                  & $1.0508$        & $1.1474$        & $1.03948$    & $1.011223$\\
\hline\hline
\multicolumn{5}{|l|}{dimensionless parameters of the data-sets}\\\hline
$E_{ADM}\Omega$      &  $0.08813$      &  $0.08040$      & $0.01652$       & $0.00279$   \\
$J_{ADM}/E_{ADM}^2$  &  $0.9007$       &  $0.8767$       & $1.1174$    & $1.8408$\\
$S/M^2$              &  $0.1779$       &  $0.1629$       & $0.033287$      & $0.0055982$ \\
$m\Omega$            &  $0.0897$       &  $0.0820$       & $0.016648$      & $0.0027991$ \\
$J_{ADM}/\mu m$      &  $3.4816$       &  $3.3730$       & $4.40100$       & $7.32737$   \\
$E_b/\mu$            &  $-0.0679$      & $-0.0770$       & $-0.03073$      & $-0.00976$  \\
\hline\hline
\multicolumn{5}{|l|}{Comparison to post-Newtonian expansions}\\\hline
J-Newton  &  $0.642$  &  $0.682$  &  $0.8899$  &  $0.9684$ \\
J-PN-1    &  $0.899$  &  $0.937$  &  $0.9890$  &  $0.9991$ \\
J-PN-2    &  $0.968$  &  $1.004$  &  $1.0025$  &  $1.0005$ \\
E-Newton  &  $1.475$  &  $1.225$  &  $1.0609$  &  $1.0172$ \\
E-PN-1    &  $1.247$  &  $1.047$  &  $1.0076$  &  $1.0017$ \\
E-PN-2    &  $1.135$  &  $0.962$  &  $0.9964$  &  $1.0006$ \\
\hline\hline
\multicolumn{5}{|l|}{relevant time-scales}\\\hline
$2\pi/\Omega\equiv P$ & $71  E_{ADM}$ & $78 E_{ADM}$ & $380 E_{ADM}$   & $2300 E_{ADM}$\hspace*{-0.5cm} \\
$1/|\partial_t\ln\CF|_\infty$ 
          & $9 P$     & $66 P$     & $68P$      & $52 P$ \\
$\frac{|K_{ij}|_\infty}{|\partial_tK_{ij}|_\infty}$ 
          & $0.2 P$   & $0.6 P$   & $0.8 P$     & $0.7 P$ \\
$\frac{\sqrt{|(K_{ij})^2|_\infty}}{\sqrt{|(\partial_tK_{ij})^2|_\infty}}$  
         & $0.2 P$    & $0.5P$    & $0.9P$      & $0.7P$ \\
\hline
\end{tabular}
}
\end{centering}
\end{table}

The first block of Table~\ref{tab:BBH-QE} lists the two parameters $s$
and $\Omega$, on which the solution depends, as well as several raw
numbers extracted from the computed initial data set.  The second
block in Table~\ref{tab:BBH-QE} lists dimensionless quantities in
commonly used scalings.  

In order to make contact with post-Newtonian theory, e.g.
Eqs.~(\ref{eqn:PN_Eb}) and (\ref{eqn:PN_J}), we need to accommodate
the fact that the quasi-equilibrium solutions are {\em corotating}.
From the initial data set, we know $\Omega$ (which equals the angular
frequency of the horizon), and the apparent horizon mass
$M_{AH}=\sqrt{A_{AH}/16\pi}$.  However, in order to use
Eq.~(\ref{eqn:PN_J}), we need the full mass $M$ (including the
rotational contribution) of each black hole as well as its spin $S$.
For a Kerr black hole, irreducible mass and angular frequency of the
horizon are given by \cite{Misner-Thorne-Wheeler:1973}
\begin{align}
M_{irr}&=\frac{1}{2}\sqrt{\left(1+\sqrt{1-\left(S/M^2\right)^2}\right)^2
+\left(S/M^2\right)^2},\\
M\Omega&=\frac{S/M^2}{2+2\sqrt{1-\left(S/M^2\right)^2}}.
\end{align}
Approximating $M_{irr}\approx M_{AH}$, and inverting these two
relations, we find the desired quantities, 
\begin{align}
\label{eq:M-physical}
M&=\frac{M_{AH}}{\sqrt{1-\left(2M_{AH}\Omega\right)^2}},\\
\frac{S}{M^2}&=4M_{AH}\Omega\sqrt{1-\left(2M_{AH}\Omega\right)^2}.
\end{align}
The total mass and the reduced mass are $m=2M$ and $\mu=M/2$,
respectively.  The black hole spins are parallel to the orbital
angular momentum, so that Eq.~(\ref{eqn:PN_J}) reduces to
\begin{align}
\left(\frac{J}{\mu m}\right)^2
&=\omg^{-2/3}
\Bigg\{1+\frac{4S}{M^2}\,\omg^{1/3}
+\left(\frac{37}{12}+\frac{4S^2}{M^4}\right)\omg^{2/3} \\
&\hspace{2.7cm   }\mbox{}
+\frac{S}{3M^2}\,\omg
	+\left(\frac{143}{18}-\frac{29\,S^2}{3M^4}\right)
\omg^{4/3}\Bigg\}.\nonumber
\end{align}
We define the quantities
\begin{subequations}
\begin{align}
\mbox{J-Newton}&\equiv
\frac{\omg^{-1/3}}{J/\mu m},\\
\mbox{J-PN-1}&\equiv
\frac{\omg^{-1/3}}{J/\mu m}
\Bigg\{1+\frac{4S}{M^2}\,\omg^{1/3}+\left(\frac{37}{12}+\frac{4S^2}{M^4}\right)\omg^{2/3}
\Bigg\}^{1/2}, \\
\mbox{J-PN-2}&\equiv
\frac{\omg^{-1/3}}{J/\mu m}
\Bigg\{1+\frac{4S}{M^2}\,\omg^{1/3}
+\left(\frac{37}{12}+\frac{4S^2}{M^4}\right)\omg^{2/3} \nonumber\\
&\qquad\qquad\qquad\quad+\frac{S}{3M^2}\,\omg
	+\left(\frac{143}{18}-\frac{29\,S^2}{3M^4}\right)
\omg^{4/3}\Bigg\}^{1/2}.
\end{align}
\end{subequations}
Furthermore, Eq.~(\ref{eqn:PN_Eb}) reduces in the present context to
\begin{equation}
\frac{E_b}{\mu}=-\frac{1}{2}\omg^{2/3}\left\{1-\frac{37}{48}\omg^{2/3}
+\frac{7\,S}{3M^2}\omg-\left(\frac{1069}{384}+\frac{S^2}{M^4}\right)\omg^{4/3}
\right\}
\end{equation}
so that we can examine the ratios
\begin{subequations}
\begin{align}
\mbox{E-Newton}&\equiv-\frac{\omg^{2/3}}{2 E_b/\mu},\\
\mbox{E-PN-1}&\equiv-\frac{\omg^{2/3}}{2 E_b/\mu}\left\{1-\frac{37}{48}\omg^{2/3}\right\},\\
\mbox{E-PN-2}&\equiv-\frac{\omg^{2/3}}{2 E_b/\mu}
\left\{1-\frac{37}{48}\omg^{2/3}
+\frac{7\,S}{3M^2}\omg-\left(\frac{1069}{384}+\frac{S^2}{M^4}\right)\omg^{4/3}
\right\}
\end{align}
\end{subequations}
Deviation of these numbers from unity will measure the disagreement of
the quasi-equilibrium solution with predictions of post-Newtonian
theory.

Table~\ref{tab:BBH-QE} shows that the post-Newtonian expansions are in
good agreement with the quasi-equilibrium calculations.  At $s=100$,
the quasi-equilibrium method and post-Newtonian expansions differ by
less than $0.1$ per cent.  For $s=10$ on a maximal slice, angular
momentum and binding energy agree to better than $1$ per cent and a
few per cent, respectively.  On the Eddington-Finkelstein slice,
deviations are larger, indicating that such a slicing might be less
suited for quasi-equilibrium initial data.

As a final consistency check of the quasi-equilibrium method, we
evaluate $\partial_t\ln\CF$ and $\partial_t\K_{ij}$.    The characteristic
time-scale on which $\CF$ changes is 
\begin{equation}
\frac{1}{\left|\partial_t\ln\CF\right|_\infty},
\end{equation}
where $|\,.\,|_\infty$ denotes the maximum norm.
For the tensor-quantity $\partial_t\K_{ij}$ it is more difficult to
define a time-scale.  We use the maximum value of any Cartesian
component of $K_{ij}$ and $\partial_tK_{ij}$, as well as the maximum
value of $K_{ij}K_{kl}g^{il}g^{jl}$ and
$\partial_tK_{ij}\partial_tK_{kl}g^{il}g^{jl}$.  Thus, we estimate the
characteristic time-scale for changes in $\K_{ij}$ by 
\begin{equation}
\frac{\left|K_{ij}\right|_\infty}{\left|\partial_tK_{ij}\right|_\infty}
\qquad\mbox{and}\qquad
\frac{\sqrt{\left|K_{ij}K_{kl}g^{il}g^{jl}\right|_\infty}}
{\sqrt{\left|\partial_tK_{ij}\partial_tK_{kl}g^{il}g^{jl}\right|_\infty}}.
\end{equation}
Table~\ref{tab:BBH-QE} compares these three time-scales to the orbital
period $P=2\pi/\Omega$.  For maximal slicing, the conformal factor
changes on a fairly long timescale, about 50 orbital periods.
However, the extrinsic curvature changes much faster, within half an
orbital period.  Both of these time scales increase approximately
linearly with the orbital period.  

While the time-scale of $\CF$ is satisfactory, the time-scale
associated with the extrinsic curvature is shorter than expected.  For
a binary in ``quasi-circular orbit,'' changes should occur on
time-scales much longer than the orbital period.  This is especially
true at large separations where the radiation reaction time-scale is
much longer than the orbital period.

A close inspection of the $s=100$ data set indicates that the
comparatively large time-derivatives have the same cause as in the
test-mass limit: The free data of the quasi-equilibrium approach is
not compatible with a boosted black hole in a comoving frame.  Because
the velocity of Keplerian orbits decays very slowly with separation,
namely as $s^{-1/2}$, this effect dominates at large separations.
When comparing a single black hole solution with appropriate nonzero shift at
infinity (cf. Figure~\ref{fig:QEtestmass-SingleHole}) with the region
close to one black hole in the binary black hole solution (with
$s=100$), we find:
\begin{enumerate}
\item Very good agreement in the form and amplitude of $\partial_t\ln\CF$.
\item Good agreement in the form and amplitude of
$\sqrt{(\partial_tK_{ij})^2}$. 
\end{enumerate}
Furthermore, very close to the horizons, the binary black hole
solution exhibits tidal effects: On the hemisphere of the horizon
which points away from the other black hole,
$\sqrt{(\partial_tK_{ij})^2}$ is {\em twice} as big as on the
hemisphere pointing toward the companion.

Given the good agreement of orbital parameters with the post-Newtonian
expansions, it seems likely that the data sets are only contaminated
locally around each black hole.

\section{Discussion}
\label{sec:QE:Discussion}

We have presented the quasi-equilibrium method and demonstrated that
it provides a viable method to construct single and binary black hole
initial data.

For single non-rotating black holes, this method is spectacularly
successful.  Only the mean curvature $\trK$ and the lapse boundary
condition remain as freely specifiable data, and any (reasonable)
choice for these results in a completely time-independent slicing
of Schwarzschild.

Binary black hole initial data was constructed with two different
choices for $\trK$.  Apart from a proposed lapse boundary condition
---which we found to be ill-posed--- no difficulties arose when
solving the coupled partial differential equations with the
quasi-equilibrium boundary conditions.  Furthermore, the condition
$E_{ADM}=M_K$ singles out a unique orbital frequency.  

We examined the test-mass limit and the limit of widely separated
black holes.  In both limits, we found time-derivatives
$\partial_t\ln\CF$, $\partial_tK_{ij}$ significantly larger than
expected based on the long inspiral time-scales in these limits.  We
identified one cause for these large time-derivatives, namely the
apparent inconsistency of our choices for the remaining free data with
a boosted black hole in co-moving coordinates.

Future work should be directed toward resolving this issue.  Indeed,
the demonstration that time-derivatives approach zero in these
limiting cases consistent with the inspiral time-scale would be a
major success.  One might also continue to search for a physically
motivated boundary condition for the lapse.  One could consider, for
example, a quantity $\kappa$ defined analogous to the surface gravity
by $\zeta\cdot\fournabla\zeta^\mu=\kappa\zeta^\mu$ [$\zeta^\mu$ is
given in Eq.~(\ref{eq:t-split})].  As $\kappa$ scales linearly with
$\zeta^\mu$, the requirement that $\kappa$ take the value appropriate
for a Kerr black hole might fix the scaling of $\zeta^\mu$, i.e.  the
lapse $\N$.  Finally, work is already underway to construct sequences
of quasi-circular orbits based on the initial data presented in this
chapter, and to determine the location of the ISCO.

Although the quasi-equilibrium initial data was found not to be
perfect, it nonetheless represents a significant improvement.  It is
superior to initial data constructed with the Bowen-York approach in
many aspects:
\begin{itemize}
\item The quasi-equilibrium approach avoids ambiguities regarding the
choice of extrinsic curvature.
\item It sets certain time-derivatives explicitly to zero, whereas
with the effective potential method no rigorous statement can be made
about time-derivatives.
\item The quasi-equilibrium method has a natural consistency check,
 evaluation of the remaining time-derivatives $\partial_t\ln\CF$ and
 $\partial_t\K_{ij}$.
\item Locating quasi-circular orbits is computationally simpler than
with the effective potential method because no minimization is
necessary.
\item The initial data solve also yields lapse and shift which can be
used in the evolution of the initial data.
\item The apparent horizon boundary condition and the shift boundary
condition guarantee that the apparent horizons initially do not move
in coordinate space.  This allows evolution with black hole excision
to proceed for some time without regridding.
\end{itemize}

Given all these advantages, I believe that one should aggressively
attempt to evolve the equal-mass initial data sets constructed in this
chapter.  This is especially true in the light of the significant
improvements in black hole evolutions over the last few years.  The
initial data based on maximal slices has somewhat smaller
time-derivatives and agrees better with post-Newtonian theory than the
data based on Eddington-Finkelstein slices.  One should therefore
evolve initial data constructed on maximal slices.

The research in this chapter illustrates that numerics has reached a
level of maturity which makes it not only feasible to implement
complex proposals for constructing initial data, but also to perform
consistency checks and guide further analytical work.

\section{Appendix A: Detailed quasi-equilibrium calculations}
\label{QE:appendix}

We define the extrinsic curvature of $\S$ along the outward-going null
rays by
\begin{equation}
\Sigma_{\mu\nu}
=\frac{1}{2}h_\mu^{\,\alpha} h_\nu^{\,\beta}\Lie_k\gfour_{\alpha\beta}
=h_\mu^{\,\alpha} h_\nu^{\,\beta}\;\fournabla_{(\alpha}k_{\beta)}, 
\end{equation}
where $\fournabla$ is the spacetime derivative.  The trace and
trace-free (symmetric) parts of $\Sigma_{\mu\nu}$ are expansion and
shear, which we want to compute.  Use $k^\mu=(n^\mu+s^\mu)$
and insert a redundant projection operator to find
\begin{equation}\label{eq:Null-ExCurv0}
\Sigma_{\mu\nu} 
=h_\mu^\alpha h_\nu^\beta
\underbrace{\g_\alpha^{\alpha'}\g_\beta^{\beta'}
\fournabla_{(\alpha'}n_{\beta')}}_{-K_{\alpha\beta}}
+h_\mu^\alpha h_\nu^\beta\fournabla_{(\alpha}s_{\beta)}.
\end{equation}
The first term contains the extrinsic curvature of $\Sigma$, the
second term is the extrinsic curvature of $\S$ as embedded in
$\Sigma$.  All tensors in Eq.~(\ref{eq:Null-ExCurv0}) are purely
spatial, so we can switch to Latin (spatial) indices:
\begin{equation}\label{eq:Null-ExCurv}
\Sigma_{ij}=-h_i^kh_j^lK_{kl}+h_i^kh_j^l\nabla_{\!(k}s_{l)}.
\end{equation}

The trace of Eq.~(\ref{eq:Null-ExCurv}) gives the expansion,
\begin{equation}\label{eq:Null-Expansion}
\theta=\Sigma_{ij}h^{ij}= -h^{ij}K_{ij}+h^{ij}\nabla_{\!i}s_j.
\end{equation}
Use $h^{ij}\deriv_{\!i}s_j=\CF^{-4}\tilde h^{ij}\deriv_{\!i}(\CF^2\tilde
s_j)=\CF^{-2}\tilde h^{ij}\deriv_{\!i}\tilde s_j$ (because $\tilde
h^{ij}\tilde s_i=0$), and transform the derivative into conformal
space with Eq.~(\ref{eq:Gamma-scaling}):
\begin{align}
h^{ij}\nabla_{\!i}s_j&
=\CF^{-2}\tilde h^{ij}\left(\partial_i\tilde s_j-\Gamma^k_{ij}\tilde s_k\right)
\nonumber\\
&=\CF^{-2}\tilde h^{ij}\left(\partial_i\tilde
s_j-\tilde\Gamma_{ij}^k\tilde s_k
-2\left(\delta_i^k\partial_j\ln\CF+\delta_j^k\partial_i\ln\CF
-\cg_{ij}\cg^{kl}\partial_l\ln\CF\right)\tilde s_k\right)\nonumber\\
\label{eq:temp-AH-BC}
&=\CF^{-2}\left(\tilde h^{ij}\cderiv_{\!i}\tilde s_j+4\tilde
s^l\partial_l\ln\CF\right).
\end{align}
Combining Eqs.~(\ref{eq:Null-Expansion}) and (\ref{eq:temp-AH-BC}),
and setting $\theta=0$ gives
\begin{equation}
\tilde s^k\partial_k\ln\CF=-\frac{1}{4}\left(\tilde
h^{ij}\cderiv_{\!i}\tilde s_j-\CF^2h^{ij}K_{ij}\right),
\end{equation}
which is Eq.~(\ref{eq:BC-AH}) from the main text.

To compute the shear, it is easiest to return to
Eq.~(\ref{eq:Null-ExCurv}) and use the conformal thin sandwich
decomposition of $K_{ij}$, Eqs.~(\ref{eq:K-split})
and~(\ref{eq:CTS-Aij}).  The first term of (\ref{eq:Null-ExCurv})
gives (with $\cu_{ij}=0$)
\begin{align}
\nonumber
h_i^kh_j^lK_{kl}
&=
h_i^kh_j^l\left(
\frac{1}{3}g_{kl}\trK+\frac{1}{2N}(\Long\beta)_{kl}\right)\\
\label{eq:temp}
&=\frac{1}{3}h_{ij}\left(K-\frac{1}{N}\nabla_{\!k}\beta^k\right)
+\frac{1}{N}h_i^kh_j^l\nabla_{(k}\beta_{l)}.
\end{align}
Now substitute the boundary condition $\beta^k=\beta_\perp
s^k+\beta^k_\parallel$ into the second term of
Eq.~(\ref{eq:Null-ExCurv}).  It suffices that the boundary
condition holds only on $\S$, because the derivative is projected
into $\S$.  Using $h^k_is_k=0$, we find
\begin{align}
h_i^kh_j^lK_{kl}
&=\frac{1}{3}h_{ij}\left(K-\frac{1}{N}\nabla_{\!k}\beta^k\right)
+\frac{\beta_\perp}{N}h_i^kh_j^l\nabla_{(k}s_{l)}
+\frac{1}{N}D_{(i}\beta_{\parallel\,j)},
\end{align}
where $D_i$ denotes the induced two-dimensional covariant derivative
in $\S$.  Equation~(\ref{eq:Null-ExCurv}) becomes
\begin{equation}\label{eq:Null-ExCurv2}
\Sigma_{ij}
=-\frac{1}{3}h_{ij}\left(K-\frac{1}{N}\nabla_{\!k}\beta^k\right)
+\left(1-\frac{\beta_\perp}{N}\right)h_i^kh_j^l\nabla_{(k}s_{l)}
-\frac{1}{N}D_{(i}\beta_{\parallel\,j)}.
\end{equation}
The first term of~(\ref{eq:Null-ExCurv2}) is purely trace and the
second term vanishes because of the boundary condition
$\beta_\perp=N$.  Therefore, the shear is simply
\begin{equation}
\sigma_{ij}=\Sigma_{ij}-\frac{1}{2}h_{ij}h^{kl}\Sigma_{kl}
=-\frac{1}{N}\left(D_{(i}\beta_{\parallel\,j)}
                   -\frac{1}{2}h_{ij}D_k\beta^k_\parallel\right)
=-\frac{1}{2N}(L_\S\beta_\parallel)_{ij},
\end{equation}
where $L_\S$ denotes the two-dimensional longitudinal operator within
$\S$.  Thus, the shear vanishes if $(L_\S\beta_\parallel)^{ij}=0$, or,
using the conformal scaling behavior of the longitudinal operator,
Eq.~(\ref{eq:L-scaling}), if
\begin{equation}
(\tilde L_\S\beta_\parallel)^{ij}=\tilde D^i\beta^j_\parallel+\tilde
D^j\beta^i_\parallel-\tilde h^{ij}\tilde D_k\beta^k_\parallel=0,
\end{equation}
which is Eq.~(\ref{eq:BC-shift-parallel}) from the main text.

\section{Appendix B: Code Tests}
\label{sec:QE:CodeTest}

\begin{figure}[tb]
\centerline{\includegraphics[scale=0.44]{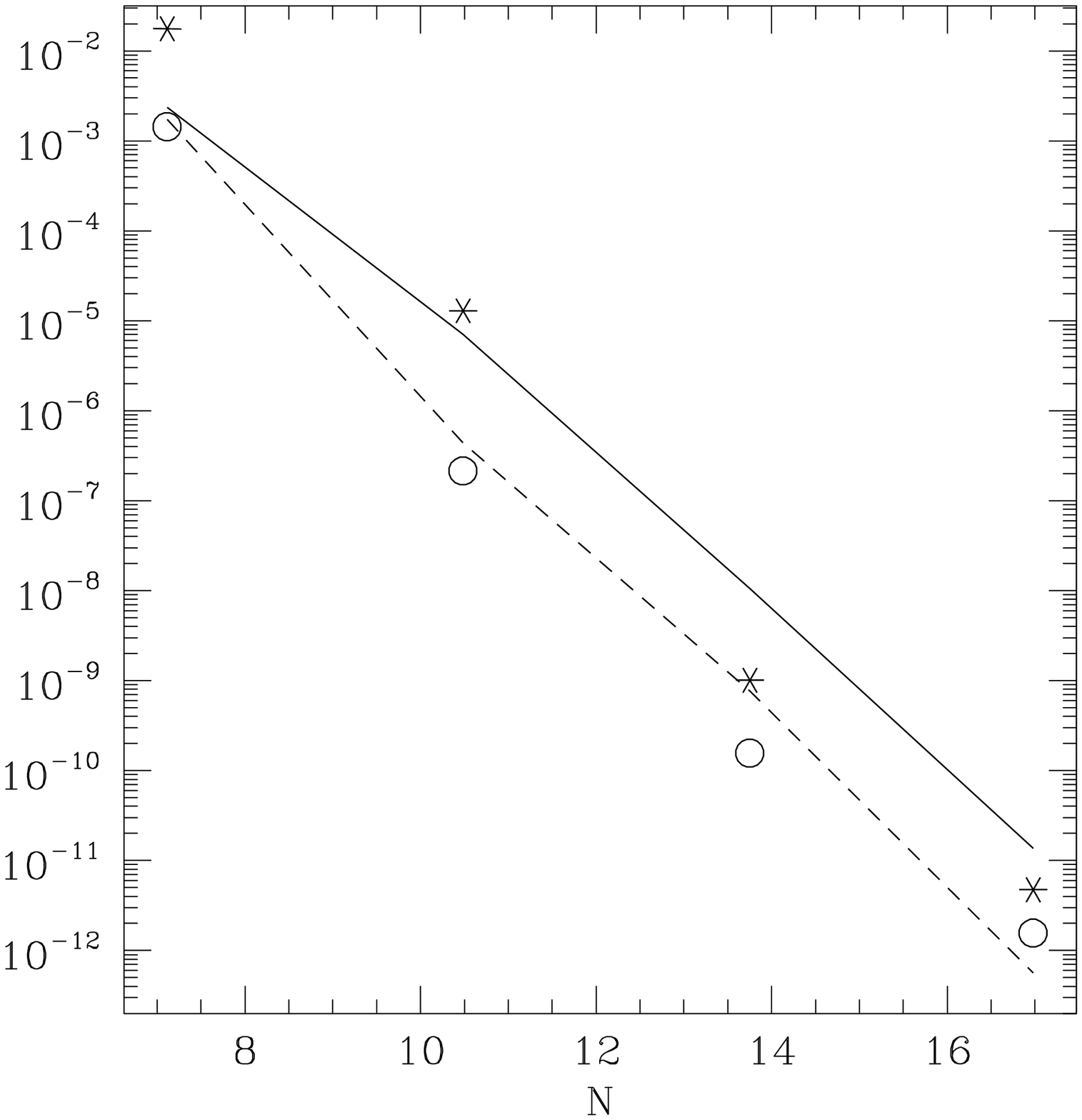}}
\CAP{Testing the quasi-equilibrium equations and the lapse
quasi-equilibrium condition on an Eddington-Finkelstein
slice.}{\label{fig:TestSandwich5QE}Residuals of the quasi-equilibrium
boundary conditions~(\ref{eq:QE}) and the lapse quasi-equilibrium
condition~(\ref{eq:BC-lapse}) for an Eddington-Finkelstein slice in
the standard coordinates (solid line) and the ``conformally flat''
coordinates (dashed line).  The stars and circles denote the violation
of the Hamiltonian constraint for both cases.  All norms are
$L_\infty$-norms.}
\end{figure}

We set the conformal metric $\cg_{ij}$ and the mean curvature $\trK$
to their values for Eddington-Finkelstein, Eqs.~(\ref{eq:gij-EF}) and
(\ref{eq:trK-EF}), and verify that the conformal factor $\CF\equiv 1$
and lapse and shift given by Eqs.~(\ref{eq:lapse-EF}) and
(\ref{eq:shift-EF}) indeed solve the quasi-equilibrium equations
(\ref{eq:QE}).  We also perform the variable transformation
(\ref{eq:EF-flat-r}) to a conformally flat metric, and verify that a
flat background metric, $\cg_{ij}=\delta_{ij}$, transformed lapse and
shift and the now non-trivial conformal factor~(\ref{eq:CF-EF-flat})
still satisfy Eqs.~(\ref{eq:QE}).  (In the latter case, one has to
change the radius of the excised sphere to $\hat r_{AH}$ as given
by~(\ref{eq:r-AH-EF})~).  Convergence of the residuals to zero is
demonstrated in Figure \ref{fig:TestSandwich5QE}.

The lapse boundary condition~(\ref{eq:BC-lapse}) is very complicated,
and was implemented only for spherically symmetric conformal metrics
$\cg_{ij}$, which includes conformal flatness, and the usual
Eddington-Finkelstein slicing~(\ref{eq:EF}).  (The {\em solutions}
$\CF$, $\beta^i$ and $\N$ are not required to be spherically
symmetric).  The lapse condition is included in the residuals shown
Figure~\ref{fig:TestSandwich5QE}, already providing a first test of
Eq.~(\ref{eq:BC-lapse}).  However, many terms in
Eqs.~(\ref{eq:BC-lapse}) and (\ref{eq:calD}) vanish identically for
spherically symmetric solutions like Eqs.~(\ref{eq:lapse-EF})
and~(\ref{eq:shift-EF}).

\begin{figure}[tb]
\centerline{\includegraphics[scale=0.44]{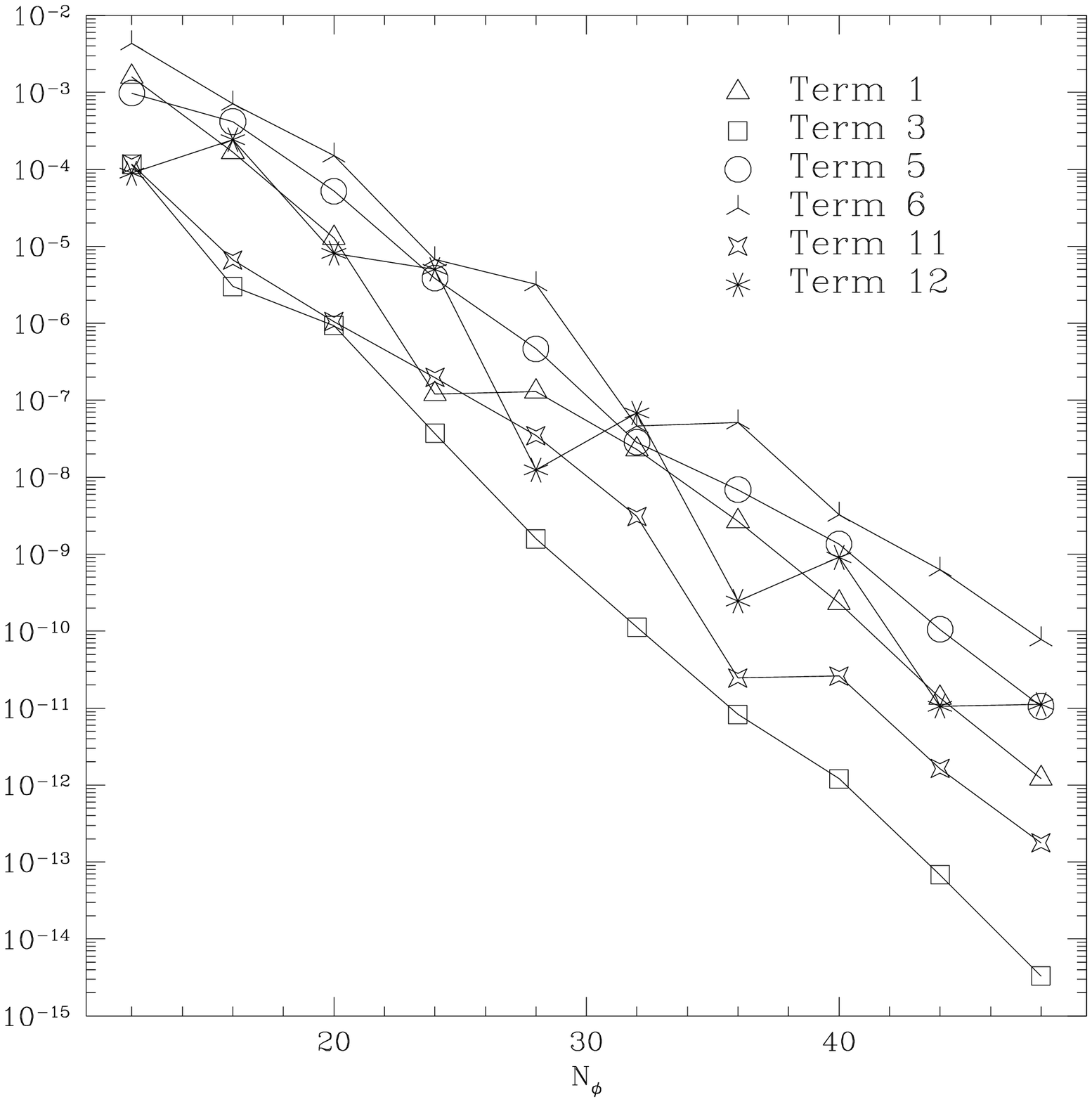}}
\CAP{Convergence of selected terms of the lapse boundary condition.}
{\label{fig:TestLapseBC}Convergence of selected terms of the
   lapse boundary condition Eq.~(\ref{eq:BC-lapse}).}
\end{figure}

We therefore test the code implementing Eq.~(\ref{eq:BC-lapse})
against a separate analytical calculation with Mathematica.  We choose
a flat background metric $\cg_{ij}\!=\!\mbox{flat}$, and set
\begin{equation}
\begin{gathered}
  \CF=1+\frac{a_0}{|\vec r-\vec r_0|},\qquad
  \vec\beta=\hat e_r\frac{1}{r}+\hat e_x\frac{a_1}{|\vec r-\vec r_1|}
  +\hat e_z\frac{a_2}{|\vec r-\vec r_2|},\\
  \N\CF=1+\frac{a_3}{|\vec r-\vec r_3|},\qquad
  K=\frac{a_4}{|\vec r-\vec r_4|},\qquad
  \tilde u_{ij}=\frac{U_{ij}}{|\vec r-\vec r_u|},
\end{gathered}
\end{equation}
with constants 
\begin{equation*}\label{eq:Parameters2}
\begin{aligned}
    a0&=-0.2,&\vec r_0&=(-0.1, 0.2, 0.3),
    &a1&= 0.3,& \vec r_1&=( 0.3, 0.0,-0.1),\\
    a2&=-0.2,& \vec r_2&=( 0.3,-0.1, 0.2),&
    a3&=-0.3,& \vec r_3&=( 0.3, 0.1,-0.2),\\
    a_4&=1,& \vec r_4&=(0.3, 0.1, 0.1),&
  U_{ij}&=0.1(i+j),& \vec r_u&=(0.3, 0, -0.1).
\end{aligned}
\end{equation*}
$\S$ was taken as a sphere centered at the origin with radius 1.2.
These data do not satisfy Eq.~(\ref{eq:BC-lapse}), of course.
However, we can nonetheless compare the computed values of various
terms in (\ref{eq:BC-lapse}) against the results obtained with
Mathematica.  Figure~\ref{fig:TestLapseBC} confirms that these results
converge exponentially with resolution, as expected.  (In the
implementation, Eq.~(\ref{eq:BC-lapse}) has been expanded into smaller
terms, see Eq.~(\ref{eq:LapseBC1}).  The numbering of terms in
Figure~\ref{fig:TestLapseBC} corresponds to the terms in the code).

\section{Appendix C: Extrapolation into the interior of the excised spheres}
\label{sec:QE:Extrapolation}

The quasi-equilibrium method yields initial data {\em up to} the
apparent horizon.  Evolutions with black hole excision, however, place
the inner boundary {\em inside} the apparent horizon, so that one
needs to extrapolate the initial data set a small distance into the
interior of $\S$.  Simple summation of the spectral series outside its
domain is very unstable as the high frequency modes blow up very
quickly.  A function $f$ given in the inner spherical shell is
extrapolated into the excised region with the following method:

\begin{enumerate}
\item For each grid point on the inner surface of the shell, compute
radial derivatives up to order $N_{extrp}$; we denote these derivatives by
$f^{(n)}_{jk}$, where $j$ and $k$ label the angular grid-points.
\item To extrapolate to a point $(r, \theta, \phi)$ with $r<r_{exc}$,
first spectrally interpolate $f^{(n)}_{jk}$ to the direction $(\theta,
\phi)$.  This is done by summing up the expansion in spherical
harmonics, and results in $f^{(n)}(\theta,\phi)$, $n=0, \ldots, N_{extrp}$.
Then use a Taylor series in radius to find
\begin{equation}
f(r, \theta,\phi)=\sum_{n=0}^{N_{extrp}}\frac{f^{(n)}(\theta,\phi)}{n!}\big(r_{exc}-r\big)^n.
\end{equation}
\end{enumerate}

By construction, this procedure leads to an extrapolation that is
smooth up to and including the $N_{extrp}$-th derivative.  Note that
this procedure is {\em not} equivalent to just truncating the spectral
series after $N_{extrp}$ terms; it rather takes polynomials up to
order $N_{extrp}$ from {\em all} terms in the spectral series.
Extrapolation by a truncated spectral series, i.e.
\begin{equation}
f(r)=\left\{
\begin{aligned}
\sum_0^N\tilde f_k\Phi_k(r),\qquad\mbox{for $r\ge r_{exc}$}\\
\sum_0^{N_{extrp}}\tilde f_k\Phi_k(r),\qquad\mbox{for $r<r_{exc}$}
\end{aligned}
\right.
\end{equation}
results in a discontinuous function,
\begin{equation}
\lim_{\varepsilon\to 0}\left(f(r_{exc})-f(r_{exc}-\varepsilon)\right)
=\sum_{N_{extrp}+1}^N\tilde f_k\Phi_k(r_{exc})\neq 0.
\end{equation}

Two considerations influence the choice of $N_{extrp}$.  If it is too
big, then the Taylor series is dominated by the very fast growing
highest radial derivatives, which in turn depend on high-order
spectral coefficients.  These coefficients are small and have a
comparatively large (relative) error so that the extrapolation becomes
inaccurate.  On the other hand, using a Taylor series up to order
$N_{extrp}$ implies that the $(N_{extrp}+1)$--st derivative of the
extrapolated function will be discontinuous at the location of the
inner boundary.  If $N_{extrp}$ is too small, this discontinuity will
destroy exponential convergence.

In practice, we found that $N_{extrp}=10$ is satisfactory.  For
example, evolutions run for roughly the same evolution
time~\cite{Kidder:2003} with completely analytic Eddington-Finkelstein
initial data, or with initial data that is the analytic
Eddington-Finkelstein just {\em up to} the horizon, while the
grid-points inside the horizon are filled by extrapolation.


\chapter{Initial data with superposed gravitational waves}
\label{chapter:GW}

\section{Introduction}

The goal of this chapter is to construct initial data sets that
contain a strong gravitational wave where one can control certain
characteristics like shape, location and direction of propagation.

Spacetimes containing some component of gravitational radiation are
very interesting for several reasons.  One can, for example,
investigate perturbed black holes~\cite{Price:1972,Papadopoulos:2002},
or examine critical collapse to a black
hole~\cite{Choptuik:1993,Abrahams-Evans:1993}.  Moreover, an initial
data set with gravitational radiation is non-stationary and is
therefore a good test for numerical evolution codes and gravitational
wave extraction in a setting computationally much simpler than a full
binary black hole evolution.

Numerical work on perturbed black holes is either done on spacelike
hypersurfaces (the formalism we use throughout this thesis), or on
null surfaces, with so-called characteristic formulations.
Characteristic formulations are very well adapted to studying
gravitational radiation, since gravitational radiation moves along
null-cones which coincide with coordinate surfaces.  Bishop et
al~\cite{Bishop-Gomez-etal:1997} demonstrate their characteristic code
by scattering a strong ingoing pulse of radiation on a Schwarzschild
black hole.  Papadopoulos~\cite{Papadopoulos:2002} perturbs a
Schwarzschild black hole with an outgoing pulse centered on the
potential barrier at $r=3M$ and examines the power emitted into
different modes.  Characteristic formulations, however, are limited
by formation of caustics; in particular, a binary black hole
coalescence most likely cannot be computed with a characteristic
formulations alone (see for example the discussion in
\cite{Gomez-Lehner-etal:1998}).  Therefore we now turn out attention
back to the initial value problem on spacelike surfaces, where we will
recover some aspects of work already done using the characteristic
formulation.

Perturbed spacetimes represented on {\em spacelike surfaces} dates
back to Brill~\cite{Brill:1959}.  He pointed out that, for
time-symmetry, a certain axisymmetric form of the metric,
\begin{equation}\label{eq:gij-Brill}
ds^2= \CF^4\big[e^{2q}\left(d\rho^2+dz^2\right)+\rho^2d\phi^2\big],
\end{equation}
allows the Hamiltonian constraint (\ref{eq:Ham4}) to be written as a
very simple {\em flat space} Laplace equation for the conformal factor
$\CF$.  Here, $(\rho, z, \phi)$ are cylindrical coordinates, and the
function $q(\rho, z)$ can be freely specified (subject to certain
regularity and fall-off conditions), encoding the perturbation of the
spacetime.  Brill
used~(\ref{eq:gij-Brill}) for a positivity of energy proof.

Since then, so-called ``Brill-waves'' with various choices for $q$
have been used.  Perturbations of {\em flat space} were obtained, for
example, with
(\cite{Eppley:1977,Abrahams-Heiderich-etal:1992,Garfinkle-Duncan:2001,
Alcubierre-Allen-etal:2000})
\begin{align}
q&=\frac{A\rho^2}{1+(r/\lambda)^n},\\
q&=Ar^2\exp\left[-\left(\frac{\rho}{a}\right)^2
-\left(\frac{z}{b}\right)^2\right],\\
q&=A\rho^2e^{-r^2}\left[1+c\frac{\rho^2}{1+\rho^2}\cos^2(m\varphi)\right],\\
\end{align}
with constants $A, \lambda, a, b, c, n$ and integer $m$.  For
perturbations around flat space, there are no singularities or inner
boundaries in the computational domain, and the conformal factor $\CF$
will be close to unity; in particular, $q\equiv 0$ yields the solution
$\CF\equiv 1$: flat space.  Shibata~\cite{Shibata:1997b} generalized
(\ref{eq:gij-Brill}) to compute three-dimensional time-symmetric
initial data sets.

{\em Black hole solutions} with superposed Brill waves can be obtained
by excising a sphere from the computational domain and applying an
isometry boundary condition on $\CF$.  Then the conformal factor will
significantly differ from unity close to the inner boundary.  This
approach has been used, for example, in
\cite{Brandt-Seidel:1996,Camarda-Seidel:1999,Brandt-Camarda-etal:2003}.
These authors also modify the conformal metric~(\ref{eq:gij-Brill})
and abandon time-symmetry to incorporate spin.  They choose the
extrinsic curvature in such a way that it satisfies the momentum
constraint, so that only the Hamiltonian constraint has to be solved
for $\CF$.  In the spirit of Brill, a function $q$ still parametrizes
the distortion, for example~\cite{Brandt-Camarda-etal:2003},
\begin{equation}
q=A\sin^n\theta\left(e^{-(\eta+b)^2/w^2}+e^{-(\eta-b)^2/w^2}\right)
\left(1+c\,\cos^2\varphi\right).
\end{equation}
Here, $\eta$ is a radial coordinate (the horizon is at $\eta=0$), and
$A, b, c, w, n$ are constants.  $b$ was set to zero in the
computations in
\cite{Brandt-Seidel:1996,Camarda-Seidel:1999,Brandt-Camarda-etal:2003},
so that the Brill distortion is located {\em on the throat}, making it
difficult to distinguish between effects of the initial perturbation
from subsequent interaction of the black hole with the gravitational
wave.

This short and non-exhaustive survey illustrates that a lot of work is
based on Brill's original idea.  The different choices for $q$ show
that there is considerable freedom in this approach.  It appears that,
generally, the function $q$ is chosen rather ad hoc, its purpose
mainly being to perturb the spacetime in {\em some} way.  Every one of
these choices for $q$ leads to a perturbed initial data set containing
a ``blob'' of gravitational radiation.  However, it is not clear what
properties the gravitational radiation has nor how to control them.

Part of the motivation to use the Brill metric (\ref{eq:gij-Brill})
was certainly that the resulting elliptic equation is relatively
simple.  As we have a robust elliptic solver for the coupled
constraint equations, computational complexity is no longer a concern.

We describe here a different approach, which is based on {\em linear}
gravitational waves.  By choosing the underlying linear wave one can
construct initial data sets with very specific properties.  This
approach is related to and generalizes work by Abrahams \&
Evans~\cite{Abrahams-Evans:1992,Abrahams-Evans:1993}.  Their
axisymmetric work sets a certain component of the extrinsic curvature
(namely $K^r_{\;\theta}$ in spherical coordinates) equal to the value
appropriate for the linear wave, and solves the momentum constraints
for the remaining components of $\K_{ij}$; this procedure singles out
a preferred coordinate system.  Our method here, in contrast, is
three-dimensional, and covariant with respect to spatial coordinate
transformations.  Moreover, we use the conformal thin sandwich
formalism and not an extrinsic curvature decomposition.

\section{Method}

We first recall some properties of linear gravitational waves.
In linear gravity, one writes the spacetime metric as
\begin{equation}\label{eq:gfour-linwave}
\gfour_{\mu\nu}=\eta_{\mu\nu}+A\,h_{\mu\nu},
\end{equation}
where $\eta_{\mu\nu}$ is the Minkowski-metric, $A\ll 1$ a constant,
 and $h_{\mu\nu}={\cal O}(1)$ the linear gravitational wave.  (We
 separate the amplitude $A$ from $h_{\mu\nu}$ for later convenience.)
 In the {\em transverse-traceless}
 gauge~\cite{Misner-Thorne-Wheeler:1973}, $h_{\mu\nu}$ is purely
 spatial, $h_{\mu0}=0$, transverse with respect to Minkowski space,
 $\nabla^ih_{ij}=0$, and traceless, $\eta^{ij}h_{ij}=0$.  To first
 order in the amplitude $A$, Einstein's equations reduce to
\begin{equation}\label{eq:box-hij}
\square h_{ij}=0.
\end{equation}
The 3+1 decomposition of the metric (\ref{eq:gfour-linwave}) in
transverse-traceless gauge is
\begin{subequations}
\label{eq:3+1-linwave}
\begin{align}
\label{eq:3+1-linwave-gij}
g_{ij}&=f_{ij}+A\,h_{ij}\\
\beta^i&=0,\\
\N&=1,
\end{align}
where $f_{ij}$ denotes the flat metric.  The evolution equation for
$\g_{ij}$, Eq.~(\ref{eq:dtgij1}), yields the extrinsic curvature
\begin{equation}\label{eq:3+1-linwave-Kij}
K_{ij}=-\frac{A}{2}\dot h_{ij}.
\end{equation}
\end{subequations}

The spacetime metric (\ref{eq:gfour-linwave}) satisfies Einstein's
equations to first order in $A$.  Consequently, $(g_{ij}, K^{ij})$
from Eqs.~(\ref{eq:3+1-linwave-gij}) and (\ref{eq:3+1-linwave-Kij})
will satisfy the constraints to linear order in $A$, also.  Since we
will increase $A$ to order unity later on, this is not sufficient, and
we must solve the constraint equations.  But which formalism and what
free data should we use?

Given a linear wave $h_{ij}$ it is straightforward to compute the
three-dimensional {\em metric} $\g_{ij}=f_{ij}+A h_{ij}$ and its
{\em time-derivative} $A\dot h_{ij}$.  In the conformal thin sandwich
decomposition (section \ref{sec:IVP:CTS}), we are free to specify a
conformal {\em metric} $\cg_{ij}$ and its {\em time-derivative}
$\partial_t\cg_{ij}=\cu_{ij}$.  This suggests using the conformal
thin sandwich equations with the choices
\begin{subequations}\label{eq:GW-flatspace}
\begin{align}
\label{eq:cgij-linwave1}
\cg_{ij}&=f_{ij}+A\, h_{ij},\\
\label{eq:cuij-linwave1}
\cu_{ij}&=A\dot h_{ij}-\frac{1}{3}\cg_{ij}\cg^{kl}A\dot h_{kl}.
\end{align}
The second term in (\ref{eq:cuij-linwave1}) ensures that $\tilde
u_{ij}$ is tracefree with respect to $\cg_{ij}$.  Because $h_{ij}$ and $\dot h_{ij}$ are both traceless, Eq.~(\ref{eq:3+1-linwave-Kij}) suggests the choice
\begin{equation}\label{eq:trK-linwave}
 \trK=0.
\end{equation}
The free data Eq.~(\ref{eq:CTS-freedata2}) is completed by setting
\begin{equation}
\dot\trK=0.
\end{equation}
\end{subequations}

We now have a prescription to construct a slice through perturbed flat
 space with an arbitrarily strong gravitational wave: Pick a
 gravitational wave in TT gauge, $h_{ij}$, and solve the conformal
 thin sandwich equations with free data given by Eqs.~(\ref{eq:GW-flatspace}).

We generalize this prescription to an arbitrary background spacetime
as follows:

Let $\g_{ij}^0$ and $\trK^0$ be the 3-metric and mean curvature of a
slice through a stationary spacetime (for example flat space or a
Kerr black hole).  Solve the conformal thin sandwich equations with
the free data
\begin{subequations}
\label{eq:linwave-freedata}
\begin{align}\label{eq:cgij-linwave2}
\cg_{ij}&=\g_{ij}^0+A\,h_{ij},\\
\cu_{ij}&=A\,\dot h_{ij}-\frac{1}{3}\cg_{ij}\cg^{kl}A\dot h_{kl},\\
\trK&=\trK^0,\\
\dot\trK&=0.
\end{align}
\end{subequations}
We consider a few limiting cases
\begin{itemize}
\item
$A\to 0$ recovers the underlying stationary spacetime in
time-independent coordinates (if appropriate boundary conditions were
used).
\item For $A\ll 1$ with linear wave ``far away'' where the spacetime
is approximately flat, $g_{ij}^0\approx f_{ij}$, $\trK^0\approx 0$ the
linear theory is valid and the properties of the linear wave are well
defined and well known\footnote{If $h_{ij}$ is a linear wave on the
curved background $g_{ij}^0$, then the $A\ll 1$ case is well-defined
even for a wave in the strong curvature region.  However, we consider
here only the simpler case of $h_{ij}$ being a flat space wave.}.
\item
Finally, for large $A$ we will have a nonlinearly perturbed spacetime,
our primary interest.  Due to the nonlinearity of Einstein's
equations, the properties of such a strongly perturbed spacetime will
differ from the linear wave, however, we would expect that the
qualitative properties still agree.
\end{itemize}

We remark that this formalism is covariant; any suitable three
dimensional coordinate system can be used.  Furthermore, the conformal
thin sandwich equations have to be solved in a general conformal
background, cf. Eqs.~(\ref{eq:cgij-linwave1}) and
(\ref{eq:cgij-linwave2}), which requires an elliptic solver capable of
handling this situation.

\section{Quadrupole waves}

We use for $h_{ij}$ quadrupole waves as described by
Teukolsky~\cite{Teukolsky:1982}.  The even parity outgoing wave
is~\cite{Teukolsky:1982}
\begin{equation}\label{eq:TeukolskyWave}
\begin{aligned}
ds^2=-dt^2&+(1+Af_{rr})dr^2+(2Bf_{r\theta})rdrd\theta
+(2Bf_{r\theta})r\sin\theta drd\phi\\
&+\left(1+Cf_{\theta\theta}^{(1)}+Af_{\theta\theta}^{(2)}+\right)r^2d\theta^2
+\left[2(A-2C)f_{\theta\phi}\right]r^2\sin\theta d\theta d\phi\\
&+\left(1+Cf_{\phi\phi}^{(1)}
+Af_{\phi\phi}^{(2)}\right)r^2\sin^2\theta d\phi^2.
\end{aligned}\end{equation}
with radial dependence given by
\begin{align}
A&=3\left[\frac{F^{(2)}}{r^3}+\frac{3F^{(1)}}{r^4}+\frac{3F}{r^5}\right],
\\
B&=-\left[\frac{F^{(3)}}{r^2}+\frac{3F^{(2)}}{r^3}+\frac{6F^{(1)}}{r^4}+\frac{6F}{r^5}\right],\\
C&=\frac{1}{4}\left[\frac{F^{(4)}}{r}+\frac{2F^{(3)}}{r^2}+\frac{9F^{(2)}}{r^3}+\frac{21F^{(1)}}{r^4}+\frac{21F}{r^5}\right],\\
\label{eq:F-deriv}
F^{(n)}&\equiv \left[\frac{d^nF(x)}{dx^n}\right]_{x=t-r}.
\end{align}
$F(x)=F(t-r)$ describes the shape of the wave.  The functions $f_{rr},
\ldots, f^{(2)}_{\phi\phi}$ depend only on angles $(\theta,\phi)$;
they are given explicitly in Ref.~\cite{Teukolsky:1982} for azimuthal
quantum number $m=-2, \ldots, 2$.  {\em Ingoing} quadrupole waves are
obtained by replacing $F(t-r)$ with a function of $t+r$, and reversing
the signs in front of odd derivatives of $F$ in Eq.~(\ref{eq:F-deriv}).
Reference~\cite{Teukolsky:1982} gives also the metric for odd parity
waves.  From Eq.~(\ref{eq:TeukolskyWave}), one can easily extract
$h_{ij}$ and $\dot h_{ij}$ in spherical coordinates, which we
transform to the Cartesian coordinates used in the code.

\section{Results}

\subsection{Flat space with ingoing pulse}

We first consider a perturbation of flat space,
$g_{ij}^0=f_{ij}$, $\trK^0=0$.  We choose the even parity, $m=0$
ingoing mode.  The shape of the pulse is taken as a Gaussian
\begin{equation}\label{eq:linwave-Gaussian}
F(x)=e^{-(x-x_0)^2/w^2}
\end{equation}
of width $w=1$ and with an initial radius of $x_0=20$.

\begin{figure}
\centerline{\includegraphics[scale=0.4]{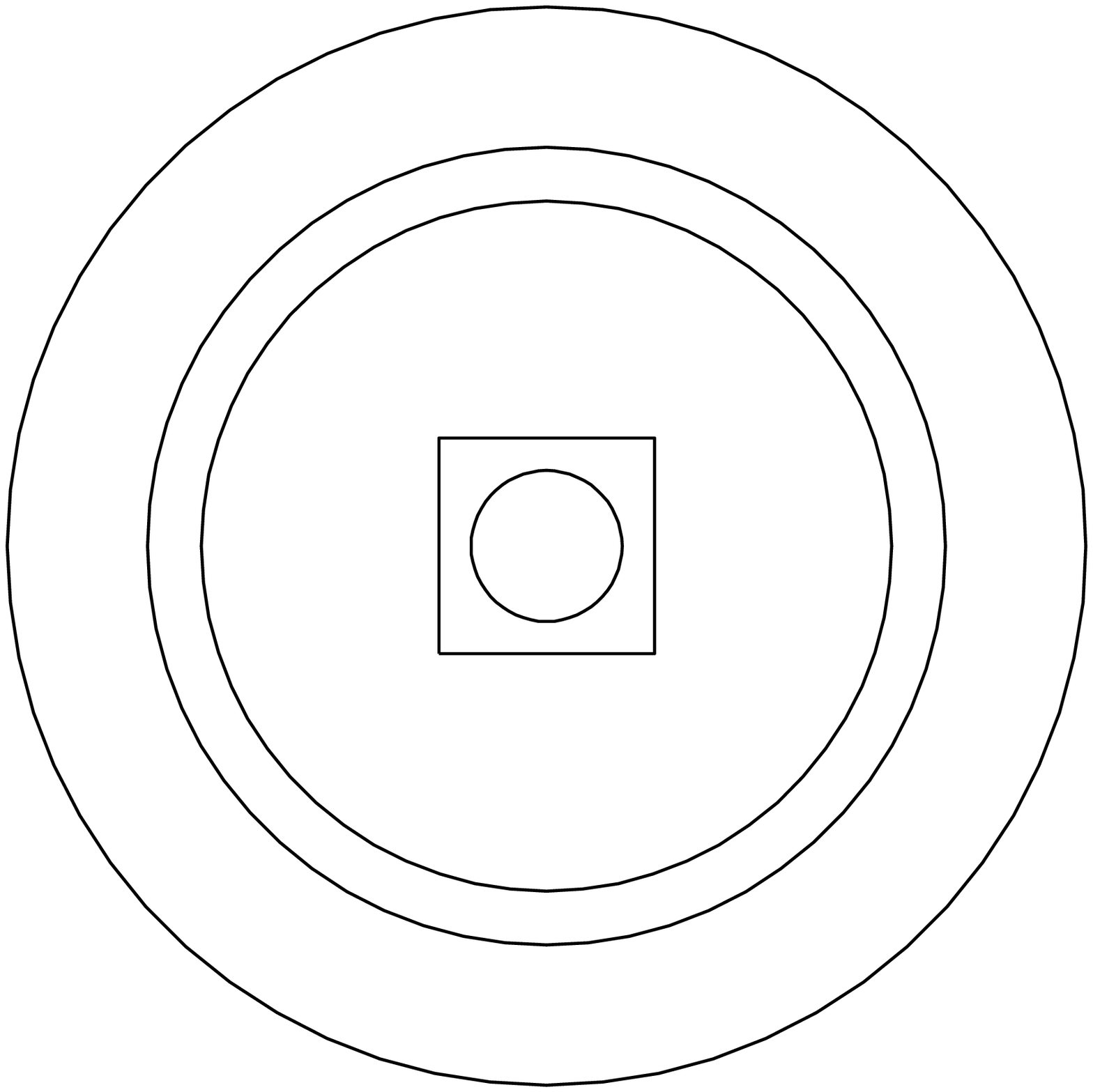}} \CAP{Domain
decomposition used for elliptic solves in full $R^3$.}
{\label{fig:GW-domain}Domain decomposition used for elliptic solves in
full $R^3$.  A cube covers the central region which is not covered by
the spherical shells.}
\end{figure}

The domain decomposition used during the solve is shown in
Figure~\ref{fig:GW-domain}.  We use three spherical shells with
boundaries at $r=1.5, 16, 24, 10^6$.  The middle shell is centered on
the gravitational wave and has higher radial resolution than the other
shells.  The inner shell does not extend down to the origin, since the
regularity conditions at the origin of a sphere are not implemented.
Instead, we place a cube on the origin which overlaps with the
innermost spherical shell.

\begin{figure}
\centerline{\includegraphics[scale=0.4]{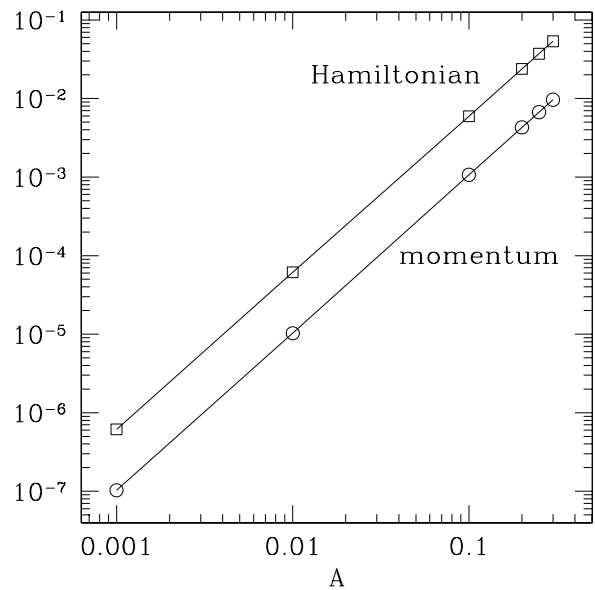}}
\caption{\label{fig:GW-violation}Constraint violation of linear
gravitational wave in flat background.}
\end{figure}

As a first step, we verify that such a linear wave does satisfy the
constraints to first order in $A$.  Figure~\ref{fig:GW-violation}
shows the constraint violation of $(\g_{ij}, \K_{ij})$ given by
Eqs.~(\ref{eq:3+1-linwave-gij}) and (\ref{eq:3+1-linwave-Kij}), or,
equivalently to linear order in $A$, by the free data
Eqs.~(\ref{eq:linwave-freedata}) {\em without} solving the conformal
thin sandwich equations.

\begin{figure}
\centerline{\includegraphics[scale=0.4]{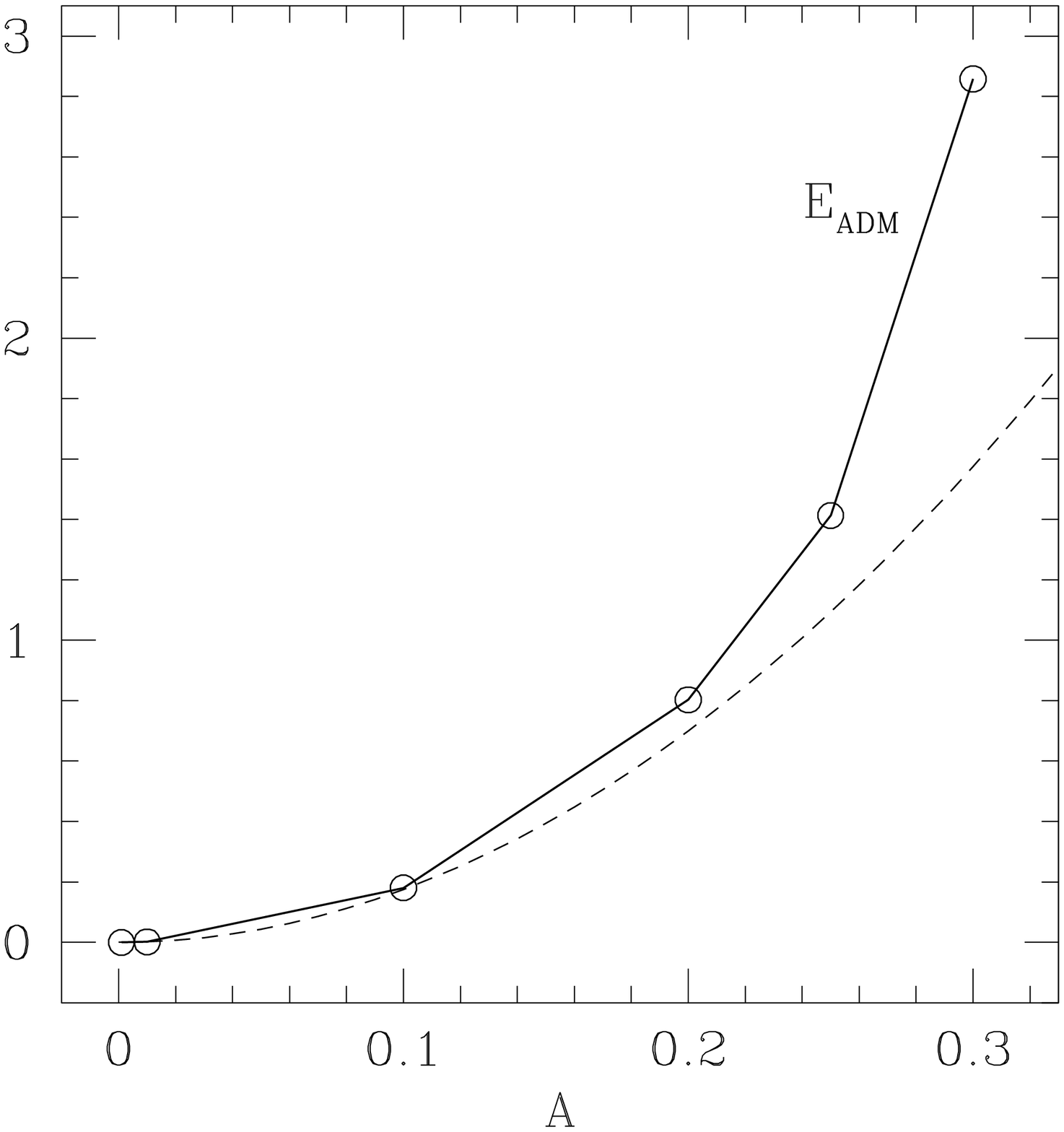}}
\CAP{ADM energy of an ingoing Gaussian pulse in flat
space.}{\label{fig:GW-flatspace-E}ADM energy of an ingoing Gaussian
pulse in flat space.  The dashed line indicates the low-amplitude
quadratic behavior.}
\end{figure}

We now solve the conformal thin sandwich equations with the free
data~(\ref{eq:linwave-freedata}) for different $A$.  For small $A$, we
find that $\CF-1$ is proportional to $A^2$.  This is expected, because
$\CF-1$ corrects the conformal metric to satisfy the Hamiltonian
constraint.  As the constraint violation is proportional to $A^2$, so
is this correction. Figure~\ref{fig:GW-flatspace-E} presents the
ADM-energy of the resulting hypersurfaces as a function of $A$.  One
can clearly achieve initial data sets with a significant energy
content.  At low amplitudes, $E_{ADM}$ is proportional to $A^2$, as
one expects given that $\CF-1$ is proportional to $A^2$.  At high
amplitudes, however, $E_{ADM}$ grows faster than $A^2$, indicating
that the non-linear regime with self-interaction is reached.
For $A>0.3$, the elliptic solver fails to converge.

\begin{figure}
\centerline{\includegraphics[scale=0.4]{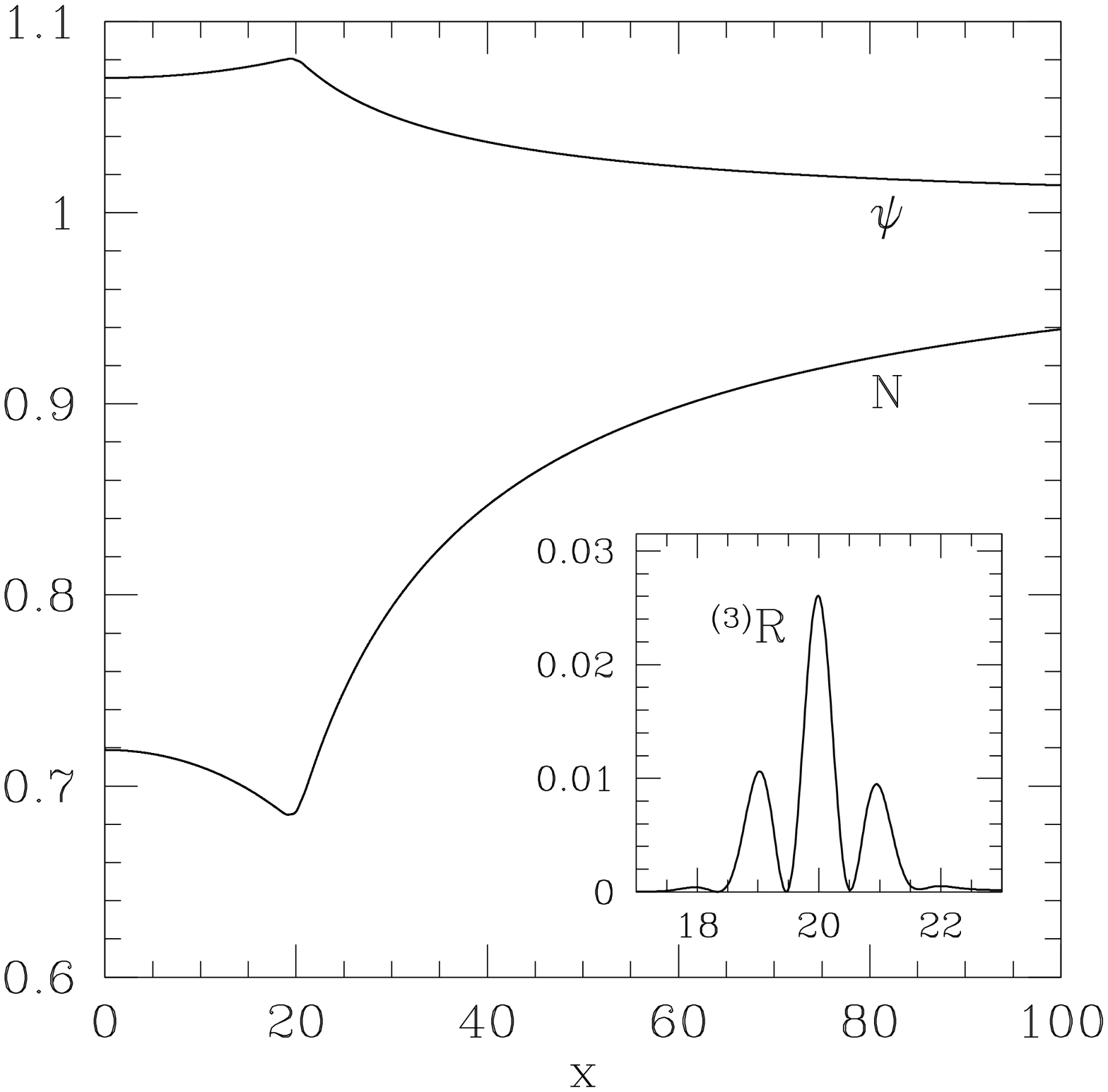}}
\CAP{Cuts through the equatorial plane of the $A\!=\!0.3$ data set of
Fig.~\ref{fig:GW-flatspace-E}.}{\label{fig:GW-flatspace-cuts}Cuts
through the equatorial plane of the $A\!=\!0.3$ data set of
Fig.~\ref{fig:GW-flatspace-E}. The large plot shows lapse and
conformal factor, the insert shows the scalar curvature of the
3-metric.}
\end{figure}

We now examine the highest amplitude data set with $A=0.3$, having
$E_{ADM}=2.858$.  Figure~\ref{fig:GW-flatspace-cuts} presents cuts
through the conformal factor $\CF$, lapse $\N$ and the scalar
curvature of the physical 3-metric, $^{(3)}\R$.  Conformal factor and
lapse deviate significantly from unity confirming that the solution is
indeed deep in the nonlinear regime.  The scalar curvature is
virtually zero everywhere except within a spherical shell with $18\lesssim
r\lesssim 20$.

The gravitational wave is concentrated in a spherical shell of width
$w=1$.  The underlying {\em linear} wave is purely ingoing, so it
seems reasonable that the gravitational perturbation in the physical,
{\em nonlinear} spacetime is also predominantly ingoing.  Neglecting
dispersion, the wave will concentrate in a sphere centered at the
origin with radius $r\sim w$.  Black holes usually form for systems
with mass to size ratio of order unity; here, $E_{ADM}/w\approx 2.8$,
so that black hole formation appears very likely once the pulse is
concentrated at the origin.

These data sets could be used to examine critical collapse to a black
hole, repeating Abrahams \& Evans~\cite{Abrahams-Evans:1993} and
extending it to genuinely three-dimensional collapse by choosing
$m\neq 0$ in the underlying quadrupole wave.  These datasets also
provide a testbed for evolution codes in situations where the {\em
topology} of the horizons changes.

\subsection{Black hole with gravitational wave}
\label{sec:GW:BH}

As a second example of the flexibility of our method, we now superpose
a gravitational wave on a black hole background.  For the metric
$\g_{ij}^0$ and mean curvature $\trK^0$ in
Eqs.~(\ref{eq:linwave-freedata}), we use a Schwarzschild black hole in
Eddington-Finkelstein coordinates, Eqs.~(\ref{eq:gij-EF}) and
(\ref{eq:trK-EF}).  (We do not apply the radial coordinate
transformation to make it conformally flat; instead, we use the
deviation of the conformal factor from unity as a diagnostic).

We choose an odd, ingoing $m=0$ quadrupole wave with Gaussian shape,
Eq.~(\ref{eq:linwave-Gaussian}) at location $x_0=15$ and width $w=1$.
The metric is singular at the origin, therefore we excise a inner
sphere with radius $1.5M$ inside the horizon.  At this inner boundary,
we impose simple Dirichlet boundary conditions appropriate for the
unperturbed black hole: $\CF=1$, and $\N$ and $\beta^i$ given by
Eqs.~(\ref{eq:lapse-EF}) and~(\ref{eq:shift-EF}), respectively.

\begin{figure}
\centerline{\includegraphics[scale=0.4]{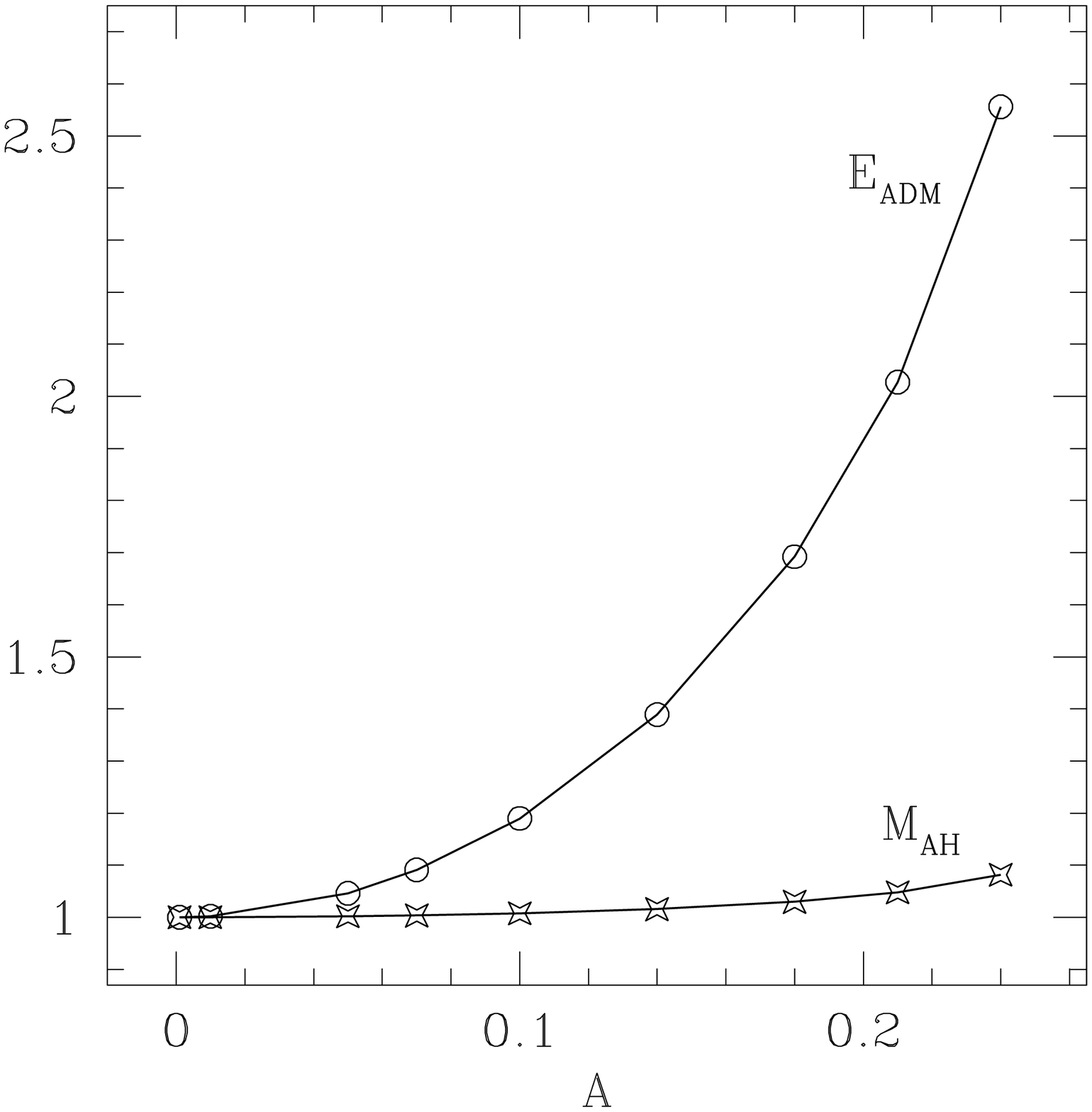}}
\CAP{Black hole with superposed gravitational wave: ADM-energy and
apparent horizon mass}{\label{fig:GW-blackhole-E}Black hole with
superposed gravitational wave.}
\end{figure}

Figure~\ref{fig:GW-blackhole-E} shows the ADM-energy and the apparent
horizon mass of the central black hole as a function of the amplitude
of the gravitational wave.  The apparent horizon mass changes only
slowly with $A$ indicating that the central black hole is only
slightly perturbed by the gravitational wave.  However, the ADM-energy
changes drastically; for large amplitudes it is over twice as large as
the apparent horizon mass, indicating a huge amount of energy in the
gravitational wave outside the black hole.

\begin{figure}
\centerline{\includegraphics[scale=0.4]{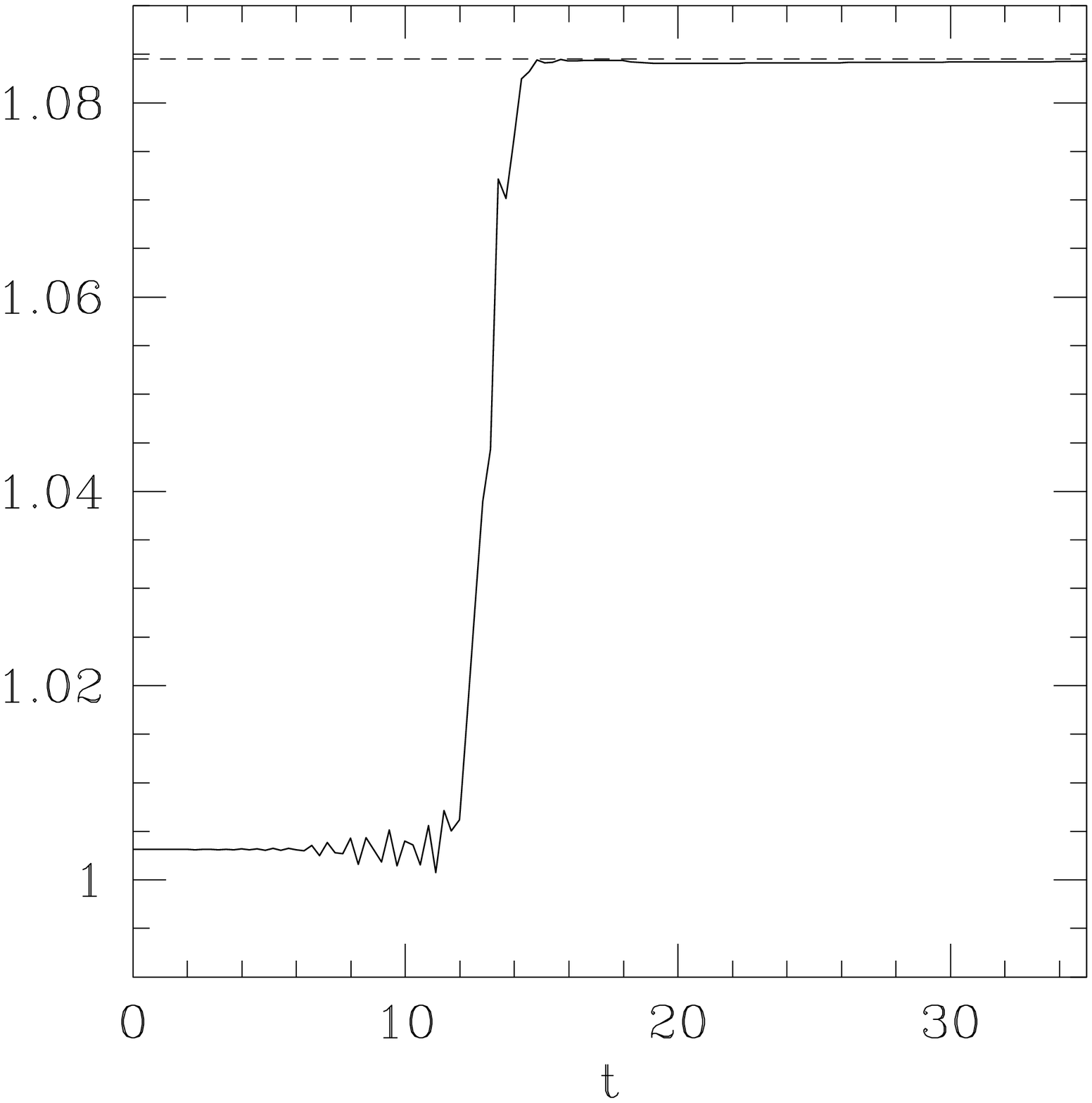}} \CAP{Apparent
horizon mass during an evolution of a perturbed black hole
spacetime.}{\label{fig:GW-AHmass}Apparent horizon mass during an
evolution of a perturbed black hole spacetime.  The dashed line
indicates $E_{ADM}$ from the initial data set.}
\end{figure}

We finally present an evolution of a black hole with superposed
ingoing gravitational wave~\cite{Kidder-Pfeiffer-Shoemaker:2003}.  The
initial data for the evolution is identical to the data sets used in
Figure~\ref{fig:GW-blackhole-E} with the one exception that the
gravitational wave has even parity.  Figure~\ref{fig:GW-AHmass} shows
the apparent horizon mass as a function of evolution time.  For
$t\lesssim 12$, $M_{AH}$ is constant, its value being that from the
initial data set.  Between $12\lesssim t\lesssim 16$, $M_{AH}$
increases rapidly to an asymptotic value of $M_{AH}\approx 1.084$.
The ADM-energy of the initial data set was $E_{ADM}=1.0845$.
Apparently, the ingoing gravitational wave outside the black hole
falls into it, increasing the area of the apparent horizon.  The final
apparent horizon mass is very close to the ADM-energy, and the growth
of $M_{AH}$ happens during a time-interval of roughly the width of the
initial pulse.  Thus it appears that a large fraction of the wave is
coherently ingoing and falls into the black hole.

\section{Discussion}

We propose a new method to construct spacetimes containing
gravitational radiation.  Since our method is based on a {\em linear}
wave, it allows for easy physical interpretation of the generated data
set in the limit of small amplitude.  For strong amplitudes, the
initial data sets are very nonlinear, but we expect that the solutions
still retain qualitatively the properties of the underlying linear
wave.

Two examples illustrate that the method is capable of generating
initial data surfaces containing a large amount of gravitational
energy.

There are many possible uses for these initial data sets: For example,
data analysis of gravitational wave experiments would benefit from
knowing in how many different frequencies a black hole radiates
significantly, i.e., whether a nonlinearly perturbed black hole
settles down into the lowest frequency eigenmode very quickly, or
whether higher frequency eigenmodes are important for an
intermediate period of time.  This question could be addressed by
hitting a black hole with one of the very strong pulses presented in
section \ref{sec:GW:BH}.  

The example in section \ref{sec:GW:BH} can also be generalized to
spinning black holes, off-centered gravitational waves, or
gravitational waves with $m\neq 0$.  Interesting questions would
include, whether one can impart linear or angular momentum on the
black hole, and which fraction of the gravitational wave is scattered
at the black hole potential and reaches infinity.

Numerically, these initial data sets provide test-beds of evolution
codes in situations far away from stationarity.  The mass of a central
black hole changes when a large gravitational wave falls into it; can
todays gauge conditions handle this situation?  If a gravitational
wave collapses to a black hole, horizons appear, and evolution codes
using black hole excision must accommodate this change.  Furthermore,
such spacetimes are ideal test-beds for gravitational wave extraction.

\cleardoublepage

\bibliographystyle{thesis_page}
\bibliography{thesis}

\end{document}